\documentclass[11pt,oneside,a4paper]{article}

\usepackage{authblk}
\usepackage[usenames, dvipsnames]{color} 
\usepackage{graphics} 
\usepackage[format=hang]{caption}
\usepackage[linktocpage, colorlinks, citecolor=blue, linkcolor=blue, urlcolor=blue]{hyperref}
\usepackage{amsfonts, amssymb, amsthm, mathtools, mathrsfs}
\usepackage{scalerel, stackengine}
\usepackage[applemac]{inputenc}
\usepackage{empheq}
\usepackage{bbm}
\usepackage{accents}
\usepackage{enumerate}
\usepackage{dsfont}
\usepackage{scalerel}[2016/12/29]
\usepackage{verbatim} 
\usepackage{float}
\usepackage{MyCommands} 
\usepackage[T1]{fontenc}
\usepackage[makeroom]{cancel}
\usepackage{calc}
\usepackage[record, symbols, nomain, stylemods={tree}, shortcuts=other]{glossaries-extra}
\usepackage{titlesec}
\usepackage{subcaption}
\usepackage[hang,flushmargin]{footmisc}
\usepackage[default]{gillius}

\titleformat{\section}{\normalfont \Large \bfseries}
{Chapter\ \thesection:}{2.3ex plus .2ex}{} 
\titlespacing{\subsection}{2em}{*1}{*1}

\newcommand{\asection}[2]{
\setcounter{section}{#1}
\addtocounter{section}{-1}
\section{#2}
}

\glsxtrsetgrouptitle{manifolds}{Manifolds and other Spaces}
\glsxtrsetgrouptitle{metrics}{Metrics}
\glsxtrsetgrouptitle{tensors}{Tensor Fields}
\glsxtrsetgrouptitle{groups}{Groups and Algebras}
\glsxtrsetgrouptitle{operators}{Derivative Operators}
\glsxtrsetgrouptitle{acronyms}{Acronyms}
\glsxtrsetgrouptitle{other}{Other Symbols}

\newglossarystyle{longgroup}{
    \setglossarystyle{long}
  
}

\newtheoremstyle{ExerciseStyle}
  {0.2cm}{1cm}
  {}
  {0cm}
  {\bfseries}{ }
  {0cm}
  {\thmname{#1}\thmnumber{ #2}\hfill \thmnote{ \textcolor{blue}{$\to$ \hyperref[#3]{Solution}}}\newline}           

\newtheoremstyle{SolutionStyle}
  {0.2cm}{1cm}
  {}
  {0cm}
  {\bfseries}{ }
  {0cm}
  {\thmname{#1}\thmnumber{ #2}\hfill \thmnote{ \textcolor{blue}{$\to$ \hyperref[#3]{Back to Exercise}}}\newline}           

\theoremstyle{ExerciseStyle}
\newtheorem{Exercise}{Exercise}[section]

\theoremstyle{SolutionStyle}

\theoremstyle{definition}

\renewcommand{\thesubsection}{\thesection.\Alph{subsection}}

\numberwithin{equation}{section}

\title{
\textbf{Gravitational Waves in Full, Non-Linear General Relativity}}
\date{}

\author[1]{Fabio D'Ambrosio\textcolor{blue}{\thanks{ \href{mailto:fabioda@phys.ethz.ch}{fabioda@phys.ethz.ch}}}$^{\textsf{, }}$}

\author[2]{Shaun D.~B. Fell\textcolor{blue}{\thanks{\href{mailto:fell@thphys.uni-heidelberg.de}{fell@thphys.uni-heidelberg.de}}}$^{\textsf{, }}$}

\author[2, 1]{Lavinia Heisenberg\textcolor{blue}{\thanks{\href{mailto:lavinia.heisenberg@phys.ethz.ch}{lavinia.heisenberg@phys.ethz.ch}}}$^{\textsf{, }}$}

\author[2]{David Maibach\textcolor{blue}{\thanks{\href{mailto:d.maibach@thphys.uni-heidelberg.de}{d.maibach@thphys.uni-heidelberg.de}}}$^{\textsf{, }}$}

\author[1]{Stefan Zentarra\textcolor{blue}{\thanks{\href{mailto:szentarra@phys.ethz.ch}{szentarra@phys.ethz.ch}}}$^{\textsf{, }}$}

\author[1]{Jann Zosso\textcolor{blue}{\thanks{\href{mailto:jzosso@phys.ethz.ch}{jzosso@phys.ethz.ch}}}$^{\textsf{, }}$}

\affil[1]{Institute for Theoretical Physics, ETH Zurich, Wolfgang-Pauli-Strasse 27, 8093 Zurich, Switzerland}

\affil[2]{Institut f\"{u}r Theoretische Physik, Philosophenweg 16, 69120 Heidelberg, Germany}

\newcommand\linesep{4pt}

\newentry{Mhat}{
	name={$\ \ \hatM$},
	description={Physical spacetime manifold; four-dimensional}\vspace{\linesep},
	group={manifolds},
	type={symbols}
}

\newentry{M}{
	name={$\ \ \M$},
	description={Conformal completion of physical spacetime, $\M = \hatM\cup\scri$}\vspace{\linesep},
	group={manifolds},
	type={symbols}
}

\newentry{2Sphere}{
	name={$\ \ \bbS^2$},
	description={Topological $2$-sphere}\vspace{\linesep},
	group={manifolds},
	type={symbols}
}

\newentry{Scris}{
	name={$\ \ \scri^{\pm}$, $\scri$},
	description={Future/past null infinity (``scri plus/minus''), null infinity ($\scri := \scrip\cup\scrim$, ``scri'')}\vspace{\linesep},
	group={manifolds},
	type={symbols}
}

\newentry{DeltaScris}{
	name={$\ \ \Delta\scri^{\pm}$},
	description={Finite portion of $\scri^{\pm}$}\vspace{\linesep},
	group={manifolds},
	type={symbols}
}

\newentry{ipm0}{
	name={$\ \ i^{\pm}$, $i^0$},
	description={Future/past timelike infinity ($i^{+}$/$i^{-}$), spacelike infinity ($i^0$)}\vspace{\linesep},
	group={manifolds},
	type={symbols}
}

\newentry{Cross_Section}{
	name={$\ \ \C$},
	description={cross-section of $\scri$ with topology of $\bbS^2$}\vspace{\linesep},
	group={manifolds},
	type={symbols}
}

\newentry{SpaceOfGenerators}{
	name={$\ \ \bbG$},
	description={Space of generators of $\scri$ (see Chapter~\ref{Chap4})}\vspace{\linesep},
	group={manifolds},
	type={symbols}
}

\newentry{Spacetime_Metric}{
	name={$\ \ g_{ab}$},
	description={Metric on spacetime $\M$; signature $(-,+,+,+)$}\vspace{\linesep},
	group={metrics},
	type={symbols}
}

\newentry{Minkowski_Metric}{
	name={$\ \ \eta_{ab}$},
	description={Minkowski metric; signature $(-,+,+,+)$}\vspace{\linesep},
	group={metrics},
	type={symbols}
}

\newentry{Degenerate_Metric}{
	name={$\ \ q_{ab}$},
	description={Pullback of $\eta_{ab}$ or $g_{ab}$ to $\scri$; signature is $(0,+,+)$ if $\scri$ is null}\vspace{\linesep},
	group={metrics},
	type={symbols}
}

\newentry{2Sphere_Metric}{
	name={$\ \ s_{ab}$},
	description={Pullback of $q_{ab}$ to cross-section $\C$ of $\scri$; signature $(+,+)$}\vspace{\linesep},
	group={metrics},
	type={symbols}
}

\newentry{2Sphere_Metric_projected}{
	name={$\ \ \underline{s}_{ab}$},
	description={Metric on the space of generators; $\pi^*\underline{s}_{ab} = q_{ab}$, with projector $\pi:\scri\to\bbG$ (see Chapter~\ref{Chap4})}\vspace{\linesep},
	group={metrics},
	type={symbols}
}

\newentry{NPEMScalars}{
	name={$\ \ \Phi_i$},
	description={Electromagnetic Newman-Penrose scalars, $i\in\{0,1,2\}$; complex functions}\vspace{\linesep},
	group={tensors},
	type={symbols}
}

\newentry{NPWeylScalars}{
	name={$\ \ \Psi_i$},
	description={Newman-Penrose Weyl scalars, $i\in\{0,1,2,3,4\}$; complex functions}\vspace{\linesep},
	group={tensors},
	type={symbols}
}

\newentry{Maxwell_2Form_Component}{
	name={$\ \ F_{ab}, F$},
	description={Components of Maxwell $2$-form ($F_{ab} = -F_{ba}$); Maxwell $2$-form $F := F_{ab}\,\dd x^{a}\wedge \dd x^{b}$}\vspace{\linesep},
	group={tensors},
	type={symbols}
}

\newentry{Weyl_Tensor}{
	name={$\ \ C_{abcd}, \hat{C}_{abcd}$},
	description={Weyl tensor of the conformally completed spacetime; Weyl tensor of the physical spacetime}\vspace{\linesep},
	group={tensors},
	type={symbols}
}

\newentry{Rescaled_Weyl_Tensor}{
	name={$\ \ K_{abcd}$},
	description={Asymptotic Weyl tensor, $K_{abcd} := \Omega^{-1} C_{abcd}$}\vspace{\linesep},
	group={tensors},
	type={symbols}
}

\newentry{Energy_Momentum_Tensor}{
	name={$\ \ T_{ab}$},
	description={Energy-momentum tensor of the conformally completed spacetime}\vspace{\linesep},
	group={tensors},
	type={symbols}
}

\newentry{Rescaled_Energy_Momentum_Tensor}{
	name={$\ \ \hat{T}_{ab}$},
	description={Energy-momentum tensor of the physical spacetime}\vspace{\linesep},
	group={tensors},
	type={symbols}
}

\newentry{BMSGroup}{
	name={$\ \ \B, \mathfrak b$},
	description={Bondi-Metzner-Sachs (BMS) group, BMS Lie algebra (fraktur b)}\vspace{\linesep},
	group={groups},
	type={symbols}
}

\newentry{LorentzGroup}{
	name={$\ \ \lie, \mathfrak l$},
	description={Lorentz group, Lorentz Lie algebra (fraktur l)}\vspace{\linesep},
	group={groups},
	type={symbols}
}

\newentry{PoincareGroup}{
	name={$\ \ \boxdot$},
	description={This symbol denotes the Poincar\'e group as well as its Lie algebra}\vspace{\linesep},
	group={groups},
	type={symbols}
}

\newentry{SupertranslationGroup}{
	name={$\ \ \S, \mathfrak s$},
	description={Group of angle-dependent supertranslations, Lie algebra of supertranslation (fraktur s)}\vspace{\linesep},
	group={groups},
	type={symbols}
}

\newentry{TranslationGroup}{
	name={$\ \ \T, \mathfrak t$},
	description={Group of spacetime translations, Lie algebra of spacetime translations (fraktur t)}\vspace{\linesep},
	group={groups},
	type={symbols}
}

\newentry{eth}{
	name={$\ \ \eth$},
	description={Angular derivative which acts on scalars $f$ of spin weight $s$ (see definition~\eqref{eq:eth}); pronounced ``eth''}\vspace{\linesep},
	group={operators},
	type={symbols}
}

\newentry{ethbar}{
	name={$\ \ \bar{\eth}$},
	description={Conjugate of the angular derivative $\eth$ (see definition~\eqref{eq:ConjugateEth});}\vspace{\linesep},
	group={operators},
	type={symbols}
}

\newentry{Lie}{
	name={$\ \ \lie_v$},
	description={Lie derivative along vector field $v$}\vspace{\linesep},
	group={operators},
	type={symbols}
}

\newentry{Exterior_Derivative}{
	name={$\ \ \dd$},
	description={Exterior derivative which acts on differential forms}\vspace{\linesep},
	group={operators},
	type={symbols}
}

\newentry{nabla}{
	name={$\ \ \nabla_a$},
	description={Unique metric-compatible and torsion-free covariant derivative of $\eta_{ab}$ or $g_{ab}$}\vspace{\linesep},
	group={operators},
	type={symbols}
}

\newentry{CovD_q}{
	name={$\ \ \D_a$},
	description={Pullback of $\nabla_a$ to $\scri$ (see Chapters~\ref{Chap6} and~\ref{Chap7})}\vspace{\linesep},
	group={operators},
	type={symbols}
}

\newentry{CovD_s}{
	name={$\ \ \mathfrak D_a$},
	description={Unique metric-compatible and torsion-free covariant derivative of $s_{ab}$ (obtained by pulling back $\D_a$ to cross-section $\C$ of $\scri$)}\vspace{\linesep},
	group={operators},
	type={symbols}
}

\newentry{equalhat}{
	name={$\ \ \hat{=}$},
	description={Equality holds on $\scri$}\vspace{\linesep},
	group={other},
	type={symbols}
}

\newentry{unequalhat}{
	name={$\ \ \hat{\cancel{=}}$},
	description={Not equal on $\scri$}\vspace{\linesep},
	group={other},
	type={symbols}
}

\newentry{definition}{
	name={$\ \ :=, =:$},
	description={left side defined by right side, right side defined by left side}\vspace{\linesep},
	group={other},
	type={symbols}
}

\newentry{identity}{
	name={$\ \ \equiv$},
	description={Identity}\vspace{\linesep},
	group={other},
	type={symbols}
}

\newentry{isomorphic}{
	name={$\ \ \simeq$},
	description={isomorphic to}\vspace{\linesep},
	group={other},
	type={symbols}
}

\newentry{hat}{
	name={$\ \ \hat{T}_{a_1\dots a_p}{}^{b_1\dots b_p}$},
	description={Tensor fields with a hat are defined on the physical spacetime manifold $\hatM$}\vspace{\linesep},
	group={other},
	type={symbols}
}

\newentry{underleftarrow}{
	name={$\ \ \pb{T\du{a_1\dots a_p}{b_1\dots b_q}}$},
	description={Pullback of $T_{a_1\dots a_p}{}^{b_1\dots b_q}$ to $\scri$}\vspace{\linesep},
	group={other},
	type={symbols}
}

\newentry{underline}{
	name={$\ \ \underline{T_{a_1\dots a_p}{}^{b_1\dots b_q}}$},
	description={Pullback of $T_{a_1\dots a_p}{}^{b_1\dots b_q}$ to cross-section $\C$ of $\scri$}\vspace{\linesep},
	group={other},
	type={symbols}
}

\newentry{Hodge_Dual}{
	name={$\ \ \prescript{\star}{}{\mathbf{\omega}}$},
	description={Hodge dual of differential form $\mathbf{\omega}$ on $\M$}\vspace{\linesep},
	group={other},
	type={symbols}
}

\begin{document}

\maketitle
\begin{abstract}
    \noindent These notes provide a student-friendly introduction to the theory of gravitational waves in full, non-linear general relativity (GR). We aim for a balance between physical intuition and mathematical rigor and cover topics such as the Newman-Penrose formalism, electromagnetic waves, asymptotically Minkowski spacetimes, the peeling theorem, the universal structure of null infinity, the Bondi-Metzner-Sachs group, and the definition of radiative modes in linear as well as in non-linear GR. Many exercises and some explicitly calculated examples complement the abstract theory and are designed to help students build up their intuition and see the mathematical machinery at work.
\end{abstract}

\pagenumbering{gobble}
\clearpage

\newpage
\pagenumbering{roman}
\section*{\Huge Preface}\addcontentsline{toc}{section}{Preface}
These notes are based on a lecture series by Prof. Abhay Ashtekar, which can be found on the YouTube channel of the Institute for Gravitation and the Cosmos at Penn State~\cite{Ashtekar:2019YT}.

In 2021, the authors of these notes founded the \textit{Gravitational Waves Working Group} at ETH Zurich, with the purpose of studying and discussing recent advances in the field of gravitational waves. Our intention has been to learn as much as possible about different aspects of this highly interesting and active field of research --- from observational questions, detectors, data analysis all the way to mathematical foundations. 

Right from the start, our aim was to create a document of high didactic value. Each chapter is complemented by a number of exercises, qualitative arguments often foreshadow results which will be derived, and we have provided examples to illustrate certain aspects of the formalism. More examples and exercises, including solutions, will follow in a forthcoming update of these notes. 

Furthermore, we aimed for a balance between mathematical rigor, intuition, and qualitative reasoning. In the hope of having succeeded in this effort, we believe that these notes offer students an easy introduction into a range of topics of mathematical relativity and it could be useful to junior researchers, who wish to contribute to this field.

$$ \textsf{\textbf{Acknowledgements}} $$
It is our pleasure to thank Prof. Ashtekar for his lectures.
Furthermore, we would like to thank Neev Khera, who kindly presented recent advances~\cite{AshtekarII:2020,Ashtekar:2020} on gravitational waves in full, non-linear general relativity to our group. Last but not least, we would like to thank Tommaso De~Lorenzo for sparking our interest, expanding our horizon, and introducing us to the beautiful subject of gravitational waves in full general relativity.
LH is supported by funding from the European Research Council (ERC) under the European Unions Horizon 2020 research and innovation programme grant agreement No 801781 and by the Swiss National Science Foundation grant
179740.

\newpage
\tableofcontents
\newpage

\printunsrtglossary[type=symbols, style=longgroup, title={List of Symbols}]
\newpage

\section*{\Huge Conventions}\addcontentsline{toc}{section}{Conventions}
Throughout these notes, spacetime is assumed to be four-dimensional and the spacetime metric has signature $(-,+,+,+)$. The curvature tensor is defined via $2\nabla_{[a}\nabla_{b]}k_c =: R\du{abc}{d}k_d$, $R_{ab} := R\du{amb}{m}$, and $R := g^{ab}R_{ab}$. We work in geometric units, where $c = G = 1$.
\newpage

\pagenumbering{gobble}
\pagenumbering{arabic}

\asection{1}{Electromagnetic Waves}\label{Chap1}

\subsection{Why the Notion of Radiation is Non-Trivial}
We already know what electromagnetic waves are --- at least we think we know. It is a simple exercise to derive the electromagnetic wave equation from Maxwell's equations, and we can even write down the formal solution in the presence of sources. From simple examples, which we can work out in detail, we know that the radiation field, described by the vector potential $A^\mu$, has three characteristic properties: It oscillates, it is transversal (this is actually true in general), and it decays as $\frac{1}{r}$ as we move away from the source which generates the field.

But now suppose there is a source $J^\mu := (\rho, \vec{j})^\transpose$ which generates a vector potential $A^\mu$. An observer is tasked with determining by local measurements whether this source generates a \textit{radiation} field. How could the observer achieve this? How do we know whether a source, which we may not directly access, generates electromagnetic waves? Are there observables we can theoretically compute and then compare with observations?

\begin{figure}[htb!]
	\centering
	\includegraphics[width=0.45\linewidth]{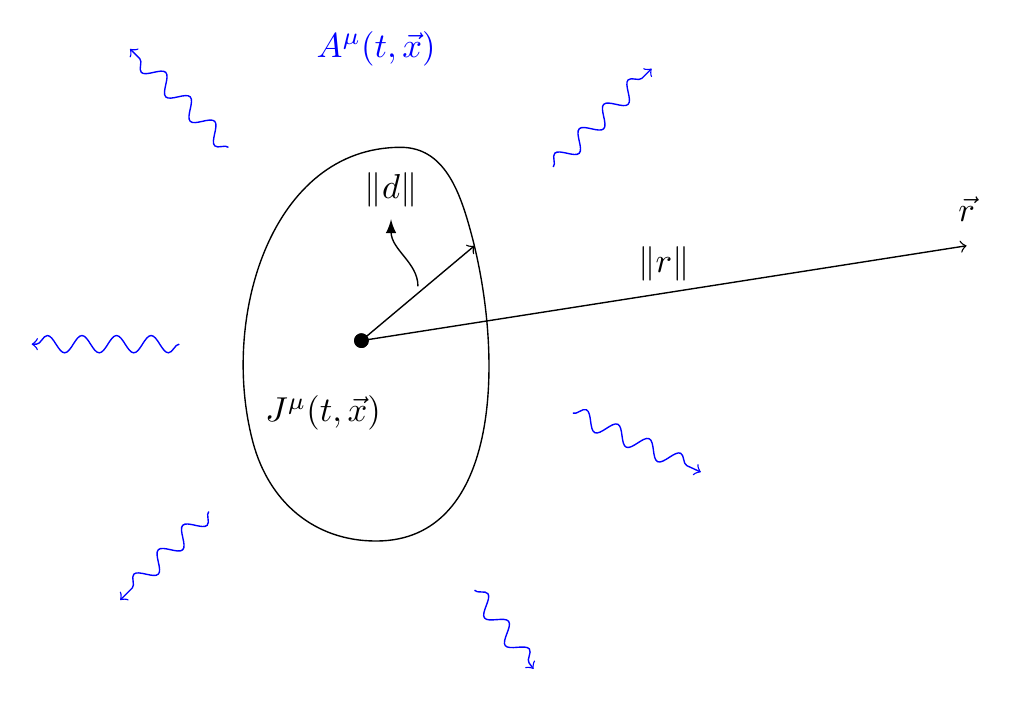}
	\caption{\textit{An electromagnetic source $J^\mu$ with finite, characteristic spatial extent $d$ producing a field $A^\mu$. The vector $\vec{r}$ indicates the position of the observer.}}
	\label{fig:ChargeDistrib}
\end{figure}

These questions seem na\"{i}ve, but we will shortly see that the notion of electromagnetic waves is less trivial than we think. In fact, we will see that we can only determine whether a given source generates radiation if we go to ``infinity'', in an appropriate sense. This will lead us to a reformulation of Maxwell's equations, and in particular the theory of electromagnetic waves, in a language which is also suitable to describe the gravitational field. This will provide us with a guideline to develop the theory of gravitational waves in full, non-linear general relativity (GR). So let us return to the question of what electromagnetic waves are and how we can determine whether a given source generates such waves. Given that we know the equations which govern electromagnetic waves, it should be easy to answer these questions. For concreteness, consider the situation shown in Figure~\ref{fig:ChargeDistrib}, where an electromagnetic source $J^\mu$ is confined to a finite spatial region of characteristic extension $d$.
The field generated by this source is assumed to satisfy Maxwell's equations and we can therefore immediately write down the formal solution (see for instance~\cite{JacksonBook}, chapter 9)
\begin{equation}\label{eq:formal_solution}
	A^\mu(t, \vec{x}) = \frac{1}{4\pi} \int_\Omega\dd^3 x'\int_\bbR\dd t'\,\frac{J^\mu(t', \vec{x}')}{\|\vec{x}-\vec{x}'\|}\,\delta\left(t'+\|\vec{x}-\vec{x}'\|-t\right).
\end{equation}
Notice that the fact that this expression solves Maxwell's equations does not help us in determining whether there is radiation or not. This information is implicitly contained in $J^\mu$, but it is bunched together with a lot of other information and we may not directly have access to $J^\mu$. The source may have static parts which only produce coulombic fields, it may have currents, and it may have radiating contributions. The point is that everything is entangled and we do not yet know how to disentangle the different contributions. But we can try the following: We assume there is radiation and that it has a wavelength $\lambda = \frac{2\pi}{\omega}$. Moreover, we assume an observer is located at the radial distance $r$ from the source (cf. Figure~\ref{fig:ChargeDistrib}). The assumption that there is radiation of wavelength $\lambda$ can be translated into
\begin{align}
	J^\mu(t, \vec{x}) &= J^\mu_\textsf{rad}(t, \vec{x}) + J^\mu_\textsf{stat}(\vec{x})\notag\\
	&= \e^{-i \omega t} J^\mu_{0}(\vec{x})+ J^\mu_\textsf{stat}(\vec{x}).
\end{align}
In words, we can decompose our source into static contributions, $J^\mu_\textsf{stat}(\vec{x})$, and radiative contributions, $J^\mu_\textsf{rad}(t, \vec{x})$. The latter can be assumed to oscillate like $\e^{-i \omega t} J^\mu_0(\vec{x})$, without loss of generality.\footnote{In general, we would have to represent the source by its Fourier transform, $J^\mu_\textsf{rad}(t, \vec{x}) = \int_\bbR\dd \omega\, J^\mu_0(\omega, \vec{x})\, e^{-i\omega t}$. Hence, we would have to carry around an integral in all expressions. It is just simpler to do things for one Fourier mode at a time, as we do in the main text.} By inserting this ansatz into the formal solution~\eqref{eq:formal_solution}, we obtain
\begin{align}\label{eq:ansatz_A}
	A^\mu(t, \vec{x}) &= \int_\Omega\dd^3 x'\, J^\mu_0(\vec{x}') \frac{\e^{ i\omega \|\vec{x}-\vec{x}'\|}}{\|\vec{x}-\vec{x}'\|}\e^{-i\omega t} + A^\mu_\textsf{stat}(\vec{x}),
\end{align}
where $A^\mu_\textsf{stat}(\vec{x})$ contains the static contributions. We have not yet taken into account the position of the observer relative to the source. There are three different zones we can distinguish:
\begin{enumerate}
	\item The near zone: $d\ll r\ll \lambda$
	\item The transition zone: $d\ll r \simeq \lambda$
	\item The far/radiation zone: $d\ll \lambda \ll r$
\end{enumerate}
The behavior of the vector potential is very different in the three zones and this directly impacts the observer's ability to infer the existence of electromagnetic radiation from local measurements compared with theoretically expected properties. We will see this explicitly for the near and the far zone. In the former case, the condition $r\ll \lambda$ allows us to expand the exponential in~\eqref{eq:ansatz_A} and we find, by also applying an expansion of~\eqref{eq:ansatz_A} in spherical harmonics,
\begin{align}
	A^\mu(t, \vec{x}) &= \sum_{l=0}^\infty\sum_{|m|\leq l}\frac{\e^{-i\omega t}}{2l+1}\frac{Y_{lm}(\theta,\phi)}{r^{l+1}}\int_\Omega\dd^3 x' J^\mu_0(\vec{x}')\, r'^{l} Y^*_{lm}(\theta',\phi') + A^\mu_\textsf{stat}(\vec{x}) & \text{(for $r\ll \lambda$)}.
\end{align}
Observe that this expression is time-dependent, but it is not the dependence one would expect from a wave. In fact, fields which oscillate like $\e^{-i\omega t}$ are called \textbf{quasi-static}~\cite{JacksonBook}. Moreover, the field does not drop like $\frac{1}{r}$, but rather there is a sum over terms which go like $\frac{1}{r^{l+1}}$. This is not the behavior we expect from a radiation field and even though the source is not static, the vector potential in the near zone is quasi-static. Hence, in this region, an observer would not be able to see any electromagnetic waves!

We omit a discussion of the transition zone, which is more complicated, but also irrelevant for our purposes, and we directly move to the far zone. In this case, we implement the condition $\lambda\ll r$, which means we have to expand $\|\vec{x}-\vec{x}'\|$ as
\begin{equation}
	\|\vec{x}-\vec{x}'\| \approx r - \vec{n}\cdot \vec{x}',
\end{equation} 
where $\vec{n}$ is a unit vector in the direction of $\vec{x}$. Using this approximation, we find that~\eqref{eq:ansatz_A} assumes the form
\begin{equation}\label{eq:FarZone}
	\lim_{r\to \text{far zone}} A^\mu(\vec{x}, t) = \frac{1}{4\pi} \frac{e^{i \omega (r-t)}}{r}\int_\Omega\dd^3 x'\, J^\mu_0(\vec{x}')\,\e^{-i \omega\vec{n}\cdot \vec{x}'} + A^\mu_\textsf{stat}(\vec{x}).
\end{equation}
The first term in the above expression has the expected properties: It oscillates, it decays like $\frac{1}{r}$, and it is transversal. This is a genuine radiation field!

The moral of the story is that the observer has to be far enough away from the source to perform his or her measurements. Too close, and the vector potential is quasi-static and the observer can therefore not infer whether or not there is electromagnetic radiation. But even if the observer is far enough away from the source, there is the problem that, in~\eqref{eq:FarZone}, static and radiative contributions are mixed up. So the question is whether the observer can perform a measurement which disentangles the different contributions and ultimately isolates the radiative part.

Since electromagnetic waves carry energy and momentum, it is natural to attempt to measure the flux of energy and momentum through some small spatial region. Is this sufficient to determine whether a given source produces electromagnetic radiation? Because the energy flux of an electromagnetic field is described by the Poynting vector, it is natural to study its behavior. However, in doing so, we will soon find that the Poynting vector alone does not suffice in order to tell radiation and other field contributions apart! We need one more ingredient.

Let us dive right in and consider the Poynting vector $\vec{S}:=\vec{E}\times\vec{B}$ with its associated Poynting flux, $\oint_{\bbS^2}\vec{S}\, \dd^2\sigma$, where $\bbS^2$ is a $2$-sphere and $\dd^2\sigma = r^2\, \sin\theta\,\dd\theta\,\dd\phi$. Since the flux carries energy and momentum, is it true that if it is non-zero there must be electromagnetic radiation? The answer is no: You can have a non-zero Poynting vector even when there is no electromagnetic radiation. Partially, this is due to the fact that the Poynting vector is not a Lorentz invariant quantity and it therefore depends on a choice of reference frame. As an example, consider the Coulomb solution, i.e., the field of a point charge for an observer in the rest frame of the particle. Clearly, for such an observer the magnetic field is zero and consequently the Poynting vector vanishes as well. 
But now let us consider this point charge from the point of view of a boosted observer. This observer will see a current, rather than a static charge. From elementary electromagnetism we know that a current produces an electric and a magnetic field. We also know that these fields are orthogonal to each other. This implies that the boosted observer sees a non-zero Poynting vector, which leads to the conclusion
\begin{equation}
	\vec{S}_\textsf{rest frame} = 0 \neq \vec{S}_\textsf{boosted}.
\end{equation}
It follows that this na\"{i}ve approach of using the Poynting vector to determine whether there is radiation or not is not viable: The observer in the rest frame of the particle measures a zero flux and therefore concludes there is no radiation, while the boosted observer measures a non-zero flux and therefore erroneously concludes there is radiation.

Luckily, the situation is not quite so hopeless. The Poynting flux \textit{does} carry information about electromagnetic radiation, but we need to take a certain limit in order to extract it. In fact, an explicit computation for the above example shows that the boosted observer sees a Poynting vector which decays like $\frac{1}{r^4}$. Since $\vec{S}$ in the flux integral is multiplied by $r^2$ (this factor stems from the area element $\dd^2\sigma = r^2\sin\theta\,\dd\theta\,\dd\phi$), we find that the Poynting flux of the boosted observer vanishes at infinity. That is, we obtain
\begin{equation}
	\lim_{r\to\infty}\int_{\bbS^2} \vec{S}_\textsf{rest frame}\,\dd^2\sigma = 0 = \lim_{r\to\infty}\int_{\bbS^2} \vec{S}_\textsf{boosted}\,\dd^2\sigma.
\end{equation}
Both observers now agree that there is no electromagnetic radiation! Of course, it could just be a coincidence that in this example the two fluxes at infinity give the same result. Furthermore, it is not immediately clear why we should take that particular limit. However, we will now show that this is not a coincidence and that ``going to infinity'' always acts as a ``filter'' which only lets through the radiative parts of a field. More precisely, we will show that the Poynting flux of static contributions vanishes at infinity while the Poynting flux of electromagnetic waves is non-zero.

The technical tool we need for this is the multipole expansion of the scalar $A^0$ and the vector potential~$A^{i}$, respectively. Of course, if we talk about a scalar and a vector potential, this means we explicitly break Lorentz covariance because we need to pick a reference frame to define and distinguish the two potentials. This is something to keep in mind and we will return to this point later on.

From elementary electrodynamics we recall that the two multipole expansions can be written as
\begin{align}
	A^0(\vec{x}) &= \sum_{l=0}^\infty\sum_{|m|\leq l}\frac{1}{2l+1} q_{lm} \frac{Y_{lm}(\theta,\phi)}{r^{l+1}}\quad \text{with }  q_{lm} = \int\dd^3 x'\, Y^*_{lm}(\theta',\phi')\, r'^{l}\, \rho_\textsf{stat}(\vec{x}')\notag\\
	A^{i}(\vec{x}) &= \frac{1}{4\pi} \frac{\vec{x}}{\|\vec{x}\|^3}\cdot\int\dd^3 x'\, j_\textsf{stat}^{i}(\vec{x}')\,\vec{x}' + \textsf{higher order multipoles}.
\end{align}
We see that the lowest term in the expansion of the scalar potential is the monopole, which scales as $\frac{1}{r}$, while the lowest term in the expansion of the vector potential is the dipole, which scales as $\frac{\vec{x}}{\|\vec{x}\|^3}$. This is an important observation: On physical grounds we know that there are electric monopoles, but no magnetic monopoles. Hence, the lowest terms in the two expansions will \textit{always} be given by a monopole and a dipole. In particular, this is independent of the reference frame we use to define the scalar and the vector potential. Moreover, these two contributions, when computing the Poynting vector, combine to give a vector which falls \textit{faster} than $\frac{1}{r^2}$. Again, this is true in every reference frame since no Lorentz transformation can change the fact that there are no magnetic monopoles. It follows from these simple considerations that the Poynting flux of static sources \textit{always} vanishes at infinity.

What remains to be shown is that the flux of the radiation field does not vanish at infinity. To that end, we need to look at the first term in~\eqref{eq:FarZone}. Of course, to get the electric and magnetic fields from this expression we need to take derivatives. But even after taking derivatives there will always be a term which goes like $\frac{1}{r}$. Hence, the electric and magnetic fields of electromagnetic waves decay like $\frac{1}{r}$ (to leading order) and the Poynting vector consequently behaves like $\frac{1}{r^2}$ (to leading order). We therefore find that the Poynting flux at infinity is given by
\begin{equation}\label{eq:RadiationFlux}
	\lim_{r\to\infty}\oint_{\bbS^2}\left(\vec{E}_\textsf{rad}+\vec{E}_\textsf{stat}\right)\times\left(\vec{B}_\textsf{rad}+\vec{B}_\textsf{stat}\right)\,\dd^2\sigma = \lim_{r\to \infty}\oint_{\bbS^2}\vec{E}_\textsf{rad}\times\vec{B}_\textsf{rad}\,\dd^2\sigma,
\end{equation}
where it follows from the above considerations that $\vec{E}_\textsf{rad}\times \vec{B}_\textsf{stat}$, $\vec{E}_\textsf{stat}\times\vec{B}_\textsf{rad}$, and $\vec{E}_\textsf{stat}\times\vec{B}_\textsf{stat}$ do not contribute to the flux at infinity. The only contribution comes from the radiation field. It can be shown that this contribution is indeed non-zero (this has to be expected, since this simply means that the electromagnetic wave carried energy and momentum to infinity) and hence we reach the conclusion that the right hand side of~\eqref{eq:RadiationFlux} is a good quantity to measure in order to determine whether there is an electromagnetic wave or not.

These considerations can be summarized as follows: Not only do we need to be far enough away from the sources in order to detect electromagnetic radiation, it is actually convenient to go infinitely far away in order to disentangle the radiation field from the other electromagnetic field components.

\begin{figure}[htb!]
	\centering
	\includegraphics[width=0.68\linewidth]{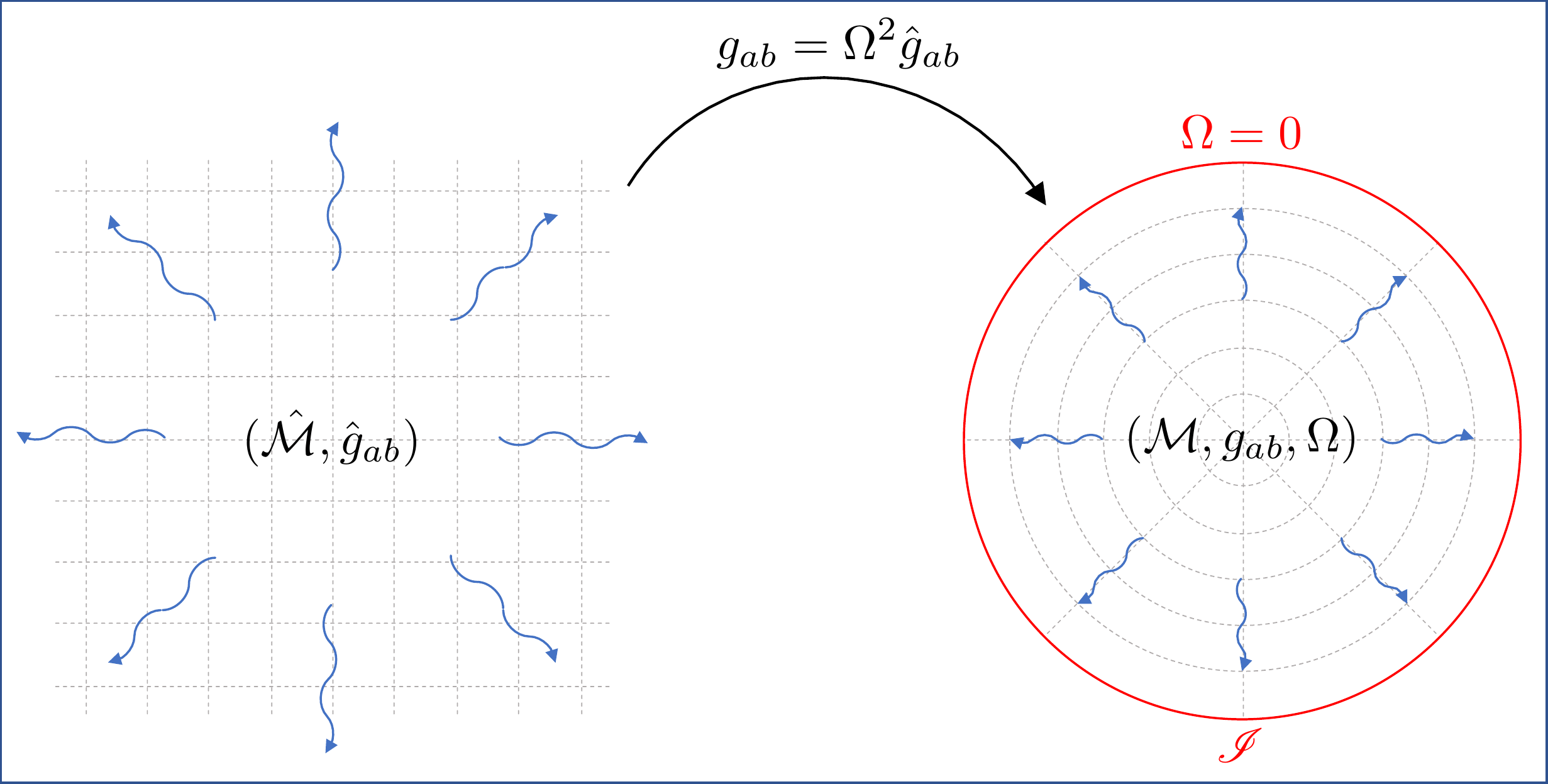}
	\caption{\textit{Representation of a conformal completion for an arbitrary spacetime $(\hatM, \hat{g}_{ab})$. The blue lines represent radiation emanating from a source. The boundary $\scrip$ acts as a ``screen'' which collets radiative information.}}
	\label{fig:ConformalCompletion}
\end{figure}

\begin{figure}[htb!]
	\centering
	\includegraphics[width=0.28\linewidth]{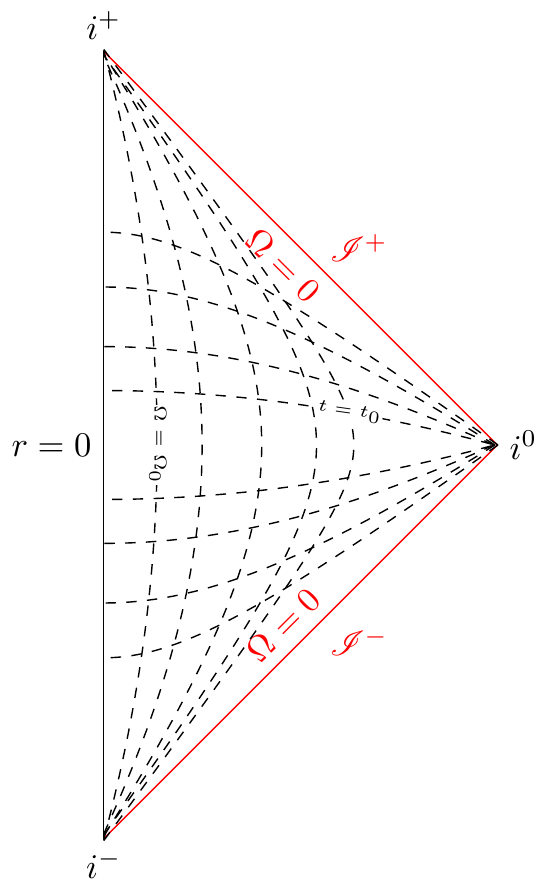}
	\caption{\textit{Carter-Penrose diagram of Minkowski space in coordinates $\{t,\Omega,\theta,\phi\}$. The horizontal lines are lines of constant $t$, while vertical lines represent lines of constant $\Omega$.}}
	\label{fig:PenroseDiag}
\end{figure}

The discussion thus far was certainly hand-wavy in parts, but the general strategy can be made rigorous, as we will show in what follows. First, we need to make the idea of ``going infinitely far away'' more precise. This is achieved by a conformal completion of spacetime. The idea is a very simple one: The physical spacetime is modelled by a manifold $\hatM$  which is endowed with a Minkowski metric $\hat\eta_{ab}$. (We always use hats to indicate physical quantities. The reason for this will become clear during the first few chapters). The manifold has an infinite extension, but we can bring ``infinity'' to a finite distance by means of a conformal transformation. That is, we introduce a conformal factor $\Omega>0$ and we define the conformal metric $\eta_{ab} := \Omega^2\hat\eta_{ab}$. In the case of the Minkowski line element, which in outgoing Eddington-Finkelstein coordinates is given by
\begin{align}
	\dd \hat{s}^2 = -\dd u^2 - 2 \dd u\,\dd r + r^2\dd \omega^2,
\end{align}
we would choose $\Omega = \frac{1}{r}$ and the conformally rescaled line element would thus read
\begin{align}
	\dd s^2 = \Omega^2 \dd \hat s^2 = -\Omega^2\,\dd u^2 + 2 \dd u\, \dd \Omega+\dd \omega^2.
\end{align} 
While the components of the physical line element diverge as $r\to \infty$, we find that the components of the rescaled line element are well-behaved in the $r\to\infty$ limit. Moreover, we can regard $\Omega$ as a new coordinate and the $r\to\infty$ limit is equivalent to the $\Omega\to 0$ limit. The advantage of this conformal rescaling is thus that we end up with a metric which is well-behaved in the asymptotic region of Minkowski space. Additionally, we can go one step further and complete the spacetime by adding a boundary to it. The ``point'' $r=\infty$ (or, equivalently, the ``point'' $\Omega = 0$) is not part of the original manifold $\hatM$. But if we work with the conformally rescaled metric, we can add the $3$-manifold described by $\Omega=0$ to~$\hatM$. This is the conformal completion (cf. Figure~\ref{fig:ConformalCompletion}). 
It allows us to talk about the asymptotic region of Minkowski space as a genuine manifold which possesses a well-behaved metric, namely the conformally rescaled metric. We can thus do differential geometry in the asymptotic region and it is more convenient to work with the mathematical model $(\M, \eta_{ab})$, rather than with the physical spacetime $(\hatM, \hat\eta_{ab})$. The model is defined by $\M := \hatM\cup\scri$ and $\eta_{ab} := \Omega^2\hat\eta_{ab}$, where $\scri$ (read ``scri'') is the $3$-manifold defined by $\Omega=0$. See also Figure~\ref{fig:PenroseDiag} for a graphical representation of the conformal completion in terms of a Penrose diagram.

It is important to point out that $(\M, \eta_{ab})$ is a purely mathematical construct. But it is a very powerful one, as we will see, and we can always relate results obtained in $(\M, \eta_{ab})$ to the physical spacetime~$(\hatM, \hat\eta_{ab})$ by means of a conformal transformation. We will make extensive use of this fact in the following subsections. 

Specifically, in subsection~\ref{ssec:NPFormalism}, we will introduce the Newman-Penrose null tetrad formalism, which will further facilitate the discussion of electromagnetic waves and, later on, of gravitational waves in the asymptotic region $\scri$. In subsection~\ref{ssec:PeelingTheorem}, we will reap the first fruits of our efforts and prove the so-called \textbf{Peeling Theorem}. This theorem, which only relies on the conformal invariance of Maxwell's theory, describes how the different components of the Maxwell $2$-form decay, or ``peel off'', at different rates. This will allow us to disentangle the radiative modes from the coulombic modes (in a much more rigorous way than we did in this section) and in Chapter~\ref{Chap2}, we will be able to compute the flux of energy and momentum of electromagnetic waves through regions of $\scri$. 

The tools and techniques introduced for electromagnetism can be carried over to GR. This is a task which we initiate in Chapter~\ref{Chap3}, where we introduce a special class of (curved) spacetimes $(\hatM, \hat g_{ab})$. We will see in subsequent chapters that the Peeling Theorem also holds for the gravitational field, which will ultimately lead us to the identification of radiative modes in full, non-linear GR. Moreover, we will be naturally led to discover an asymptotic symmetry group, the so-called \textbf{BMS group}, which has far-reaching consequences and applications.
\bigskip

\subsection{Newman-Penrose Null Tetrad Formalism}\label{ssec:NPFormalism}
The Newman-Penrose null tetrad formalism plays a crucial role in what follows. For the time being, we are interested in Minkowski space and in keeping things simple. We will thus introduce explicit expressions for the Newman-Penrose null tetrad. This formalism, however, is much more general and it can be applied to generic curved backgrounds, as we will see in Chapter~\ref{Chap3}. 

To start with, we define the null tetrad in the physical spacetime $(\hatM, \hat{\eta}_{ab})$. That is, we introduce the $1$-forms
 \begin{align}\label{eq:DefinitionNullTetrad}
 	\hat\ell_a &:= -\frac{1}{\sqrt{2}}\left(\nabla_a t-\nabla_a r\right),& \hat n_a &:= -\frac{1}{\sqrt{2}}\left(\nabla_a t+\nabla_a r\right),&  \hat m_a &:= \frac{r}{\sqrt{2}}\left(\nabla_a\theta + i\sin\theta\,\nabla_a\phi\right),
 \end{align}  
 where $t$, $r$, $\theta$, and $\phi$ are the spherical Minkowski coordinates, $\nabla_a$ denotes the covariant derivative operator, and $i$ is the imaginary unit. One can now easily check that these $1$-forms are null with respect to the physical Minkowski metric. That is, these $1$-forms satisfy
 \begin{align}
     \hat\eta^{ab} \hat\ell_a\hat\ell_b \equiv \hat\ell^{a}\hat \ell_a &= 0, &  \hat\eta^{ab} \hat n_a \hat n_b \equiv \hat n^{a}\hat n_a  &= 0 && \text{and} &  \hat\eta^{ab} \hat m_a \hat m_b \equiv \hat m^{a} \hat m_{a}= 0.
 \end{align}
 We remark that indices on objects with a hat are raised and lowered with a metric with a hat and that the complex conjugate tetrad $\hat{\bar{m}}_a$ is of course also null with respect to $\hat\eta^{ab}$. Moreover, a few quick computations reveal that the following cross-normalization relations hold:
  \begin{align}
 	\hat  \ell_a \hat n^{a} &= -1 &\text{and}& & \hat m_a\hat{\bar{m}}^a &= 1.
 \end{align}
 All other contractions between $\hat\ell_a$, $\hat n_a$, $\hat m_a$, and $\hat{\bar{m}}_a$ vanish. Since a tetrad carries the same information as the metric, it is no surprise that the Minkowski metric can be expressed in terms of the null tetrad and one easily finds (see Exercise~\ref{ex:MetricInNPForm})
  \begin{equation}
 	\hat{\eta}_{ab} = -2\hat \ell_{(a} \hat n_{b)} + 2\hat m_{(a} \hat{\bar{m}}_{b)}.
 \end{equation}
 We point out that the factor of $r$ in the definition of $\hat m_a$ is required in order to obtain the $r^2$ in the spherical part of the physical metric, $\dd\hat{\omega}^2 = r^2\left(\dd\theta^2 + \sin^2\theta\dd\phi^2\right)$, and that $\hat m_{(a} \hat{\bar{m}}_{b)}$ is real despite the fact that $\hat m_{a}$ is complex. Using the definitions given in~\eqref{eq:DefinitionNullTetrad}, we can easily determine that the co-tetrad is explicitly given by
 \begin{align}\label{ContravariantTetrads}
 	\hat \ell^{a} &= \frac{1}{\sqrt{2}}\left(\hat t^{a} + \hat r^{a}\right),& \hat n^{a} &= \frac{1}{\sqrt{2}}\left(\hat t^{a}-\hat r^{a}\right), &\hat  m^{a} &= \frac{1}{\sqrt{2} r}\left(\hat \theta^{a} + \frac{i}{\sin\theta}\,\hat \phi^{a}\right),
 \end{align}
 where $\hat t^{a}$ and $\hat r^{a}$ are unit timelike and spacelike vectors, respectively, i.e., $\hat t_{a} \hat t^{a} = -1$ and $\hat r_{a} \hat r^{a} = 1$, and $\hat \theta^{a}\partial_a := \PD{}{\theta}$ and $\hat \phi^{a}\partial_a := \PD{}{\phi}$. These equations and relations completely define the formalism in the physical spacetime $(\hatM, \hat{\eta}_{ab})$. Concretely, this means that everything that can be done using a metric can now also be done using the null tetrad and we can think of the spacetime $(\hatM, \hat{\eta}_{ab})$ as being equivalently described by $(\hatM, \hat \ell_a, \hat n_a, \hat m_a, \hat{\bar{m}}_a)$. The advantage of this point of view will become apparent shortly.
 
 At this point, we recall that we wish to work with a conformally completed spacetime where $\eta_{ab} = \Omega^2 \hat\eta_{ab}$, with $\Omega=\frac{1}{r}$, and $\M = \hatM\cup\scri$. Our task is therefore to infer how the null tetrad of the physical spacetime transforms under a conformal rescaling. This will give us the null tetrad of the conformally completed spacetime $(\M, \eta_{ab})$.

Let us begin with the tetrad $\hat\ell^{a}$ and let us work in outgoing Eddington-Finkelstein coordinates $(u, \Omega, \theta, \phi)$, with $u:=t-r$ and $\Omega=\frac{1}{r}$. In terms of these coordinates, $\scrip$ is the $\Omega = 0$ hypersurface and $(u, \theta, \phi)$ are well-defined coordinates on all of $\scrip$. We can then rewrite $\ell^{a}$ in terms of the conformally rescaled metric (notice that the inverse metric of the conformal completion satisfies $\hat\eta^{ab} = \Omega^2\eta^{ab}$, as shown in Exercise~\ref{ex:RelationPhysicalMetricToConformal}):
\begin{align}
    \hat\ell^{a} &= -\frac{1}{\sqrt{2}}\hat\eta^{ab}\left(\nabla_b t - \nabla_b r\right) = -\frac{\Omega^2}{\sqrt{2}}\eta^{ab}\left(\nabla_b \left(u+\frac{1}{\Omega}\right) - \nabla_b \frac{1}{\Omega}\right)\notag\\
    &= -\frac{\Omega^2}{\sqrt{2}}\eta^{ab}\nabla_b u.
\end{align}
From this we immediately deduce that $\hat\ell^{a}$ has a smooth limit to $\scrip$, which is given by
\begin{equation}
    \hat\ell^{a} \equalhat 0,
\end{equation}
where we recall that the symbol `$\equalhat$' stands for ``equality on $\scri$''. We can compute the limit of $\hat n^{a}$ in a similar fashion. First, we rewrite this tetrad in outgoing Eddington-Finkelstein coordinates and in terms of the rescaled metric:
\begin{align}
    \hat n^{a} = -\frac{1}{\sqrt{2}} \hat{\eta}^{ab}\left(\nabla_b t + \nabla_b r\right) = -\frac{\Omega^2}{\sqrt{2}}\eta^{ab} \left(\nabla_b u - \frac{2}{\Omega^2}\nabla_b\Omega\right).
\end{align}
If we take the limit to $\scri$ of this expression, the first term vanishes because it is proportional to $\Omega^2$, but the second term is finite:
\begin{equation}
    \hat n^{a}\equalhat \sqrt{2}\eta^{ab}\nabla_b\Omega.
\end{equation}
Not only is $\eta^{ab}\nabla_b\Omega$ finite at $\scri$, it also has a very simple geometric interpretation: The vector
\begin{equation}
    \tilde{n}^{a} := \eta^{ab}\nabla_b\Omega
\end{equation}
is the normal vector to $\scrip$. This follows from the fact that normal vectors to hypersurfaces described by $\Phi(x^{a}) = 0$ are given by $\tilde{n}_a \propto \left.\nabla_a\Phi\right|_{\Phi=0}$. In our case, $\scrip$ is defined by $\Omega = 0$ and $\tilde{n}_a$ is therefore normal to $\scrip$. We can even say a little bit more than that: $\tilde{n}^{a}$ is a null normal to $\scrip$ which means that $\scrip$ itself is a null hypersurface. It is left as an exercise (see Exercise~\ref{ex:nIsNull}) to show that $\tilde{n}^{a}$ is a null vector, i.e., that it satisfies $\tilde{n}_a\tilde{n}^{a} = 0$.

Hence, both vectors $\hat r^{a}$ and $\hat t^{a}$ have a smooth limit to $\scrip$. From this we can immediately deduce the limit of $\hat n^{a}$ and $\hat \ell^{a}$ to $\scrip$ from the defining equations~\eqref{ContravariantTetrads}. We easily find
\begin{align}
	\hat n^{a} &\equalhat \sqrt{2}\,\tilde{n}^{a} &\text{and} & & \hat \ell^{a} &\equalhat 0.
\end{align}
This means that $\hat n^{a}$ and $\hat \ell^{a}$ have a smooth limit to $\scrip$.

\begin{mysidenote}{An intuitive geometric reason for why $\hat{t}^{a}$ becomes null on $\scri$}{TimetranslationBecomingNull}
    Notice that $\hat t^{a}$ is timelike in the physical spacetime, but it becomes null when we move it to $\scri$ (i.e., to the boundary which we add to the physical spacetime). There is a geometric reason for this: $\hat t^{a}$ is the time-translation Killing vector field of Minkowski space. In particular, this means it describes an isometry of the metric. But since $\scri$ is determined by the metric, and $\hat t^{a}$ cannot change the metric because it is a Killing vector field, $\hat t^{a}$ cannot ``move'' or change $\scri$. Hence, it must be tangential to $\scri$. However, since $\scri$ is a null surface, this is only possible if $\hat t^{a}$ is itself a null or a spacelike vector at $\scri$. Since $\hat t^{a}$ is a smooth timelike vector in the bulk of spacetime, it cannot suddenly ``jump'' across the light cone and become spacelike. In the limit to $\scri$, it can only become null. This is precisely what we found in our computation.
\end{mysidenote}

Let us now return to the task of inferring how the physical null tetrad transforms under conformal rescaling. Our goal is to obtain a null tetrad which is well-defined on the whole conformally completed spacetime $(\M, \eta_{ab})$, such that we can represent this spacetime equivalently as $(\M, \ell^{a}, n^{a}, m^{a}, \bar{m}^{a})$. To that end, it is convenient to impose that the conformally rescaled null tetrad satisfies the same cross-normalization conditions (with respect to the rescaled metric $\eta_{ab}$) as the physical null tetrad. That is, we impose that
\begin{align}
    \eta_{ab}\ell^{a} n^{a} &= -1 & \text{and} & &  \eta_{ab}m^{a} \bar{m}^{b} &= 1,
\end{align}
while all other contractions are zero. Moreover, we impose that the physical and the rescaled null tetrads are related by the transformation law
\begin{equation}
    (\ell^{a}, n^{a}, m^{a}, \bar{m}^{a}) = (\Omega^{s_1} \hat\ell^{a}, \Omega^{s_2}\hat n^{a}, \Omega^{s_3} \hat m^{a}, \Omega^{s_4} \hat{\bar{m}}^{a}),
\end{equation}
where $s_1$, $s_2$, $s_3$, and $s_4$ are real numbers which need to be determined. Since we have seen that the physical null tetrad $\hat n^{a}$ has a well-defined limit to $\scrip$, we \textit{choose} the conformally rescaled $n^{a}$ to be equal to the physical one. That is, we set
\begin{equation}
	n^{a} := \hat{n}^{a},
\end{equation} 
which is tantamount to setting $s_2 = 0$. This is a convenient choice because $n^{a}$ has the interpretation of being the normal vector to $\scrip$. We cannot define the rescaled $\ell^{a}$ as being the limit of the physical $\hat{\ell}^{a}$ because the latter one vanishes on $\scrip$ and this would lead to a degenerate null tetrad. However, we can exploit the fact that $\hat{n}^{a}$ and $\hat{\ell}^{a}$ are cross-normalized as $\hat \eta_{ab}\hat\ell^{a}\hat n^{b} = -1$ and that we demanded that this cross-normalization shall be preserved under the conformal completion. This leads us to the condition
\begin{equation}\label{eq:ConditionOnell}
	\hat \eta_{ab} \hat\ell^{a}\hat n^{b} = \Omega^2 \eta_{ab}\hat\ell^{a} n^{b} = \Omega^{2+s_1}\underset{= -1 }{\underbrace{\eta_{ab}\ell^{a} n^{b}}} \overset{!}{=} -1.
\end{equation}
This is obviously solved by $s_1 = -2$ and hence we conclude that $\ell^{a}$ is obtained from the physical $\hat \ell^{a}$ via the relation
\begin{equation}
    \ell^{a} = \Omega^{-2} \hat\ell^{a}.
\end{equation}

Notice that $\ell^{a}$ is a tetrad which is well-defined on the whole conformally completed spacetime. Hence, we can express $\hat\ell^{a}$ in terms of $\ell^{a}$ and take the limit to $\scrip$, which results in
\begin{equation}
	\lim_{\,\,\to \scrip}\hat\ell^{a} = \lim_{\Omega\to 0}\left(\Omega^2\ell^{a}\right) = 0.
\end{equation}
This is a nice consistency test and it also tells us that the physical $\hat\ell^{a}$ decays as $\frac{1}{r^2}$ as it approaches $\scrip$. This can also easily be verified by a direct computation.

Now let us turn to the rescaling behavior of $\hat{m}^{a}$. From its definition, equation~\eqref{ContravariantTetrads}, we see that it can be written as
\begin{equation}
	\hat{m}^{a} = \frac{\Omega}{\sqrt{2}}\left(\hat{\theta}^{a} + \frac{i}{\sin\theta}\,\hat{\phi}^{a}\right).
\end{equation}
We simply define the term multiplied by $\Omega$ to be the rescaled $m^{a}$ and we thus obtain
\begin{equation}
	m^{a} = \Omega^{-1}\hat{m}^{a},
\end{equation}
where $m^{a}$ is again well-defined on the whole conformally completed spacetime. Notice that this rescaling preserves the cross-normalization $\hat\eta_{ab}\hat{m}^{a}\hat{\bar{m}}^{b} = 1$:
\begin{equation}
	\eta_{ab} m^{a} \bar{m}^{b} = \Omega^2 \hat\eta_{ab} \Omega^{-1}\hat{m}^{a} \Omega^{-1}\hat{\bar{m}}^{b} = 1.
\end{equation}
In summary, we have found that the Newman-Penrose null tetrad of the conformally completed spacetime is related to the physical Newman-Penrose null tetrad by
\begin{equation}
\boxed{\begin{aligned}
    n^{a} &= \hat{n}^{a}, & \ell^{a} &= \Omega^{-2}\hat\ell^{a}, & m^{a} &= \Omega^{-1}\hat{m}^{a}
\end{aligned}}
\end{equation}
and they satisfy
\begin{align}
	\eta_{ab}\ell^{a}n^{a} &= -1, &\text{and} & & \eta_{ab}m^{a}\bar{m}^b = 1.
\end{align}
All other contractions vanish identically. It should also be noted that these rescaling properties hold in full generality. That is, they do not only hold in Minkowski space, they hold for any curved background.

We can now immediately derive an interesting result from these rescaling properties which goes by the name of the \textbf{Peeling Theorem}. In short, this theorem tells us at which rate the components of the Maxwell $2$-form $F_{ab}$ decay as one approaches $\scrip$. From the decaying behavior, one can in turn extract information about coulombic modes, radiative modes, and multipole moments. Proving the Peeling Theorem will be the main goal of the next subsection.\bigskip

\subsection{The Peeling Theorem for Electrodynamics}\label{ssec:PeelingTheorem}
As we will see shortly, the Peeling Theorem is a consequence of the smoothness of the Maxwell $2$-form $F_{ab}$, which in turn is a consequence of the conformal invariance of electrodynamics. Nothing else is required or assumed in order to prove the theorem.

This will change in the case of gravity, where the smoothness of the Weyl tensor ---the analogue of Maxwell's $2$-form in the gravitational context--- has to be assumed and cannot be traced back to some fundamental property of GR. It is nevertheless instructive to see how to prove the Peeling Theorem for electrodynamics, as the main steps can be carried over to GR. Also, we will gain some first intuition and familiarity with the Newman-Penrose formalism.

To begin with, we note that in the conformally completed spacetime $(\M, \eta_{ab})$, the fields $n^{a}$, $\ell^{a}$, $m^{a}$ are smooth and possess a well-defined limit to $\scrip$. The $2$-form $F_{ab}$ is smooth as well, which is, as mentioned above, due to the conformal invariance of Maxwell's theory. We now introduce the following definitions:
\begin{align}
	\Phi_2 &:=  F_{ab}  n^{a} \bar{m}^b\notag\\
	\Phi_1 &:= \frac12  F_{ab}\left( n^{a} \ell^{b} +  m^{a}\bar{m}^{b}\right)\notag\\
	\Phi_0 &:=  F_{ab}  m^{a}  \ell^{b}.
\end{align}
These are just definitions without any underlying meaning. In tensorial language, they look a little bit awkward, but they are completely natural in a spinorial language. Nevertheless, what we achieve through these definitions is a representation of the six components of $F_{ab}$ in terms of three complex functions. In other words, the scalars $\Phi_i = \Phi_i(u, \Omega, \theta,\phi)$ contain the same information as the $2$-form $F_{ab}$.

\begin{mysidenote}{A cautionary remark on Newman-Penrose scalars}{RemarkOnNPScalars}
    These functions are called the Newman-Penrose scalars. A word of caution though: This does not imply that these functions are some kind of invariant! They clearly represent components of a $2$-form. In fact, the functions $\Phi_i$ are clearly scalars with respect to coordinate transformations. But they do depend on a choice of tetrad and they will change when we choose a different tetrad to work with. This is akin to choosing different reference frames and getting different expressions for the electric and magnetic fields.
\end{mysidenote}

We now proceed and introduce functions at $\scrip$, which are simply defined as
\begin{align}
	\Phi^\circ_i(u,\theta,\phi) &:= \left.\Phi_i(u, \Omega,\theta,\phi)\right|_{\scrip} & \text{for } i\in\{0,1,2\} .
\end{align}
These are functions\footnote{To be more precise: These are spin-weighted functions. We will introduce the concept of spin-weighted fields in subsection~\ref{Chap1.D}.} of the coordinates $(u,\theta,\phi)$ and $\Phi^\circ_i$ capture the leading order behavior of~$\Phi_i$ at~$\scrip$. This can also be seen by performing a Taylor expansion of $\Phi_i$ around $\Omega=0$, which gives us
\begin{equation}
	\Phi_i = \Phi^\circ_i + \left.\frac{\dd\Phi_i}{\dd \Omega}\right|_{\Omega=0}\Omega + \O(\Omega^2).
\end{equation}
Our goal is now to relate the Newman-Penrose scalars $\Phi_i$ of the conformally completed spacetime $(\M, \eta_{ab})$ to the physical scalars $\hat{\Phi}_i$ of the physical spacetime $(\hatM, \hat\eta_{ab})$. This is straightforward since we only need the rescaling behavior of the Newman-Penrose null tetrad, which we have established in the previous subsection. We immediately obtain 
\begin{align}\label{eq:PeelingMaxwell}
	\hat \Phi_2 &= \hat F_{ab} \hat n^{a} \hat{\bar{m}}^{b} = \frac{1}{r} F_{ab}  n^{a} {\bar m}^{b} = \frac{\Phi_2}{r} = \frac{\Phi^\circ_2(u, \theta, \phi)}{r} + \O(r^{-2})\notag\\
	\hat \Phi_1 &= \frac{{\Phi}^\circ_1(u,\theta,\phi)}{r^2} + \O(r^{-3})\notag\\
	\hat \Phi_0 &=  \frac{{\Phi}^\circ_0(u,\theta,\phi)}{r^3} + \O(r^{-4}).
\end{align}
The computations for $\hat\Phi_1$ and $\hat\Phi_2$ are explicitly done in Exercise~\ref{ex:PeelingMaxwell}.\footnote{It is useful to notice that the fall-off property of the physical scalar $\hat\Phi_i\propto\frac{1}{r^n}$ can be remembered from the equation $i+n=3$.} Observe what we have achieved: The functions $\hat\Phi_i$ are the physical Newman-Penrose scalars which carry the same information as the physical Maxwell $2$-form $\hat{F}_{ab}$ and these scalars fall-off in a characteristic manner. Or one could say that as one approaches $\scrip$, the components of $\hat F_{ab}$ are ``peeled off'' at different rates. This is the Peeling Theorem and it offers us a first clue that $\Phi_2(u, r,\theta,\phi)$ encodes the radiative modes while $\Phi_1(u,r,\theta,\phi)$ carries information about coulombic modes. This is because we know that the radiation field decays like $\frac{1}{r}$ in the radiation zone while the Coulomb field behaves like $\frac{1}{r^2}$. But there are of course more reasons, as we will see in the next sections. 

Before we can explore the consequences of the Peeling Theorem, we need to develop some mathematical tools. This is the main task of the next subsection.\bigskip

\subsection{Spin-weighted Fields}\label{Chap1.D}
In the following chapters, the $2$-sphere will play an essential role as we will explore certain hypersurfaces for which the metric has the form of a $2$-sphere metric. Hence, as a mathematical interlude, we want to consider fields defined on the $2$-sphere. It is easy to see that the unit $2$-sphere is parametrized by the vectors $m^{a}$ and $\bar{m}^{a}$, which satisfy $\eta_{ab} m^{a} \bar{m}^b = 1$. In fact, in one of the exercises (see Exercise~\ref{ex:2form}), it is shown that the metric on the unit $2$-sphere is given by $2m_{(a}\bar{m}_{b)}$, while the area element is given by $2m_{[a}\bar{m}_{b]}$. What is interesting about that is that the Newman-Penrose formalism introduces a $U(1)$ gauge freedom into the description of $2$-spheres, which can also be seen directly from $\eta_{ab} m^{a} \bar{m}^b = 1$. This gauge freedom is characterized by
\begin{equation}\label{eq:GaugeTrafo}
	 m^{a}\, \longrightarrow\, \e^{i\alpha(\theta,\phi)} m^{a}
\end{equation}
and it clearly leaves the cross-normalization invariant. Moreover, the metric and the area element, 
\begin{align}
	s_{ab} &= 2 m_{(a}\bar{m}_{b)} & &\textsf{(metric on unit $2$-sphere)}\notag\\
	\varepsilon_{ab} &= 2 m_{[a}\bar{m}_{b]} & &\textsf{(area form on unit $2$-sphere)},
\end{align}
are also both invariant under the gauge transformation~\eqref{eq:GaugeTrafo}. It is now natural to ask how $1$-forms behave under the transformation~\eqref{eq:GaugeTrafo}, since any other field can be constructed from $1$-forms and the metric.\\

Since $m_a$ and $\bar m_a$ provide a basis on the $2$-sphere, it is natural to expand the $1$-form $v_a$ in this basis: $v_a := \bar{f}\, m_a + f\, \bar{m}_a$. If $m^{a}$ changes under the $U(1)$ gauge transformation, then $v_a$ will seemingly also change under this transformation. However, $v_a$ is a real $1$-form and it does not know anything about the complex basis we used to expand it in or the gauge freedom we have introduced through our formalism. In other words, it should not change. This implies that the expansion functions $f$ have to transform as well such that $v_a$ remains invariant under the transformation~\eqref{eq:GaugeTrafo}. This is a sensible requirement and we find that 
\begin{equation}
	f\, \longrightarrow\, \e^{i\alpha(\theta,\phi)}f
\end{equation}
does the job. Quantities which transform in this way are called \textbf{functions of spin weight $\mathbf{1}$}. More generally, if 
\begin{equation}
	h\,\longrightarrow\, \e^{i\, s\alpha(\theta,\phi)} h\quad\text{for }s\in\bbZ,
\end{equation}
then $h$ is said to be of \textbf{spin weight $\mathbf{s}$}. Notice that spin weight is a notion which is defined for \textit{functions}, but we make an exception for $m^{a}$, which is said to have spin weight~$1$, and for $\bar{m}^{a}$, which has spin weight~$-1$.

What these definitions show, is that spin weighted objects are just a way of talking about components of tensors on $2$-spheres. This follows from the fact that  $f$ and $\bar f$ represent the two components of the $1$-form $v_a$ with respect to the basis $\{m^{a}, \bar{m}^{a}\}$. Similarly, $h=T_{a_1\cdots a_s}m^{a_1}\cdots m^{a_s}$ represents a component of the tensor $T_{a_1\cdots a_s}$ and it has spin weight $s$. It is left as an exercise to show that $\Phi_2$, $\Phi_1$, and $\Phi_0$, i.e., the components of $F_{ab}$, have spin weight $-1$, $0$, and $1$, respectively (see Exercise~\ref{ex:SpinWeight}).\bigskip

Finally, we want to consider the differential calculus of spin weighted objects. This leads us to introducing the angular derivative operator $\eth$ (pronounced ``eth''). This operator acts on functions of spin weight $s$ via
\begin{align}\label{eq:eth}
	\eth f_s &= \frac{1}{\sqrt{2}} m^{a} \underline{D}_a f_s - \frac{s}{\sqrt{2}} \cot\theta f_s\notag\\
	&= \frac12 \left(\partial_\theta f_s + \frac{i}{\sin\theta}\partial_\phi f_s - s\,\cot\theta f_s\right),
\end{align}
where $\underline{D}_a$ is the covariant derivative operator on the $2$-sphere. This looks like a messy definition, but it has a simple origin: Suppose you are given the $1$-form $v_a$ and you take the derivative $m^{a} m^{b} \left(\underline{D}_a v_b\right)$. By Leibniz's rule,  this is equal to
\begin{equation}
	m^{a} m^{b} \left(\underline{D}_a v_b\right) = m^{a} \underline{D}_a\left(v_b m^{b}\right) - m^{a} \left(\underline{D}_a m^{b}\right)v_{b}.
\end{equation}
But $v_a m^{a} = f_1$, where we added the index $1$ to emphasize that this is a spin weight $1$ function. Hence, we obtain
\begin{equation}
	m^{a} m^{b} \left(\underline{D}_a v_b\right) = m^{a} \underline{D}_a f_1 - m^{a} \left(\underline{D}_a m^b\right)v_{b}.
\end{equation}
To proceed, all we need to do is to compute the following derivatives:
\begin{align}
    \frac{1}{\sqrt{2}}m^{a}\underline{D}_a f_1 &= \frac12\left(\partial_\theta f_1 + \frac{i}{\sin\theta}\,\partial_\phi f_1\right) \notag\\
    -\frac{1}{\sqrt{2}}m^{a} \left(\underline{D}_a m^b\right)v_{b} &= -\frac12 \cot\theta\, f_1.
\end{align}
We have introduced a factor of $\frac{1}{\sqrt{2}}$ in order to obtain nicer expressions and we have used the fact that $\underline{D}$ is the covariant derivative with respect to the Levi-Civita connection defined by the metric $s_{ab} = 2 m_{(a} \bar{m}_{b)}$. Hence, we can now define
\begin{equation}\label{eq:SW1}
    \eth f_1 := \frac{1}{\sqrt{2}}m^{a} m^{b} \left(\underline{D}_a v_{b}\right) = \frac12\left(\partial_\theta f_1 + \frac{i}{\sin\theta}\partial_\phi f_1 - \cot\theta\, f_1\right).
\end{equation}
Observe that this is precisely what follows from~\eqref{eq:eth} for $s=1$. Similarly, we define $\eth f_{-1}$ as
\begin{equation}\label{eq:SW-1}
    \eth f_{-1} := \frac{1}{\sqrt{2}} m^{a} \bar{m}^{b} \left(\underline{D}_a v_b\right) = \frac12\left(\partial_\theta f_{-1} + \frac{i}{\sin\theta} \partial_\phi f_{-1} + \cot\theta\, f_{-1}\right),
\end{equation}
where the right hand side follows from similar straightforward computations as before. Finally, the action of $\eth$ on a spin weight $0$ function is defined as
\begin{equation}\label{eq:SW0}
    \eth f_{0} := m^{a} \underline{D}_a f_0 = \frac12 \left(\partial_\theta f_0 + \frac{i}{\sin\theta} \partial_\phi f_0\right).
\end{equation}
Now that we know how the angular derivative operator acts on spin weight $-1$, $0$, and $1$ functions, we can easily determine how it acts on spin weight $s$ functions. Let us first make the qualitative observation that  all three definitions contain an $m^{a}$. This has the effect of \textit{increasing} the spin weight by one. The effect of $m^{b}$ in~\eqref{eq:SW1} and of $\bar{m}^{b}$ in~\eqref{eq:SW-1} is to ``isolate'' the spin weight $1$ and $-1$ parts of $v_b$, respectively. 

Now let us see how to generalize these observations to spin weight $s$ functions. Without loss of generality, we set
\begin{equation}
    f_s = T_{a_1\cdots a_p b_1\cdots b_{q}} m^{a_1}\cdots m^{a_p}\bar{m}^{b_1}\cdots \bar{m}^{b_q}\qquad\text{with } p-q = s.
\end{equation}
Let us also introduce the abbreviation
\begin{equation}
    P^{a_1\cdots a_p b_1\cdots b_{q}} := m^{a_1}\cdots m^{a_p}\bar{m}^{b_1}\cdots \bar{m}^{b_q}.
\end{equation}
We can then define the angular derivative of a spin weight $s$ function as
\begin{align}
    \eth f_{s} :=&\ \frac{1}{\sqrt{2}} m^{a} P^{a_1\cdots a_p b_1\cdots b_{q}}\left(\underline{D}_a T_{a_1\cdots a_p b_1\cdots b_{q}}\right)\notag\\
    =&\ \frac{1}{\sqrt{2}} m^{a} \left(\underline{D}_{a} f_s\right) - \frac{1}{\sqrt{2}} m^{a} \left(\underline{D}_a P^{a_1\cdots a_p b_1\cdots b_{q}}\right) T_{a_1\cdots a_p b_1\cdots b_{q}}.
\end{align}
Clearly, the first term in the second line always gives us
\begin{equation}
    \frac{1}{\sqrt{2}} m^{a} \left(\underline{D}_a f_s\right) = \frac12\left(\partial_\theta f_s + \frac{i}{\sin\theta} \partial_\phi f_s\right).
\end{equation}
The second term is more interesting. What we need to use is the fact that
\begin{align}
    m^{a} \left(\underline{D}_a m^{b}\right) &= \frac{1}{\sqrt{2}}\cot\theta\, m^{b} & \text{and} && m^{a}\left(\underline{D}_a \bar{m}^{b}\right) &= -\frac{1}{\sqrt{2}}\cot\theta\, \bar{m}^b.
\end{align}
It then follows from a repeated application of Leibniz's rule that
\begin{align}
   m^{a}\left(\underline{D}_a P^{a_1\cdots a_p b_1\cdots b_{q}}\right) &= \frac{p}{\sqrt{2}}\cot\theta \, P^{a_1\cdots a_p b_1\cdots b_{q}} - \frac{q}{\sqrt{2}}\cot\theta\, P^{a_1\cdots a_p b_1\cdots b_{q}} \notag\\
   &= \frac{s}{\sqrt{2}} \cot\theta\, P^{a_1\cdots a_p b_1\cdots b_{q}},
\end{align}
where in the last step we have used $p - q = s$. Since, $P^{a_1\cdots a_p b_1\cdots b_{q}} T_{a_1\cdots a_p b_1\cdots b_{q}} = f_s$, by definition, we finally obtain
\begin{equation}
    -\frac{1}{\sqrt{2}} m^{a} \left(\underline{D}_a P^{a_1\cdots a_p b_1\cdots b_{q}}\right)T_{a_1\cdots a_p b_1\cdots b_{q}} = -\frac{s}{2}\cot\theta\, f_s
\end{equation}
and hence we have shown that
\begin{equation}\label{eq:SWs}
    \eth f_s = \frac12\left(\partial_\theta f_s + \frac{i}{\sin\theta}\partial_\phi f_s - s\, \cot\theta\, f_s\right).
\end{equation}
Let us conclude with the remark that $P^{a_1\cdots a_p b_1\cdots b_{q}}$ acts like a ``projector'' which isolates the spin weight $s$ part of the tensor $T$, while the $m^{a}$ which contracts the $\underline{D}_a$ has again the effect of \textit{increasing} the spin weight by one. Thus, in full generality, it holds true that if $f_s$ has spin weight $s$, then $\eth f_s$ has spin weight~$s+1$.

We conclude this chapter by stating that the action of the angular derivative operator can also by written as (see Exercise~\ref{ex:EquivalenceAngularDerivative} for a proof that this is equivalent to~\eqref{eq:SWs})
\begin{equation}\label{eq:AlternativeAngularDeriv}
    \eth f_s = \frac{1}{2}(\sin \theta)^s\left(\frac{\partial}{\partial\theta}+\frac{i}{\sin \theta}\frac{\partial}{\partial\phi}\right)(\sin\theta)^{-s}f_s \;
\end{equation}
and that a conjugate angular derivative operator, $\bar{\eth}$, can be defined starting from $\bar{m}^{a} \bar{m}^{b} \left(\underline{D}_a v_b\right)$. It is explicitly given by
\begin{align}\label{eq:ConjugateEth}
	\bar{\eth}f_s := \frac12\left(\partial_\theta f_s - \frac{i}{\sin\theta}\partial_\phi f_s + s\, \cot\theta\, f_s\right) \equiv \frac{1}{2}(\sin \theta)^{-s}\left(\frac{\partial}{\partial\theta}-\frac{i}{\sin \theta}\frac{\partial}{\partial\phi}\right)(\sin\theta)^{s} f_s
\end{align}
and it \textit{lowers} the spin weight of $f_s$ by one.

\newpage
\subsection{Exercises}
\renewcommand{\thesection}{\arabic{section}}

\begin{Exercise}[]\label{ex:NormalsToNullSurface}
    Let $\Sigma$ be a null surface which is defined by the constraint $\Phi(x^{a}) = 0$ with $\nabla_a \Phi(x) \neq 0$ for all $x\in \Sigma$. Prove the following claims.
    \begin{itemize}
    	\item[a)] The normal vector $n^a := -g^{ab}\nabla_b \Phi$ to $\Sigma$ is also tangential to $\Sigma$.
    	\item[b)] Show that if the non-null vector $s^{a}\neq 0$ is tangential to $\Sigma$, it has to be a spacelike vector.
    	\item[c)] Let $v^{a}$ be a null vector which is tangential to $\Sigma$. Show that $v^{a}\propto n^{a}$.
    	\item[d)] Let $\ell^{a}$ be a vector which is null but which is \textit{not} tangential to $\Sigma$. Show that it is always possible to normalize the vector such that $\ell_a n^{a} = -1$
    \end{itemize}
\end{Exercise}

\begin{Exercise}[]\label{ex:MetricInNPForm}
    Show that the spacetime metric $g_{ab}$ can be expressed in terms of the Newman-Penrose tetrad as
	\begin{equation*}
		g_{ab} = -2\ell_{(a}n_{b)} +2m_{(a}\bar{m}_{b)}.
	\end{equation*}
\end{Exercise}

\begin{Exercise}[]\label{ex:nIsNull}	
    Use $r^{a} = -\eta^{ab}\nabla_b\Omega$ and the definition $\tilde{n}^{a} := \lim_{\Omega\to 0} r^{a}$ to show that $\tilde{n}^{a}$ is a null vector, i.e., that it satisfies $\tilde{n}_a \tilde{n}^{a} = 0$.
\end{Exercise}

\begin{Exercise}[]\label{ex:RelationPhysicalMetricToConformal}
    The physical metric $\hat g_{ab}$ and the conformally rescaled metric $g_{ab}$ are related to each other by $\hat{g}_{ab} = \Omega^2 g_{ab}$. Show that the inverse of the physical metric satisfies $\hat{g}^{ab} = \Omega^2 g^{ab}$.
\end{Exercise}

\begin{Exercise}[]\label{ex:TetradIdentities}
	Prove the following identities for the tetrad $\ell := \ell_a\dd x^{a}$, $n:=n_a \dd x^{a}$, $m:=m_a\dd x^{a}$:
	\begin{itemize}
		\item[a)]  $\ell\wedge n\wedge m\wedge \bar{m} = i\,r^2\,\sin\theta\,\dd t\wedge\dd r\wedge \dd \theta\wedge\dd\phi$
		\item[b)]  $\epsilon^{abcd}\ell_a n_b m_c \bar{m}_d = i$
		\item[c)]  $\epsilon_{abcd} = -4!\, i\, \ell_{[a}n_b m_c \bar{m}_{d]}$
	\end{itemize}

	\textit{Hint: For the last exercise, use the fact that $T_{[\mu_1\dots\mu_n]} := \frac{1}{n!} \epsilon_{\mu_1\dots\mu_n}\epsilon^{\nu_1\dots\nu_n}T_{\nu_1\dots\nu_n}$.}
\end{Exercise}

\begin{Exercise}[]\label{ex:PeelingMaxwell}
    Use the rescaling properties of the Newman-Penrose null tetrad and the conformal invariance of the Maxwell $2$-form to derive the Peeling Properties of $\hat\Phi_1$ and $\hat\Phi_2$.
\end{Exercise}

\begin{Exercise}[]\label{ex:SpinWeight}
    Show that $\Phi_2$, $\Phi_1$, and $\Phi_0$ have spin weight $-1$, $0$, and $1$, respectively. 
\end{Exercise}

\begin{Exercise}[]\label{ex:EquivalenceAngularDerivative}
    Show that the angular derivative operator~\eqref{eq:SWs} can equivalently be written as~\eqref{eq:AlternativeAngularDeriv}.
\end{Exercise}

\newpage

\asection{2}{Electromagnetic Waves and Null Infinity}\label{Chap2}
In the first chapter, we introduced the Newman-Penrose scalars, which are defined as
\begin{align}
	\Phi_0 &:=  F_{ab}  m^{a} \ell^{b} & &\text{(spin weight $+1$)}\notag\\
	\Phi_1 &:= \frac12 F_{ab}\left(m^{a}\bar{m}^{b} - \ell^{a} n^{b} \right) & &\text{(spin weight $\phantom{-}0$)}\notag\\
	\Phi_2 &:=  F_{ab}  n^{a} {\bar{m}}^b & &\text{(spin weight $-1$)},
\end{align}
and we proved the Peeling Theorem. This theorem tells us how the physical Newman-Penrose scalars, which carry the same information as the Maxwell $2$-form, decay as one approaches $\scrip$. In particular, we have seen that $\hat\Phi_2$ decays like $\frac{\Phi^\circ_2}{r}$, and we are therefore tempted to call $\hat\Phi_2$ the radiation field, while $\hat\Phi_1 = \frac{\Phi^\circ_1}{r^2}$ is thought to encode the Coulomb field. There are indeed stronger reasons for believing this, which we will explore in this and the following subsections. We begin our exploration with a study of Maxwell's equations in the Newman-Penrose formalism and move then, in the next subsection, to a discussion of charges and energy-momentum carried by electromagnetic waves. In subsection~\ref{Chap2:Examples}, we discuss the Coulomb field and the linear dipole antenna within the Newman-Penrose formalism, in order to gain some familiarity with the formalism and demonstrate how it works in practice.

First of all, on physical grounds, we are interested in sources with compact spatial support. This means that near $\scrip$, the electromagnetic sources vanish and Maxwell's equations take the form
\begin{align}
	\nabla_{[a}{F}_{bc]} &= 0 &\text{and} &&  \nabla_{[a}\prescript{\star}{}{F}_{bc]} &= 0.
\end{align}
These are eight equations which we wish to explore on $\scrip$, i.e., on the hypersurface defined by $\Omega = 0$. To do so, we first need to translate the above equations into the Newman-Penrose formalism. In principle, this is an easy task: Express $F_{ab}$ and $\prescript{\star}{}{F}_{ab}$ in terms of the null tetrad and the Newman-Penrose scalars (see Exercises~\ref{ex:Maxwell2FormInNPForm} and~\ref{ex:FIdentity}). Then, contract the equations with the null tetrad $\{n^{a}, \ell^{a}, m^{a}, \bar{m}^{a}\}$ in order to generate four complex scalar equations.

This procedure is carried out in detail in Appendix~\ref{App:A1}. What is of interest to us here, is that the resulting equations can be divided into two groups: There are two (complex) equations which contain derivatives with respect to the retarded time coordinate $u$ and therefore tell us something about the dynamical behavior of the Newman-Penrose scalars, and there are two (complex) constraint equations which carry no dynamical information. When we take the $\Omega\to 0$ limit, i.e., when we restrict ourselves to $\scrip$, these equations take the form
\begin{align}\label{eq:FourIntrinsic}
	&\textsf{Dynamical equations} &  &\textsf{Constraints} \notag\\
	&\partial_u{\Phi}^\circ_1(u,\theta,\phi) = \eth {\Phi}^\circ_2(u,\theta,\phi) &  	&\bar{\eth}\Phi^\circ_1(u,\theta,\phi) = 0 \notag\\
	&\partial_u{\Phi}^\circ_0(u,\theta,\phi) = \eth {\Phi}^\circ_1(u,\theta,\phi) & &\bar{\eth}\Phi^\circ_0(u,\theta,\phi) = 0.
\end{align}
Observe that these equations only contain derivatives intrinsic to $\scrip$, as should be expected, and that the total spin weight of the right hand side matches the spin weight of the left hand side, because the angular derivative operator $\eth$ increases the weight by one. 

\begin{figure}[htb!]
	\centering
	\includegraphics[width=0.5\linewidth]{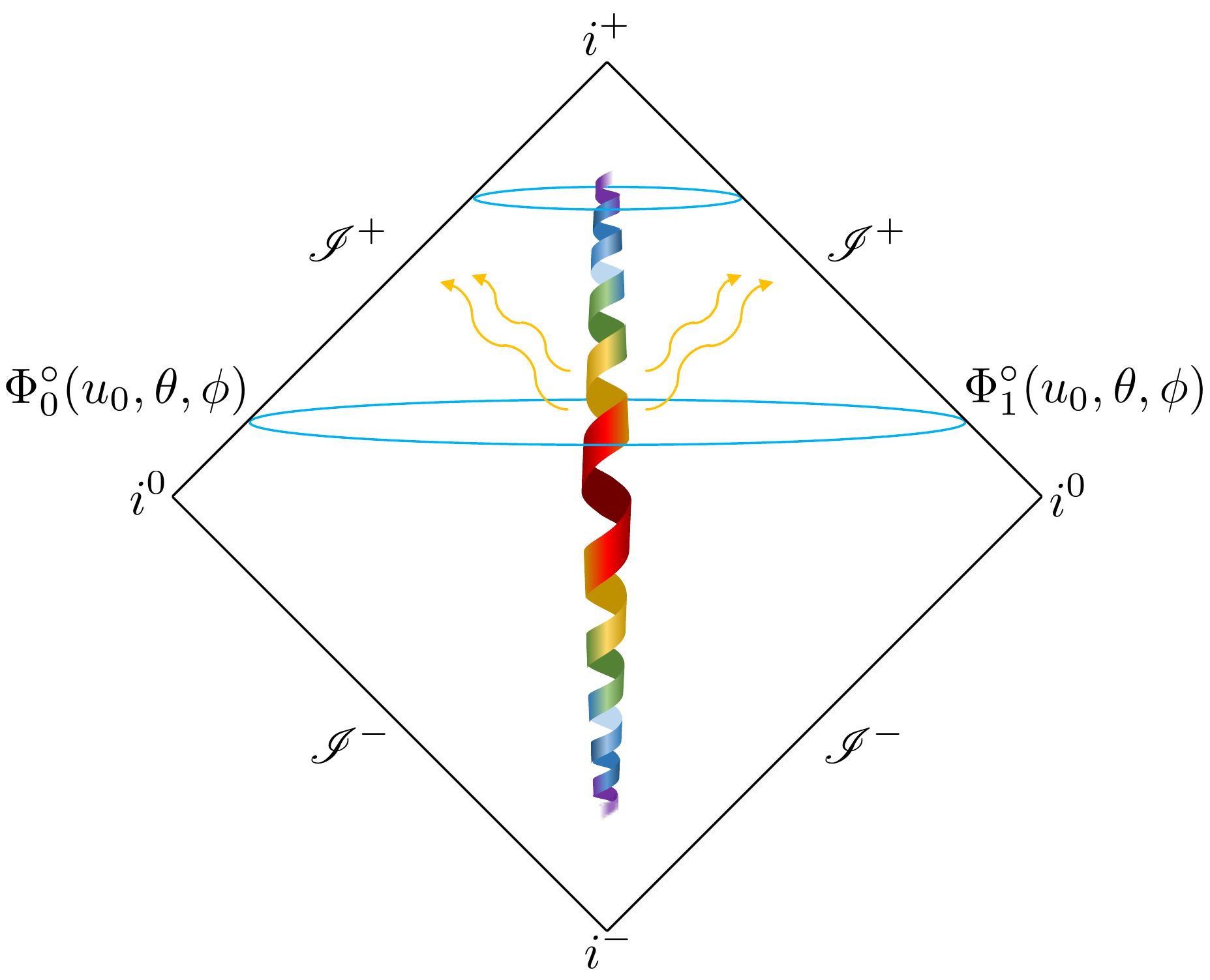}
	\caption{\textit{The field $\Phi^\circ_2$ is freely specifiable and encodes the radiative modes of the Maxwell field. The field~$\Phi^\circ_1$ needs to be specified on a Cauchy surface, here represented by a large blue ring labeled by the data $\Phi^\circ_1(u_0,\theta,\phi)$, and its dynamics depends on the behavior of the radiative field $\Phi^\circ_2$. It is therefore thought to capture the coulombic information of the electromagnetic field.}}
	\label{fig:InitialValueFormulation}
\end{figure}

Furthermore, observe that $\Phi^\circ_2$ only appears in the first equation of~\eqref{eq:FourIntrinsic} and it does so without a ``time'' derivative, i.e., there is no term of the form $\partial_u\Phi^\circ_2$ in any equation. This means that Maxwell's equations do \textit{not} determine the dynamics of $\Phi^\circ_2$. We are thus left with the following situation:

In order to solve Maxwell's equations, which are first order equations for $\Phi^\circ_0$ and $\Phi^\circ_1$, we need to specify initial data at some initial ``time'' $u=u_0$ (a typical choice is $u=-\infty$). Let this initial data be $\Phi^\circ_0(u_0,\theta,\phi)$ and $\Phi^\circ_1(u_0, \theta, \phi)$ (see Figure~\ref{fig:InitialValueFormulation}). However, this is \textit{not} sufficient to determine a unique solution to Maxwell's equations because the angular derivatives of $\Phi^\circ_2$ appear on the right hand side of the first equation in~\eqref{eq:FourIntrinsic}. Hence, we need to specify $\Phi^\circ_2$ everywhere on $\scrip$ by hand! Once we have done that, we can solve the first equation in~\eqref{eq:FourIntrinsic} for $\Phi^\circ_1$ and then use this solution to solve the second equation for $\Phi^\circ_0$. After having solved these equations, we know the electromagnetic field $F_{ab}$ everywhere on~$\scrip$. 

Now comes the main observation: Since $\Phi^\circ_2$ is not determined by the equations themselves, it has to represent the radiative modes! Since this claim is not at all obvious, let us clarify:
\begin{itemize}
    \item[1.]  Given a set of covariant field equations, what we first of all need to do in order to solve them is to perform a $3+1$ decomposition. That is, we single out one coordinate as ``the'' evolution coordinate (typically we think of this coordinate as representing time and we call it $t$, but this labeling is not necessary) and we describe the dynamics of the fields with respect to that coordinate. Furthermore, we need to specify initial data on an initial value surface, i.e., a $t=\textsf{const}.$ surface. Typically, when $t$ is indeed a \textit{timelike} coordinate, the initial value surface is spacelike and we can think of the initial data on that surface as representing our knowledge, gathered by measurements and observation, of the field configuration at a given instant of time, throughout all of space.
    
    For completeness, we mention that in gauge theories we also need to perform a gauge fixing and ensure that the initial data satisfies the constraints on the initial value surface. Once we have picked a gauge, specified initial data, and made sure that the data satisfies the constraints, the field equations should tell us how the fields evolve off the $t=\textsf{const}.$ surface, either into the future or into the past of $t=\textsf{const}$. However, and this is the main point here, not every $t=\textsf{const}.$ surface is adequate for determining a solution! In other words, not every choice of $t$ allows us to determine a solution to the field equations! In yet other words, some choices of $t$ lead to ``bad'' initial value surfaces ($t=t_0$ surfaces), which do not allow us to determine the future or the past using the field equations. Such surfaces are known as \textbf{characteristic surfaces} in the theory of partial differential equations. Appendix~\ref{App:A2} provides a self-contained introduction to basic notions of the theory of partial differential equations and clarifies this issue. 
    \item[2.] In relativistic field theories, we find that the characteristic surfaces are null surfaces. In particular, it can be shown that Maxwell's equations admit unique solutions (up to gauge transformations) when the initial value surfaces are spacelike. However, timelike and null surfaces are ``bad'', or characteristic, surfaces and therefore do not allow us to determine the electromagnetic field in the rest of spacetime. This is also shown in Appendix~\ref{App:A2}. Furthermore, it is shown in Appendix~\ref{App:A3} that what remains undetermined by the field equations are precisely the radiative modes. Once the radiative modes are known, it is possible to solve the equations. This is precisely what we found here: Once $\Phi^\circ_2$ is given, it is possible to solve the equations. Physically, this is akin of saying ``once we know what the radiation field is doing, we can determine what all the charges and the other fields in the spacetime are doing''.
    \item[3.] The fact that we can not determine a solution to Maxwell's equations from the equations~\eqref{eq:FourIntrinsic}, or, more generally, from the point of view of null surfaces, does \textit{not} mean that these equations are not solved by solutions to Maxwell's equations. Put differently, we can always perform the $3+1$ decomposition with respect to a spacelike initial value surface, solve Maxwell's equations, and then plug these solutions into Maxwell's equations decomposed with respect to a null initial value surface. In either case, the equations will be satisfied. What we cannot do, however, is \textit{solve} Maxwell's equations when we decompose them with respect to a null initial value surface.
\end{itemize}

Let us summarize the situation thus far: Firstly, the Newman-Penrose formalism neatly separated the components of the Maxwell $2$-form which asymptotically fall off like $\frac{1}{r^3}$, $\frac{1}{r^2}$, and $\frac{1}{r}$ to leading order. Secondly, when we analyze Maxwell's equations using the Newman-Penrose formalism, which introduces a decomposition with respect to a null surface, we find that $\Phi^\circ_2$ is freely specifiable, i.e., it is not determined by Maxwell's equations. Hence, the theory of partial differential equations tells us that $\Phi^\circ_2$ has to represent the radiative modes. This is further corroborated by the fact that $\hat\Phi_2$ asymptotically falls off like $\frac{1}{r}$. We call $\Phi^\circ_2$ the \textbf{radiation field}. Furthermore, because $\Phi^\circ_1$ falls off like $\frac{1}{r^2}$ and because its dynamical behavior is only determined once the radiation field $\Phi^\circ_2$ has been determined, we call $\Phi^\circ_1$ the \textbf{coulombic degrees of freedom}.

This is a rather intuitive picture and it reproduces in parts results which we expect from the well-known multipole expansion of standard electrodynamics (though it must be stressed that the language used here is more flexible and it can be generalized to the context of gravitational waves). However, can we see the physics more directly? What can we compute once we know the radiative and coulombic degrees of freedom?\bigskip

\subsection{Flux of Energy-Momentum carried by Electromagnetic Waves}
\begin{figure}[htb!]
	\centering
	\includegraphics[width=0.5\linewidth]{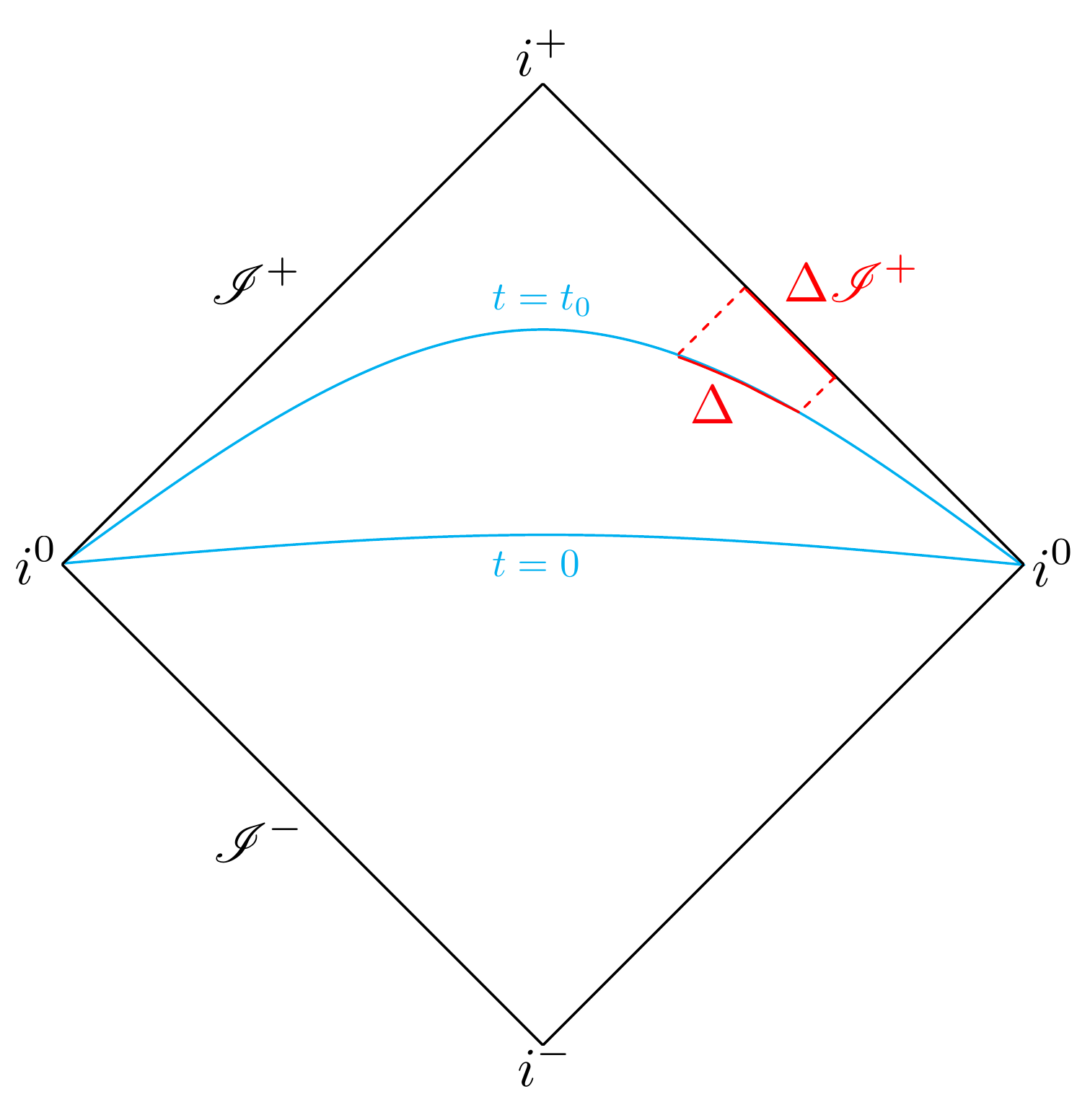}
	\caption{\textit{We consider a family of spacelike hypersurfaces, characterized by $t=\textsf{const}.$, and then compute the flux of energy and momentum through an element $\Delta$ of these surfaces. There is no problem with taking the limit of these surfaces to $\scrip$. Those ultimately allows us to compute the flux of energy and momentum through the element $\Delta\scrip$ or even the total flux through all of $\scrip$.}}
	\label{fig:EnergyMomentumFlux}
\end{figure}

We now wish to determine the flux of energy and momentum carried by electromagnetic waves through some area element. To that end, we consider $t=\text{const.}$ slices (cf. Figure~\ref{fig:EnergyMomentumFlux}) and a finite volume element~$\Delta$ on the $t=t_0$ slice. Then we take the limit to $\scrip$ in order to determine what an observer at~$\scrip$ would measure. If we take $\Delta$ to represent the whole $t=t_0$ surface and extend it to $\scrip$, we get the total amount of energy and momentum carried to $\scrip$ by electromagnetic waves.

In the physical spacetime $(\hatM, \hat\eta_{ab})$, the flux of energy-momentum through $\Delta$ is simply defined as
\begin{equation}\label{DefEnergyMomentumFlux}
	(\P\cdot k)(\Delta) := \int_\Delta \hat T_{ab} \,\hat k^{a}\, \hat \tau^b\,\dd^3 v,
\end{equation}
where $\hat k^{a}$ is the timelike Killing vector field of spacetime translations and $\hat \tau^b$ is the unit normal vector to the surface $\Delta$ (while $\dd^3v$ represents its volume form). Clearly, if $\hat k^{a}$ is the Killing vector field of time translations, $(\P\cdot k)(\Delta)$ simply measures the flux of energy through $\Delta$, and if $\hat k^{a}$ is the Killing vector field of spatial translations, $(\P\cdot k)(\Delta)$ measures the flux of momentum through $\Delta$.

Let us now take the limit of~\eqref{DefEnergyMomentumFlux} to $\scrip$. To that end, we need the results derived in Chapter~\ref{Chap1}. In particular we need that the limit to $\scrip$ of the Killing vector field $t^{a}\partial_a = \PD{}{t}$ is given by $\tilde n^{a} \equalhat \eta^{ab}\nabla_b\Omega$. In order to determine the flux of momentum, we also need to determine the limits of
\begin{align}
	\hat x^{a} &:= \hat{\eta}^{ab}\nabla_b x, & \hat y^{a} &:=\hat{\eta}^{ab}\nabla_b y, & \hat z^{a} &:= \hat{\eta}^{ab}\nabla_b z,
\end{align}
which are the Killing vector fields of spatial translations, to $\scrip$. This is easily achieved by first expressing the above vectors in the chart $(u, r, \theta, \phi)$, and then replacing $r$ with $\Omega$. Taking the limit $\Omega\to 0$ then yields (see Exercise~\ref{ex:KillingFieldsAtScri})
\begin{equation}\label{eq:TranslationsOnScri}
\boxed{
\begin{aligned}
	\hat x^{a} &\equalhat \sin\theta\, \cos\phi\,  \tilde{n}^{a},& && && && \hat y^{a} &\equalhat \sin\theta\,\sin\phi\, \tilde{n}^{a},& && && && \hat z^{a} &\equalhat \cos\theta\,\tilde{n}^{a}\;.
\end{aligned}
}
\end{equation}
Hence, all Killing vector fields which generate spacetime translations have well-defined and smooth limits to $\scrip$ and the limits are of the form $\alpha\,\tilde{n}^{a}$, where $\alpha$ is either equal to $1$ or one of the first three spherical harmonics ($Y_{\ell,m}$ functions). Notice that it had to be expected that the translational Killing vector fields are all proportional to $\tilde{n}^{a}$. The reasoning is the same as the one presented in Side Note~\ref{sn:TimetranslationBecomingNull}.

Since $\tilde \tau^{a}$ in equation~\eqref{DefEnergyMomentumFlux} is normal to $t=t_0$, its limit is easily seen to be given by $\tilde n^{a}$. Hence, when we take the limit of $\Delta$ to $\scrip$, we obtain
\begin{align}
	(\P\cdot k)(\Delta\scrip) &\equalhat \int_{\Delta\scrip}\alpha\,\sin\theta\, r^2\, \hat T_{ab}\,\tilde n^{a} \tilde{n}^b\,\dd u\,\dd \theta\,\dd \phi,
\end{align} 
where we have used that $\dd^3 v$ on $\scrip$ simply becomes $r^2\sin\theta\,\dd u\, \dd \theta\, \dd \phi$. Of course, the above expression can be further simplified by using $r^2 = \Omega^{-2}$ and by expressing $\hat T_{ab}$ in terms of the Maxwell $2$-form. For the latter one, we obtain
\begin{align}\label{eq:TabMaxwell}
	\hat T_{ab} &= \hat F_{am}\hat F_{bn}\hat{\eta}^{mn}-\frac14\hat \eta_{ab} \hat F_{mn} \hat F_{pq} \hat \eta^{mp}\hat \eta^{nq}\notag\\
	&= \Omega^{2}\left( F_{am}  F_{bn}  \eta^{mn}-\frac14   \eta_{ab}   F_{mn}   F_{pq}  \eta^{mp}  \eta^{nq}\right),
\end{align}
where the second line is obtained by replacing $F_{ab} \to \hat F_{ab}$ (since the Maxwell $2$-form is conformally invariant) and by using the fact that the metric and its inverse scale as $\hat \eta_{ab} = \Omega^{-2} \eta_{ab}$ and $\hat \eta^{ab} = \Omega^{2} \eta^{ab}$, respectively. Observe that the $\Omega^2$ in~\eqref{eq:TabMaxwell} precisely cancels the $r^2=\Omega^{-2}$ from the integration measure. Putting everything together one finally finds
\begin{equation}\label{MaxwellEnergyMomentumFlux}
	(\P\cdot k)(\Delta\scrip) \equalhat \int_{\Delta\scrip} \alpha\, \sin\theta\,\left( F_{am}  F_{bn}  \eta^{mn}-\frac14   \eta_{ab}   F_{mn}   F_{pq}  \eta^{mp}  \eta^{nq}\right)\tilde{n}^{a}\tilde{n}^{b}\,\dd u\, \dd\theta\,\dd\phi.
\end{equation}
Notice that all quantities under the integral are well-defined on $\scrip$. It can now be shown (see Exercise~\ref{ex:EnergyMomentumFlux}) that the above expression for the flux can be expressed more compactly in terms of the leading order Newman-Penrose scalar $\Phi^\circ_2$ as
\begin{equation}\label{NPEnergyMomentumFlux}
	(\P\cdot k)(\Delta\scrip) \equalhat \int_{\Delta\scrip} \alpha\,\left| \Phi_2^\circ\right|^2\sin\theta\, \dd u\, \dd\theta\, \dd \phi.
\end{equation}
This expression clearly shows that there are no coulombic or other contributions to the flux of energy and momentum through a region of $\scrip$. Hence, the above integral only counts the energy and momentum carried by electromagnetic waves through $\Delta\scrip$.

\begin{figure}[htb!]
	\centering
	\includegraphics[width=0.5\linewidth]{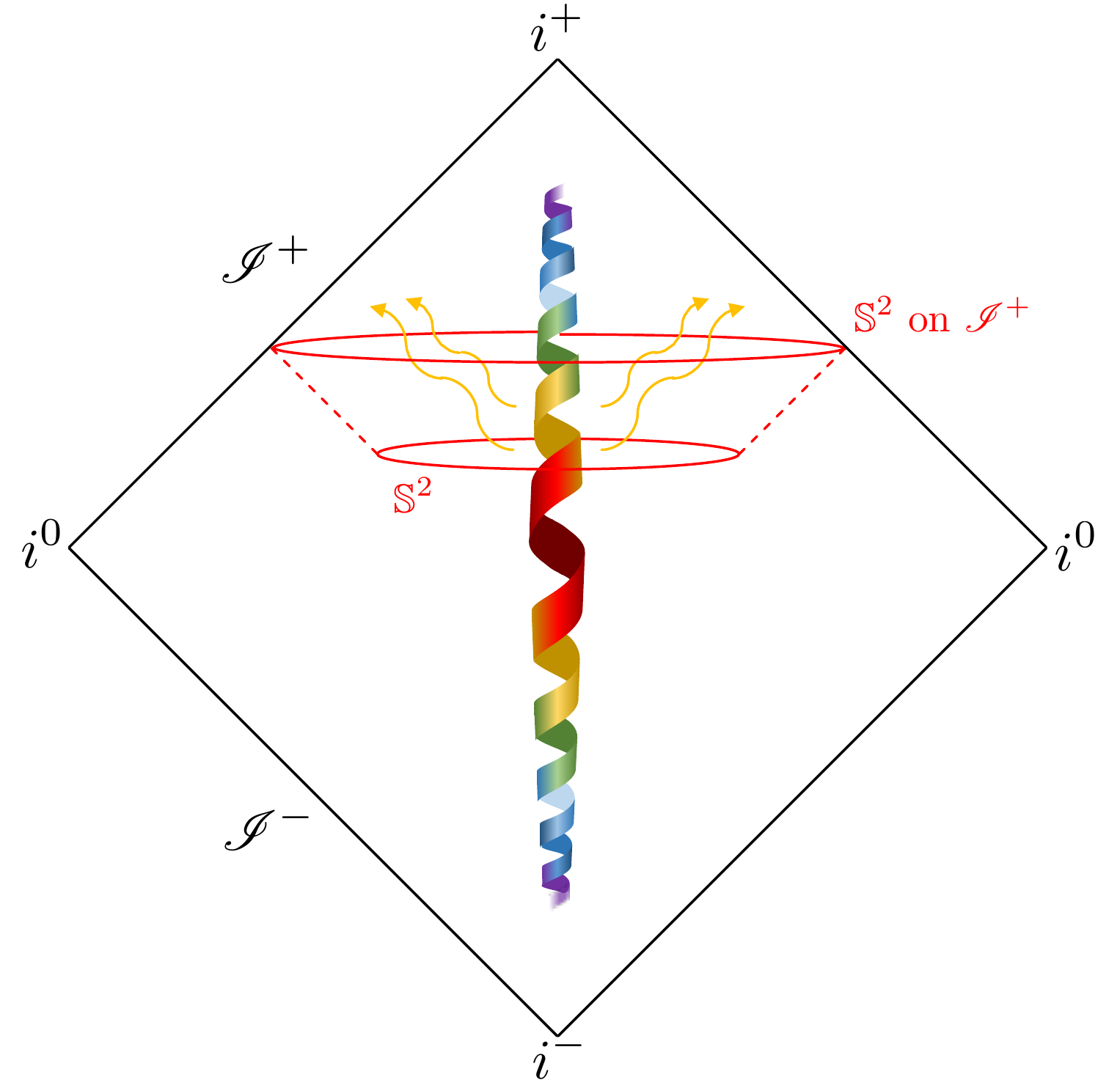}
	\caption{\textit{Any topological $2$-sphere which completely surrounds the sources, here represented by the colorful spiral, can be used for computing the electric and magnetic charges, $Q$ and $\prescript{\star}{}{Q}$, respectively. Because of the independence of $Q$ and $\prescript{\star}{}{Q}$ on the choice of $\bbS^2$, we can always blow up, or project, $\bbS^2$ to $\scrip$. This allows us to define and compute the electric and magnetic charges directly on~$\scrip$, and to show that this coulombic information is indeed captured by $\Phi^\circ_1$.}}
	\label{fig:Charges}
\end{figure}

Is there anything we can say about the coulombic mode? Yes, we can look at the electric charge $Q$. In the physical spacetime, we can define $Q$ in the usual way using Gauss's law,\footnote{See Exercise~\ref{ex:Charges} for a derivation of this expression for $Q$.}
\begin{equation}
	Q := \frac{1}{4\pi}\oint_{\bbS^2}\prescript{\star}{}{\hat F},
\end{equation}
where $\bbS^2$ is any topological $2$-sphere surrounding the sources. Since it does not matter which $2$-sphere we take (as long as it contains all sources), we can take the limit to $\scrip$ (cf. Figure~\ref{fig:Charges}), thus obtaining
\begin{equation}\label{eq:ElectricCharge}
	Q = -\frac{1}{2\pi} \oint_{\bbS^2 \text{ on }\scrip}\Re{\Phi^\circ_1}\sin\theta\,\dd\theta\,\dd\phi,
\end{equation}
which further reinforces the notion that $\Phi^\circ_1$ encodes information about the coulombic modes of the Maxwell field. It is again left as an exercise (see Exercise~\ref{ex:Charges}) to derive~\eqref{eq:ElectricCharge}. If we like, we can also define a magnetic charge, which we denote by $\prescript{\star}{}{Q}$, via the expression
\begin{equation}\label{eq:MagneticCharge}
	\prescript{\star}{}{Q} := \frac{1}{4\pi} \oint_{\bbS^2} \hat F \equalhat -\frac{1}{2\pi} \oint_{\bbS^2 \text{ on } \scrip}\Im{ \Phi^\circ_1}\sin\theta\,\dd\theta\,\dd\phi.
\end{equation}
The last equality is also shown in Exercises~\ref{ex:Charges}. Notice that the magnetic charge is of course zero if we assume that there is a global vector potential, i.e., if there is a $1$-form $A$ such that $F= \dd A$, where $A$ is globally defined and smooth. If this is true, then the first integral is just an integral over $\dd A$, which can be turned into an integral over the boundary of $\bbS^2$ via Stokes' theorem. However, $\partial\bbS^2 = \emptyset$ and therefore the magnetic charge vanishes, as we would expect. 

With this, we conclude our abstract discussion of electromagnetic waves. Recall that the purpose of our discussion was to introduce new concepts which not only describe electromagnetic waves, but also gravitational waves. This is the Newman-Penrose formalism and, more importantly, the conformal completion in terms of $\scri$. The usefulness of the $\scri$-formalism is, qualitatively, that it allows us to follow radiation from its source all the way to infinity, where it can be studied more easily and disentangled from other phenomena which might occur in the bulk of spacetime. In our discussion of gravitational waves, we will make extensive use of $\scri$ and its properties, so it is useful to recall them. This will be the subject of Chapter~\ref{Chap3}.

Before completely closing the discussion on electromagnetic waves, we will have a look at more practical matters. In the next subsection, we will use the Newman-Penrose formalism to study the Coulomb field and the linear dipole antenna.\bigskip

\subsection{Examples of the Newman-Penrose Formalism for Electromagnetism}\label{Chap2:Examples}
So far we have discussed the Newman-Penrose formalism for electromagnetism, albeit on a rather abstract level. Thereby we have seen that the Newman-Penrose scalars $\hat\Phi_i$ obey the Peeling Theorem, and we have seen that physical quantities, such as electric and magnetic charges as well as energy and momentum of the radiation field,\footnote{Notice the distinction: We can only express energy and momentum of the radiative modes in terms of $\Phi^\circ_2$. We have not derived an expression for the energy and momentum of the ``full'' electromagnetic field.} can be expressed in terms of the leading order functions $\Phi^\circ_1$ and $\Phi^\circ_2$, respectively.

In this subsection, we will actually apply this formalism to simple examples which we know and understand very well. This has the purpose to further familiarize ourselves with the formalism and to learn how to actually apply it to physical problems.

Specifically, we will consider the Coulomb field and the linear dipole antenna. We recall that one of the virtues of the Newman-Penrose formalism is that it is able to tell us, given some electromagnetic field as input, whether or not the field contain radiative modes. These modes are encoded in $\Phi^\circ_2$. Hence, we will explicitly check whether $\Phi^\circ_2$ is zero or non-zero for the two examples mentioned above. We will also compute the other Newman-Penrose scalars and explicitly check whether they satisfy the Peeling Theorem and the Maxwell equations~\eqref{eq:FourIntrinsic} on $\scrip$. Then, we will compute the electric and magnetic charges as well as the energy and momentum of the radiation field. At the end of the day, we will see that the Newman-Penrose formalism, quite reassuringly, reproduces precisely the results we would expect from classical electrodynamics.\bigskip

\subsubsection{The Coulomb Field}
Our goal is to compute the physical Newman-Penrose scalars $\hat\Phi_i$ for the Coulomb field generated by a charge $q$. To that end, we work in units where $4\pi\epsilon_0 = 1$ and we use the vector space basis $\{\hat t, \hat r, \hat \theta, \hat\phi\}$. This last remark is actually important, since the Coulomb field in spherical coordinates with respect to the vector space basis $\{\hat t, \hat x, \hat y, \hat z\}$ is given by
\begin{equation}
    \vec{E} = \frac{q}{r^2} \sin\theta\cos\phi \,\hat x + \frac{q}{r^2} \sin\theta\sin\phi\, \hat y + \frac{q}{r^2}\cos\theta\,\hat z = \frac{q}{r^2}
    \begin{pmatrix}
        \sin\theta\cos\phi\\
        \sin\theta\sin\phi\\
        \cos\theta
    \end{pmatrix},
\end{equation}
whereas with respect to the other basis mentioned above, we simply have
\begin{equation}
    \vec{E} = \frac{q}{r^2} \, \hat{r} + 0\, \hat\theta+ 0\,\hat\phi = \begin{pmatrix}
        \frac{q}{r^2}\\
        0\\
        0
    \end{pmatrix}.
\end{equation}
In order to compute the Newman-Penrose scalars, we need the Maxwell $2$-form and the physical Newman-Penrose tetrad. The former is simply given by
\begin{equation}
    \hat F_{ab} = 
    \begin{pmatrix}
        0 & \frac{q}{r^2} & 0 & 0\\
        -\frac{q}{r^2} & 0 & 0 & 0\\
        0 & 0 & 0 & 0\\
        0 & 0 & 0 & 0
    \end{pmatrix},
\end{equation}
while the Newman-Penrose tetrad (in the basis $\{\hat t, \hat r, \hat \theta, \hat\phi\}$) can be written in vector notation as
\begin{align}\label{eq:NPinLightCone}
    \hat \ell^{a} &= \left(-\frac{1}{\sqrt{2}}, \frac{1}{\sqrt{2}},0,0\right)^\transpose\notag\\
    \hat n^{a} &= \left(-\frac{1}{\sqrt{2}},-\frac{1}{\sqrt{2}},0,0\right)^\transpose\notag\\
    \hat m^{a} &= \left(0,0,\frac{1}{\sqrt{2}\, r}, \frac{i}{\sqrt{2}\,\sin\theta\, r}\right)^\transpose.
\end{align}
These are all the ingredients we need to compute the physical Newman-Penrose scalars. We find
\begin{align}
    \hat\Phi_0 &:= \hat F_{ab} \hat m^{a} \hat\ell^{b} = 0\notag\\
    \hat\Phi_1 &:= \frac12 \hat F_{ab} \left(\hat n^{a}\hat\ell^{b} + \hat{m}^{a} \hat{\bar{m}}^b\right) = -\frac{q}{2r^2}\notag\\
    \hat\Phi_2 &:= \hat F_{ab}\hat{n}^{a} \hat{\bar{m}}^{b} = 0.
\end{align}
First of all, we make the trivial observation that the physical scalars have the correct $r$-behavior, as predicted by the Peeling Theorem. Namely, we find that
\begin{align}
    \hat\Phi_0 &= \frac{\Phi^\circ_0}{r^3} + \O(r^{-4}) & \text{with} & & \Phi^\circ_0 &= 0\notag\\
    \hat\Phi_1 &= \frac{\Phi^\circ_1}{r^2} + \O(r^{-3}) & \text{with} & & \Phi^\circ_1 &= -\frac{q}{2}\notag\\
    \hat\Phi_2 &= \frac{\Phi^\circ_2}{r} + \O(r^{-2}) & \text{with} & & \Phi^\circ_2 &= 0.
\end{align}
It is also reassuring that $\hat\Phi_2$ is zero, which implies, as expected, that there is no radiation. This is also confirmed by equation~\eqref{NPEnergyMomentumFlux}, which tells us that the flux of energy and momentum carried through any portion of $\scrip$ is zero (because $\Phi^\circ_2 = 0$). However, $\Phi^\circ_1$ is not zero and we can therefore use this scalar to compute the electric and magnetic charges. Since $\Phi^\circ_1$ is real, the magnetic charge vanishes, according to~\eqref{eq:MagneticCharge}, and this is of course precisely as it should be. For the electric charge, on the other hand, we find
\begin{equation}
    Q = - \frac{1}{2\pi}\int_{\bbS^2 \textsf{ on }\scrip} \Re{\Phi^\circ_1} \,\sin\theta\,\dd\theta\dd\phi = \frac{q}{4\pi}\int_{0}^{\pi}\sin\theta\,\dd\theta\int_0^{2\pi}\dd\phi = q,
\end{equation}
which is precisely the expected result! Finally, we note that Maxwell's equations on $\scrip$, i.e., equations~\eqref{eq:FourIntrinsic}, are trivially satisfied. To summarize: We have investigated the Coulomb field using the Newman-Penrose formalism and we have found that
\begin{itemize}
    \item[a)] The peeling properties are satisfied;
    \item[b)] Maxwell's equations~\eqref{eq:FourIntrinsic} are satisfied;
    \item[c)] The electric charge is given by $q$, while the magnetic charge vanishes;
    \item[d)] There is no radiation and the flux of energy and momentum through any portion of $\scrip$ vanishes.
\end{itemize}
Next, we consider the slightly less trivial example of a linear dipole antenna, which provides us with the simplest\footnote{Strictly speaking, the simplest example would be electromagnetic plane waves. However, such waves only exist in media (like waveguides, for instance) and do not describe the behavior of electromagnetic waves travelling through empty space. In fact, plane waves do not possess the typical $\sim\frac{1}{r}$ behavior which would make them decay as they move away from the source. Consequently, plane waves are not described by the Newman-Penrose formalism.} example of a radiation field.\bigskip

\subsubsection{The Linear Dipole Antenna}
\begin{figure}[htb!]
	\centering
	\includegraphics[width=0.4\linewidth]{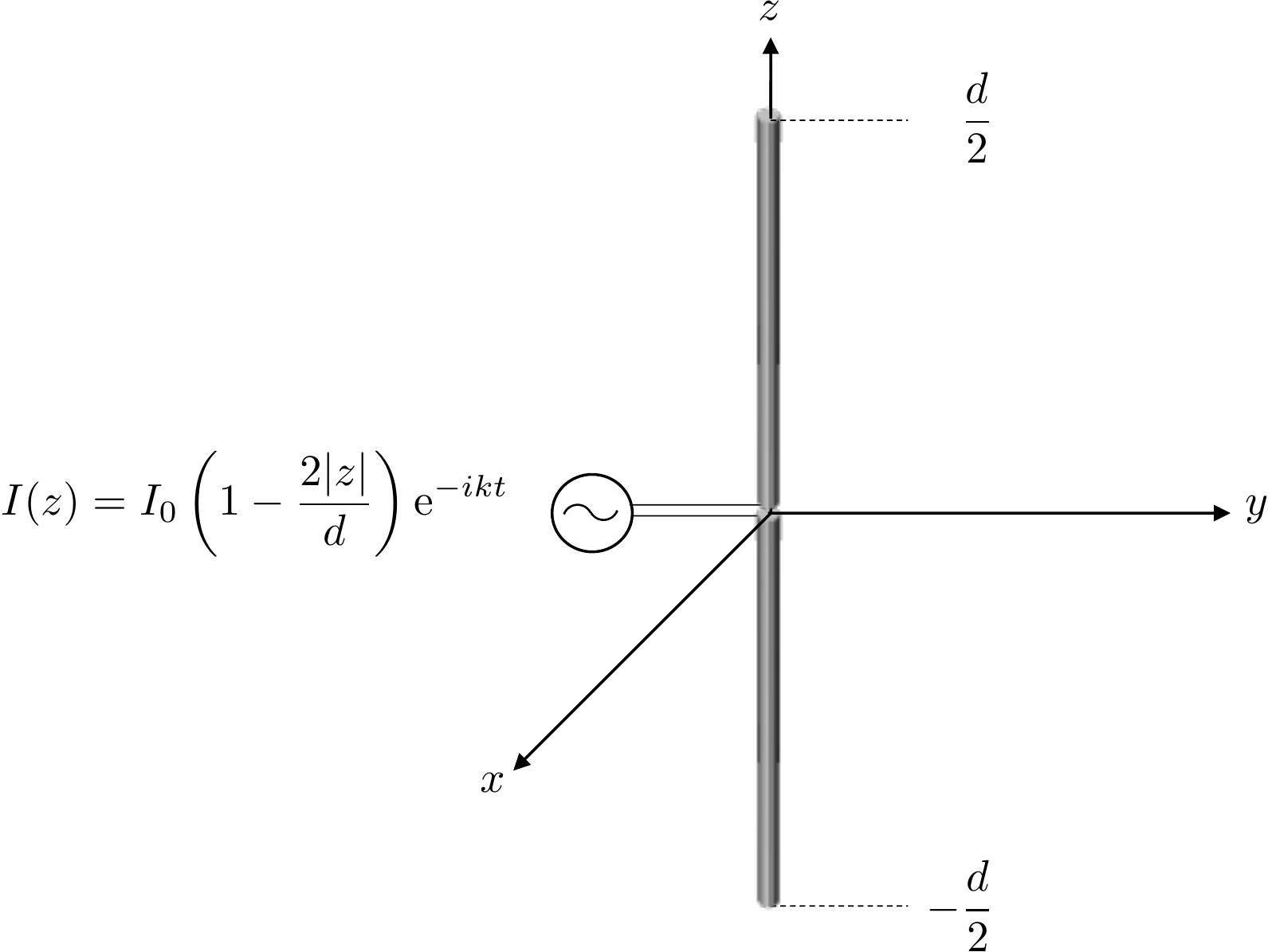}
	\caption{\textit{A linear dipole antenna consisting of two metallic rods, each of length $\frac{d}{2}$, aligned with the $z$-axis of our Cartesian coordinate system and fed by a sinusoidally varying current with maximum value $I_0$ and angular frequency $\omega$.}}
	\label{fig:LinearDipoleAntenna}
\end{figure}

A linear dipole antenna, such as the one shown in Figure~\ref{fig:LinearDipoleAntenna},  simply consists of two metallic rods separated by a gap, which are fed by an oscillating current. We assume that the rods are aligned with the $z$-axis of our Cartesian coordinate system, such that the gap is at the origin. Each rod has a length $\frac{d}{2}$ and the current is assumed to vary sinusoidally with angular frequency $\omega$. The maximum value of the current shall be $I_0$. We will, as it is commonly done in basic electrodynamics, work with a complex current of the form (see for instance~\cite{JacksonBook} chapter 9.2)
\begin{equation}
    I(t, z) = I_0 \left(1-\frac{2|z|}{d}\right) \e^{-i\omega t},
\end{equation}
and we have to remember to take the real part at the end of computations in order to get the physical result. In particular, one can then show (see again~\cite{JacksonBook}) that the vector potential for this antenna is complex and given by
\begin{equation}\label{eq:VecPot}
    \vec{A}(t,\vec{x}) = \begin{pmatrix}
    0 \\
    0 \\
    \frac{d I_0}{8\pi} \frac{\e^{i\, \omega\left(r-t\right)}}{r}
    \end{pmatrix},
\end{equation}
with $r:=\sqrt{x^2+y^2+z^2}$, since we are working in Cartesian coordinates. 

Our goal is the following: We will use the vector potential~\eqref{eq:VecPot} to compute the Maxwell $2$-form, from which  we will derive the Newman-Penrose scalars. We will then show that the Peeling properties and Maxwell's equations are satisfied, that there is no net electric charge in the spacetime, and that there is a radiation field. Moreover, we will show that the total power radiated away to infinity (computed from~\eqref{NPEnergyMomentumFlux}) is exactly equal to the expression found in~\cite{JacksonBook}.

In order to do all that, we need to express the vector potential in spherical coordinates and the resulting Maxwell $2$-form in outgoing Eddington-Finkelstein coordinates. Only then can we meaningfully take the limit to $\scrip$ in order to check the Peeling properties, compute the net charge, and determine the flux of energy and momentum.

The first task is easy to achieve. After computing $\hat F_{ab}$ in Cartesian coordinates from $A^{a} = (0, \vec{A})^\transpose$, with $\vec{A}$ given by~\eqref{eq:VecPot}, we can transform the $2$-form to outgoing Eddington-Finkelstein coordinates $(u, r, \theta, \phi)$ and we obtain 
\begin{equation}\label{eq:FabSpherical}
    \hat F_{ab} = \frac{d\, I_0}{4\pi \omega r^3}\e^{- i \omega u} 
    \begin{pmatrix}
        0 & -\left(i+ \omega r\right)\cos\theta & \frac{i}{2}r\,\left((\omega r+i)^2 - i\, \omega r\right)\, \sin\theta  & 0\\
        \left(i+ \omega r\right)\cos\theta & 0 & -\frac{i}{2}r\sin\theta & 0\\
        -\frac{i}{2}r\,\left((\omega r+i)^2 - i\, \omega r\right)\,\sin\theta & \frac{i}{2}r\sin\theta & 0 & 0\\
        0 & 0 & 0 & 0
    \end{pmatrix}.
\end{equation}
As mentioned above, the vector potential is complex and we need to take the real part of $\hat F_{ab}$ before we can compute physical quantities. Taking the real part of~\eqref{eq:FabSpherical} results in the expression
\begin{align}\label{eq:RealF}
    & \hat F_{ab}= \frac{d\, I_0}{4\pi \omega r^3} 
    \begin{pmatrix}
        0 & -\Sigma\cos\theta & -\frac{r}{2} \left(\Sigma-\omega^2 r^2\sin(\omega u)\right)\sin\theta & 0\\
        \Sigma\cos\theta & 0 & -\frac{r}{2}\sin(\omega u)\sin\theta & 0\\
        \frac{r}{2} \left(\Sigma-\omega^2 r^2\sin(\omega u)\right)\sin\theta & \frac{r}{2}\sin(\omega u)\sin\theta & 0 & 0\\
        0 & 0 & 0 & 0
    \end{pmatrix},
\end{align}
where we have introduced $\Sigma := \omega r\cos(\omega u) + \sin(\omega u)$. For the Newman-Penrose tetrad in outgoing Eddington-Finkelstein coordinates, one finds
\begin{align}\label{eq:TetradinEDCoord}
    \hat\ell^{a} &= \left(0,\frac{1}{\sqrt{2}},0,0\right)^\transpose\notag\\
    \hat{n}^{a} &= \left(\sqrt{2},-\frac{1}{\sqrt{2}},0,0\right)^\transpose\notag\\
    \hat{m}^{a} &= \left(0,0,\frac{1}{\sqrt{2}\, r}, \frac{i}{\sqrt{2}\, r\,\sin\theta r}\right)^\transpose.
\end{align}
Now we are in a position to compute the physical Newman-Penrose scalars. Using~\eqref{eq:RealF} and~\eqref{eq:TetradinEDCoord}, we find
\begin{align}
    \hat\Phi_0 &= \frac{d\, I_0}{16\pi \omega r^3}\sin(\omega u)\sin\theta\notag\\
    \hat\Phi_1 &= -\frac{d\, I_0}{8\pi r^2}\cos(\omega u)\cos\theta - \frac{d\, I_0}{8\pi \omega r^3}\sin(\omega u)\cos\theta\notag\\
    \hat\Phi_2 &= \frac{d\, I_0 \omega}{8\pi r}\sin(\omega u)\sin\theta - \frac{d\, I_0}{8\pi r^2}\cos(\omega u)\sin\theta - \frac{d\, I_0}{16 \pi \omega r^3}\sin(\omega u)\sin\theta.
\end{align}
We immediately observe that $\hat\Phi_2\neq 0$, which means there is radiation, just as it should! Let us now check whether the Peeling Theorem is satisfied. For the leading order contributions to the Newman-Penrose scalars, we find
\begin{align}
    \hat\Phi_0 &= \frac{\Phi^\circ_0}{r^3} + \O(r^{-4}) & \text{with} && \Phi^\circ_0 &= \frac{d\, I_0}{16\pi \omega}\sin(\omega u)\sin\theta \notag\\
    \hat\Phi_1 &= \frac{\Phi^\circ_1}{r^2} +\O(r^{-3}) &\text{with} && \Phi^\circ_1 &= -\frac{d\, I_0}{8\pi}\cos(\omega  u)\cos\theta\notag\\
    \hat\Phi_2 &= \frac{\Phi^\circ_2}{r} +\O(r^{-2}) &\text{with} &&  \Phi^\circ_2 &= \frac{d\, I_0 \omega}{8\pi}\sin(\omega u)\sin\theta.
\end{align}
In other words, the Peeling Theorem is satisfied! Moreover, all three leading order terms are non-vanishing and it is easy to check that they satisfy Maxwell's equations~\eqref{eq:FourIntrinsic}.

Next, we consider the electric and magnetic charges. Since $\Phi^\circ_1$ is real, the magnetic charge vanishes trivially, while for the electric charge we need to use~\eqref{eq:ElectricCharge} to find
\begin{align}
    Q &= -\frac{1}{2\pi}\int_0^{2\pi}\dd\phi\int_{0}^{\pi}\dd\theta\,\sin\theta\,\Re{\Phi^\circ_1}\notag\\
    &= \frac{d\, I_0}{8\pi}\cos(k u)\underset{=0}{\underbrace{\int_{0}^{\pi}\dd\theta\,\sin\theta\,\cos\theta}} = 0.
\end{align}
Hence, there is no net electric charge, just as had to be expected! Finally, we compute the power radiated to $\scrip$, using~\eqref{NPEnergyMomentumFlux} with $\alpha = 1$ and averaged over a period $T = \frac{2\pi}{\omega}$. We denote this average by $\left\langle P_\textsf{rad}\right\rangle$ and obtain
\begin{align}
   \left\langle P_\textsf{rad}\right\rangle &= \frac{\omega}{2\pi}\int_0^{\frac{2\pi}{\omega}}\dd u\int_{0}^{2\pi}\dd\phi\int_0^\pi\dd\theta\,\sin\theta\,|\Phi^\circ_2|^2\notag\\
   &= \frac{d^2\, I^2_0 \omega^3}{64\pi^2}\underset{=\frac{\pi}{\omega}}{\underbrace{\int_{0}^{\frac{2\pi}{\omega}}\dd u\, \sin^2(\omega u)}}\,\,\underset{=\frac{4}{3}}{\underbrace{\int_0^\pi\dd\theta\,\sin^3\theta}} = \frac{d^2\, I^2_0 \omega^2}{48\pi}.
\end{align}
This is precisely the result found in~\cite{JacksonBook} for the total power radiated by a linear dipole antenna! This completes our investigation of this simple physical system. Let us summarize:
\begin{itemize}
    \item[a)] The peeling properties are satisfied;
    \item[b)] Maxwell's equations are satisfied;
    \item[c)] The magnetic charge vanishes trivially and there is no net electric charge;
    \item[d)] There is radiation and the total power radiated to $\scrip$, as computed from~\eqref{NPEnergyMomentumFlux}, precisely reproduces the formula found in~\cite{JacksonBook}. 
\end{itemize}

\newpage
\subsection{Exercises}
\renewcommand{\thesection}{\arabic{section}}

\begin{Exercise}[]\label{ex:KillingFieldsAtScri}
Show that the Killing vector fields of spatial translation on Minkowski space, namely
\begin{align*}
    \hat x^{a} &:= \hat{\eta}^{ab}\nabla_b x, & \hat y^{a} &:=\hat{\eta}^{ab}\nabla_b y, & \hat z^{a} &:= \hat{\eta}^{ab}\nabla_b z,
\end{align*}
have well-defined limits to $\scrip$. Show that the limits are given by equation~\eqref{eq:TranslationsOnScri}.\bigskip

\textit{Hint: Use the definitions of $x,y,z$ in spherical coordinates and take the limit  $\Omega\to 0$. Use the retarded time coordinate $u$.}
\end{Exercise}

\begin{Exercise}[]\label{ex:EnergyMomentumFlux}
Show that the flux of energy and momentum, described by equation~\eqref{MaxwellEnergyMomentumFlux} in the main text, can be written as~\eqref{NPEnergyMomentumFlux}.
\end{Exercise}

\begin{Exercise}[]\label{ex:2form}
Show that $\prescript{2}{}{\hat{\varepsilon}}:= i\, \hat{m}\wedge\hat{\bar{m}}$, where $\hat{m}:= \hat{m}_a\,\dd x^{a}$, is equal to the standard area element of a $2$-sphere of radius $r$. That is, show that 
\begin{align*}
	\prescript{2}{}{\hat{\varepsilon}} = r^2\,\sin\theta \, \dd \theta\wedge\dd \phi.
\end{align*}
Compute also the Hodge dual of the area element,~$\prescript{\star\,2}{}{\hat{\varepsilon}}$.\bigskip

\textit{Hint: The components of the dual $\prescript{\star}{}{\hat\varepsilon}$ are given by $\left(\prescript{\star}{}{\hat\varepsilon}\right)_{ab} = \frac12 \epsilon_{abcd}\hat\varepsilon^{cd}$.}
\end{Exercise}

\begin{Exercise}[]\label{ex:Maxwell2FormInNPForm}
Show that the Maxwell $2$-form $F:= F_{ab}\,\dd x^{a}\wedge \dd x^{b}$ can be written as
\begin{equation*}
	F = \Phi_0\, n\wedge \bar{m} + \bar{\Phi}_0\,n\wedge m+2\left(\Re{\Phi_1}\,n\wedge \ell - i\Im{\Phi_1}\,m\wedge \bar{m}\right) -\Phi_2\, \ell\wedge m -\bar{\Phi}_2\, \ell\wedge \bar{m},
\end{equation*}
where the $1$-forms $\ell$, $n$ and $m$ are defined as $\ell := \ell_{a}\,\dd x^{a}$, $n := n_{a}\, \dd x^{a}$, and $m:=m_{a}\, \dd x^{a}$. Furthermore, show that $F$ is real despite being expressed in terms of complex functions and a complex null tetrad.
\end{Exercise}

\begin{Exercise}[]\label{ex:Charges}
Show that the electric and magnetic charges can be written as 
\begin{align*}
	Q &= \frac{1}{4\pi} \oint_{\bbS^2} \prescript{\star}{}{F} \equalhat -\frac{1}{2\pi} \oint_{\bbS^2 \text{ on } \scrip}\Re{{\Phi}^\circ_1}\sin\theta\,\dd\theta\,\dd\phi \\
	\prescript{\star}{}{Q} &= \frac{1}{4\pi} \oint_{\bbS^2} \prescript{\textcolor{white}{\star}}{}{F} \equalhat -\frac{1}{2\pi} \oint_{\bbS^2 \text{ on } \scrip}\Im{{\Phi}^\circ_1}\sin\theta\,\dd\theta\,\dd\phi.
\end{align*}
Start with proving the first equality of both equations and, subsequently, prove the second one. 
\end{Exercise}

\begin{Exercise}[]\label{ex:FIdentity}
Starting from the Maxwell $2$-form given in Exercise~\ref{ex:Maxwell2FormInNPForm}, compute the Hodge dual $\prescript{\star}{}{F}$ and show that $\Phi_1 = \frac12\left(F-i\,\prescript{\star}{}{F}\right)_{ab}m^{a}\bar{m}^{b}$.\bigskip

\textit{Hint: The components of $\prescript{\star}{}{F}$ are given by $\left(\prescript{\star}{}{F}\right)_{ab} = \frac12 \epsilon_{abcd}F^{cd}$.}
\end{Exercise}

\newpage

\asection{3}{Properties of Asymptotically Minkowski Spacetimes}\label{Chap3}
In this chapter we take the first step toward defining and studying gravitational waves in full, non-linear GR. Contrary to our discussion of electromagnetic radiation, we can no longer assume that all of spacetime is adequately modeled by Minkowski space. Rather, we consider curved spacetimes which contain sources, i.e., a non-trivial energy-momentum tensor $\hat T_{ab}$, and which are approximately Minkowskian far away from those sources.

The latter is a reasonable physical assumption, since ultimately we wish to study the emission of gravitational radiation by the coalescence and merger of compact binaries. In what follows, we will make the idea of asymptotically Minkowski spacetimes mathematically precise. Subsequently, we will demonstrate how to construct a Newman-Penrose null tetrad for curved spacetimes. This is slightly more involved than in Minkowski space, because we can no longer rely on global symmetries and globally well-defined coordinate systems. 

We then study the Riemann tensor for asymptotically Minkowski spacetimes and introduce the Newman-Penrose scalars for GR. In the final subsection of this chapter, we discuss a mathematical theorem concerning these scalars, which will then enable us to prove the Peeling Theorem for GR in Chapter~\ref{Chap4}.\bigskip

\subsection{Asymptotically Minkowski Spacetimes and their Geometric Properties}
As already alluded to at the end of Chapter~\ref{Chap2}, the concept of $\scrip$ allows us to follow radiation along its null direction from the source all the way out to infinity --- in a precise mathematical sense. We will see in later chapters, that in this asymptotic region the radiative modes disentangle from other modes, thus making it easier to study their properties. Naturally, most of our discussion about gravitational waves will take place on $\scrip$. However, in order to have this discussion, we first need to properly define the asymptotic region. Just as before, we will make use of the concept of conformal completion, because it brings ``infinity'' or ``asymptotically far away regions'' to a finite distance and it allows us to study these regions with tools of differential geometry. The first concept we introduce is the one of asymptotic flatness.\bigskip

\textbf{Definition 3.1: Asymptotic flatness}\\
A physical spacetime $(\hatM, \hat{g}_{ab})$ which satisfies Einstein's field equations with vanishing cosmological constant, $\hat{R}_{ab}-\frac12 \hat{R}\,\hat{g}_{ab} = 8\pi\, \hat{T}_{ab}$, is said to be \textbf{asymptotically flat} if
\begin{itemize}
	\item[1)] There exists a conformal completion $(M, g_{ab}, \Omega)$ such that $\M:= \M\cup\scri$ is a manifold with a boundary and the boundary has the topology $\scri\simeq \bbS^2\times \bbR$. Moreover, the conformally rescaled metric and the physical metric are related by $g_{ab} = \Omega^2\,\hat{g}_{ab}$. The conformal factor is assumed to satisfy $\Omega\equalhat 0$ and $\nabla_a\Omega \unequalhat 0$.
	\item[2)] $\Omega^{-2}\hat{T}_{ab}$ has a smooth limit to $\scri$. 
\end{itemize}

Let us briefly pause here and paint a heuristic picture. First of all, the conditions $\Omega\equalhat 0$ and $\nabla_a \Omega\unequalhat 0$ tell us that $\Omega$ is a good coordinate near $\scri$, that $\scri$ has a well-defined normal $n^{a} := \left.\nabla^a\Omega\right|_{\Omega=0}$, and that $\Omega$ is heuristically the same as $\frac{1}{r}$. Of course, in general we do not know what ``$r$'' is, but at least for simple spacetimes, such as Minkowski or Schwarzschild, this makes sense. Also, if for some reason we would choose $\Omega=\frac{1}{r^2}$ to conformally complete these spacetimes, we would get $\nabla_a\Omega \equalhat 0$, in violation of the condition spelled out in our definition of asymptotic flatness. This reinforces the interpretation of $\Omega$ being qualitatively the same as $\frac{1}{r}$.

Secondly, the condition that $\Omega^{-2}\,\hat{T}_{ab}$ has a smooth limit to $\scri$ heuristically tells us that $\hat{T}_{ab}$ falls-off at a certain rate. Namely it falls-off \textit{at least} like $\frac{1}{r^2}$. One finds that this is a condition which is satisfied by all reasonable sources.

The above definition of asymptotic flatness can be extended to describe asymptotically Minkowski spacetimes. If we look more closely at the definition of asymptotic flatness, we notice that we only say that a boundary $\scri$ has to exist for the choice of $\Omega$ we made, but we do not say anything about the ``size'' of $\scri$. It is possible to choose $\Omega$ in such a way that even for Minkowski space, we obtain only a finite portion of null infinity. This can easily be amended by introducing the concept of completeness of $\scri$. \bigskip

\textbf{Definition 3.2: Completeness of $\scri$}\\
We say that $\scri$ is \textbf{complete} if the normal vector field to $\scri$, $n^{a}:=\left.g^{ab}\nabla_b\Omega\right|_{\Omega=0}$, is complete.\bigskip

We recall that a vector field is called complete, if there exists an affine parameter $u$ such that $n^{a}\nabla_a u = 1$, which then implies $u\in(-\infty, +\infty)$. In the case of Minkowski space, it is easy to see that this condition implies that $n^{a}$ generates all of $\scri$ and not just a finite portion of it. However, this definition presents us with a new problem: Which $n^{a}$ should we use? There is no canonical choice since many different choices of conformal factor could lead to a normal vector which generates all of $\scri$. In fact, it is even possible that one person chooses a conformal factor for which $\scri$ is complete but another person might choose a factor for which $\scri$ is not complete. Both choices are acceptable as far as the conformal completion is concerned. Luckily, we can canonicalize our choice of $\Omega$ by choosing a so-called \textbf{divergence-free conformal frame}. Such a frame is defined as follows:\bigskip

\textbf{Definition 3.2: Divergence-free conformal frame}\\
Let $(\hatM, \hat{g}_{ab})$ be a physical spacetime. A \textbf{divergence-free conformal frame} consists of a conformal completion $(\M, g_{ab}, \Omega)$ for which $\nabla_a n^{a} \equalhat 0$. That is, the normal vector $n^{a}$ is divergence-free for the given choice of $\Omega$.\bigskip

It is always possible to choose a divergence-free conformal frame. To see this, we assume we are given a conformal completion with conformal factor $\Omega$ for which $n^{a}$ is \textit{not} divergence free. We can always introduce a new conformal completion $\Omega'$ defined by $\Omega' := \omega\, \Omega$, where $\omega$ is a smooth, nowhere vanishing function. Under such a conformal rescaling, the covariant derivative transforms in a particular way (see for instance appendix D of~\cite{WaldBook}) and one finds for the \textit{new} normal co-vector $n'_a$ and its divergence
\begin{align}\label{eq:Transformations}
	n'_{a} &= \omega\, n_{a} + \Omega\, \nabla_a \omega & \text{and} && \nabla'_a n'^{a} \equalhat \nabla_a n^{a} + 4n^{a} \nabla_a \omega,
\end{align}
where the second equation holds only on $\scri$. If we choose $\omega$ such that it satisfies
\begin{align}\label{eq:DivFreeCond}
	\underset{\equiv n^{a} \nabla_a\omega}{\underbrace{\lie_n \omega}} \equalhat -\frac14 \nabla_a n^{a},
\end{align}
we obtain a divergence-free normal vector $n'^{a}$ in the new conformal frame. Notice that this is a first order differential equation for $\omega$, so we are guaranteed to always find a solution (albeit not a unique one, because we have not chosen any initial-value conditions). It follows that we can always find a divergence-free conformal frame by a suitable conformal rescaling $\Omega \mapsto \Omega'=\omega'\,\Omega$. From now on, we will \textit{always} work in a divergence-free conformal frame. This has the following advantage: Given a divergence-free conformal frame $\Omega$, we can change to a different divergence-free conformal frame $\Omega' := \omega\, \Omega$, provided~$\omega$ is smooth, nowhere vanishing \textit{and} provided it is Lie dragged by $n^{a}$,
\begin{equation}\label{eq:Lie_n_omega}
	\lie_n\omega \equalhat 0.
\end{equation}
It is evident from the second equation in~\eqref{eq:Transformations} that this condition preserves the divergence-freeness and it follows from equation~\eqref{eq:Lie_n_omega} that $\omega = \omega(\theta, \phi)$ on $\scri$. Moreover, if $\scri$ is complete with respect to the divergence-free conformal frame $\Omega$, then it is also complete with respect to the divergence-free conformal frame $\Omega'$. This follows from the first equation in~\eqref{eq:Transformations}, which on $\scri$ reads
\begin{align}
	n'^{a} \equalhat \omega(\theta, \phi)^{-1}\, n^{a}.
\end{align}
Hence, if $n^{a}\nabla_a u = 1$ in the first frame (this was the condition for completeness), then $n'^{a}\nabla'_a u' = 1$ with $u' = \omega(\theta,\phi)\, u$ holds in the second frame, making $n'^{a}$ complete. In other words, the affine parameter $u$, which ensures completeness, is transformed by an angle-dependent function which is nowhere vanishing. Hence, $\scri$ is also complete with respect to the divergence-free conformal frame $\Omega'$. Let us summarize the two results we have obtained so far.
\begin{itemize}
    \item[1)] Given any conformal frame $(\M, g_{ab}, \Omega)$, we can always find a conformal rescaling $\Omega\mapsto\Omega' = \omega\, \Omega$, where $\omega$ is a smooth, nowhere vanishing function on $\scri$, such that $\nabla'_a n'^{a} = 0$ holds in the new frame. That is, \textit{we can always work in a divergence-free conformal frame}.
    \item[2)] Given a divergence-free conformal frame $(\M, g_{ab}, \Omega)$, we have a residual rescaling freedom $\Omega\mapsto\Omega' = \omega\, \Omega$, where $\omega$ is a smooth, nowhere vanishing function on $\scri$, which is Lie dragged by the normal vector, $\lie_n \omega \equalhat 0$. This rescaling freedom preserves the divergence-freeness. Thus, it maps one divergence-free conformal frame to another divergence-free conformal frame.
\end{itemize}
These results allow us to \textit{canonicalize} our choice of conformal frame. We will \textit{always} choose a conformal frame which is divergence-free and with respect to which $\scri$ is complete. If we have found one such frame, we can change to a different frame without losing divergence-freeness nor completeness of $\scri$.

Having cleared up the issue of completeness and choice of frame, we are finally ready to define asymptotically Minkowski spacetimes.\bigskip

\textbf{Definition 3.3: Asymptotically Minkowski spacetimes}\\
An \textbf{asymptotically Minkowski spacetime} is an asymptotically flat spacetime which, in a divergence-free conformal frame, is also complete.\bigskip

In the next subsection, we will use the field equations to derive important properties of this class of spacetimes.\bigskip

\subsection{Physical Properties of Asymptotically Minkowski Spacetimes}\label{subsec:PhysicalPropofasympflat}
Having given a definition of asymptotically Minkowski spacetimes, we wish to extract some consequences. To that end, we will use the field equations to derive a number of results on the nature of $\scrip$ and its intrinsic metric. More precisely, we will see that the field equations imply that $\scrip$ is a null surface, that divergence-freeness implies the stronger equation $\nabla_a n_b \equalhat 0$, and that the intrinsic metric of $\scrip$ is smooth, degenerate, and only depends on two out of three coordinates.

After that, we will see that all relevant information about gravitational waves is encoded in the Weyl tensor and that this tensor has to satisfy a constraint equation at $\scrip$. This constraint forces the Weyl tensor to vanish at $\scrip$, as one might have intuitively\footnote{The vanishing of the Weyl tensor is intuitively clear: Away from sources we have $\hat{R}_{ab} = 0$ and hence $\hat{R}_{abcd} = \hat{C}_{abcd}$. But we also expect that far away from sources, in some asymptotic region, the metric approaches the Minkowski metric and hence $R_{abcd} \to 0$ which then implies $C_{abcd}\to 0$} expected for this class of spacetimes.

\subsubsection{Properties of the Asymptotic Region}
Let us begin by studying the Ricci tensor and the Ricci scalar on the physical as well as on the conformally completed spacetime. To do so, recall that on the physical spacetime we have a torsion-free, metric compatible covariant derivative $\hat\nabla$. On the conformally completed spacetime, we can also introduce a torsion-free, metric-compatible derivative operator $\nabla$. What is the relation between $\hat\nabla$ and $\nabla$? 

From differential geometry we know that the difference between any two covariant derivatives is a tensor,
\begin{equation}
    \left(\nabla_a - \hat\nabla_a\right)\omega_b = C\du{ab}{c}\omega_c,
\end{equation}
where $\omega_b$ is any $1$-form which is defined on both spacetimes. Using the fact that $\hat\nabla$ is the covariant derivative with respect to the Levi-Civita connection of the physical metric $\hat g_{ab}$, one can work out that $C\du{ab}{c}$ is given by (see for instance appendix D of~\cite{WaldBook})
\begin{equation}
    C\du{ab}{c} = -\Omega^{-1}\left(2\hat\nabla_{(a}\Omega\delta\du{b}{c} - \hat\nabla^{c} \Omega g_{ab}\right).
\end{equation}
This knowledge enables us to express the covariant derivative $\nabla$ on the conformally completed spacetime in terms of the physical derivative and the factor $\Omega$. In turn, this allows us to relate the curvature tensors on $(\hatM, \hat g_{ab})$ and $(\M, g_{ab})$ to each other. For now we only need the Ricci tensor and the Ricci scalar, for which one finds (we refer the reader again to~\cite{WaldBook} for details on the derivation)
\begin{align}\label{eq:Riccis}
    \hat R_{ab} &= R_{ab} + 2\Omega^{-1} \nabla_a\nabla_b \Omega + \left[\Omega^{-1}\nabla^c\nabla_c\Omega - 3 \Omega^{-2}\left(\nabla^c\Omega\right)\left(\nabla_c\Omega\right)\right] g_{ab}\notag\\
    \hat R &= \Omega^2\, R + 6\Omega\, \nabla^c\nabla_c\Omega - 12\left(\nabla^c\Omega\right)\left(\nabla_c \Omega\right).
\end{align}
Furthermore, we recall that in an asymptotically Minkowski spacetime, the limit $\lim_{\Omega\to 0}\Omega^{-2}\hat T_{ab}$ is smooth by definition. Qualitatively, this means that the stress-energy tensor falls-off \textit{at least} like $\frac{1}{r^2}$ and it implies that $\lim_{\Omega\to 0}\Omega^{-1}\hat T_{ab} = 0$. Using the field equation
\begin{equation}
	\hat{R}_{ab} - \frac12 \hat{R}\,\hat{g}_{ab} = 8\pi\hat{T}_{ab},
\end{equation} 
we can re-express the physical Ricci scalar and the physical Ricci tensor as
\begin{align}\label{eq:RelatingTandR}
	\hat R &= -8\pi \hat T &\text{and} & & \hat R_{ab} = 8\pi\left(\hat T_{ab}-\hat T\, \hat g_{ab}\right).
\end{align} 
These relations can be used to conclude that the left hand side of equation~\eqref{eq:Riccis} vanishes on $\scrip$. Let us now examine the equation for the Ricci scalar.

The term $\Omega^2\,R$ vanishes at $\scri$ because $\Omega\equalhat 0$ and $R$ is smooth. This leaves us with the second and the third term in the second equation of~\eqref{eq:Riccis}. Since $\nabla_c\Omega \equalhat n_c$, we find that the second term vanishes due to divergence-freeness, which finally leaves us with
\begin{equation}\label{eq:ScriIsNull}
    \boxed{n^{c}n_{c} \equalhat 0}
\end{equation}
Hence, it follows that $\scrip$ is a \textit{null} hypersurface. We emphasize that we have not assumed this property, it is a consequence of the definition of asymptotic flatness and the field equations. Moreover, if the field equations contained a non-vanishing cosmological constant, we would have found that $\scrip$ is either a spacelike or a timelike hypersurface, depending on whether $\Lambda$ is positive or negative, respectively. This is shown in Exercise~\ref{ex:ScriSpacelike}. 

The equation~\eqref{eq:ScriIsNull} also comes in handy when examining the first equation in~\eqref{eq:Riccis}. We first multiply it with $\Omega$ and then find that the first term, $\Omega R_{ab}$, vanishes on $\scrip$. That is because $\Omega\equalhat 0$ and $R_{ab}$ is smooth. This leaves us with
\begin{equation}
    2\nabla_a n_b + \left[\nabla^{c}n_c - 3 \Omega^{-1} n^{c} n_c\right] g_{ab} \equalhat 0,
\end{equation}
where we used $\nabla_a\Omega \equalhat n_a$. The first term in the square brackets vanishes because $n_c$ is divergence-free. For the second term in the bracket we can use $n^{c} n_c \equalhat 0$, but notice that we need to use L'Hospital's rule to properly compute the limit $\Omega^{-1}n^{c}n_c$. Finally, this leaves us with
\begin{equation}\label{eq:NablaN}
    \boxed{\nabla_a n_b \equalhat 0}\;.
\end{equation}
This equation is stronger than the divergence-freeness of $n^{a}$ and it allows us to derive further properties of $\scrip$. To that end, we first define the intrinsic metric $q_{ab}$ of $\scrip$ as
\begin{equation}\label{eq:IntrinsicMetric}
    q_{ab} := \pb{g}_{ab}.
\end{equation}
Because $\scrip$ is null, this metric is degenerate with signature $(0,+,+)$. In particular, this means that $n^{a}$ is an eigenvector of $q_{ab}$ with eigenvalue zero, i.e.,
\begin{equation}
    \boxed{q_{ab}n^{b} \equalhat 0}\;.
\end{equation}
This is shown in Exercise~\ref{ex:qIsDegenerate}. Using~\eqref{eq:NablaN}, we can conclude that $q_{ab}$ is not only degenerate, but that it also only depends on two of the three coordinates on $\scrip$. To see this, we consider the Lie derivative of $g_{ab}$ along $n^{a}$ and pull the expression back to $\scrip$. On the one side, this gives us $\underleftarrow{\lie_n g_{ab}} = \lie_n q_{ab}$. On the other side, $\lie_n g_{ab} = 2\nabla_{(a}n_{b)} \equalhat 0$, where we used~\eqref{eq:NablaN} in the last step. It thus follows that
\begin{equation}
    \boxed{\lie_n q_{ab} \equalhat 0}
\end{equation}
This means that the intrinsic metric is Lie dragged by the normal vector and it is hence possible to find a coordinate system such that $q_{ab}$ only depends on two out of three coordinates. We will introduce such coordinates adapted to $\scrip$ in subsection~\ref{Chap3_NullTetrad}.

\subsubsection{A Constraint Equation for the Weyl Tensor}\label{Chap3_Constraint}
Let us now turn to studying the Weyl tensor. We begin our considerations by first defining the tensor. To that end, we recall that the Riemann tensor can be decomposed into terms constructed from the Ricci tensor and a trace-free tensor $\hat C_{abcd}$, which has the same symmetries as the Riemann tensor:
\begin{equation}\label{DecompositionRiemann}
	\hat R_{abcd} = \hat C_{abcd} + \hat g_{a [c}\hat S_{d]b} - \hat g_{b[c}\hat S_{d]a}.
\end{equation}
The tensors $\hat C_{abcd}$ and $\hat S_{ab}$ are called \textbf{Weyl tensor} and \textbf{Shouten tensor}, respectively. The latter is defined as
\begin{align}
	\hat S_{ab} &:= \hat R_{ab} - \frac16 \hat R\, \hat g_{ab},
\end{align}
i.e., it is constructed from the Ricci tensor and its trace, while the Weyl tensor is implicitly defined by equation~\eqref{DecompositionRiemann}. The motivation for this decomposition is that the field equations of GR relate the Ricci tensor and the Ricci scalar to the stress-energy tensor via equation~\eqref{eq:RelatingTandR}. Hence, we can heuristically think of these two tensors as being given \textit{by} and encoding information \textit{about} the matter content of spacetime. In turn this means that the Shouten tensor is completely determined by the matter content. The Weyl tensor, on the other hand, can be thought of as encoding information about the gravitational field even when there are no matter fields. In fact, since the vacuum field equations of GR are simply given by $\hat R_{ab}=0$, the Shouten tensor vanishes for such field configurations and we are left with
\begin{equation}
	\hat R_{abcd} = \hat C_{abcd}.
\end{equation}
Hence, all information about the gravitational field $g_{ab}$ outside of matter sources is coded in the Weyl tensor. Therefore, it is relevant to further study this tensor to extract information about gravitational waves far away from matter sources.

To proceed in our analysis of the Weyl tensor, it is useful to add the following fact about smooth functions to our toolbox.
\begin{mysidenote}{A Lemma on smooth functions}{Lemma}
    Let $f$ be a function which is smooth in a neighborhood of $\scrip$ and which satisfies $\left.f\right|_{\scrip} = 0$. Its Taylor expansion looks like
\begin{equation}
	f = \left.\PD{f}{\Omega}\right|_{\Omega=0}\Omega + \O(\Omega^2).
\end{equation}
This implies that $\Omega^{-1} f$ is also \textit{smooth} and its limit is given by
\begin{equation}
    \lim_{\Omega\to 0}\Omega^{-1}f = \left.\PD{f}{\Omega}\right|_{\Omega = 0}.
\end{equation}
\end{mysidenote}
Let us now consider the transformation behavior of the Shouten tensor under conformal rescaling (cf. appendix D of~\cite{WaldBook}):
\begin{equation}\label{eq:RescalingShouten}
    \hat S_{ab} = S_{ab} + 2\Omega^{-1}\nabla_a n_b -\Omega^{-2} g_{ab} n_c n^{c},
\end{equation}
where $n_c := \nabla_c\Omega$. We multiply this equation by $\Omega^2$ and take the limit $\Omega\to 0$. Since the Shouten tensor can be expressed in terms of the energy-momentum tensor,
\begin{equation}
    \hat S_{ab} = 8\pi \left(\hat T_{ab} - \frac13 \hat g_{ab} \hat T\right),    
\end{equation}
we can conclude that $\lim_{\Omega\to 0}\Omega^2 \hat S_{ab} = 0$ because $\hat S_{ab}$ is smooth. The smoothness of $\hat S_{ab}$ also implies that the right hand side of equation~\eqref{eq:RescalingShouten} is smooth. This is only possible if every term is smooth by itself. Thus, it follows that $\Omega^2 S_{ab} \to 0$ for $\Omega\to 0$. The second term vanishes simply because $\nabla_a n_b \equalhat 0$, as we know from~\eqref{eq:NablaN}. The only term which is left is $g_{ab}n_c n^{c}$. Of course, we already know that $n^{a}$ is null. But what this computation shows is that $n_c n^{c}$ can be written as
\begin{equation}
    n_c n^{c} = \Omega^2\, \alpha,
\end{equation}
where $\alpha$ is a smooth function. Thus, $n^{a}$ is in general not a null vector in the bulk of spacetime, but it becomes null in the limit $\Omega\to 0$. It  does so, qualitatively, at a rate of at least $\frac{1}{r}$.

Let us now return to equation~\eqref{eq:RescalingShouten}, where we replace $n_c n^{c}$ by the smooth function $\Omega^{2}\, \alpha$. Next, we multiply the whole equation by $\Omega$, take the covariant derivative $\nabla_a$ and anti-symmetrize in the first and second indices. This procedure turns the term $\nabla_b n_c$ into $\nabla_{[a}\nabla_{b]}n_c$, which can be re-expressed in terms of the Riemann tensor. All in all, we find
\begin{equation}
	\nabla_{[a}\Omega\hat S_{b]c} = \nabla_{[a}\Omega S_{b]c} + \underset{R_{abc}{}^d n_d}{\underbrace{2\nabla_{[a}\nabla_{b]}n_c}} - \big(\nabla_{[a}\Omega\,\alpha\big)g_{b]c}.
\end{equation}
Next, we use the decomposition~\eqref{DecompositionRiemann} of the Riemann tensor into the Shouten and Weyl tensors, which leads us to
\begin{equation}
    \nabla_{[a}\Omega\hat S_{b]c} = \nabla_{[a}\Omega S_{b]c} + C_{abcd}n^{d} + g_{a [c} S_{d]b}n^{d} -  g_{b[c} S_{d]a}n^{d} - \big(\nabla_{[a}\Omega\,\alpha\big)g_{b]c}.
\end{equation}
Using what we have learned thus far, we see that every term vanishes when we take the limit $\Omega\to 0$. Only the Weyl tensor remains and we obtain the constraint equation
\begin{equation}\label{eq:WeylConstraint}
	\boxed{C_{abc}{}^d n_d \equalhat 0}
\end{equation}

Let us briefly pause here and compare the situation with the electromagnetic theory described by Maxwell's $2$-form $F_{ab}$. This is the electromagnetic analogue of the Weyl tensor. We can think that way because in both cases, in regions where there are no sources, all the curvature resides in $C_{abcd}$ (spacetime curvature) and $F_{ab}$ (curvature of the connection $A^{a}$), respectively. Moreover, these two fields satisfy structurally similar field equations:
\begin{align}
	 &\text{Electromagnetism}  & &  \text{Gravity}\notag\\
	&\nabla^{a} F_{ab} = 0 & &\nabla^{a} C_{abcd} = 0\notag\\
	&\nabla_{[a}F_{bc]} = 0 & &\nabla_{[a}C_{bc]de} = 0
\end{align}
However, the key difference is that electromagnetism is conformally invariant and $F_{ab}$ is not forced to vanish at $\scrip$, while gravity is \textit{not} a conformally invariant theory and the Weyl tensor has to satisfy a constraint at $\scrip$. This constraint, equation~\eqref{eq:WeylConstraint}, does not imply that the full Weyl tensor vanishes at~$\scrip$. We will elaborate more on this issue in the last subsection of this chapter. Here, we only remark that eight of the ten components are forced to be zero, while the remaining two components are unconstrained. To understand this, it is instructive to consider an analogous situation in electromagnetism (see Exercise~\ref{ex:DecomposingFwrtSigma}). We also point out that the situation is different in the presence of a non-zero cosmological constant (Exercise~\ref{ex:DecomposingFwrtSigma} also helps in understanding this claim).

\begin{mysidenote}{Behavior of the Weyl tensor when $\Lambda\neq 0$}{}
    Let $E_a := F_{ab}n^b$ and $B_a := \prescript{\star}{}{F}_{ab}n^b$ be the electric and magnetic fields. It is shown in Exercise~\ref{ex:DecomposingFwrtSigma} that if $n^{a}$ is a timelike or spacelike vector, then $E_a$ and $B_a$ carry exactly the same information as $F_{ab}$. Moreover, $E_a = 0 = B_a$ implies $F_{ab}=0$. This is no longer true when $n^{a}$ is a null vector. Then it is possible to have $E_a = 0 = B_a$, but a non-vanishing $F_{ab}$. 
    
    In analogy to it, we define the \textbf{electric and magnetic parts of the Weyl tensor} as 
 \begin{align}
	E_{ab} &:= C_{ambn}n^{m} n^{n} & \text{and} && B_{ab} &:= \prescript{\star}{}{C}_{ambn} n^{m} n^{n}.
\end{align}
    If $\scrip$ is spacelike ($\Lambda>0$) or timelike ($\Lambda<0$), then $C_{abcd}n^{d}\equalhat 0$ implies that the electric and magnetic parts of the Weyl tensor are zero. In turn, this implies $C_{abcd}\equalhat 0$, just as in electromagnetism. Thus, we have
\begin{equation}\label{eq:WeylZeroCosmo}
	\boxed{C_{abcd}\equalhat 0\quad\text{for }\Lambda\neq 0}
\end{equation}
This conclusion does \textit{not} hold when $n^{a}$ is a null vector (i.e., when the cosmological constant vanishes) because the electric and magnetic parts of the Weyl tensor are \textit{not} independent. The electric and magnetic parts are still zero, but they do not completely determine the Weyl tensor and therefore we can \textit{not} conclude that $C_{abcd}\equalhat 0$. This is in perfect analogy with electromagnetism.
\end{mysidenote}

Even though the constraint~\eqref{eq:WeylConstraint} does not immediately imply that $C_{abcd} \equalhat 0$ when $\Lambda = 0$, we can still arrive at this conclusion by a careful analysis of the Bianchi identities. Moreover, we point out that $C_{abcd} \equalhat 0$ is a consequence of the topology of $\scrip$ being $\bbS^2\times\bbR$. We will sketch a proof in the last subsection of this chapter. In the next subsection, we introduce a coordinate system adapted to $\scrip$ and we discuss the construction of the Newman-Penrose null tetrad for asymptotically Minkowski spacetimes. The null tetrad will also be helpful in the proof of the last subsection.\bigskip

\subsection{Construction of a Newman-Penrose Null Tetrad for Curved Spacetimes}\label{Chap3_NullTetrad}
In the previous subsection we have introduced the concepts of asymptotic flatness and asymptotic Minkowski spaces. In order to work out physical consequences and construct a theory of gravitational radiation in full, non-linear GR, it is convenient to introduce a Newman-Penrose null tetrad. The basic idea is the same as for electromagnetic radiation. What we seek, is a way to follow radiation along null geodesics to infinity. 

Unlike electromagnetism, where we worked in Minkowski space, in GR we can no longer rely on global symmetries and globally defined coordinate systems to define a null tetrad. We have to carry out a more elaborate construction, where the strategy is to first install a null tetrad on a cross-section of $\scrip$. Then we parallel transport this ``reference'' tetrad along $\scrip$, in order to obtain a null tetrad on all of $\scrip$. Finally, we extend the null tetrad into the bulk of spacetime. 

\begin{figure}[htb!]
	\centering
	\includegraphics[width=0.5\linewidth]{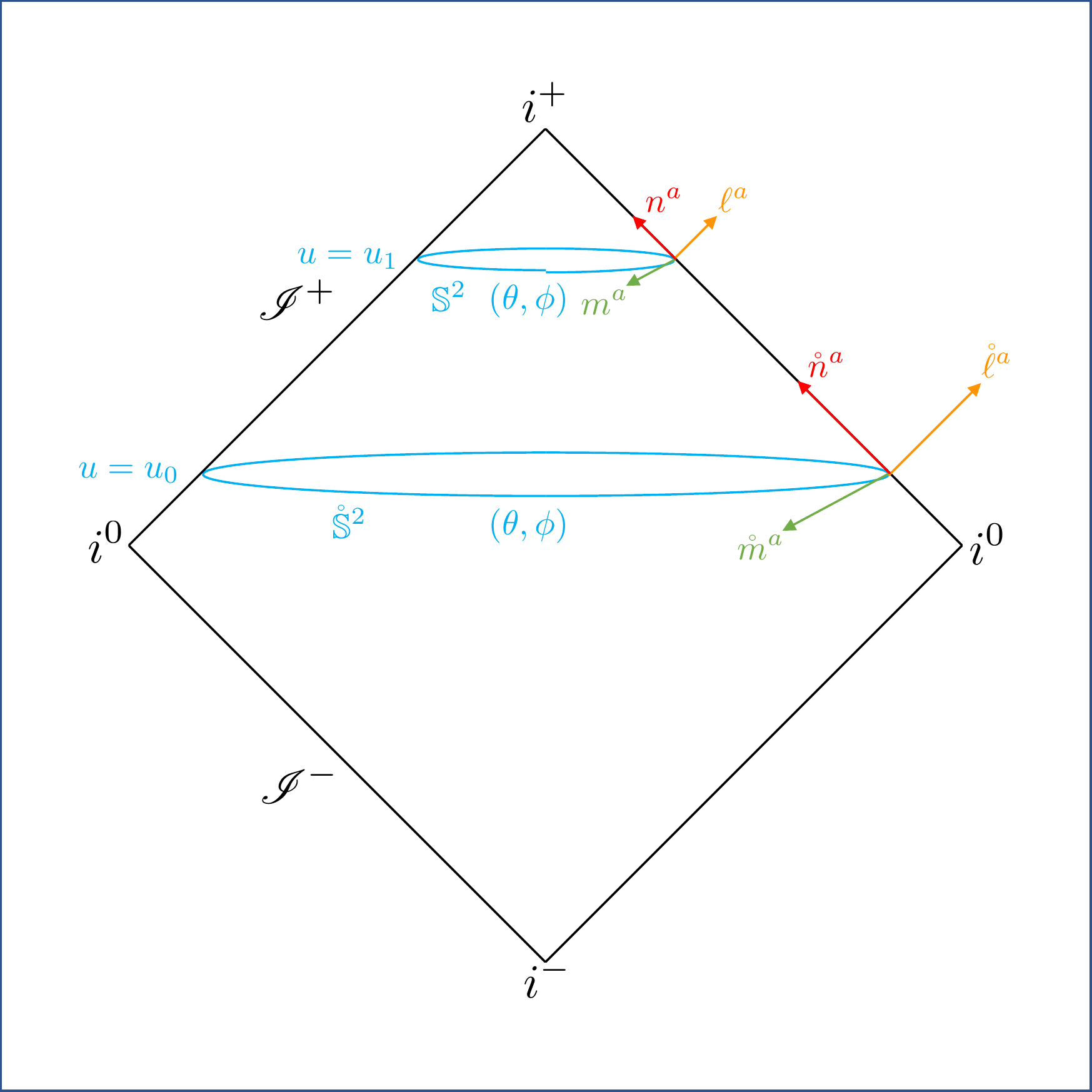}
	\caption{\textit{Schematic construction of a Newman-Penrose null tetrad for a generic asymptotically Minkowski spacetime. We install coordinates $(u, \theta, \phi)$ which are defined on all of $\scrip$ and define a reference null tetrad $\{\mathring{\ell}^{a}, \mathring{n}^{a}, \mathring{m}^{a}, \mathring{\bar{m}}^{a}\}$ on a cross-section $\mathring{\bbS}^2$. This reference null tetrad is extended to a null tetrad on all of $\scrip$ by Lie dragging it along $n^{a}$. Finally, the null tetrad on $\scrip$ is extended to a null tetrad in a neighborhood of $\scrip$ by Lie dragging it along $\ell^{a}$ into the bulk of $\M$.}}
	\label{fig:Construction}
\end{figure}

Let $(\M, g_{ab}, \Omega)$ be a divergence-free conformal frame of an asymptotically Minkowski spacetime. This means that $n^{a}:=g^{ab}\nabla_a\Omega$ is a well-defined vector field which is normal no $\scri$. Moreover, it is complete and thus generates all of $\scri$ (cf. Figure~\ref{fig:Construction}). We choose this vector field to be the first element of our null tetrad. Because we are in a divergence-free conformal frame, we can affinely parametrize the integral curves of $n^{a}$, i.e., we impose the equation $n^{a}\nabla_a u = 1$, where $u\in(-\infty, +\infty)$ is an affine parameter.

Next, we consider a cross-section of $\scrip$ which is defined by $u=u_0$, where $u_0$ is some constant. Since the cross-section is transverse to $\scrip$ (it is nowhere tangential to it), it is a spacelike surface and, moreover, it has the topology of a $2$-sphere (this follows from $\scri\simeq\bbS^2\times\bbR$, as required by the definition of asymptotic flatness). Therefore, we call this cross-section $\mathring{\bbS}^2$ (cf. Figure~\ref{fig:Construction}). The cross-section $\mathring{\bbS}^2$ can be parallel transported by $n^{a}$ along $\scrip$, as indicated in Figure~\ref{fig:Construction}, and we can therefore foliate $\scrip$ in terms of $2$-spheres corresponding to $u=$ const. surfaces. On each $2$-sphere, we can introduce coordinates $\theta$ and $\phi$, which are required to satisfy the equations $n^{a}\nabla_a \theta = 0$ and $n^{a}\nabla_a \phi = 0$. These equations simply guarantee that $\theta$ and $\phi$ are coordinates which are independent of the affine parameter $u$. With this we achieve the installment  of a coordinate system $(u,\theta,\phi)$ on $\scrip$.
Let us pause and contrast this with the situation we encountered in Minkowski space in Chapter~\ref{Chap1}. There, we started with vector fields in the interior of spacetime and took the limit to $\scrip$. This was easily possible due to the symmetries of Minkowski space and because there is a global coordinate system which allows us to introduce null coordinates. Here, we consider a more general class of spacetimes, which do not necessarily possess symmetries or globally well-defined coordinate systems. All we can rely upon, is the existence of $\scrip$ and its geometric properties. This is what we did so far.

The next step in our construction is the introduction of a null vector $\mathring{\ell}^{a}$, which is future-directed and normal to the cross-sections $u=u_0$. We can always introduce such a vector because given a $2$-sphere in a four-dimensional Lorentzian manifold, there are precisely two null normals. But notice that we have the freedom to normalize our vector and we choose $\mathring{n}^{a}\mathring{\ell}_{a} = -1$, where $\mathring{n}^{a}:=\left.n^{a}\right|_{u=u_0}$. Moreover, we can extend $\mathring{\ell}^{a}$ to $\ell^{a}$ by imposing the condition
\begin{align}\label{condition1}
	0 \overset{!}{=} \lie_n \ell^{a} = n^b\nabla_b \ell^{a} - \ell^a\nabla_b n^{b} &= n^b\nabla_b \ell^{a} & \textsf{with} && \left.\ell^{a}\right|_{u=u_0} = \mathring{\ell}^{a}.
\end{align}
We used $\nabla_b n^{b} = 0$ because we are in a divergence-free conformal frame. Hence, we are left with $n^b\nabla_b \ell^{a} \overset{!}{=}0$, which is the parallel transport equation of $\ell^{a}$ along $n^{a}$. 

Next, we introduce two complex null vectors, $\mathring{m}^{a}$ and $\mathring{\bar{m}}^{a}$, on the cross-section $\mathring{\bbS}^2$, which satisfy the cross-normalization condition $\mathring{m}^{a} \mathring{\bar{m}}_a = 1$. Because we have installed coordinates $\theta$ and $\phi$ on $\scrip$, we can represent the vector $m^{a}$ more explicitly like in Minkowski space:
\begin{equation}
	\mathring{m}^{a} \partial_a = \frac{1}{\sqrt{2}}\left(\partial_\theta + \frac{i}{\sin\theta}\,\partial_\phi\right).
\end{equation}
Notice that this does \textit{not} mean that the co-vectors $\mathring{m}_a$ and $\bar{m}_a$ also have the same form as in Minkowski space. The reason is that $g_{ab}$, which is needed to lower the index of $\mathring{m}^{a}$, is a potentially non-trivial metric. To continue our construction, we extend $\mathring{m}^{a}$ to $m^{a}$ by demanding that it is Lie dragged by $n^{a}$, i.e.,
\begin{align}\label{condition2}
	0 \overset{!}{=} \lie_n m^{a} = n^b\nabla_b m^{a} - m^a\nabla_b n^{b} &= n^b\nabla_b m^{a} & \textsf{with} && \left.m^{a}\right|_{u=u_0} = \mathring{m}^{a}.
\end{align}
In the last step we used again that the divergence of $n^{a}$ is zero. What we achieve with this is the following: We have installed the well-defined, global coordinate system $(u, \theta, \phi)$ on $\scrip$. Furthermore, we have introduced null vectors $\mathring{n}^{a}$, $\mathring{\ell}^{a}$, and $\mathring{m}^{a}$ which satisfy $\mathring{n}_{a} \mathring{\ell}^{a} = -1$ and $\mathring{m}_a \mathring{\bar{m}}^{a} = 1$. By Lie dragging these vectors along $n^{a}$, we obtain a Newman-Penrose null tetrad $\{n^{a}, \ell^{a}, m^{a}, \bar{m}^{a}\}$. Provided $\ell^{a}$ is transverse to \textit{all} $u=$ const. cross-sections and $m^{a}$, $\bar{m}^{a}$ are tangent to \textit{all} $u=$ const. cross-sections. This is shown in Exercise~\ref{ex:NullTetradConstruction}.

The final step in our construction is to extend the null tetrad from $\scrip$ into the bulk of $\M$. This is achieved by imposing the following parallel transport equations:
\begin{equation}
	\ell^{a} \nabla_a\ell^b = 0,\quad\quad\quad \ell^{a} \nabla_a n^{b} = 0,\quad\quad\quad \ell^{a}\nabla_a m^{b} = 0.
\end{equation}
In words: We parallel transport the null tetrad along $\ell^{a}$ from $\scrip$ into $\M$, thus obtaining a null tetrad $\{n^{a}, \ell^{a}, m^{a}, \bar{m}^{a}\}$ in a neighborhood of $\scrip$ with respect to the conformally rescaled metric $g_{ab}$. (This means that these vectors are null with respect to $g_{ab}$ and the normalization conditions hold with respect to $g_{ab}$).

Let us summarize the whole strategy: First, we chose $n^{a}$ as the first null normal to $\scrip$. This vector field is defined on all of $\scrip$ and it allows us to introduce an affine parameter $u$ which foliates $\scrip$. We have then introduced $(\theta, \phi)$ coordinates on the $u=$ const. leaves of the foliation. Put together, $(u, \theta, \phi)$ provides us with a globally defined coordinate system for $\scrip$. Next, we have introduced a Newman-Penrose null tetrad $\{\mathring{\ell}^{a}, \mathring{n}^{a}, \mathring{m}^{a}, \mathring{\bar{m}}^{a}\}$ on a cross-section $\mathring{\bbS}^2$. This ``reference'' null tetrad is normalized in the usual way and it serves as ``generator'' of a null tetrad on all of $\scrip$. In fact, we can generate such a null tetrad by Lie dragging (or parallel transporting, which is the same in this context) the reference tetrad along $n^{a}$ (or along its integral lines). Finally, we have extended the null tetrad from $\scrip$ into a neighborhood of $\scrip$ by Lie dragging it along $\ell^{a}$ into the bulk of spacetime. 

Given a Newman-Penrose null tetrad $\{n^{a}, \ell^{a}, m^{a}, \bar{m}^{a}\}$ on the conformally completed spacetime, we can construct a null tetrad $\{\hat{n}^{a}, \hat\ell^{a}, \hat m^{a}, \hat{\bar{m}}^{a}\}$ on the physical spacetime. We do so by demanding that the vectors of the null tetrad transform as
\begin{equation}
    (\hat n^{a}, \hat\ell^{a} , \hat m^{a}, \hat{\bar{m}}^{a}) = (\Omega^{s_1} n^{a}, \Omega^{s_2} \ell^{a}, \Omega^{s_3}m^{a}, \Omega^{s_4} \bar{m}^{a})
\end{equation}
for some $s_1, s_2, s_3, s_4\in\bbZ$ under a conformal transformation. We choose $\hat{n}^{a} \equiv n^{a}$ and the remaining rescaling relations follow from the same logic as the one used in Chapter~\ref{Chap1}. Concretely, we find the rescaling relations
\begin{equation}\label{eq:RescalingRelations_general}
\boxed{\begin{aligned}
    \hat{n}^{a} &= n^{a}, & \hat\ell^{a} &= \Omega^{2}\,\ell^{a}, & \hat{m}^{a} &= \Omega \,m^{a}
\end{aligned}}
\end{equation}
Finally, we note that the metric can be expressed in terms of the null tetrad as (see Exercises~\ref{ex:MetricInNPForm}, whose method holds in full generality)
\begin{equation}
    \boxed{g_{ab} = -2\ell_{(a} n_{b)} + 2m_{(a} \bar{m}_{b)}}
\end{equation}
Observe that the rescaling relation~\eqref{eq:RescalingRelations_general} correctly reproduce the relation between the physical and the conformally rescaled metric, i.e., $\hat g_{ab} = \Omega^2\, g_{ab}$.
\bigskip

\subsection{The Vanishing of the Weyl Tensor at Null Infinity}
We return to studying the Weyl tensor and our goal is the proof of the claim we made at the end of subsection~\ref{Chap3_Constraint}. Namely, that $C_{abcd} \equalhat 0$. To that end, we start by introducing a Newman-Penrose null tetrad $\{\ell^{a}, n^{a}, m^{a}, \bar{m}^{a}\}$ and we consider the alternating tensor $\epsilon_{abc}$ intrinsic to $\scrip$,  defined by $\epsilon_{abc} := \epsilon_{abcd} \ell^d$. We recall from Exercise~\ref{ex:TetradIdentities} that $\epsilon_{abcd}$ can be written as $\epsilon_{abcd} = -4!\, i\, \ell_{[a} n_b m_c \bar{m}_{d]}$. Hence, the only non-zero contraction in $\epsilon_{abcd} \ell^d$ is when $\ell^d$ hits $n_d$ and we obtain $\epsilon_{abc} = 3!\, i \,\ell_{[a}m_b\bar{m}_{c]}$. Next, we define $\bbC_{ab} := \epsilon_{mna} \epsilon_{pqb} C^{mnpq}$. This tensor has the following properties:
\begin{itemize}
	\item[a)] It is real because it is constructed from real tensors;
	\item[b)] It is symmetric (this follows from the symmetry $C_{abcd} = C_{cdab}$);
	\item[c)] It is trace-free (this follows from the definition of $\epsilon_{abc}$ and $g_{ab} = -2\ell_{(a}n_{b)}+2m_{(a}\bar{m}_{b)}$);
	\item[d)] It is transverse, i.e., it satisfies $\bbC_{ab} n^{b} = 0$ (this follows again from the definition of $\epsilon_{abc}$);
	\item[e)] It satisfies the Bianchi identity $\nabla_{[a}\bbC_{b]c} =0$ (this follows from the Bianchi identity of the Weyl tensor, $\nabla_{[a}C_{bc]de} = 0$ and because $\epsilon_{abc}$ is covariantly constant, $\nabla_d \epsilon_{abc} = 0$).
\end{itemize}
We pull back the tensor $\bbC_{ab}$ to $\scrip$. Then, using property d), i.e., the transversality, and property b), i.e., the symmetry, to conclude that $\bbC_{ab}$ is tangential in both indices to $n^{a}$. That is, the tensor $\bbC_{ab}$ lies in the orthogonal complement of the vector space span$\{n^{a}\}$. This orthogonal complement is topologically $\bbS^2$, because $\scrip$ has the topology $\bbS^2\times\bbR$. Hence, we can think of $\bbC_{ab}$ as being a tensor on a manifold which has the topology of a $2$-sphere. Furthermore, it follows that $\bbC_{ab}$ can be written as
\begin{equation}\label{eq:CExpansion}
	\bbC_{ab} = \alpha\, m_{a} m_{b} + \bar{\alpha}\, \bar{m}_a\bar{m}_b + \beta\, m_{(a}\bar{m}_{b)},
\end{equation}
where $\alpha = \alpha(\theta,\phi)$ is a complex function on the $2$-sphere and $\beta = \beta(\theta,\phi)$ is real. All we have done is expand the tensor $\bbC_{ab}$ in a basis of the $2$-sphere. Notice that it is necessary to include the complex conjugate of the first term in order to ensure that $\bbC_{ab}$ is real, as required by property a). The third term is real provided $\beta$ is real.

Next, we use the fact that $\bbC_{ab}$ is trace-free to conclude
\begin{equation}
	0 = g^{ab} \bbC_{ab} = 2 m^{(a}\bar{m}^{b)}\bbC_{ab} = \beta.
\end{equation}
Hence, the last term in the expansion~\eqref{eq:CExpansion} vanishes and we are left with $\alpha\, m_{a} m_b$ and its complex conjugate. Up to this point, we have used properties a) through d) and we are left with imposing the Bianchi identities. To do so, we observe that the intrinsic metric $q_{ab}$ is only a function of the coordinates $(\theta, \phi)$ on $\bbS^2$ and that we can always perform a diffeomorphism, such that the pull back to the orthogonal complement of span$\{n^{a}\}$ is the metric of the unit $2$-sphere. Moreover, since $\bbC_{ab}$ is defined on that orthogonal complement, we have to pull back the Bianchi identities to that space. Let us denote the covariant derivative with respect to the unit $2$-sphere of the orthogonal complement by $\D$. Then, the equations to consider read
\begin{equation}
    \D_{[a}\bbC_{b]c} = \alpha\D_{[a}m_{b]}m_b - m_{[b}\left(\D_{a]}\alpha\right) m_c + \textsf{c.c.} = 0,
\end{equation}
where ``$\textsf{c.c.}$'' stands for ``complex conjugate''. Explicitly working out the covariant derivatives finally leads to the conclusion that the Bianchi identities are satisfied if and only if $\alpha = 0$. Hence, we find that $\bbC \equalhat 0$. What is left to show is that $\bbC_{ab} \equalhat 0$ forces the two components of $C_{abcd}$ to vanish which remained unconstrained by $C_{abcd}n^{d}\equalhat 0$.

This is most easily achieved by introducing Newman-Penrose scalars for the Weyl tensor. We define them as 
\begin{align}
	\underline{\Psi}_4 &:= C_{abcd}\,n^{a} \bar{m}^{b} n^{c} m^{d}\notag\\
	\underline{\Psi}_3 &:= C_{abcd}\,\ell^{a} n^{b} \bar{m}^c n^d\notag\\
	\underline{\Psi}_2 &:= C_{abcd}\,\ell^{a} m^{b} \bar{m}^c n^d\notag\\
	\underline{\Psi}_1 &:= C_{abcd}\,\ell^{a} n^{b} \ell^{c} m^{d}\notag\\
	\underline{\Psi}_0 &:= C_{abcd}\, \ell^{a} m^{b}\ell^{c} m^{d},
\end{align}
where the use of the underline will be explained further below. Notice that the scalars are complex, just as their electromagnetic analogues. Furthermore, there are five such scalars and they encode the ten independent components of the Weyl tensor. It is shown in Exercise~\ref{ex:PsisAreZero} that $C_{abcd}n^{d}\equalhat 0$ implies $\underline{\Psi}_4 \equalhat \underline{\Psi}_3 \equalhat \underline{\Psi}_2 \equalhat \underline{\Psi}_1 \equalhat 0$. In the same exercise, it is shown that $\bbC = \underline{\Psi}_0 m_a m_b + \textsf{c.c.}$, from which it finally follows that the Weyl tensor of the conformally completed spacetime vanishes at null infinity. All in all, we can state the following result.
\begin{equation}
	\boxed{C_{abcd}\equalhat \begin{cases}
		0 & \text{for }\Lambda\neq 0\\
		0 & \text{for }\Lambda = 0\text{ if the topology of $\scrip$ is $\bbS^2\times\bbR$}
	\end{cases}}
\end{equation}
This result is in contrast to Maxwell's theory where we have $F_{ab}\,\unequalhat\, 0$. This leads us to introduce the \textbf{asymptotic Weyl tensor} defined as
\begin{equation}
	K_{abcd} := \Omega^{-1} C_{abcd}.
\end{equation}
Recall from Side Note~\ref{sn:Lemma} that if a smooth tensor field vanishes at $\scrip$, then $\Omega^{-1}$ times that tensor has a smooth limit at $\scrip$. Thus, the asymptotic Weyl tensor is well-defined. Moreover, the Newman-Penrose scalars with an underline we introduced above, are mere auxiliary quantities. Their purpose was to simplify our proof of $C_{abcd} \equalhat 0$. The Newman-Penrose scalars we shall use in the remainder of these notes are defined with respect to the asymptotic Weyl tensor.
\begin{align}
	{\Psi}_4 &:= K_{abcd}\,n^{a} \bar{m}^{b} n^{c} m^{d}\notag\\
	{\Psi}_3 &:= K_{abcd}\,\ell^{a} n^{b} \bar{m}^c n^d\notag\\
	{\Psi}_2 &:= K_{abcd}\,\ell^{a} m^{b} \bar{m}^c n^d\notag\\
	{\Psi}_1 &:= K_{abcd}\,\ell^{a} n^{b} \ell^{c} m^{d}\notag\\
	{\Psi}_0 &:= K_{abcd}\, \ell^{a} m^{b}\ell^{c} m^{d} 
\end{align}
These are the Newman-Penrose scalars we will use in the remainder of these notes. In particular, in the next chapter we will see that the smoothness of $K_{abcd}$ together with the rescaling properties of the Newman-Penrose null tetrad imply the Peeling Properties for GR.\bigskip

\newpage
\subsection{Exercises}
\renewcommand{\thesection}{\arabic{section}}

\begin{Exercise}[]\label{ex:ScriSpacelike}
    Use the field equations $\hat{R}_{ab} - \frac12 \hat R\, \hat g_{ab} + \Lambda\, g_{ab} = 8\pi T_{ab}$ with $\Lambda$ to show that $\scrip$ is timelike when $\Lambda <0$ and timelike when $\Lambda >0$.
\end{Exercise}

\begin{Exercise}[]\label{ex:qIsDegenerate}
    Let $(\N, q_{ab})$ be a co-dimension one hypersurface which is embedded into the Lorentzian manifold $(\M, g_{ab})$, where $q_{ab}:=\pb{g}_{ab}$. Let $n^{a}$ be the normal vector to $\N$ and assume that $g_{ab}n^{a} n^{b}$. Show that the normal vector $n^{a}$ is an eigenvector of the intrinsic metric $q_{ab}$ with eigenvalue zero. That is, show that
    \begin{equation*}
        q_{ab}n^{b} \equalhat 0,
    \end{equation*}
    where in this context `$\equalhat$' means ``equality on $\N$''. Furthermore, show that $q_{ab}$ has signature $(0,+,\cdots, +)$ and is thus degenerate.
\end{Exercise}

\begin{Exercise}[]\label{ex:NullTetradConstruction}
    Complete the construction of the Newman-Penrose null tetrad in~\ref{Chap3_NullTetrad} by showing that $\ell^{a}$ is transverse to \textit{all} $u=$ const. cross-section, while $m^{a}$ is tangent to \textit{all} these cross-sections.
\end{Exercise}

\begin{Exercise}[]\label{ex:PsisAreZero}
    Prove the following claims:
    \begin{itemize}
        \item[a)] The constraint $C_{abcd}n^d \equalhat 0$ implies $\underline{\Psi}_1 \equalhat \underline{\Psi}_2 \equalhat \underline{\Psi}_3 \equalhat \underline{\Psi}_4 \equalhat 0$.
        \item[b)] The real, symmetric, trace-free, and transverse tensor $\bbC_{ab} := \epsilon_{mna}\epsilon_{pqb}C^{mnpq}$ is equal to $\Psi_0 m_{a} m_{b} + \bar{\Psi}_0 \bar{m}_a \bar{m}_b$. Thus, $\bbC_{ab} \equalhat 0 \Leftrightarrow \Psi_0 \equalhat 0$.
    \end{itemize}
\end{Exercise}

\begin{Exercise}[]\label{ex:AreaElement}
    In Exercise~\ref{ex:2form} we showed that $\prescript{2}{}{\hat{\varepsilon}} := i\, \hat{m}\wedge \hat{\bar{m}}$ with $\hat{m} := \hat{m}_a \dd x^{a}$ is the area element of a sphere of radius~$r$, provided we use the null tetrad for Minkowski space. Shown in full generality (i.e., for any spacetime, not just for Minkowski space) that 
    \begin{itemize}
        \item[a)] $s_{ab} := 2\hat{m}_{(a}\hat{\bar{m}}_{b)}$ is the metric on a cross-section $\C$ of $\scrip$ (the cross-section has topology $\bbS^2$ by assumption, but not necessarily the geometry of a $2$-sphere);
        \item[b)] $\prescript{2}{}{\hat{\varepsilon}} := i\, \hat{m}\wedge \hat{\bar{m}}$ is the area element of the cross-section $\C$. This means that $\textsf{Area}(\C) = \displaystyle\int_{\C} \prescript{2}{}{\hat{\varepsilon}}$.
    \end{itemize}
\end{Exercise}

\begin{Exercise}[]\label{ex:DecomposingFwrtSigma}
Let $\Sigma$ be a hypersurface which is either null or spacelike and let $n^{a}$ be the null/timelike normal no $\Sigma$. The electric and magnetic $1$-forms are then defined as 
\begin{align*}
	E &:= E_I\,\dd y^{I} = \left.F_{ab} n^{b}\, \dd x^{a}\right|_{\Sigma}    & \text{and} && B &:= B_{I}\,\dd y^{I} = \left.\prescript{\star}{}{F_{ab}} n^{b}\,\dd x^{a}\right|_{\Sigma},
\end{align*}
where $\left.T\right|_{\Sigma}$ means ``restriction of $T$ to $\Sigma$'' and $y^{I}$ with $I\in\{1,2,3\}$ are coordinates on $\Sigma$. Prove the following claims:
\begin{itemize}
	\item[a)] If $\Sigma$ is spacelike (meaning $n^{a}$ is timelike), then the components $E_I$ and $B_I$ completely determine $F_{ab}$, (i.e., the electric and magnetic fields carry the same information as $F_{ab}$).
	
	\textit{Hint: Every spacelike hypersurface can be represented as a $t=\text{const}.$ hypersurface.}
	\item[b)] If $\Sigma$ is null (meaning $n^{a}$ is null, too), then the components $E_I$ and $B_I$ do \textit{not} determine $F_{ab}$ completely (i.e., the so-defined electric and magnetic fields carry less information than $F_{ab}$). In particular, show that $E_I = 0$ and $B_I = 0$ do \textit{not} imply $F_{ab} = 0$. What information is carried by the so-defined $E$ and $B$ fields?
	
	\textit{Hint: Use the expressions for $F$ and $\prescript{\star}{}{F}$ in terms of Newman-Penrose null tetrads derived in previous exercises.}
	\item[c)] Verify that the energy density of the electromagnetic field is given by
	\begin{equation*}
		\frac12 \left(\left<E,E\right> + \left<B,B\right>\right) = \begin{cases}
			\frac12\left(\|\vec{E}\|^2 + \|\vec{B}\|^2\right) & \text{for $\Sigma$ spacelike} \\
			2|\Phi_2|^2 & \text{for $\Sigma$ null},
		\end{cases}
	\end{equation*}
	where the bilinear inner product is defined as $\left<\omega,\mu\right> := g^{ab}\omega_a\mu_b$ for any $1$-forms $\omega$ and $\mu$. Compare this to the result of Exercise~\ref{ex:EnergyMomentumFlux}.
\end{itemize}
\end{Exercise}

\newpage
\asection{4}{Peeling and Universal Structure of Null Infinity}\label{Chap4}
\subsection{The Peeling Theorem for GR}\label{ssec:PeelingTheoremGR}
In Chapter~\ref{Chap3}, we have seen that the definition of asymptotic flatness in conjunction with the validity of Einstein's field equations implies that the Weyl tensor satisfies $C_{abcd}n^{d} \equalhat 0$. Hence, from purely local considerations on $\scrip$, we find that $\underline{\Psi}_1 \equalhat \underline{\Psi}_2 \equalhat \underline{\Psi}_3 \equalhat \underline{\Psi}_4 \equalhat 0$. In other words, eight of the ten components of the Weyl tensor vanish at $\scrip$. The only component which is not immediately set to zero by the constraint $C_{abcd}n^{d} \equalhat 0$ is $\underline{\Psi}_0$.

This component only vanishes if we make an additional assumption on the topology of $\scrip$. Namely, we need to assume that $\scrip\simeq \bbS^2\times\bbR$, which is physically well-motivated. We have argued that the symmetric, traceless, and transverse tensor $\bbC_{ab}$ in conjunction with the Bianchi identities of the Weyl tensor then imply $\underline{\Psi}_0 \equalhat 0$. Hence, it follows that the Weyl tensor of the conformally completed spacetime $(\M, g_{ab})$ vanishes on $\scrip$, $C_{abcd}\equalhat 0$.

This is an important fact, because it allows us to introduce a new tensor, $K_{abcd} := \Omega^{-1}C_{abcd}$, which is smooth at $\scrip$. We call it the \textit{asymptotic Weyl tensor}. Since the asymptotic Weyl tensor is in general \textit{not} zero at $\scrip$, we can use it to extract physics from it. In fact, this allows us to derive the Peeling Theorem for GR, which we do now.

To that end, we define, in the conformally completed spacetime $(\M, g_{ab})$, the Newman-Penrose scalar
\begin{equation}\label{eq:Psi_4}
	\Psi_4 := K_{abcd} n^{a} \bar{m}^{b} n^{c} \bar{m}^d.
\end{equation}
A word on notation: Newman-Penrose scalars with an underline, such as $\underline{\Psi}_i$, are defined with respect to the Weyl tensor $C_{abcd}$. Scalars without an underline, such as $\Psi_i$, are defined with respect to the asymptotic Weyl tensor $K_{abcd}$. With this distinction out of the way, we note that we can write $\Psi_4$ in~\eqref{eq:Psi_4} equivalently as
\begin{equation}
	\Psi_4 = \Omega^{-1} C\du{abc}{d} n^{a} \bar{m}^{b} n^{c} \bar{m}_d.
\end{equation}
We have just substituted $K_{abcd}$ with $\Omega^{-1}C_{abcd}$ and we have raised the last index of the Weyl tensor. The reason for doing so is that the Weyl tensor $C\du{abc}{d}$, with the last index raised, is conformally invariant. Hence, we have $C\du{abc}{d} = \hat{C}\du{abc}{d}$, where $\hat{C}\du{abc}{d}$ is the Weyl tensor with respect to the \textit{physical} metric $\hat g_{ab}$. We therefore get
\begin{align}
	\Psi_4 = \Omega^{-1} \hat{C}\du{abc}{d} n^{a} \bar{m}^{b} n^{c} \bar{m}_d.
\end{align}
To proceed, we recall that the physical Newman-Penrose tetrad is related to the one of the conformally completed spacetime via the relations
\begin{align}
	n^{a} &= \hat n^{a}, & \ell^{a} &= \Omega^{-2}\, \hat\ell^{a},  & m^{a} = \Omega^{-1}\, \hat m^{a}.
\end{align}
After using these rescaling relations, we find that the scalar $\Psi_4$ can be written in terms of the \textit{physical} fields as
\begin{align}
	\Psi_4 &= \Omega^{-1} \hat{C}_{abc}{}^d \hat{n}^{a} \left(\Omega^{-1}\,\hat{\bar{m}}^b\right)\hat{n}^{c}\left(\Omega\,\hat{\bar{m}}_d\right)\notag\\
	&= \Omega^{-1} \hat{C}_{abc}{}^d \hat{n}^{a} \hat{\bar{m}}^b\hat{n}^{c}\hat{\bar{m}}_d,
\end{align}
where we used $\bar{m}_a = \bar{m}^{b} g_{ab} = \left(\Omega^{-1} \hat{\bar{m}}^b\right)\Omega^2 \hat{g}_{ab} = \Omega\, \hat{\bar{m}}_a$. We therefore obtain a relation between the physical scalar $\hat{\Psi}_4$ and the Weyl scalar with respect to the asymptotic Weyl tensor, $\Psi_4$:
\begin{equation}\label{eq:NPscalarpeeling}
	\Psi_4(\Omega, u, \theta, \phi) = r\, \hat{\Psi}_4(\Omega, u, \theta, \phi).
\end{equation}
This implies that the physical scalar $\hat{\Psi}_4$ decays like
\begin{equation}
    \hat{\Psi}_4(r, u, \theta, \phi) = \frac{\Psi^\circ_4(u,\theta,\phi)}{r} + \O(r^{-2}),
\end{equation}
where we have introduced $\Psi^\circ_4(u,\theta,\phi):=\left.\Psi_4(\Omega,u,\theta,\phi)\right|_{\Omega=0}$, which is the zeroth order term in the Taylor expansion of $\Psi_4$ around $\Omega=0$. This is the first peeling property and the other peeling properties follow in a similar manner (see Exercise~\ref{ex:PeelingForGR}). All in all, we find
\begin{equation}
\boxed{\begin{aligned}\label{eq:peelinginGR}
    \hat{\Psi}_4(r, u, \theta, \phi) &= \frac{\Psi^\circ_4(u,\theta,\phi)}{r} + \O(r^{-2})\\
	\hat\Psi_3(r, u, \theta, \phi) &= \frac{\Psi_3^\circ(u,\theta,\phi)}{r^2}+\O(r^{-3})\\
	\hat\Psi_2(r, u, \theta, \phi) &= \frac{\Psi_2^\circ(u,\theta,\phi)}{r^3}+\O(r^{-4})\\	
	\hat\Psi_1(r, u, \theta, \phi) &= \frac{\Psi_1^\circ(u,\theta,\phi)}{r^4}+\O(r^{-5})\\
	\hat\Psi_0(r, u, \theta, \phi) &= \frac{\Psi_0^\circ(u,\theta,\phi)}{r^5}+\O(r^{-6})
\end{aligned}}
\end{equation}
The scalar $\Psi_i^\circ$ always denotes the value of $\Psi_i$ at $\scrip$, i.e., it always stands for the zeroth order of the Taylor expansion of $\Psi_i$ (the scalar with respect to the asymptotic Weyl tensor) around $\Omega=0$.

With this, we have proven the Peeling Theorem for GR. The theorem tells us that the different components of the physical Weyl tensor $\hat{C}_{abcd}$ decay, or ``peel off'', at different rates\footnote{It is again easy to remember at which rate $\hat\Psi_i\propto\frac{1}{r^n}$ decays. The correct behavior follows from $i+n=5$.} as one approaches $\scrip$. This peeling leads again, just as in the case of electromagnetism, to a neat separation of different modes. In fact, the peeling property of $\hat\Psi_4$ suggests that it encodes the radiation field, since it decays like $\frac{1}{r}$. Conversely, one can guess that $\hat\Psi_2$ encodes the ``coulombic'' information of the gravitational field (i.e., the mass of the source which generates the field). This is motivated by the following simple example: In the Schwarzschild spacetime, we have $\hat g_{ab} = \hat\eta_{ab} + \O(r^{-1})$, which then implies $\hat C_{abcd}\propto \frac{1}{r^3}$. This is a first indication that $\hat\Psi_2\propto\frac{1}{r^3}$ carries information about the mass of the source and the gravitational force. This is what we call the ``coulombic'' information of the gravitational field, using an obvious analogy with electromagnetism.

To actually extract physical information from the Weyl scalars $\hat\Psi_i$ such as energy (mass), momentum, and angular momentum of the source, or whether or not there is gravitational radiation, requires more work and the introduction of new ideas. The reason is that unlike in special relativistic theories (think of Maxwell, Yang-Mills fields, the Klein-Gordon field, etc.) we do \textit{not} have an energy momentum tensor for the gravitational field and, related to that issue, in GR we do not have a preferred symmetry group. Special relativistic theories are invariant under Poincar\'e transformations because they are defined on a fixed background manifold equipped with a fixed metric -- the Minkowski metric. In GR, on the other hand, the manifold and the metric are determined from within the theory as a solution of Einstein's equations. Moreover, the theory is generally covariant. Hence, no preferred coordinate systems and no preferred coordinate transformations between these systems exist with respect to which we could define notions such as energy and momentum.

What we will see in this and the next chapter, however, is that in asymptotically Minkowski spacetimes, there is a way of singling out preferred coordinates and preferred coordinate transformations, which will ultimately enable us to define conserved quantities.

Before going into more details, in the next subsection we will illustrate the Peeling Theorem for the Schwarzschild and the Kerr-Newman family of spacetimes. 
\bigskip

\subsection{Illustrating the Newman-Penrose Formalism using Black Hole Solutions}
We have shown that the Newman-Penrose scalars of GR $\hat\Psi_i$ obey the Peeling Theorem. Now we want to put the theory to a test and explicitly illustrate the Peeling Theorem! For technical reasons, we focus on simple and exact solutions, which means we will not be talking about spacetimes containing radiation. Rather, we will show that the spacetimes under consideration possess a vanishing $\Psi_4^\circ$, consistent with the fact that they are devoid of gravitational waves. Moreover, we check whether the ``coulombic'' part, i.e., $\Psi^\circ_2$, has the correct dependencies on spacetime parameters and behaves in a manner we would expect. We motivate the reader to redo the computations of the examples we provide here.

Concretely, we consider the Schwarzschild solution as a warm-up exercise in~\ref{ssec:Schwarzschild} and then we move to the Kerr-Newman solution in subsection~\ref{ssec:KerrNewman}, which describes a charged and rotating black hole. Since these solutions are stationary, we expect $\Psi^\circ_4 = 0$, while the ``coulombic'' part should be proportional to the mass, $\Psi_2^\circ\propto M$. Contrary to the electrodynamic examples we studied in~\ref{Chap2:Examples}, we will not compute quantities such as the energy and momentum associated with the gravitational field. Introducing such notions for the gravitational field is a subtle issue and we will elaborate more on this in Chapters~\ref{Chap6} and~\ref{Chap7}. \bigskip

\subsubsection{The Schwarzschild Black Hole}\label{ssec:Schwarzschild}
Our aim is to compute the physical Newman-Penrose scalars $\hat\Psi_i$ for the Schwarzschild solution of Einstein's field equations. Thus, physically speaking, we consider a non-rotating, uncharged black hole of mass $M$. In the Schwarzschild chart $\{t,r,\theta,\phi\}$, the line element takes the well-known form
\begin{align}
    \dd \hat{s}^2 = -\left(1-\frac{2M}{r}\right)\dd t^2 + \frac{1}{1-\frac{2M}{r}}\dd r^2 + r^2 \dd \theta^2 + r^2 \sin{\theta}\, \dd \phi^2.
\end{align}
In order to be able to use the Newman-Penrose formalism, our first step has to be to rewrite the above line element in outgoing Eddington-Finkelstein coordinates $\{u,\Omega,\theta,\phi\}$, with $u := t - r - 2M\ln\left(\frac{r}{2M}-1\right)$, for $r>2M$, and $\Omega := \frac{1}{r}$. In the new chart, the above line element reads
\begin{align}\label{eq:schwarzschildmetric}
    \dd \hat{s}^2 = -\left(1-2M\Omega\right)\dd u^2 + 2\Omega^{-2}\dd u\,\dd \Omega + \Omega^{-2} \dd \theta^2 + \Omega^{-2} \sin{\theta}\, \dd \phi^2.
\end{align}
From this line element, we can read off the physical metric $\hat{g}_{ab}$ in the chart $\{u,\Omega,\theta,\phi\}$ and consequently calculate the Newman-Penrose null tetrad $\{\hat n^{a}, \hat\ell^{a}, \hat m^{a}, \hat{\bar{m}}^{a}\}$, the Weyl tensor $\hat C_{abcd}(\hat g)$ and, finally, the corresponding $\hat\Psi_i$. 

Let us begin with the null tetrad. In Chapter~\ref{Chap3}, we have provided a recipe for constructing the null tetrad for a curved spacetime and we encourage the reader to follow the steps of that recipe. Here, we will simply sketch the procedure: Based on $\hat n^a=\nabla^a\Omega$ in the coordinate chart $\{u,\Omega,\theta,\phi\}$ we chose $\hat n^a\propto\delta\du{1}{a}$ where the proportionality coefficient can be chosen for convenience. We go for $\hat n^a=\sqrt{2}\delta\du{1}{a}$. Then, via $\hat n^a\hat \ell_a=-1$ we find the condition $\sqrt{2}\hat\ell^0/\Omega^2=-1$ which determines $\hat \ell^0$. Further, from $\hat \ell^a\hat \ell_a=0$ and with the knowledge that in this coordinate chart $\hat \ell^a$ has no components in $\theta,\phi$-direction, we deduce $\hat\ell^0\hat\ell^0\left(-1+2M\Omega\right)+4\hat\ell^0\hat\ell^1/\Omega=0$. With the latter equation we find $\hat\ell^1$ and, thus, have a full description of $\hat\ell^a$. In similar fashion, we obtain the vectors $\hat m^a, \hat{\bar{m}}^a$. Overall, this results in
\begin{align}
    \hat{n}^{a} &= \sqrt{2}\,\delta\du{1}{a}, \notag\\
    \hat\ell^{a} &= -\frac{1}{\sqrt{2}}\Omega^2\, \delta\du{0}{a} + \frac{1}{2\sqrt{2}} \Omega^4 (-1 + 2 M \Omega)\, \delta\du{1}{a}, \notag\\
    \hat{m}^{a} &= \frac{1}{\sqrt{2}} \Omega\, \delta\du{2}{a} + \frac{i}{\sqrt{2} \sin\theta}\Omega\, \delta\du{3}{a}.
\end{align}
Next, we compute the physical Weyl tensor and we find the following non-zero components (we do not display components which can be obtained from the components listed below by symmetries of the Weyl tensor):
\begin{align}
    \hat C_{0101} &= -2M\Omega^{-1}         \notag\\
    \hat C_{0202} &= -\left(2 M \Omega - 1\right)M \Omega \notag\\
    \hat C_{0212} &= -M\Omega^{-1}      \notag\\
    \hat C_{0303} &=  -(2 M\Omega -1 )M \Omega\, \sin^2{\theta} \notag\\
    \hat C_{0313} &= -M\Omega^{-1} \, \sin^2{\theta}   \notag\\
    \hat C_{2323} &= 2 M\Omega^{-1} \, \sin^2{\theta}.
\end{align} 
We now have all the ingredients needed to compute the Newman-Penrose scalars $\hat{\Psi}_i$ and we find
\begin{align}
    \hat{\Psi}_4(u, \Omega, \theta, \phi) = \hat{C}\du{abc}{d}\hat{n}^{a}\hat{\bar{m}}^{b}\hat{n}^c\hat{m}_d &=\ 0   \notag\\
    \hat{\Psi}_3(u, \Omega, \theta, \phi) = \hat{C}\du{abc}{d}\hat{\ell}^{a}\hat{n}^{b}\hat{\bar{m}}^{c}\hat{n}_d &=\ 0\notag\\
    \hat{\Psi}_2(u, \Omega, \theta, \phi) = \hat{C}\du{abc}{d}\hat{\ell}^{a}\hat{m}^{b}\hat{\bar{m}}^{c}\hat{n}_d &=\  -M\,\Omega^3 \notag\\
    \hat{\Psi}_1(u, \Omega, \theta, \phi) = \hat{C}\du{abc}{d}\hat{\ell}^{a}\hat{n}^{b}\hat{\ell}^{c}\hat{m}_d &=\  0 \notag\\
    \hat{\Psi}_0(u, \Omega, \theta, \phi) = \hat{C}\du{abc}{d}\hat{\ell}^{a}\hat{m}^{b}\hat{\ell}^{c}\hat{m}_d &=\ 0,
\end{align}
from which we can simply read off the leading order contributions $\Psi^\circ_i$,
\begin{align}
    \Psi^\circ_4 = \Psi^\circ_3 = \Psi^\circ_1 = \Psi^\circ_0 &= 0 &\text{and} && \Psi^\circ_2 = -M,
\end{align}
which immediately confirms the Peeling Theorem for the Schwarzschild spacetime. As anticipated, $\Psi^\circ_2$ is not zero since it encodes the ``coulombic'' information, i.e., it carries information about the mass which is sourcing the field. Also, because $\Psi^\circ_4$ vanishes, we find, reassuringly, that there is no gravitational radiation in this spacetime.

This was a rather simple example and intended to be more of a proof of concept for the Newman-Penrose formalism. In the next subsection, we consider the computationally more complicated case of the Kerr-Newman solution and explicitly confirm the validity of the Peeling Theorem also for this family of spacetimes.\bigskip

\subsubsection{The Kerr-Newman Family of Black Holes}\label{ssec:KerrNewman}
Let us now turn our attention towards the Kerr-Newman black hole solution. That is, we consider a black hole of mass $M$, charge $Q$, and angular momentum $J$. In the metric, the angular momentum is encoded through $a:=J/M$ and in outgoing Eddington-Finkelstein coordinates $\{u,r,\theta,\phi\}$, the Kerr-Newman line element reads
\begin{align}\label{eq:KerrNewmanmetric}
    \dd \hat{s}^2 =& -\left(1-\frac{2Mr-Q^2}{r^2+a^2\cos^2\theta} \right)\dd u^2-2\dd u\, \dd r - \frac{2a\sin^2\theta}{r^2+a^2\cos^2\theta}(2Mr-Q^2)\dd u\, \dd \phi + 2a\sin^2\theta \dd r\, \dd\phi \notag\\
    & +(r^2+a^2\cos^2\theta)\dd \theta^2 + \sin^2\theta \left(r^2+a^2+\frac{a^2\sin^2\theta}{r^2+a^2\cos^2\theta}(2Mr-Q^2)\right)\dd \phi^2.
\end{align}
We perform a conformal completion by introducing the new coordinate $\Omega := \frac{1}{r}$ and multiplying the physical line element by $\Omega^2$.\footnote{Strictly speaking, the conformal completion is not necessary. In order to compute the Newman-Penrose scalars $\hat{\Psi}_i$ and verify their Peeling properties, we do not need to work in the conformally completed spacetime. However, we do need to know in which ``direction'' to go in order to find $\scrip$ and check that the $\hat{\Psi}_i$ fall off as predicted by the Peeling Theorem. For this purpose, it is useful to work in the $\{u,\Omega, \theta, \phi\}$ chart.} In the chart $\{u, \Omega, \theta, \phi\}$, the physical line element becomes
\begin{align}\label{eq:NullKNLineElement}
    \dd \hat{s}^2 =& -\left(1-\frac{2M\Omega^{-1}-Q^2}{\Omega^{-2}+a^2\cos^2\theta} \right)\dd u^2+2\Omega^{-2}\dd u\, \dd \Omega - \frac{2a\sin^2\theta}{\Omega^{-2}+a^2\cos^2\theta}(2M\Omega-Q^2)\dd u\, \dd \phi \notag\\
    &+ 2a\sin^2\theta\, \dd \Omega\, \dd\phi  +(\Omega^{-2}+a^2\cos^2\theta)\,\dd \theta^2 \notag\\ &+ \sin^2\theta \left(\Omega^{-2}+a^2+\frac{a^2\sin^2\theta}{\Omega^{-2}+a^2\cos^2\theta}(2M\Omega^{-1}-Q^2)\right)\dd \phi^2.
\end{align}
It can  easily be confirmed that the above line element reduces to the Schwarzschild expression if one takes the limits $Q\to 0$ and $a\to 0$. As one can imagine, based on this rather lengthy expression for the line element, the construction of the Newman-Penrose tetrads is more involved than in the Schwarzschild case. The recipe is the same, though, and we encourage the reader to verify that in the end one finds, in the chart  $\{u,\Omega,\theta,\phi\}$, the following null tetrad:
\begin{align}\label{eq:KNnullTetrad}
    \hat{n}^{a} &= \sqrt{2}\,\delta\du{1}{a}\notag\\
    \hat\ell^{a} &= -\frac{\Omega^2}{\sqrt{2}\left(\Omega^{-2}+a^2\cos^2\theta\right)}\left((\Omega^{-2}+a^2)\,\delta\du{0}{a}\,+\,\Omega^2\frac{\Omega^{-2}+a^2+Q^2-2M\Omega^{-1}}{2}\delta\du{1}{a}+a\,\delta\du{3}{a}\right)\notag\\
    \hat{m}^{a} &= \frac{1}{\sqrt{2}(\Omega^{-1}+i\,a\,\cos \theta)}\left(i\,a\,\sin \theta \delta\,\du{0}{a} + \delta\du{2}{a} + \frac{i}{\sin\theta} \delta\du{3}{a}\right).
\end{align}
As a consistency check, observe that for $Q\to 0$ and $a\to 0$ we recover the null tetrad of the Schwarzschild spacetime. Moreover, notice that other choices for the null tetrad are possible which nevertheless satisfy the normalization and cross-normalization properties.

At this point, we need to compute the Weyl tensor. Because of the length and complexity of the resulting expressions, we do not display the individual components here. Rather, we move directly to computing the Newman-Penrose scalars, for which we obtain
\begin{align}
    \hat\Psi_4(u, \Omega,\theta,\phi) &= 0 \notag\\
    \hat\Psi_3(u, \Omega,\theta,\phi) &= 0 \notag\\
    \hat\Psi_2(u, \Omega,\theta,\phi) &= \frac{\Omega^3 \left(M-\left(Q^2+i\,Ma\cos\theta\right)\Omega\right)}{\left(a\cos\theta\,\Omega-i\right)^3\left(a\cos\theta\,\Omega+i\right)}\notag\\
    \hat\Psi_1(u, \Omega,\theta,\phi) &= 0 \notag\\
    \hat\Psi_0(u, \Omega,\theta,\phi) &= 0.
\end{align}
As expected, we find that $\hat{\Psi}_4$ vanishes, while the ``coulombic'' part of the Weyl tensor is non-trivial. We can expand $\hat{\Psi}_2$ in powers of $\Omega$ and we find that the lowest order term is given by
\begin{align}
    \hat{\Psi}_2(u,\Omega, \theta,\phi) &= -M\, \Omega^3 + \O(\Omega^4) &\Longrightarrow &&  \Psi^\circ_2(u, \theta, \phi) &= -M.
\end{align}
This is precisely the same result as for the Schwarzschild spacetime and, moreover, we have confirmed that the Peeling properties are satisfied.

This is not the end of the story, though. As the Kerr-Newman black hole is charged, it also generates an electromagnetic field and we should be able to determine, via the Newman-Penrose formalism, whether it emits electromagnetic radiation. Of course, we expect that the Kerr-Newman black hole simply represents a Coulomb charge in a curved background, but we would like to confirm that via the formalism.

First of all, we need the Maxwell $2$-form for the Kerr-Newman spacetime. In the chart $\{u,\Omega,\theta,\phi\}$, it is given by (see for instance~\cite{Adamo:2016})
\begin{align}
    F_{01} &= -F_{10} = -\frac{Q\left(1-a^2\cos^2\theta\,\Omega^2\right)}{\left(1+a^2\cos^2\theta\,\Omega^2\right)^2} \notag\\
    F_{02} &= -F_{20} = -\frac{2 a^2 Q \, \cos\theta\,\sin\theta\, \Omega^3}{\left(1+a^2\cos^2\theta\,\Omega^2\right)^2} \notag\\
    F_{13} &= -F_{31} = -\frac{a\, Q\,\sin^2\theta\left(1-a^2\cos^2\theta\, \Omega^2\right)}{\left(1+a^2\cos^2\theta\,\Omega^2\right)^2}\notag\\
    F_{23} &= -F_{32} = -\frac{2a\, Q\, \cos\theta\,\sin\theta\left(1+a^2\Omega^2\right)\Omega}{\left(1+a^2\cos^2\theta\,\Omega^2\right)^2}.
\end{align}
To convince oneself that this is the correct expression for the Maxwell $2$-form, one can use symbolic manipulation software, such as \texttt{Mathematica}, to verify that (a) the trace of $T_{ab} := g^{cd}F_{ac}F_{bd}-\frac14 g_{ab}F_{cd} F_{ef} g^{ce} g^{df}$ vanishes, as it should, and that (b) the equation $G_{ab} = \kappa\, T_{ab}$ is satisfied. 

Using the null tetrad~\eqref{eq:KNnullTetrad}, we can calculate the physical Newman-Penrose scalars $\hat{\Phi}_i$ for the electromagnetic field. We obtain
\begin{align}
    \hat\Phi_2(u,\Omega,\theta,\phi) & =0 \notag\\
    \hat\Phi_1(u,\Omega,\theta,\phi) &= \frac{Q\,\Omega^2}{2 (i - a \cos
\theta\, \Omega)^2}\notag\\
    \hat\Phi_0(u,\Omega,\theta,\phi) &= 0.
\end{align}
Since $\hat{\Phi}_2=0$, we can confirm that there is no electromagnetic radiation. By expanding $\hat{\Phi}_1$ in $\Omega$ around $\Omega = 0$, we find 
\begin{align}
    \hat{\Phi}_1(u,\Omega,\theta,\phi) &= -\frac{Q}{2} \Omega^2 + \O(\Omega^3) & \Longrightarrow&& \Phi^\circ_1(u,\theta,\phi) &= -\frac{Q}{2}.
\end{align}
This is in perfect agreement with the Peeling Theorem and, moreover, we find exactly the same result we found in~\ref{Chap2:Examples} for the Coulomb charge. This further confirms our intuition that the Kerr-Newman black hole describes a Coulomb charge in a curved background. Indeed, we find for the total charge in the spacetime, unsurprisingly,
\begin{align}
    \text{Total charge} = -\frac{1}{2\pi} \oint_{\bbS^2 \text{ on }\scrip}\Re{\Phi^\circ_1}\sin\theta\,\dd\theta\,\dd\phi = Q.
\end{align}
To summarize, in this example we have illustrated that the Kerr-Newman family of spacetimes 
\begin{itemize}
    \item[a)] obey the Peeling properties of the gravitational field;
    \item[b)] also obey the Peeling properties of the electromagnetic field;
    \item[c)] has a non-vanishing ``coulombic'' part of the gravitational Newman-Penrose scalars which encodes the mass of the black hole and it has a non-vanishing coulombic part of the electromagnetic Newman-Penrose scalars which encode the charge of the black hole.
\end{itemize}
\bigskip

\subsection{Extracting Physics}   
So far we have seen that, with a little more effort than for electromagnetism, the Peeling Theorem also holds for GR. This theorem tells us at what rates the different Weyl scalars decay as one approaches~$\scrip$. This theorem also confirms our intuition, that the gravitational field becomes asymptotically flat and only differs by terms of order $\O(r^{-1})$ from the Minkowski metric.

Now we would like to go a step further and extract some actual physics. For instance, we know that the coalescence of compact objects is caused by a loss of energy due to gravitational waves. We also know that when two compact objects merge, the sudden emission of gravitational waves can cause a ``kick''. In principle, such a ``kick'' could eject the remnant of the merger from the galaxy. From observations we can also learn that the rest mass of the remnant is less than the rest masses of the bodies which coalesced. Supposedly, the difference in mass was converted into energy and radiated away by gravitational waves.

In all these examples we made implicit or explicit use of some notion of energy and momentum. Not just of material bodies, but of the gravitational field itself. Or, if not about the gravitational field in general, then about gravitational waves which have traveled far from their source. Can we make these implicit notions mathematically precise? 

To answer the question, we will again take inspiration from Maxwell's theory. In that case, spacetime symmetries play an important role for the definition of energy and momentum of the field. More precisely, we make use of the Poincar\'{e} group, which is generated by time translation, spatial translations, rotations, and boosts. What we call energy-momentum and angular momentum of the field are then quantities associated with the invariance of the action functional under spacetime translations and rotations. Alternatively, we can also defined energy-momentum and angular momentum as the generators of spacetime translations and rotations.

The reason we can make use of the Poincar\'{e} group is Noether's theorem and the fixed background metric. Having a fixed metric provides us with a universal background structure which is present in all special relativistic theories (think of Klein-Gordon fields, Dirac fields, Yang-Mills fields, etc.). Ultimately, it is the invariance of the background structure under certain coordinate transformations which determines the symmetry (or isometry) group. To be more precise, the Poincar\'{e} group is formed by generators $\mathfrak p^{a}$, which satisfy $\lie_{\mathfrak p}\eta_{ab} = 0$. The latter is the mathematical statement that the background structure (here, the Minkowski metric) is invariant under transformations generated by $\mathfrak p^{a}$.

Our strategy for GR is to emulate what we learned from electrodynamics: We start by looking for a universal structure which is common to all asymptotically Minkowski spacetimes -- this is the class of spacetimes we are interested in. That is, we look for a geometric structure which is common to all these spacetimes and then we ask which generators of ``infinitesimal'' coordinate transformations leave this structure invariant.

Of course, in GR we do not have a preferred subgroup of the diffeomorphism group in general. However, we are not interested in generic spacetimes. Rather, we only consider the sector of GR which consists of asymptotically Minkowski spacetimes. For this sector, we can reasonably expect the existence of a universal structure. The intuition is quite simple: At $\scrip$, the spacetime metric tends to the Minkowski metric, which is the universal background structure we seek. However, we will see that because the asymptotic Minkowski metric can be obtained in many different ways, the isometry group will \textit{not} be the Poincar\'{e} group, as one may intuitively expect, but a much larger group which is known as the Bondi-Metzner-Sachs (BMS) group.

Our task in the remainder of this chapter is to determine the universal structure. The BMS group will then be the subject of the next chapter.\bigskip

\subsection{Universal Structure of Asymptotically Minkowski Spacetimes}\label{susec:UniversalStructure} 
\begin{figure}[htb!]
     \centering
     \begin{subfigure}[b]{0.45\textwidth}
         \centering
         \includegraphics[width=\textwidth]{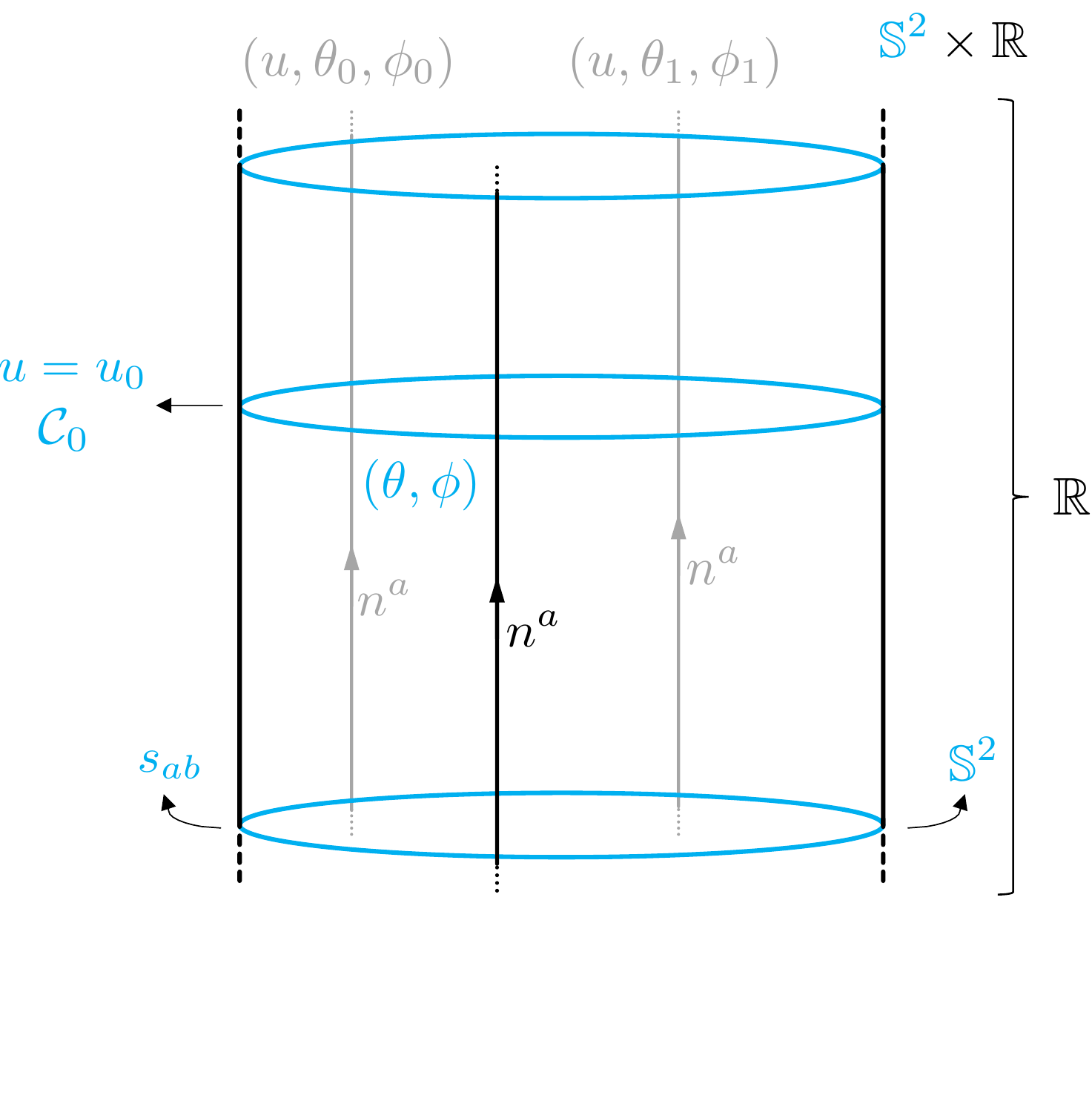}
         \caption{Null infinity as a cylinder}
         \label{fig:ScriAsCylinder}
     \end{subfigure}
     \hfill
     \begin{subfigure}[b]{0.45\textwidth}
         \centering
         \includegraphics[width=\textwidth]{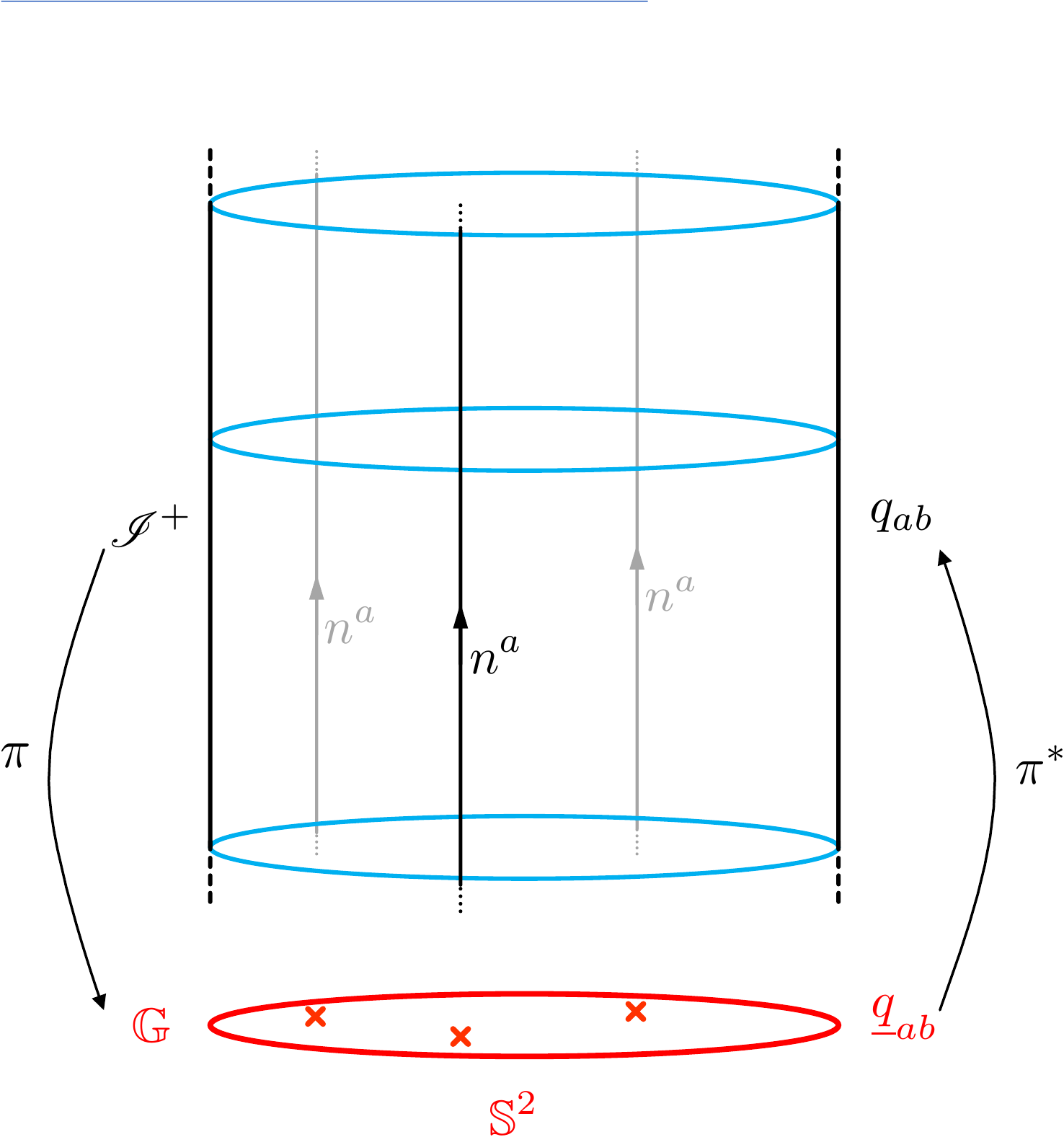}
         \caption{The space of generators}
         \label{fig:BaseSpaceScri}
    \end{subfigure}
	\caption{\textit{Panel (a) shows a visualization of $\scrip$ as a cylinder. Null infinity is ruled by the integral lines of $n^{a}$, which are of the form $(u, \theta_0, \phi_0)$. Each cross-section $\C$ of constant $u$ has the topology of $\bbS^2$ and is equipped with the same metric $s_{ab}$. Panel (b) shows the projection from $\scrip$ to the space of generators $\bbG$. This space has the topology $\bbS^2$ and each integral line of $n^{a}$ is projected to a point in $\bbG$. There is a unique, non-degenerate metric $\underline{q}_{ab}$ on $\bbG$, which carries the same information as the degenerate metric $q_{ab}$ and whose pullback to $\scrip$ generates $q_{ab}$.}}
	\label{fig:ScriAndBaseSpace}
\end{figure}

What we seek is a structure which is common to all spacetimes under consideration. Since we only consider the sector of asymptotically Minkowski spacetimes, we can take the following properties for granted:
\begin{enumerate}
	\item The spacetime admits a boundary, which is $\scri$. This boundary has the topology $\bbS^2\times\bbR$.
	\item The boundary is a null $3$-manifold described by $(q_{ab},n^{a})$, where $n^{a}$ is the null normal to $\scri$, $q_{ab}$ is the degenerate metric of signature $(0,+,+)$, and $n^{a}$ is also the null direction on $\scri$, i.e., $q_{ab}n^{b} \equalhat 0$.
	\item We can work in a divergence-free conformal frame. This means we can make use of the property $\nabla_a n_b \equalhat 0$, which in turn implies $\lie_n g_{ab} \equalhat 0$. Pulling back the latter equation to $\scri$ results in $\lie_n q_{ab} \equalhat 0$. This is a property which is intrinsic to $\scri$.  
\end{enumerate}
What these properties tell us is the following: First of all, we can think of $\scrip$ as being a cylinder (cf. Figure~\ref{fig:ScriAsCylinder}), which is ``ruled'' by null geodesics of the form\footnote{This means there is one line, parametrized by $u\in\bbR$, for each value of $(\theta_0, \phi_0)\in[0,\pi]\times[0,2\pi)$.} $(u,\theta_0, \phi_0)$. These null geodesics are the integral curves of the null normal $n^{a}$ (see again Figure~\ref{fig:ScriAsCylinder}).
On this cylinder, we have an intrinsic (degenerate) metric $q_{ab}$ which is preserved under the Lie drag along~$n^{a}$; $\lie_n q_{ab} \equalhat 0$.  This means that the intrinsic metric does not change in the $n^{a}$ direction (i.e., along the $u$ coordinate) and therefore~$q_{ab}$ can only depend on $\theta$ and $\phi$. 

Furthermore, every cross-sections $\C$ of $\scrip$, determined by $u=$const.,  has the topology~$\bbS^2$. Each cross-section is equipped with a metric $s_{ab}$, which is obtained by pulling back~$q_{ab}$ from $\scrip$ to~$\C$,
\begin{equation}
    s_{ab}:=\pb{q_{ab}}.
\end{equation}
Because $q_{ab}$ is degenerate, there is at least one null vector.\footnote{A null vector in the sense of linear algebra: A vector $v\neq 0$ which satisfies $M v = 0$, where $M$ is a square matrix. In other words, an eigenvector with eigenvalue zero.} This null vector is $n^{a}$ because we know from point~2 that $q_{ab} n^{b}\equalhat 0$. Moreover, the metric $q_{ab}$ does not change as we move in the vertical direction of the cylinder shown in Figure~\ref{fig:ScriAndBaseSpace}, i.e., when we move along $n^{a}$. That is because $\lie_n q_{ab}\equalhat0$. From this we learn that $s_{ab}$ looks exactly the same on each cross-section $\C$ and that all metric information is actually contained in~$s_{ab}$. It also follows thatthe metric $q_{ab}$ is necessarily of the form
\begin{equation}
	q_{ab} = \begin{pmatrix}
		0 & 0 & 0\\
		0 & s_{22} & s_{23}\\
		0 & s_{23} & s_{33}
	\end{pmatrix}
\end{equation}
with respect to the basis $\{n^{a}, m^{a}, \bar{m}^{a}\}$. Also, $s_{ab}$ is only a function of the coordinates in the plane perpendicular to $n^{a}$. Since $q_{ab}$ does not change in the direction $n^{a}$ (it only changes in the two directions perpendicular to $n^{a}$), we can imagine that on $\scrip$, we are stacking one metric $s_{ab}$ on top of another, since on each cross-section $\C$ we have the same metric. This is a highly redundant description of the intrinsic geometry of~$\scrip$! 

Therefore, it makes sense to introduce the \textbf{projector} $\pi:\scrip\to \bbG$, as illustrated in Figure~\ref{fig:BaseSpaceScri}. The space~$\bbG$ is called the \textbf{base space} or the \textbf{space of orbits} or \textbf{the space of generators} -- the terminology in the literature varies, but it all means the same. Namely, the projector maps every generator (i.e., every vertical line of the cylinder) of $\scrip$ to a single point in $\bbG$ (see Figure~\ref{fig:BaseSpaceScri}). Since every generator is mapped to a \textit{different} point in $\bbG$ (meaning that $\pi$ is \textit{injective}), we can think of every point in $\bbG$ as representing precisely one generator. Therefore, in these notes, we refer to it as the \textbf{space of generators}. Let us also note that because $\scrip$ has the topology $\bbS^2\times \bbR$, the space of generators necessarily has the topology of $\bbS^2$, since we project along the $\bbR$ direction. Moreover, the projector $\pi$ allows us to introduce a metric on $\bbG$, which is free of all the redundancies which plague $q_{ab}$. In fact, the metric $\underline{q}_{ab}$ on $\bbG$ is implicitly defined via a pullback induced by $\pi$:
\begin{equation}
    \pi^* \underline{q}_{ab} := q_{ab}.
\end{equation}
We summarize the situation in simple words: The projector ``collapses'' the infinite tower of stacked-up cross-sections $\C$, which are all equipped with the same metric, into a single space. This space has the topology $\bbS^2$ and it is equipped with a unique, non-degenerate metric, $\underline{q}_{ab}$. In a sense, the projector isolates all the information contained in $q_{ab}$, thus providing us with the simplest possible description of that information.

What we have described in this subsection so far is the universal structure of $\scrip$. To be more precise, we $(q_{ab}, n^{a})$ \textit{in a given divergence-free conformal frame} the \textbf{universal structure} of $\scrip$. However, recall from Chapter~\ref{Chap3} that in a conformal frame we have the freedom of perform a rescaling transformation of the form $\Omega\mapsto \Omega' = \omega\, \Omega$, where $\omega$ is a smooth, nowhere zero function with $\lie_n \omega \equalhat 0$. Under this rescaling, the intrinsic metric and the null normal transform as
\begin{equation}
    \boxed{
    \begin{aligned}
        q'_{ab} &= \omega^2\, q_{ab} & \text{and}&& n'^{a} &= \omega^{-1} n^{a}
    \end{aligned}
    }
\end{equation}
This transformation does not affect any aspect of the universal structure we have discussed. Thus, the pair $(q'_{ab}, n'^{a})$ is an equally admissible description of the universal structure, provided that $(q_{ab}, n^{a})\mapsto (q'_{ab} = \omega^2\, q_{ab}, n'^{a} = \omega^{-1}\, n^{a})$. In other words, the universal structure is described by an equivalence class of pairs $(q_{ab}, n^{a})$ under the conformal rescaling operation discussed here. 

In the next chapter, we will see that this universal structure allows us to introduce an asymptotic symmetry group for the class of asymptotically Minkowski spacetimes.\bigskip

\newpage
\subsection{Exercises}
\renewcommand{\thesection}{\arabic{section}}

\begin{Exercise}[]\label{ex:PeelingForGR}
    Use the asymptotic Weyl tensor and the rescaling relations of the null tetrad to derive the Peeling properties of the physical Newman-Penrose scalars $\hat{\Psi}_i$.
\end{Exercise}

\begin{Exercise}[]\label{ex:NPScalarsAndStrain}
    In this exercise, we glimpse at what expects us toward the end of these notes, namely the relation between $\Psi_4=C_{abcd}n^a\bar{m}^bn^c\bar{m}^d$ and the gravitational wave strain fields $h_\times$ and $h_+$. We will not go into detail here (which would be massive spoiling), however keeping this exercise in our minds is helpful for understanding the overall guideline of these notes. \\
    
    Consider the perturbed metric $g_{ab}=\eta_{ab}+h_{ab}$, where $\eta_{ab}$ is the Minkowski metric and $|h_{ab}|\ll 1$ are small perturbations $h_{ab}$. Show the following by direct calculations.
    \begin{itemize}
        \item[a)] The linearized Riemann tensor is given by
        \begin{align*}
            R_{abcd}=\frac12\left(\partial_c\partial_bh_{ad}+\partial_a\partial_dh_{bc}-\partial_b\partial_dh_{ac} -\partial_c\partial_ah_{bd}  \right).
        \end{align*}
        \item[b)] Compute the Weyl tensor, which is defined as
        \begin{align*}
            C_{abcd} = R_{abcd} - \frac{1}{2} \left(R_{ac} g_{bd} - R_{ad} g_{bc} + R_{bd} g_{ac} - R_{bc} g_{ad}\right) + \frac{1}{6} \left(g_{ac} g_{bd} - g_{ad} g_{bc} \right) R.
        \end{align*}
        \textit{Hint: Use the Einstein field equations to simplify the Ricci part.}
        \item[c)] Using the Newman-Penrose null tetrad
    \begin{align*}
        &\ell^a = \frac{1}{\sqrt{2}}\left(\hat{t}+\hat{r}\right) && n^a = \frac{1}{\sqrt{2}}\left(\hat{t}-\hat{r}\right) & m^a = \frac{1}{\sqrt{2}}\left(\hat{\theta}+i\hat{\phi}\right)
    \end{align*}
    and the transverse-traceless gauge for $h_{ab}$, (i.e., $h_{0a} = 0$ and $h_{\theta\theta} = -h_{\phi\phi}$), show that
    \begin{align*}
        \Psi_4=\frac{1}{2}\left(\Ddot{h}_{\hat{\theta}\hat{\theta}}-\Ddot{h}_{\hat{\phi}\hat{\phi}} \right)+ i\Ddot{h}_{\hat{\theta}\hat{\phi}} =: \Ddot{h}_\times-i\Ddot{h}_+,
    \end{align*}
    \textit{Hint: For simplicity, assume that the propagation of the gravitational wave is in the $\hat{r}$ direction, i.e., $h_{ra} = 0$.}
    \end{itemize}
\end{Exercise}

\newpage
\asection{5}{The Bondi-Metzner-Sachs Group}\label{Chap5}
The universal structure of asymptotically Minkowskian spacetimes provides us with a compact mathematical description of the most fundamental attributes of this class of spacetimes. A natural first step is to find all transformations which leave this universal structure invariant. In other words, we are seeking the asymptotic isometry group of spacetime metrics which are asymptotically Minkowskian. Na\"ively, we would expect to recover the Poincar\'e group, i.e., the isometry group of Minkowski space. However, we will see that this is not the case. The class of asymptotically Minkowski spacetimes admits a \textit{larger} group which contains the Poincar\'e group as a subgroup. This enlargement is actually what allows us to account for gravitational radiation. As we will see later on, the asymptotic symmetry group plays a crucial role in the study of conserved charges on $\scrip$ and are therefore of direct relevance for extracting physics.\bigskip

Qualitatively, a symmetry group consists of diffeomorphisms which leave the structure to be studied invariant. In our case, the structure to be preserved is the universal structure of $\scrip$, which is characterized, as we recall, by the following properties:
\begin{itemize}
    \item[1)] $\scrip$ has the topology $\bbS^2\times \bbR$.
    \item[2)] $\scrip$ is endowed with a degenerate metric $q_{ab}:=\pb{g}_{ab}$ which satisfies $q_{ab}n^{a} \equalhat 0$, where $n^{a}$ is the null normal to $\scrip$.
    \item[3)] Moreover, in a divergence free conformal frame,\footnote{Fixing a conformal frame is akin to fixing a gauge. It is not a necessary step and everything can be done without choosing a special conformal frame, but this choice drastically simplifies computations.} the degenerate metric satisfies $\lie_n q_{ab} \equalhat 0$.
\end{itemize}
A pictorial representation of the universal structure is provided by Figure~\ref{fig:ScriAndBaseSpace}. However, recall that working in a divergence-free conformal frame does not completely fix the conformal completion. There is a residual rescaling freedom and any given spacetime can be described by an equivalence class of conformal completions. That is, the pairs $(q_{ab}, n^{a})$ and $(q'_{ab}, n'^{a})$ describe the same universal structure of $\scrip$, provided they are related to each other as $(q'_{ab}, n'^{a}) = (\omega^2 q_{ab}, \omega^{-1}\,n^{a})$, where $\omega$ has to satisfy $\lie_n \omega \equalhat 0$ and $\omega \neq 0$, and where the two conformal factors are related by $\Omega' = \omega\,\Omega$.

Hence, the asymptotic symmetry group has to be defined in the following way: A diffeomorphism $d\in\textsf{Diff}(\scrip)$ is element of the \textbf{BMS group} $\B$ if there exists some $\omega$ with $\lie_n\omega = 0$ such that
\begin{align}\label{eq:DiffeosOfBMS}
    d(q_{ab}) &\equalhat \omega^2 q_{ab}\notag\\
    d(n^{a}) &\equalhat \omega^{-1}\, n^{a}.
\end{align}
Put differently, the symmetry group of the universal structure of $\scrip$ consists of all diffeomorphisms $d$ which do not necessarily leave the pair $(q_{ab}, n^{a})$ invariant (as this would be too restrictive), but which map  $(q_{ab}, n^{a})$ to another pair which generates \textit{the same} structure.

\begin{figure}[!tb]
\centering
\includegraphics[scale=1.4]{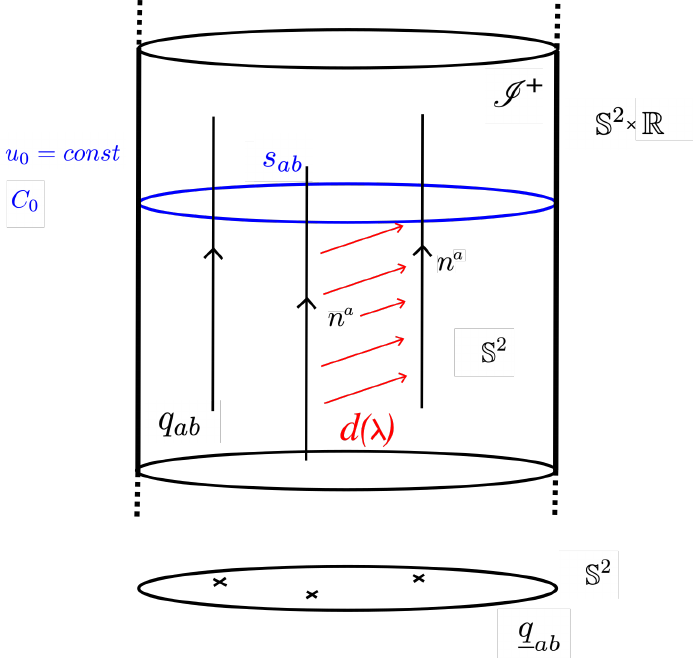}
    \caption{\textit{Visualization of the diffeomorphism which maps one integral line to an other one. This diffeomorphism clearly preserves the ruling of $\scrip$ and thus its universal structure.}}
\label{fig:DiffeomorphismsBMS}
\end{figure}

As an example, consider the diffeomorphism $d$ which maps one generator of $\scrip$ (i.e., an integral line) to another generator, as shown in Figure~\ref{fig:DiffeomorphismsBMS}. Notice that, because of the second equation in~\eqref{eq:DiffeosOfBMS}, $n^{a}$ is mapped to a new vector proportional to $n^{a}$ and hence the ruling of $\scrip$ is preserved! 

The asymptotic symmetry group $\B$ is known as the \textbf{BMS group}, which is short for \textbf{Bondi-Metzner-Sachs group}. As alluded to in the introduction to this section, the BMS group is \textit{larger} than the Poincar\'e group, which we would na\"ively have expected to appear as the symmetry group of asymptotically Minkowski spacetimes, but instead it contains the Poincar\'e group as a subgroup. This fact is not at all obvious at this point, so let us investigate the structure of the BMS group and explicitly show its relation to the Poincar\'e group.

Note that in what follows, we do not write '$\equalhat$' explicitly anymore as the computations which follow are only meaningful on $\scrip$ itself. We also adopt the following notation:
\begin{align}
    \B&:\,\text{Symmetry group which preserves universal structure of $\scrip$ (BMS group)}\notag\\
    d(\lambda)&:\,\text{$1$-parameter family of diffeomorphisms which belong to $\B$ and with $d(0) = \textsf{id}$.}\notag
\end{align}
Our next task is to study the ``infinitesimal'' generators of the BMS group. Recall that every $1$-parameter family of diffeomorphisms can be generated by a vector field. Let $b := b^{a}\partial_a := \left.\frac{\dd}{\dd\lambda}d(\lambda)\right|_{\lambda = 0}$ be the generating vector field of $d(\lambda)$ on $\scrip$. These are the ``infinitesimal'' generators we seek to better understand. Moreover, observe that under a $1$-parameter family of diffeomorphisms $d(\lambda)$, the metric and the normal vector change according to\footnote{A comment on notation: The function $\omega$ depends on the coordinates $\theta$, $\phi$ and parameter $\lambda$, i.e., $\omega(\theta,\phi,\lambda)$. We suppress the dependence on the angles and only display the dependence on the parameter $\lambda$, which is induced by the $1$-parameter family of diffeomorphisms, $d(\lambda)$.}
\begin{align}\label{eq:ActionOfd}
    \pb{q}_{ab} &:= d(\lambda)^* q_{ab} = \omega(\lambda)^2\, q_{ab}\notag\\
    \pf{n}^{a} &:= d(\lambda)_* n^{a} = \omega(\lambda)^{-1}\, n^{a}.
\end{align}
Here, $d(\lambda)^*$ and $d(\lambda)_*$ denote the pullback and pushforward operations, respectively, and the right hand side of~\eqref{eq:ActionOfd} follows from the definition of the BMS group. Hence, if we wish to study the generators $b^{a}$, we need to compare $\pb{q}_{ab}$ and $\pf{n}^{a}$ to $q_{ab}$ and $n^{a}$, respectively, for $\lambda$ close to zero (which translates into $d(\lambda)$ being close to the identity). To that end, it is convenient to introduce
\begin{equation}
	\alpha(\theta, \phi) := \left.\frac{\dd}{\dd \lambda}\omega(\lambda)\right|_{\lambda = 0},
\end{equation}
where it follows from $\lie_n\omega \equalhat 0$ that $\alpha$ is only a function of $(\theta, \phi)$ on $\scrip$. The ``infinitesimal'' action of $d(\lambda)$ on the metric is then defined as
\begin{align}
    \underset{= \lie_{b}q_{ab}}{\underbrace{\left.\frac{\dd}{\dd\lambda}\left(d(\lambda)^*q-q\right)_{ab}\right|_{\lambda = 0}}} &= \underset{= 2\alpha\, q_{ab}}{\underbrace{\left.\frac{\dd}{\dd\lambda}\left(\omega(\lambda)^2\, q-q\right)_{ab}\right|_{\lambda = 0}}},
\end{align}
where we have used the definition of the Lie derivative. Similarly, for the normal vector one finds
\begin{equation}
    \underset{= \lie_{b}n^{a}}{\underbrace{\left.\frac{\dd}{\dd\lambda}\left(d(\lambda)_* n-n\right)^{a}\right|_{\lambda=0}}} = \underset{= - \alpha\, n^{a}}{\underbrace{\left.\frac{\dd}{\dd\lambda}\left(\omega(\lambda)^{-1}\, n-n\right)^{a}\right|_{\lambda=0}}}.
\end{equation}
Thus, the ``infinitesimal'' diffeomorphisms of the BMS group are characterized as being generated by vector fields $b$ which satisfy the conditions
\begin{equation}\label{eq:BMSAlgebraConditions}
\boxed{\begin{aligned}
	\lie_b q_{ab} &= 2 \alpha\, q_{ab}, \\
	\lie_b n^{a} &= -\alpha\, n^{a},
\end{aligned}}
\end{equation}
for some function $\alpha = \alpha(\theta, \phi)$. Let us denote the set of all vectors $b$ which satisfy the above conditions by $\mathfrak{b}$ (fraktur b) and the Lie bracket of vector fields by $[\cdot, \cdot]$. Is $(\mathfrak b, [\cdot, \cdot], \bbR)$ a Lie algebra?

The answer is in the affirmative. All we need to show is that $\mathfrak b$ is a real vector space and that $[\cdot, \cdot]:\mathfrak b\times\mathfrak b\to \mathfrak b$, i.e., that the Lie bracket maps any two vectors from $\mathfrak b$ to some other vector in $\mathfrak b$. To show that $\mathfrak b$ is a real vector space, all we need to show is that if $b_1, b_2\in\mathfrak{b}$, then it follows that $\lambda_1 b_1 + \lambda_2 b_2\in\mathfrak b$ for all $\lambda_1,\lambda_2\in\bbR$. Indeed, we easily find that
\begin{align}
    \lie_{\lambda_1 b_1 + \lambda_2 b_2}q_{ab} &= \alpha\,q_{ab}\notag\\
    \lie_{\lambda_1 b_1 + \lambda_2 b_2}n^{a} &= -\alpha\,n^{a},
\end{align}
with $\alpha = \alpha_1\lambda_1 + \alpha_2\lambda_2$. Thus, $\mathfrak b$ is a real vector space. Next, we need to show that the Lie bracket maps vectors from $\mathfrak b$ on vectors in $\mathfrak b$, i.e., that it is an endomorphism on $\mathfrak b$. This translates into the question whether $[b_1, b_2]$ satisfies the conditions~\eqref{eq:BMSAlgebraConditions}, provided $b_1$ and $b_2$ satisfy them. Using $\lie_{[X,Y]}T = \lie_X\lie_Y T - \lie_Y\lie_X T$, which holds for any vector fields $X$, $Y$ and any tensor field $T$, we immediately find
\begin{align}
    \lie_{[b_1, b_2]}q_{ab} &= 2\alpha\,q_{ab}\notag\\
    \lie_{[b_1, b_2]}n^{a} &= -\alpha\, n^{a},
\end{align}
with $\alpha = \lie_{b_1}\alpha_2-\lie_{b_2}\alpha_1$. Hence, we have verified that $[\cdot,\cdot]:\mathfrak b\times\mathfrak b\to\mathfrak b$ and we finally conclude that $(\mathfrak b, [\cdot,\cdot], \bbR)$ forms a Lie algebra. Our next task is to better understand the structure of this Lie algebra and to provide an interpretation of the transformations it generates.\bigskip

\subsection{Supertranslations}
We have derived the abstract conditions~\eqref{eq:BMSAlgebraConditions}, which characterize the ``infinitesimal'' generators $b$, and we have shown that the set of these vectors, $\mathfrak b$, forms a real Lie algebra with respect to the Lie bracket $[\cdot,\cdot]$. Can we deduce a more explicit form for these generators? The answer is yes. An educated guess is that some of these generators are given by $b^{a} = \beta\, n^{a}$, where $\beta = \beta(u, \theta, \phi)$ is some function on~$\scrip$. This guess is motivated by the following observations: Minkowski space possesses four global Killing vector fields which correspond to spacetime translations and whose limit to $\scrip$ has the form $\beta(\theta,\phi)\, n^{a}$ (see Chapter~\ref{Chap2} equations~\eqref{eq:TranslationsOnScri} and see also Side Note~\ref{sn:TimetranslationBecomingNull}). Hence, these fields are also symmetries of $\scrip$ and necessarily satisfy the conditions~\eqref{eq:BMSAlgebraConditions}. More generally, we know that the universal structure of $\scrip$ demands that $\lie_n q_{ab} \equalhat 0$. Hence, $b^{a} = \beta\, n^{a}$ is a good candidate for an ``infinitesimal'' symmetry generator. Indeed, by direct computation we find that $b^{a} = \beta\, n^{a}$ satisfies the first condition in~\eqref{eq:BMSAlgebraConditions}:
\begin{align}
	\lie_{\beta n} q_{ab} &=  \beta\lie_n q_{ab}  + q_{ac}n^{c}\left(\D_{b} \beta\right) + q_{bc}n^{c}\left(\D_{a} \beta\right) = 0,
\end{align}
where we have used that the universal structure of $\scrip$ demands that $\lie_n q_{ab} = 0$ and $q_{ab}n^{a} = 0$. Notice also that $\D$ is an arbitrary\footnote{We cannot use the Levi-Civita connection on $\scrip$ because the metric $q_{ab}$ is degenerate and hence the Levi-Civita connection, which requires the inverse of $q_{ab}$, is not defined. Hence, there is not a canonical choice of connection on~$\scrip$, but any choice is admissible here since neither the covariant derivative of scalars nor the Lie derivative depend on the connection.} covariant derivative operator on $\scrip$. It follows that $b^{a} = \beta\, n^{a}$ satisfies the first condition in~\eqref{eq:BMSAlgebraConditions} with $\alpha = 0$, in agreement with our educated guess.

Next, let us consider the second condition in~\eqref{eq:BMSAlgebraConditions}, using the fact that $\alpha = 0$, which we have just derived. A direct computation yields
\begin{align}
	\lie_{\beta n} n^{a} &= \beta n^b\D_bn^a-n^c\D_c\left(\beta n^a\right) \notag\\
	&= \beta n^b \D_b n^a- n^cn^a\D_c \beta -n^c\beta \D_c n^a\notag\\
	&=-n^cn^a\D_c\beta \overset{!}{=} 0.
\end{align}
We end up with the condition that $n^{a}\D_a \beta \overset{!}{=} 0$, which obviously tells us that $\beta$ is independent of $u$. Thus, we have shown that $b^{a} = \beta\, n^{a}$, with $\beta = \beta(\theta, \phi)$, is an ``infinitesimal'' generator of the BMS group. In fact, we have found infinitely many generators since $\beta$ is an arbitrary function. This makes $\mathfrak b$ an \textit{infinite-dimensional} Lie algebra. 

Let us consider what happens if we change the conformal frame. That is, we assume that $(q_{ab}, n^{a})$ satisfy the conditions~\eqref{eq:BMSAlgebraConditions} for some $b^{a} = \beta\, n^{a}$ and then we perform a conformal rescaling $(q_{ab}, n^{a}) \mapsto (q'_{ab}, n'^{a}) = (\omega^2 q_{ab}, \omega^{-1} n^{a})$. It is easy to show that $b'^{a} = \beta'\, n'^{a}$ is an ``infinitesimal'' generator for the conformally rescaled pair $(q'_{ab}, n'^{a})$. However, if we want our considerations to be \textit{independent} of the conformal frame we are operating in, we need to demand that
\begin{align}
    \beta\, n^{a} = b^{a} \overset{!}{=} b' = \beta'\, n'^{a} = \beta'(\omega^{-1}n^a),
\end{align}
which implies that $\beta' = \omega\beta$. In other words, $\beta$ is not a ``true'' function. Rather, it is a scalar quantity which, under the residual conformal rescaling in a divergence-free conformal frame, transforms as $\beta' = \omega\, \beta$. We say that $\beta$ has \textbf{conformal weight} $\mathbf{+1}$, while $n^{a}$ is said to have \textbf{conformal weight} $\mathbf{-1}$. This ensures that $b'^{a} = b^{a}$, i.e., that the symmetry generators are invariant under conformal rescaling.\bigskip

Let us continue exploring the generators of the form $b^{a} = \beta\, n^{a}$. We already know that $\mathfrak{b}$, the set of \textit{all} generators, is a Lie algebra. It is thus natural to ask whether the set $\mathfrak s$ (fraktur s) of all $b^{a} = \beta\, n^{a}$ forms a subalgebra of $\mathfrak b$. Let us recall that a subalgebra $\mathfrak s\subseteq \mathfrak b$ is a subspace (in the sense of vector spaces), which is closed under the action of the Lie bracket. The latter requirement means that for all $s_1,s_2\in\mathfrak s$, we have $[s_1,s_2]\in\mathfrak s$. 

To emphasize that we are considering a special set of vectors, let us change notation and denote $\beta\, n^{a}$ by $s^{a} := \beta\,n^{a}$, where $\beta=\beta(\theta,\phi)$. Furthermore, let $\mathfrak s$ be the set of all these vectors $s^{a}$. It is easy to see that when $s_1\in\mathfrak s$ and $s_2\in\mathfrak s$, then it follows that $\lambda_1 s_1 + \lambda_2 s_2\in\mathfrak s$ for all $\lambda_1,\lambda_2\in\bbR$. Hence, $\mathfrak s$ forms a vector space and this is a subspace of the vector space $\mathfrak b$. For the Lie bracket, we find for $s_1, s_2\in\mathfrak s$
\begin{align}
    [s_1,s_2]^{a} &= s^{b}_1\D_b s^{a}_2 - s^{b}_2 \D_b s^{a}_1 = \beta_1 n^{b} \D_b(\beta_2 n^{a}) - \beta_2 n^{b} \D_b(\beta_1 n^{a})\notag\\
    &= \underbrace{\beta_1\beta_2 n^{b}\D_b n^{a} - \beta_1 \beta_2 n^{b}\D_b n^{a}}_{=0} + \beta_1 n^{a} \underbrace{n^{b} \D_b \beta_2}_{=0} - \beta_2 n^a \underbrace{n^b\D_b\beta_1}_{=0}\notag\\
    &= 0,
\end{align}
where we have used that $n^{a}\D_{a}\beta = 0$. Because $0\in\mathfrak s$, it follows that $\mathfrak s$ is closed under the action of the Lie bracket. Thus, $\mathbf\mathbf{\mathfrak s}$ \textit{is a subalgebra of} $\mathbf\mathbf{\mathfrak b}$. Moreover, because $[s_1, s_2] = 0$ for all $s_1,s_2\in\mathfrak s$, it is an \textit{abelian Lie algebra}.\bigskip

Let us briefly summarize the situation thus far: The set of ``infinitesimal'' generators of the BMS group, i.e., the vectors $b$ which satisfy~\eqref{eq:BMSAlgebraConditions}, form a real, infinite-dimensional Lie algebra $\mathfrak b$. This Lie algebra admits a real, infinite dimensional abelian Lie subalgebra $\mathfrak s$, which is defined by
\begin{align}
    s^{a}:=\beta\, n^{a}\in\mathfrak s\quad\text{with}\quad \lie_n\beta = 0.
\end{align}
We call this subalgebra the algebra of \textbf{supertranslations} (hence the use of the letters $s$ and $\mathfrak s$). This is motivated by the fact that our educated guess ---that $\beta\, n^{a}$ are ``infinitesimal'' generators--- originated from the observation that the translational Killing vector fields of Minkowski space have the form $\beta(\theta,\phi) n^{a}$ when we take their limit to $\scrip$. Moreover, we also know that the Poincar\'e Lie algebra admits an abelian subalgebra and this algebra is precisely the algebra of spacetime translations. However, there is an important difference: In the Poincar\'e case, the subalgebra is \textit{finite-dimensional}, while we found an \textit{infinite-dimensional} subalgebra in the BMS case. This is the reason why we call it the algebra of \textit{super}translations and we will see that this enlargement of the algebra ---from ``ordinary'' translations to supertranslations--- has important consequences. 

We will shortly see that there are further parallels between the Poincar\'e Lie algebra and the Lie algebra of the BMS group. To make this more precise, let us denote the Lie algebra of the Poincar\'e group by the symbol\footnote{In French, \textit{point} means point and \textit{carr\'e} means square. Put together, \textit{point-carr\'e} sounds like \textit{Poincar\'e} and therefore we use the symbol $\boxdot$ to represent this Lie algebra.}~$\boxdot$, and the subalgebra of translations by $\mathfrak t$ (fraktur t). It turns out that $\mathfrak t$ is not just a subalgebra of $\boxdot$, it is a so-called \textbf{ideal}. This means that the Lie bracket between any element $p\in\boxdot$ and any element $t\in\mathfrak t$ lies again in $\mathfrak t$, i.e., $[p, t]\in\mathfrak t$ for all $p\in\boxdot$ and all $t\in\mathfrak t$. As we will explain in more detail in the next subsection, this property guarantees that the quotient of $\boxdot$ and $\mathfrak t$, i.e., the space $\boxdot/\mathfrak t$, is (a) well-defined and (b) again a Lie algebra. In fact, one finds that
\begin{equation}
    \boxdot/\mathfrak t \simeq \mathfrak l,
\end{equation}
where $\mathfrak l$ (fraktur l) is the six-dimensional Lie algebra of Lorentz transformations (three rotations and three boosts). Given this fact, it is natural to inquire whether the supertranslations $\mathfrak s$ also form an ideal of $\mathfrak b$. This would then allow us to construct the quotient space $\mathfrak b/\mathfrak s$ and investigate its relation with the Lie algebra of Lorentz transformations. To check whether $\mathfrak s$ is an ideal of $\mathfrak b$, we only need to verify that $[b, s]\in\mathfrak s$ for all $b\in\mathfrak b$ and all $s\in\mathfrak s$. A direct computation yields
\begin{align}\label{5.10}
    [b,\beta\, n]^a = \lie_b(\beta n^a) = n^a \lie_b \beta + \beta \underbrace{\lie_b n^a}_{= -\alpha\,n^{a}} = n^a \lie_b \beta - \beta \alpha\, n^a = \underbrace{(\lie_b \beta - \beta \alpha)}_{=:\beta'}\, n^a =: s'^a,
\end{align}
where we used condition~\eqref{eq:BMSAlgebraConditions} to rewrite the Lie derivative of $n^a$ along $b$. We find that the final output of the above computation is a vector in $\mathfrak s$, that is, we find a supertranslation vector again. Hence, the \textit{supertranslations form an ideal of} $\mathfrak b$. In the next subsection, we will briefly review some important mathematical concepts, which will help us in fully appreciating the importance of the result we just derived. Readers familiar with equivalence relations, kernels, Lie algebra homomorphisms, kernels, and quotient groups can skip the mathematical interlude or read it as a reminder of certain definitions.
\bigskip

\subsubsection{Interlude: Important Mathematical Concepts}
Our goal is to understand how the ideal $\mathfrak i$ of a Lie algebra $\mathfrak g$ lead to a well-defined quotient space $\mathfrak g/\mathfrak i$ and how this space has to be understood. To that end, we recall some important mathematical concepts.\bigskip

\underline{\textbf{Equivalence Relations:}}\\[5pt]
Let $S$ be a set. A \textbf{relation} $\sim$ on $S$ is some way of relating elements of $S$ to each other. For instance, if~$S$ is the set of your family members, $x\sim y$ could mean ``$x=y$ ($x$ and $y$ are the same person) or $x$ is a brother or sister or $y$'' and $x\approx y$, a different relation defined on $S$, might mean ``$x$ is the mother of~$y$''. Both $x\sim y$ and $x\approx y$ define a relation on $S$.

Here, we are only interested in a special class of relations. We say that a relation $\sim$ is an \textbf{equivalence relation} if it possesses the following properties for all $x, y, z$ in $S$:
\begin{align*}
    &\textsf{1) } x\sim x &&\textsf{(Reflexivity)}\\
    &\textsf{2) } \textsf{If $x\sim y$, then $y\sim x$} &&\textsf{(Symmetry)}\\
    &\textsf{3) } \textsf{If $x\sim y$ and $y\sim z$, then $x\sim z$} && \textsf{(Transitivity)} &&&& &&&& &&&& &&&& &&&& &&&& &&&& 
\end{align*}
It is easy to check that the relation $x\sim y$ defined above is an equivalence relation, while $x\approx y$ is \textit{not} an equivalence relation.

From here on forward, $\sim$ will \textit{always} stand for an equivalence relation. Once such an equivalence relation~$\sim$ on $S$ has been declared, $S$ can be divided up into \textbf{equivalence classes}. An \textbf{equivalence class} is a set of the form $[x]:=\{y\in S\, |\,  y\sim x\}$ and $x$ in $[x]$ is called the \textbf{representative} of the equivalence class. Notice that if $x\sim y$, then $[x]=[y]$. It follows that (a) we can choose any element of~$[x]$ we want to represent the equivalence class and (b) that two different equivalence classes are always disjoint, i.e., $[x]\cap [y] = \emptyset$ if $x\not\sim y$. Finally, we denote the \textbf{set of all equivalence classes} of $S$ (with respect to the equivalence relation $\sim$) by $S/\sim$ (read $S$ mod tilde). This is also called the \textbf{quotient space}.\bigskip

\newpage

\underline{\textbf{Non-injective maps between sets and the kernel of a map:}}\\[5pt]
Let $S$ and $T$ be sets and define a map $f:S\to T$. Then there is a natural way to define an equivalence relation $\sim$ on $S$: We say $x,y\in S$ are equivalent, $x\sim y$, if and only if $f(x) = f(y)$. Notice that if $f(x) = f(y)$ holds for $x\neq y$, this means that $f$ is \textit{not injective}. In other words, $f$ maps \textit{different} elements of $S$ onto \textit{the same} element in $T$. 

At this point it is convenient to introduce the concept of a \textbf{kernel}, which is defined as the following set:
\begin{equation}
    \ker f := \left\{(x,y)\in S\times S\, | \, f(x) = f(y)\right\}.
\end{equation}
Qualitatively speaking, this set measures to which degree $f$ fails to be an injective map. Also, notice that~$f$ is injective if and only if $\ker f = \{(x,x)\, |\, x\in S\}$.
It is common to say that $\ker f$ \textit{is} an equivalence relation and to denote the set of all equivalence classes of $S$ (with respect to the equivalence relation induced by $\ker f$) as $S/\ker f$. To be more precise, the equivalence relation induced by $\ker f$ is $x\sim y$ if and only if $f(x) = f(y)$, which is equivalent to $x\sim y$ if and only if $(x,y)\in\ker f$.
\bigskip

\underline{\textbf{Lie Algebra Homomorphisms:}}\\[5pt]
Let $(\mathfrak g, [\cdot,\cdot]_{\mathfrak g})$ and $(\mathfrak h,[\cdot,\cdot]_{\mathfrak h})$ be real Lie algebras. A \textit{linear} map $\phi:\mathfrak g\to\mathfrak h$ is called a \textbf{Lie algebra homomorphism} when $[\phi(x), \phi(y)]_{\mathfrak h} = \phi([x,y]_{\mathfrak g})$ for all $x,y\in\mathfrak g$. This requirement ensures that the Lie algebra structure of $\mathfrak g$ is preserved under the map $\phi$.

We are interested in the kernel of $\phi$. So let us assume that $x,y\in\ker\phi$, which simply means that $\phi(x) = \phi(y)$. Because Lie algebra homomorphisms are linear, we find that this can equivalently be written as $\phi(x-y) = 0$. Now observe that $\ker\phi$ is a \textit{vector space}. Hence, it follows that $x-y$ is also an element of $\ker\phi$. More generally, one can show that if $x,y\in \ker\phi$, then $\lambda_1\, x+\lambda_2\, y\in\ker\phi$ as well, for any $\lambda_1,\lambda_2\in\bbR$. But this then implies that $\phi(x) = 0$ for any $x\in\ker\phi$. Hence, we have shown that any element of the kernel is mapped to zero in the space $\mathfrak h$. 

We can draw a further conclusion from the fact that $\phi$ is a Lie algebra homomorphism: Assume that $x,y\in\ker\phi$. Then we obtain from the condition that $\phi$ is a Lie algebra homomorphism
\begin{align}
    \phi([x,y]_{\mathfrak{g}}) = \underbrace{[\phi(x), \phi(y)]_{\mathfrak h}}_{=0\, \textsf{because }\phi(x) = \phi(y)} = 0.
\end{align}
Thus, $[x,y]_{\mathfrak g}$ is also an element of the kernel, provided $x,y\in\ker\phi$. Let us now consider $\phi([x+i, y+i]_{\mathfrak{g}})$, where $x,y\in\mathfrak g$ and where $i\in\ker\phi$. Using the linearity of $\phi$ and the condition that it is a Lie algebra homomorphism, we find
\begin{align}
    \phi([x+i, y+i]_{\mathfrak{g}}) &\underset{\phantom{\textsf{\tiny{homomorphism}}}}{\overset{{\textsf{\tiny{linearity}}}}{=}} \phi([x,y]_{\mathfrak g}) + \phi([i,y]_{\mathfrak g}) + \phi([x,i]_{\mathfrak g}) \notag\\
    &\overset{\textsf{\tiny{homomorphism}}}{=} [\phi(x+i), \phi(y+i)]_{\mathfrak{h}}\notag\\
    &\underset{\phantom{\textsf{\tiny{homomorphism}}}}{\overset{\textsf{\tiny{linearity}}}{=}} [\phi(x),\phi(y)]_{\mathfrak h} + [\underbrace{\phi(i)}_{=0},\phi(y)]_{\mathfrak h} + [\phi(x),\underbrace{\phi(i)}_{=0}]_{\mathfrak h}.
\end{align}
Because the homomorphism property of $\phi$ implies $\phi([x,y]_{\mathfrak g}) - [\phi(x),\phi(y)]_{\mathfrak h} = 0$, we are finally left with
\begin{equation}
    \phi([i,y]_{\mathfrak g}) + \phi([x,i]_{\mathfrak g}) \overset{!}{=} 0,\qquad\forall x,y\in\mathfrak g.
\end{equation}
Since this condition has to hold for \textit{all} $x,y\in\mathfrak g$, we conclude that $\phi([i,x]_{\mathfrak g}) = 0$ and $\phi([i, y]_{\mathfrak g}) = 0$ separately, which means that both $[x,i]_{\mathfrak g}$ and $[i,y]_{\mathfrak g}$ are elements of the kernel of $\phi$.

So, in conclusion, we find that the kernel of a Lie algebra homomorphism, $\ker\phi$, is a subspace of $\mathfrak g$ which is closed under the action of the Lie bracket, i.e., $[x,y]_{\mathfrak g}$ is also an element of $\ker\phi$, provided $x,y\in\ker\phi$. This means that $\mathbf{\ker\phi}$ \textit{is a subalgebra of} $\mathfrak g$. Moreover, we have seen that $[i,x]_{\mathfrak g}\in\ker\phi$. In other words, the Lie bracket between an element of the kernel of $\phi$ and any element of the Lie algebra $\mathfrak g$ is again an element in the kernel, i.e., it is again an element of the subalgebra $\ker\phi$. This last property finally leads us to the concept of an \textit{ideal} of a Lie algebra.\bigskip

\underline{\textbf{Ideals and Quotients of Lie Algebras:}}\\[5pt]
A subalgebra $\mathfrak i\subseteq \mathfrak g$ of the Lie algebra $\mathfrak g$ is called an \textbf{ideal} if $[x,y]\in\mathfrak i$ for all $x\in\mathfrak g$ and all $y\in\mathfrak i$. This means that the ideal is  invariant under the action of the Lie bracket. Equivalently, we could say that the Lie bracket acts on ideals as $[\cdot,\cdot]:\mathfrak g\times\mathfrak i\to\mathfrak i$.

This definition should ring a bell: We have just seen that the kernel of a Lie algebra homomorphism is a subalgebra and that this subalgebra is invariant under the action of the Lie bracket. In other words, \textit{the kernel $\ker\phi$ is an ideal}!

This raises the following question: Given an ideal $\mathfrak i$ of $\mathfrak g$, is it the kernel of some Lie algebra homomorphism~$\phi$? The answer is in the affirmative. In fact, we will now see that $\mathfrak i$ is the kernel of the quotient map $\phi:\mathfrak g\to \mathfrak g/\mathfrak i$. First of all, two elements $x,y\in\mathfrak g$ are equivalent, $x\sim y$, if and only if $x-y \in\mathfrak i$. Hence, the equivalence classes have the form
\begin{equation}
    [x] = \left\{x+i\,|\, i\in\mathfrak i \right\}.
\end{equation}
This is often denoted as
\begin{equation}
    [x] = x + \mathfrak i.
\end{equation}
Although this is just notation, it makes it immediately obvious that $\mathfrak i$ is the kernel of $\phi$. In fact, we know that $x=0$ is in the kernel of $\phi$ and that $\phi(0) = [0]$. If we take any element $i\in\mathfrak i$, we obtain $\phi(i) = [i] = i+\mathfrak i = \mathfrak i = [0]$. Thus, the ideal $\mathfrak i$ corresponds to the equivalence class $[0]$ and is therefore the kernel of the quotient map $\phi$.

By defining $\lambda [x] = [\lambda x]$ for all $\lambda \in\bbR$ and $[x]+[y] = [x+y]$, we can turn the quotient space into a vector space. Furthermore, we can define $[x+\mathfrak i, y+\mathfrak i]_{\mathfrak g/\mathfrak i} := [x, y]_{\mathfrak g} + \mathfrak i$ for all $x,y\in\mathfrak g$. Thus, we obtain a well-defined Lie algebra on the quotient space $\mathfrak g/\mathfrak i$.
\bigskip

\subsection{Quotient Group of the BMS Group}
Before the mathematical interlude, we saw that the Lie algebra of supertranslations $\mathfrak s$ forms an abelian and ideal subalgebra of the BMS Lie algebra $\mathfrak b$. This is analogous to the Poincar\'e case, where the algebra of translations $\mathfrak t$ is also abelian and ideal. Furthermore, for the Poincar\'e group we obtain that the quotient space is the Lie algebra of Lorentz transformations,
\begin{equation}
    \boxdot/\mathfrak t = \mathfrak l.
\end{equation}
What is the quotient space of the BMS Lie algebra with the algebra of supertranslations? To answer this, consider Figure~\ref{fig:GeneratorsBMS} below. We can define equivalence classes of general elements $b^a$ of the algebra using the prescription $b^a\sim b'^a$ if $b'^a-b^a\in \mathfrak{s}$, or equally $b'^a-b^a= \beta n^a$, for some $\beta$ with $\lie_n\beta=0$. Each equivalence class $[b^a]$ is unambiguously characterized by its projection to the space of generators $\bbG$. As the figure illustrates, two distinct elements $b^{a}_1$, $b^{a}_2$ of an equivalence class $[b^a]$ are projected to the same vector in $\bbG$, i.e., $\underline{b}^a_1=\underline{b}^a_2$. Hence, the projection down to $\bbG$ allows us to find a one-to-one correspondence between $[b^a]$ and the elements $\underline{b}^a$. It follows that $[b^a]\in \mathfrak{b}/\mathfrak{s}$. Intuitively, we can imagine the projection onto $\bbG$ as dividing out supertranslations from the general elements $b^a$ of the BMS Lie algebra.
\begin{figure}[!tb]
\centering
\includegraphics[scale=1.2]{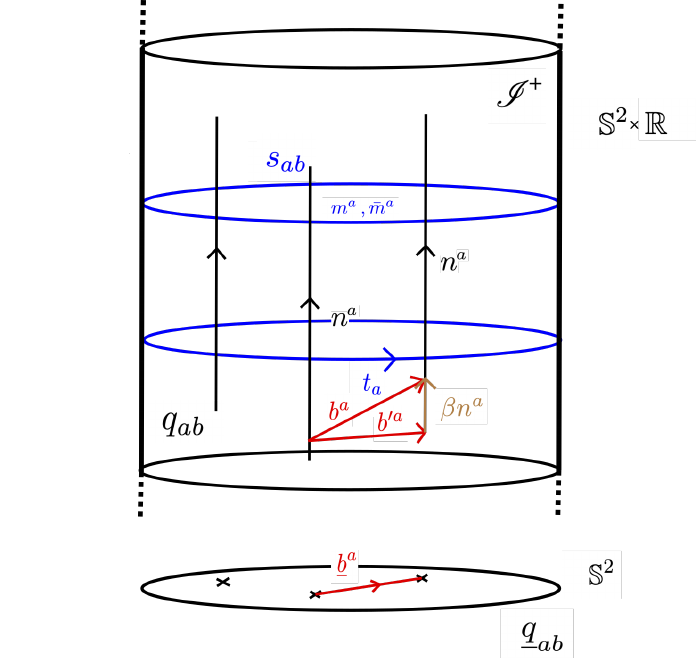}
\caption{\textit{Two vectors $b^{a}$, $b'^{a}$ on $\scrip$ which differ by $\beta(\theta,\phi)n^{a}$ are mapped to the same vector $\underline{b}^{a}$ on the space of generators $\bbG$.}}
\label{fig:GeneratorsBMS}
\end{figure}
The quotient space can be further characterized by projecting down equation~\eqref{eq:BMSAlgebraConditions}. This leads us to
\begin{align}
    \lie_{\underline{b}}\, \underline{s}_{ab}=2\underline{\alpha}\,\underline{s}_{ab},
\end{align}
where $\underline{s}_{ab}$ is the metric on $\bbG$ with signature $(+,+)$. Furthermore, recall that $\bbG$ has the topology of a $2$-sphere. Hence, the above equation is telling us that $\underline{b}$ is a \textit{conformal} Killing vector field of $(\bbG, \underline{s}_{ab})$. In other words, the quotient Lie algebra is the Lie algebra of conformal Killing vector fields on the $2$-sphere. It can be shown that this conformal algebra is isomorphic to the Lorentz Lie algebra $\mathfrak l$. Thus, we obtain
\begin{equation}
    \mathfrak b/\mathfrak s \simeq \mathfrak l.
\end{equation}
This is a remarkable result! Notice that both, the BMS Lie algebra as well as the Lie algebra of supertranslations are infinite-dimensional. By constructing the quotient space of these two algebras, we find the Lie algebra of the Lorentz group, which is finite dimensional. In conclusion, we can say that the Lie algebras of the Poincar\'e and the BMS groups are structurally very similar. However, the translational subalgebra of the BMS group is much larger than its Poincar\'e counterpart. In the next subsection we will investigate this enlargement, which so far is a purely mathematical consequence, using an example in order to gain some intuition. \bigskip

\subsection{The Enlargement of the Poincar\'e Group to the BMS Group}
We have seen that the enlargement of the Poincar\'e group to the BMS group came about as a mathematical consequence of our definition of asymptotically Minkowski spacetimes. This is a surprising result since one would expect that the asymptotic symmetry group of such spacetimes is simply the Poincar\'e group. Is it possible to understand this enlargement on a more qualitative level?

Let us piece together the lessons we have learned so far. In Chapter~\ref{Chap3}, where we first introduced the notion of asymptotically Minkowski spacetimes, we worked out that this definition implies $\hat{C}_{abcd}\equalhat 0$. Due to Einstein's field equations, the Ricci part of the Riemann tensor also vanishes as one approaches~$\scrip$ and therefore the Riemann tensor as a whole is zero for $\Omega\to 0$. 

From this we can conclude that our definition of asymptotically Minkowski spacetimes implies that
\begin{equation}\label{eq:AsymptoticExpansion}
    \hat{g}_{ab} = \hat{\eta}_{ab} + \O(r^{-1})
\end{equation}
in a neighborhood of $\scrip$. Let us have a closer look at the Minkowski metric in this expression. In the chart $\{t, x, y, z\}$, the line element corresponding to this Minkowski metric can be written as
\begin{equation}\label{eq:MinkLinElementEx}
    \dd\hat{s}^2 = -\dd t + \dd x^2 + \dd y^2 + \dd z^2.
\end{equation}
However, we have the freedom to perform angle-dependent translations. So let us consider the transformation
\begin{align}
    t\ &\mapsto\ t' = t + f(\theta, \phi)\notag\\
    x\ &\mapsto\ x' = x\notag\\
    y\ &\mapsto\ y' = y\notag\\
    z\ &\mapsto\ z' = z.
\end{align}
If $f$ was a constant, this would be a simple time-translation and the Minkowski line element~\eqref{eq:MinkLinElementEx} would be invariant. However, we want to consider the case where $f$ is \textit{not} a constant and this leads to a transformed line element. Of course, the argument of $f$ is implicitly a function of the spatial coordinates $\{x,y,z\}$, i.e., $(\theta, \phi) = (\theta(x,y,z), \phi(x,y,z))$. If one takes this into account, one can show that
\begin{align}
    \dd\hat{s}'^2 &= \hat{\eta}'_{ab}\dd x^{a}\, \dd x^{b} =  -\dd t'^2 + \dd x'^2 + \dd y'^2 + \dd z'^2 \notag\\
    &= \hat{\eta}_{ab}\dd x^{a} \, \dd x^{b} + \O(r^{-1}),
\end{align}
with $r:=\sqrt{x^2+y^2+z^2}$. This means that the transformed metric and the original metric only differ by terms of order $\O(r^{-1})$. Thus, the asymptotic expansion~\eqref{eq:AsymptoticExpansion} can also be written as
\begin{align}
    \hat{g}_{ab} &= \hat{\eta}_{ab} + \O(r^{-1})\notag\\
    &= \hat{\eta}'_{ab} + \O(r^{-1})
\end{align}
We conclude that the spacetime metric $\hat{g}_{ab}$ can approach \textit{different} Minkowski metrics. That is, the metric $\eta$ and $\eta'$ will generally not be equal but differ in some terms which decay as $1/r$. These terms are included in $\O(r^{-1})$ and, hence, instead of having a canonical choice for the asymptotic region, we have infinitely many choices. Each comes with its individual Poincar\'e group. As a result, including all possible choices of asymptotically Minkowski metrics, we obtain an ``infinite sum'' over all Poincar\'e groups. This sum constitutes the BMS group. It is worth emphasizing that the imposed fall-off condition is solely due to the presence of gravitational waves as we know that radiation dies off as $1/r$. In this sense, the presence of gravitational waves forbids the spacetime structure to have 'a single' Poincar\'e group as the asymptotic symmetry group. Instead, we find infinitely many. Finally, we end this chapter by summarizing useful properties of the BMS group in form of its generating fields which follow from the properties discussed in the previous subsections. In a coordinate chart $(u,\theta,\phi)$ of a given conformal frame, we can define $\ell_a=\D_a u$ and impose the normalization condition $n^a \ell_a=-1$. The BMS vector fields $b^a$ can be decomposed into vertical ($v^{a}\propto n^{a}$) and horizontal ($h^{a} \perp n^{a}$) parts,
\begin{align}
    b^a:=v^a+h^a=\left(\beta(\theta, \phi) + u\,\alpha(\theta,\phi)\right) n^a+h^a.
\end{align}
Furthermore, it holds that
\begin{align}
    h^a n_a &=0, &\lie_n\beta &=0, &\lie_\xi n^a &=-\alpha n^a, &\lie_\xi q_{ab} &= 2\alpha q_{ab}.
\end{align}
We can then classify the BMS fields as follows.
\begin{align}
    &\text{Supertranslations:} &\alpha &=0, &h^a &=0\notag\\
    &\text{Rotations:} &\alpha &=0, &\beta & = 0, &\lie_h q_{ab} &=0, &\lie_h n^a &=0\notag\\
    &\text{Boosts:} &\alpha &\neq0, &\beta &=0, &\lie_h q_{ab} &=2\alpha q_{ab}, &\lie_h n^a &=0\notag
\end{align}
This decomposition is handy for certain calculations involving BMS vector fields. Moreover, it clearly separates types of symmetry generators so that conserved quantities are more intuitively approachable. For a more in-depth treatment of the BMS group, including mathematical details, we refer to~\cite{Alessio:2018}.

\newpage
\subsection{Exercises}
\renewcommand{\thesection}{\arabic{section}}

\begin{Exercise}[]\label{ex:BMSalgebra}
This exercise serves the purpose of filling in some gaps in our proof that $\mathfrak b$ is a Lie algebra. To that end, fix a divergence-free conformal frame and let $(q_{ab}, n^{a})$ be a universal structure of $\scrip$. Furthermore, let $\mathfrak b$ be the set of all vector fields which satisfy
\begin{align*}
    \lie_n q_{ab} &= 2\alpha \, q_{ab}\\
    \lie_n n^{a} &= -\alpha\, n^{a},
\end{align*}
for some smooth function $\alpha = \alpha(\theta, \phi)$.
\begin{itemize}
    \item[a)] Show that 
    \begin{align*}
        \lie_{\lambda_1 b_1 + \lambda_2 b_2}q_{ab} &= \alpha\, q_{ab}\\
        \lie_{\lambda_1 b_1 + \lambda_2 b_2} n^{a} &= -\alpha\, n^{a}
    \end{align*}
    for all $b_1, b_2\in\mathfrak b$ and all $\lambda_1, \lambda_2\in\bbR$, where $\alpha = \alpha_1\lambda_1 + \alpha_2 \lambda_2$. Conclude that $\mathfrak b$ forms a real vector space.
    \item[b)] Show that
    \begin{align*}
        \lie_{[b_1, b_2]}q_{ab} &= 2\alpha\, q_{ab}\\
        \lie_{[b_1, b_2]} n^{a} &= -\alpha\, n^{a},
    \end{align*}
    for all $b_1, b_2\in\mathfrak b$, where $\alpha = \lie_{b_1}\alpha_2 - \lie_{b_2}\alpha_1$. Conclude that $\mathfrak b$ is a real Lie algebra.\bigskip
    
    \textit{Hint: $\lie_{[X,Y]}T = \lie_X\lie_Y T - \lie_Y\lie_X T$}
\end{itemize}
\end{Exercise}

\begin{Exercise}[]\label{ex:BMSI}
    The Bondi-Metzner-Sachs group is defined by the following transformations in a coordinate chart $(u,\theta,\phi)$:
    \begin{align*}
        u&\mapsto u'=\Omega(\theta, \phi)[u-\alpha(\theta,\phi)]\notag\\
        \theta&\mapsto \theta'=\theta'(\theta, \phi) \notag\\
        \phi &\mapsto \phi'=\phi'(\theta, \phi),\notag
    \end{align*}
    where $(\theta,\phi) \mapsto (\theta',\phi')$ is a transformation of a $(\theta,\phi)$-sphere into itself, $\Omega$ is the conformal factor given by
    \begin{align*}
        \dd \theta'^2 + \sin^2\theta'\, \dd\phi'^2 = \Omega\left(\dd\theta^2 + \sin^2 \theta\, \dd\phi^2\right),
    \end{align*}
    and $\alpha$ is a smooth real function on the sphere. Transformations with $\theta' = \theta$ and $\phi' = \phi$ are called supertranslations. Show that in the case of supertranslations, we can express $\alpha$ in terms of spherical harmonics. Show that we can extract the translations from this expansion and find them to be described by four parameters.
\end{Exercise}

\newpage
\asection{6}{Conserved Charges and Derivative Operators at Null Infinity}\label{Chap6}
In the previous chapter, we introduced the BMS group as the subgroup of diffeomorphisms which leaves the universal structure of $\scrip$ invariant. We have also explored its Lie algebra, $\mathfrak b$, and we have seen that it admits an infinite-dimensional, abelian, and ideal subalgebra: the algebra of supertranslations~$\mathfrak s$. Because $\mathfrak s$ is ideal, the quotient algebra $\mathfrak b/\mathfrak s$ is well-defined and we have argued that it is precisely the six-dimensional Lie algebra of Lorentz transformations.

These findings can be lifted to the level of groups: The BMS group $\B$ is infinite-dimensional and it admits an infinite-dimensional, abelian, and normal subgroup: the group of supertranslations $\S$. Moreover, the quotient group $\B/\S$ is well-defined and one finds
\begin{equation}
    \B/\S \simeq \lie,
\end{equation}
where $\lie$ denotes the six-dimensional group of Lorentz transformations. These results can also be stated in a different fashion: The BMS group is the semidirect product of the group of supertranslations with the group of Lorentz transformations:
\begin{equation}
    \B = \S \rtimes \lie.
\end{equation}
This is an important result because the structure of the BMS group closely resembles the structure of the Poincar\'e group. This gives the elements of the BMS group a natural interpretation and, moreover, it opens the door to defining energy and momentum of the gravitational field as generators of spacetime translations. Let us therefore deepen our understanding of the Lie algebra of supertranslations.\bigskip

\subsection{The Translation Subalgebra}
It is a well-known fact that in special relativistic theories, the translation subgroup of the Poincar\'e group can be used to define energy and momentum of particles and fields. Is it possible to mimic the usual procedures employed in special relativistic theories to arrive at a definition of gravitational energy and momentum? 

Since we are concerned with asymptotically Minkowski spacetimes, it might be possible to define these notions on $\scrip$ using the BMS group.  However, the ``translation'' part of the BMS group is infinite-dimensional and consists of supertranslations. Can we identify translations which, in some adequate sense, represent the ``ordinary'' translations we know from the Poincar\'e group?

Intuitively, we would argue that this should be possible because in Chapter~\ref{Chap5} we used precisely the translational Killing vector fields of Minkowski space to motivate our educated guess. This guess led us to discover the full space of ``infinitesimal'' generators of $\mathfrak s$. Let us recall that on $\scrip$, the four translational Killing vector fields of Minkowski space have the form $\beta(\theta, \phi)\, n^{a}$, where $\beta$ is an element of the set
\begin{equation}
    \left\{1, \sin\theta\cos\phi, \sin\theta\sin\phi, \cos\theta\right\}.
\end{equation}
We arrived at this conclusion in Chapter~\ref{Chap2} by working in an inertial frame. We also pointed out that this set represents the first four spherical harmonics, $\{Y_{0,0}, Y_{1,1}, Y_{1,-1}, Y_{1,0}\}$. Let us be more precise: We fix an inertial frame $(t, x, y, z)$ and consider the Minkowski line element of the physical spacetime,
\begin{equation}
    \dd \hat{s}^2 = -\dd t^2 + \dd x^2 + \dd y^2 + \dd z^2.
\end{equation}
A change to spherical coordinates and a conformal completion yields the conformally rescaled line element
\begin{equation}\label{eq:2spheremetriconscri}
    \dd s^2 = \Omega^{-2}\dd \hat{s}^2 \equalhat \dd\theta^2 + \sin^2\theta\,\dd \phi^2.
\end{equation}
The last equality holds, as indicated by the symbol `$\equalhat$', on $\scrip$. Hence, on $\scrip$ the conformally rescaled line element reduces to the one of the unit $2$-sphere.\footnote{The metric $q_{ab}$ is of course the metric of three-dimensional space. However, since it is degenerate, one dimension is ``lost'' and we can get away with this imprecise terminology.} In particular, this sphere has scalar curvature equal to $2$ and we can use it to define spherical harmonics~$Y_{l, m}$.

With respect to the inertial frame $(t, x, y, z)$, the translational Killing vector fields are defined as
\begin{align}
    \hat{t}^{a} &:= \hat{\eta}^{ab}\nabla_b t, & \hat{x}^{a} &:= \hat{\eta}^{ab}\nabla_b x, & \hat{y}^{a} &:= \hat{\eta}^{ab}\nabla_b y, & \hat{z}^{a} &:= \hat{\eta}^{ab}\nabla_b z.
\end{align}
Their limit to $\scrip$ is given, as mentioned above, by spherical harmonics. More precisely, one finds
\begin{align}
    \hat t^{a} &\equalhat Y_{0,0}\, n^{a}, & \hat x^{a} &\equalhat Y_{1,1}\, n^{a}, & \hat y^{a} &\equalhat Y_{1,-1}\, n^{a}, & \hat z^{a} &\equalhat Y_{1,0}\, n^{a}.
\end{align}
The spherical harmonics which appear in these expressions are the same $Y_{lm}$ functions one obtains from the unit $2$-sphere metric defined by~\eqref{eq:2spheremetriconscri}. This seems like a trivial observation. However, what happens if we apply a Poincar\'e transformation and change from $(t, x, y, z)$ to the inertial frame $(t', x', y', z')$? 

Translations and rotations are length-preserving transformations and thus trivially do not change the \textit{asymptotic} line element. Hence, the only interesting case are boosts. In the boosted frame, the line element reads
\begin{equation}
    \dd \hat{s}'^2 = -\dd t'^2 + \dd x'^2 + \dd y'^2 + \dd z'^2.
\end{equation}
After applying the same procedure as before to this boosted inertial frame, we obtain the line element of a \textit{new} unit $2$-sphere metric,
\begin{equation}
    \dd s'^2 \equalhat \dd\theta' + \sin^2\theta' \, \dd \phi'^2.
\end{equation}
The crucial observation is that the two spherical metrics are \textit{different} but \textit{related} to each other. Let us call the first metric $q_{ab}$ and the second one $q'_{ab}$. Then one finds that they are related by
\begin{align}
    q'_{ab} &= \omega^2\, q_{ab} &\text{with} && \omega &= \frac{1}{\gamma\left(1-\vec{v}\cdot\hat{\vec{x}}\right)},
\end{align}
where $\vec{v}$ is the constant velocity vector of the boost, $\hat{\vec{x}} = (\sin\theta \cos\phi, \sin\theta\sin\phi, \cos\theta)^\transpose$ is the unit radial vector, and $\gamma$ is the Lorentz factor. What this tells us, is that we can start with any inertial frame and produce a $3$-parameter family of unit $2$-sphere metrics. All we need to do is apply boosts and follow the procedure discussed above. In each inertial frame, we can take the limit of translational Killing vector fields to $\scrip$ and for each $2$-sphere metric we can compute spherical harmonics. The relations between these limits and the spherical harmonics are always the same. However, what is more interesting, is the relation between the spherical harmonics of $q_{ab}$ and those of $q'_{ab}$. As it turns out, the first four spherical harmonics of $q'_{ab}$ are simply given by a linear combination of the spherical harmonics of $q_{ab}$. 

Qualitatively, we can understand this if we consider what happens to the translational Killing vector fields when we apply a boost. The Killing vector fields of the boosted inertial frame will simply be linear combinations of the Killing vector fields of the original frame. Thus, what we have just established is a relationship between translational Killing vector fields of different inertial frames on $\scrip$ and spherical harmonics obtained from different $2$-sphere metrics.

This observation lies at the core of a definition which allows us to identify a \textit{unique} translation subgroup~$\T$ within the group of supertranslations $\S$: We say that that a divergence-free conformal frame $(q_{ab}, n^{a})$ is a \textbf{Bondi frame}, if $q_{ab}$ is a unit $2$-sphere metric. As we know very well by now, a divergence-free conformal frame leaves us with a rescaling freedom of the form $(q_{ab}, n^{a})\mapsto (\omega^2\, q_{ab}, \omega^{-1}\, n^{a})$. Can we find rescaling transformations which map one Bondi frame to a new Bondi frame? 

To answer this question, we impose the condition that the rescaled metric $\omega^2\, q_{ab}$ is also a unit $2$-sphere metric. This boils down to demanding that
\begin{equation}
    R(\omega^2\, q_{ab}) \overset{!}{=} 2,
\end{equation}
where $R$ is the scalar curvature. Using the fact that $R(q_{ab}) = 2$, it is possible to show that this equation admits a $3$-parameter family of solutions, which can be written as
\begin{equation}
    \omega = \frac{1}{\gamma\left(1-\vec{v}\cdot\hat{\vec{x}}\right)}.
\end{equation}
This should not come as a surprise. The normal vector $n^{a}$ can be interpreted as the limit of a timelike vector $\tau^{a}$ to $\scrip$. Any such timelike vector singles out an inertial frame (in a neighborhood of $\scrip$, where the spacetime metric is sufficiently well approximated by a Minkowski metric). The rescaling transformation can then be interpreted as asymptotically relating one inertial frame to another one via $n'^{a} = \omega^{-1}\, n^{a}$. As we have seen, boosts map $2$-sphere metrics to other $2$-sphere metrics and the conformal factor $\omega$ has therefore to be the one which is generated by an asymptotic boost.

The importance of this result is that it allows us to single out a unique algebra of translations, denoted by $\mathfrak t$, from the Lie algebra of supertranslations. Given a Bondi frame, we call 
\begin{equation}
    \mathfrak t := \textsf{span}\left\{n^{a}, \sin\theta\cos\phi\, n^{a}, \sin\theta\sin\phi\, n^{a}, \cos\theta\, n^{a}\right\} \subset \mathfrak s
\end{equation}
the \textbf{subalgebra of translations}. Notice that changing from one Bondi frame to another Bondi frame merely amounts to a change of basis of $\mathfrak t$. That is because Bondi frames are related by asymptotic boosts which simply map the basis of $\mathfrak t$ to a different basis of $\mathfrak t$. As a final remark, notice that the subalgebra of translations is four-dimensional, abelian, and ideal. Just as its Poincar\'e counterpart.\bigskip

\subsection{Flux of Momentum and Supermomentum of the Gravitational Field}\label{susec:FluxandSupermomentum}
Gravitational waves carry energy, momentum, and angular momentum. This can be inferred from theoretical considerations, such as the sticky bead argument, and, more importantly, from direct observations. For instance, the observed orbital decay of the Hulse-Taylor binary pulsar gave the first evidence that a bound system can lose energy due to gravitational radiation.

However, giving a precise mathematical definition of energy, momentum, and angular momentum is no easy task. In this subsection, we will refrain from giving mathematical derivations and instead content ourselves with just describing the basic framework and stating the definition of Bondi $4$-momentum and supermomentum.

We start our considerations with the pair $(q_{ab}, n^{a})$ in a fixed Bondi frame. In any such frame, the following is true:
\begin{itemize}
    \item[1)] Translations at $\scrip$ are represented by $t^{a} := \alpha(\beta, \phi)\, n^{a}$ with
    \begin{equation}
        \alpha(\theta, \phi) = \alpha_0\, Y_{0,0} + \sum_{|m|\leq 1}\alpha_m\, Y_{1,m}(\theta, \phi). 
    \end{equation}
    for some constants $\alpha_0, \alpha_m$.
    \item[2)] The vector $n^{a}$ is ``the'' time-translation vector field in the chosen Bondi frame and the vector fields $t^{a}_{(\vec{\alpha})} := \sum_{|m|\leq 1}\alpha_m\, Y_{1,m}\, n^{a}$ are spatial translations in that frame.
    \item[3)] Supertranslations are given by $s^{a} := \beta(\theta, \phi)\, n^{a}$, where $\beta$ is any smooth function on the $2$-sphere.
\end{itemize}
Recall that in special relativistic theories, energy and momentum arise as the Hamiltonian generators of canonical transformations which correspond to spacetime translations. For asymptotically Minkowski spacetimes, we have found that there is a unique translation subalgebra. Hence, asymptotically we can properly speak of translations and this opens the door to mimicking the procedure of special relativistic theories in GR. The idea is to construct a phase space $\Gamma_{\textsf{rad}}$ of radiative modes\footnote{At this stage in the notes we do not yet know what the radiative modes are. Thus, we would not be able to carry out this construction.} on $\scrip$~\cite{Ashtekar:1982}. One can then show that the BMS translations and supertranslations, as defined above, induce canonical transformations on $\scrip$. Finally, the last step is to compute the Hamiltonians associated with these canonical transformations~\cite{Ashtekar:1981}. This procedure then leads to the definition of \textbf{total flux of $\mathbf{4}$-momentum $\mathbf{\F_{(\alpha)}}$ across $\scrip$}:
\begin{equation}\label{eq:FluxGR}
    \F_{(\alpha)} := \frac{1}{4\pi}\int_{\scrip}\dd u\, \dd^2 \omega\, \alpha(\theta, \phi)\left(|\dot{\sigma}^\circ|-\Re{\eth\dot{\bar{\sigma}}^\circ}\right)(u,\theta, \phi).
\end{equation}
Here, $\dd^2 \omega$ is the area element of the $2$-sphere, $\eth$ is the angular derivative operator we introduced in Chapter~\ref{Chap2} (cf. definition~\eqref{eq:eth}), a dot indicates differentiation with respect to $u$, and $\sigma^\circ$ is the so-called \textbf{asymptotic shear}.\footnote{We will encounter this function again in Chapter~\ref{Chap7}, where we will study it in more detail.} It is defined as
\begin{equation}
    \sigma^\circ(u, \theta, \phi) := -\lim_{\Omega\to 0} \left(\Omega^{-1}m^{a} m^{b} \nabla_a\ell_b\right).
\end{equation}
An analogous expression can be derived for the \textbf{total flux of supermomentum} $\F_{(\beta)}$. One just has to replace $\alpha$ with $\beta$ in the integral~\eqref{eq:FluxGR}. In fact, we can always use the flux of supermomentum since it contains the $4$-momentum expression as a special case. 

Having said that, one can show that $\F_{(\beta)}$ is an integral over an exact $3$-form~\cite{Ashtekar:1982}. This allows us to rewrite $\F_{(\beta)}$ as the difference of two $2$-sphere integrals performed over the ``$u=-\infty$'' and the ``$u=\infty$'' spheres (these sphere represent spacelike infinity, $i^0$, and future timelike infinity, $i^{+}$). The explicit expression is
\begin{equation}
    \F_{(\beta)} = \lim_{u_0\to -\infty}\left. P_{(\beta)}\right|_{u=u_0} - \lim_{u_0\to \infty} \left.P_{(\beta)}\right|_{u=u_0},
\end{equation}
where
\begin{equation}\label{eq:supermomentum}
    \left.P_{(\beta)}\right|_{u=u_0} := -\frac{1}{4\pi}\oint_{u=u_0}\dd^2 \omega\, \beta(\theta, \phi)\, \Re{\Psi^\circ_2 + \bar{\sigma}^\circ \dot{\sigma}^\circ}(\theta, \phi)
\end{equation}
is the $\mathbf{\beta}$\textbf{-component of the supermomentum} evaluated at the retarded time $u=u_0$. More details on this supermomentum and its origins can be found in~\cite{Sachs:1962,Newman:1968}. We conclude with the observation that~\eqref{eq:supermomentum} contains the Newman-Penrose scalar which corresponds to ``coulombic'' modes. It therefore reinforces the notion that it contains information about masses which are present in the spacetime. This is certainly a sensible property for a quantity which measures energy and momentum of the gravitational field. Moreover, we will show in Chapters~\ref{Chap7} and~\ref{Chap8} that the asymptotic shear can be expressed in terms of the radiative modes. Hence, the supermomentum~\eqref{eq:supermomentum} also carries information about gravitational waves. In principle, knowing $\left.P_{(\beta)}\right|_{u=u_0}$ in the distant past and in the distant future allows us to determine the total flux of energy and momentum carried to infinity by gravitational waves. For interesting recent applications of this result, see~\cite{AshtekarII:2020, Ashtekar:2020, Mitman:2020}.

Finally, we remark that angular momentum has so far remained completely unmentioned. There is a good reason for this. In this subsection, we have left out mathematical details concerning the derivation of the energy and momentum fluxes, because of their complexity. When it comes to angular momentum, the situation is even worse because of the so-called \textbf{supertranslation ambiguity}. In the next subsection, we briefly discuss the origin of this ambiguity.\bigskip

\subsection{On Subtleties regarding the Definition of Angular Momentum} \label{subsec:angmomet}
It is instructive to first study the notion of angular momentum in special relativity. To that end, let $(\M,\eta_{ab})$ be Minkowski spacetime\footnote{In this subsection we drop the hats on physical quantities for notational simplicity.} endowed with coordinates $X^{a}$ with respect to a fixed coordinate origin $O$. As we know very well, there are precisely ten Killing vector fields, collectively denoted by $K^a$, which pertain to the Minkowski metric. These fields can be written as
\begin{align}\label{eq:10Killing}
    K^a = T^a + F^{ab}X_b,
\end{align}
where $T^a$ denotes the Killing vector field of spacetime translations, while $F^{ab}$ is a constant, antisymmetric tensor which encodes Lorentz transformations (three rotations and three boosts).

It is intuitively clear that spacetime translations are well-defined without having to define a point of origin first: Moving two steps to the right or moving clock handles one hour ahead can easily be achieved without reference to an origin of space nor an origin of time. Moving two steps to the right simply means moving in that direction by that amount from our \textit{current} position. The same holds true for moving clock handles: We move them relative to their \textit{current} position. However, the situation changes when we consider rotations or boosts. These operations single out a special point in space and time: The point of origin $O$.

To be more precise, there is exactly one point in space which is left invariant by \textit{all} $SO(3)$ rotations. Similarly, there is precisely one point in time which is left invariant by \textit{all} boost transformations. Together, rotations and boosts constitute the Lorentz group $\lie$, which thus leaves precisely one point in spacetime invariant. This point is the point of origin of our coordinate system with respect to which rotations and boosts are defined (and thus, with respect to which $\lie$ is defined). Changing the point of origin results in a change of the Lorentz group. For instance, if we define rotations in space with respect to point $O$ and then translate this point to $O'$, we find a new $SO(3)$ group which describes rotations around $O'$. However, the two $SO(3)$ groups are \textit{related} to each other by a \textit{translation}.

Mathematically, this dependence on the point of origin is also reflected in equation~\eqref{eq:10Killing}. Because the Killing vectors associated with rotations and boosts depend on $X^{a}$, which is the vector which measures the position of objects relative to $O$. Moreover, due to Noether's theorem, the conserved charges associated with rotations and boosts inherit this dependence on a choice of origin. 

This observation lies at the core of the angular momentum ambiguity at $\scrip$ in GR. However, before elaborating more on this issue, let us consider angular momentum in the special relativistic theory of a point particle of rest mass $M_0$. With the help of its energy-momentum tensor $T_{ab}$, the ten Killing vector fields of the Poincar\'e group define ten conserved quantities: The $4$-momentum $P^{a}$ and six quantities encoded in the antisymmetric tensor $M_{ab}$, from which angular momentum and center of mass can be extracted. Concretely, these conserved quantities can be expressed as
\begin{align}\label{eq:ConservedQuantities}
    P_a T^a + M_{ab} F^{ab} := \int_\Sigma T_{ab} K^b \dd S^a,
\end{align}
where the integral is performed over a Cauchy surface $\Sigma$ of Minkowski space. Let us introduce a rest frame via $P^a = -M_0 t^a$, where $t^a$ is a timelike vector. Under a displacement $O\rightarrow O'$ of the origin we find, using~\eqref{eq:ConservedQuantities}, the following transformations:
\begin{align}\label{transprop}
    P_a &\rightarrow P_a  &\text{and}&&  M_{ab} &\rightarrow M_{ab} + M_0\, t_{[a}d_{b]}.
\end{align}
Here, $d_a$ measures the displacement, i.e., the position of $O'$ with respect to $O$. At this point we recall that the center of mass, which is the conserved quantity conjugate to Lorentz boosts, is given by the three components $M_{i0}$, with $i\in\{1,2,3\}$, while angular momentum is defined by $J^{i} := \epsilon^{abci}t_a M_{bc}$. It therefore follows from equation~\eqref{transprop} that the three center of mass components $M_{i0}$ can be transformed away by a suitable change of the point of origin. However, the same is not true for the angular momentum components because $\epsilon^{abci}t_a\, t_{[b}d_{c]} = 0$. Hence, even though we have ten conserved quantities, the entire physical information is encoded in the $4$-momentum and the angular momentum $3$-vector $J^{i}$.

This is the situation in special relativity and it raises the question, whether we can define quantities analogous to $M_{ab}$ and $J^{i}$ for GR on $\scrip$. The first difference is that because gravitational waves carry angular momentum, these quantities have to be time-dependent. Since every cross-section of $\scrip$ represents an instant of retarded time $u$, the question is whether we can find a consistent definition of angular momentum on each such cross-section. The second difference is that the BMS group is infinite dimensional, while the Poincar\'e group is ten-dimensional. This is the main obstruction to defining angular momentum. To see this, choose a $u=u_0$ cross-section $\C$ of $\scrip$. On this cross-section, we can identify a preferred Lorentz group by demanding that boosts and rotations are tangential to $\C$. Let us denote this preferred Lorentz group by $\lie_{\C}$. Furthermore, the supermomentum $\left.P_{(\beta)}\right|_{u=u_0}$ provides us with a rest frame for that cross-section. This is the analogue of selecting a rest frame via $P^{a} = -M_0 t^{a}$ in special relativity. Given the Lorentz group $\lie_{\C}$, we can define a general relativistic analogue of the tensor $M_{ab}$. However, there is a problem when we change from the cross-section $\C$ to another cross-section $\C'$.

In fact, the Lorentz group $\lie_{\C}$ as well as the rest frame determined by $\left.P_{(\beta)}\right|_{u=u_0}$ change! If the cross-sections $\C$ and $\C'$ are related by a BMS translation, as defined in the previous subsection, we find that $\lie_{\C}$  and $\lie_{\C'}$ are also related by such a translation. This is analogous to what happens in special relativity and this is no reason for concern. However, the momentum of special relativity is invariant under translations while the BMS supermomentum changes! Hence, in going from $\C$ to $\C'$ we do not only change the Lorentz group, we also change the rest frame. The consequence is that comparing the angular momentum on~$\C$ to the angular momentum on $\C'$ becomes a meaningless operation. It would be like comparing the special relativistic $J_z \equiv M_{xy}$ at a given instant of time to $M_{xy}+M_{zt}$ at a different instant of time!

This was the ``best case'' scenario. The situation is even worse when $\C$ and $\C'$ are related by a supertranslation. Since the translation subgroup of the BMS group is four dimensional, the supermomentum corresponding to these translations has only four components. However, the supermomentum corresponding to supertranslations can be interpreted as an object with infinitely many ``components''. The reason is that on any given cross-section there is one supermomentum per supertranslation, and there are infinitely many such supertranslations. Consequently, the analogue of $M_{ab}$ would become a tensor with infinitely many ``components''. This is the infamous \textbf{supertranslation ambiguity} and it prevents us, in general, from constructing the analogue of angular momentum in GR.

However, there has been some recent progress where the ambiguity could be reduced for a class of physically interesting spacetimes. We refer the reader to~\cite{Ashtekar:2020} for these new developments and for more details on the definition of angular momentum. The presentation in this subsection was largely based on that reference.\bigskip

\subsection{Towards identifying Radiative Degrees of Freedom at Null Infinity}\label{Chap6D}
Let us return to the structure of $\scrip$ and our main task in these notes: Identifying the radiative degrees of freedom of the gravitational field. So far we studied the universal structure of asymptotically Minkowski spacetimes and the diffeomorphisms which leave it invariant. We emphasize again that this structure is common to \textit{all} asymptotically Minkowski spacetimes. This includes Minkowski space itself and the Kerr-Newman family of black hole solutions. In other words, spacetimes which are devoid of gravitational radiation. In fact, so far we have not seen a single trace of radiative modes in the universal structure and we should not expect to. Where then do radiative modes reside?

The answer might by surprising: The radiative modes are encoded in the covariant derivative operator on $\scrip$. Let us first try to understand this statement on a qualitative level.

Loosely speaking, in standard GR we introduce two geometric structures. First, we introduce a manifold and a metric on it. This is simply the spacetime $(\hatM, \hat g_{ab})$ and it constitutes what we call the first order structure. In a sense, this provides the kinematical arena of the theory. It is a description of ``where'' the physics takes place.
Secondly, we introduce a covariant derivative operator $\hat\nabla$. This enables us to address dynamical questions and learn more about how objects (particles and fields) evolve. This is the second order structure. On a purely mathematical level, there is a lot of freedom in choosing a covariant derivative operator. However, in GR we choose $\hat\nabla$ to be the covariant derivative associated with the Levi-Civita connection. It is thus a very particular operator. In fact, if we know the metric $\hat g_{ab}$, we have complete knowledge of the geometry. Because the metric determines the Levi-Civita connection and in turn the connection determines the Riemann curvature tensor.

The situation for $\scrip$ differs in a subtle way from what we described above. Asymptotically Minkowski spacetimes are described by a manifold $\scrip$, which is endowed with a universal structure $(q_{ab}, n^{a})$. This is the first order structure. However, unlike $\hat g_{ab}$, the intrinsic metric $q_{ab}$ is degenerate! The spacetime metric $\hat g_{ab}$, or, equivalently, the conformally rescaled metric $g_{ab}$ contain all the geometric information of a spacetime. In particular, they contain the information whether or not there is gravitational radiation.
When pulling $g_{ab}$ back to $\scrip$, it becomes degenerate and, in a sense, loses information. What sounds like a bad thing is actually good. If this would not happen, there would not be a universal structure. How would we otherwise account for the fact that spacetimes with and without radiation all share the same properties and symmetries described by $(q_{ab}, n^{b})$?

The information contained in $g_{ab}$ is of course not completely lost. Recall that the metric determines the derivative operator $\nabla$ and we have not yet (properly) introduced such an operator on $\scrip$. Intuitively, what we need to do is to pull back $\nabla$ to $\scrip$ in order to define a covariant derivative $\D$ on $\scrip$. In doing so, we are ``transferring'' more information from $g_{ab}$ to $\scrip$. Carrying out the construction of $\D$ and studying its properties is the main objective of this subsection. We will find that the qualitative idea of ``information transfer'' from $g_{ab}$ to $\scrip$ is correct, in the sense that the covariant derivative operator $\D$ encodes information which is not present in $q_{ab}$. As we have emphasized, the metric $q_{ab}$ does not allow us to distinguish between spacetimes with radiation and spacetimes without radiation. However, the operator $\D$ is capable of doing precisely this. Given two asymptotically Minkowskian spacetimes, their derivative operators will in general be different.

A proper definition of $\D$ on $\scrip$ is no trivial matter but, as we have hopefully motivated, an important and fruitful endeavor. To do this, we take a mathematically slightly broader perspective than is usually done in standard GR. We will build up the mathematical framework step by step and recall some important facts about covariant derivatives and connections. \bigskip

\textbf{\underline{Covariant derivatives on a general manifold:}}\\
Let $\hatM$ be a differentiable manifold. What is a covariant derivative in the absence of a metric? A covariant derivative is an operator $\hat\nabla$ which generalizes the concept of a directional derivative. Its action on a scalar~$\hat f$ is thus required to satisfy
\begin{equation}\label{eq:DirectDeriv}
    \hat v^{a}\hat\nabla \hat f = \lie_{\hat v} \hat f = \hat v^{a}\partial_a \hat f,
\end{equation}
for all vector fields $\hat v$ on $\hatM$. Furthermore, it is required to be a linear map, to satisfy Leibniz's rule, and to map $(p,q)$ tensor fields to $(p+1,q)$ tensor fields. Its action is completely determined by \textit{specifying} how it acts on vector fields and $1$-forms,
\begin{align}
    \hat{\nabla}_a \hat{v}^{b} &= \partial_a \hat{v}^{b} + \hat{\Gamma}\ud{b}{ac}\hat{v}^{c}\notag\\
    \hat{\nabla}_a \hat{\omega}_b &= \partial_a \hat{\omega}_b - \hat{\Gamma}\ud{c}{ab} \hat\omega_c.
\end{align}
On the right hand side, we have introduced the \textbf{connection} $\hat\Gamma\ud{a}{bc}$. Notice that specifying the action of~$\hat\nabla$ is the same as \textit{choosing} a connection. In a coordinate chart, this amounts to choosing $64$ functions~$\hat\Gamma\ud{a}{bc}$. Any such choice defines a notion of parallel transport on~$\hatM$. In turn, parallel transport defines, in the absence of a metric, two  geometric quantities:
\begin{align}
    \hat{T}\ud{a}{bc} &:= 2\hat\Gamma\ud{a}{[bc]} & &\textsf{(Torsion)}\notag\\
    \hat{R}\du{abc}{d} &:= 2\partial_{[b}\hat{\Gamma}\ud{d}{c]a} + 2\hat{\Gamma}\ud{d}{[b|e}\hat{\Gamma}\ud{e}{c]a} & &\textsf{(Curvature)}
\end{align}
We call the pair $(\hatM, \hat\Gamma\ud{a}{bc})$ an \textbf{affine geometry}. These geometries can be classified by whether or not they have torsion or curvature (or both). \bigskip

\textbf{\underline{Covariant derivatives on $(\hatM, \hat{g}_{ab})$:}}\\
Let $(\hatM, \hat{g}_{ab})$ by a differentiable manifold, endowed with a smooth metric $\hat g_{ab}$. We can introduce a covariant derivative just as before. The only difference is that a metric gives rise to a new fundamental tensor, which can be constructed solely from $\hat g_{ab}$ and $\hat\Gamma\ud{a}{bc}$:
\begin{align}
    \hat{Q}_{abc} &:= \hat{\nabla}_a \hat{g}_{bc} = \partial_a \hat{g}_{bc} - 2\hat{\Gamma}\ud{d}{a(b}\hat{g}_{c)d} & &\textsf{(Non-metricity)}
\end{align}
We call $(\hatM, \hat{g}_{ab}, \hat\Gamma\ud{a}{bc})$ a \textbf{metric-affine geometry}. Unsurprisingly, metric-affine geometries can be classified by whether or not they have torsion, curvature or non-metricity (or any combination of those).\footnote{Minkowski space is the only space for which all three tensors vanish identically.}\bigskip

\textbf{\underline{The covariant derivative in GR:}}\\
In GR we select a very special metric-affine geometry by imposing two geometric postulates:
\begin{itemize}
    \item[1)]  Vanishing torsion: $\hat{T}\ud{a}{bc} \overset{!}{=} 0$
    \item[2)] Metric-compatibility: $\hat\nabla_a \hat g_{bc} \overset{!}{=} 0$.
\end{itemize}
It is a well-known result of differential geometry and commonly taught in classes on general relativity, that these two postulates uniquely determine the connection $\hat\Gamma^{a}{}_{bc}$. In fact, $\hat\Gamma^{a}{}_{bc}$ has to be the Levi-Civita connection, which is completely determined by the first order derivatives of the metric and the inverse metric,
\begin{equation}\label{eq:LCConnection}
    \hat\Gamma^{a}{}_{bc} = \frac12 \hat{g}^{ad}\left(\partial_b \hat g_{cd} + \partial_c \hat g_{bd} - \partial_d \hat g_{bc}\right).
\end{equation}
The discussion thus far was completely general and $\hat\nabla$ symbolized any covariant derivative operator. However, from now on, it will be understood that $\hat\nabla$ denotes the covariant derivative with respect to the Levi-Civita connection.\footnote{We have thus restored the meaning $\hat\nabla$ had in previous chapters.} Moreover, since $\hat\nabla$ is completely determined by the metric, we will denote the metric-affine geometry $(\hatM, \hat{g}_{ab}, \hat\Gamma\ud{a}{bc})$ simply by $(\hatM, \hat{g}_{ab})$ and call it the physical spacetime. This is the standard notation in mathematical relativity.\bigskip

\textbf{\underline{The covariant derivative on the conformally completed spacetime:}}\\
Let us fix a conformal factor $\Omega$ and perform a conformal completion of the physical spacetime to $(\M = \hatM\cup\scri, g_{ab} = \Omega^{2}\hat g_{ab})$. Because the connection of the physical spacetime is completely determined by the metric, the connection of the conformally completed spacetime will be determined by the metric $g_{ab}$, or, equivalently, by $\hat g_{ab}$ and $\Omega$. The precise transformation of the connection under conformal transformations is not relevant for us and we refer the interested reader to appendix D of~\cite{WaldBook}. What is important, however, is that under a conformal transformation we have $\hat\nabla\mapsto\nabla$, where $\nabla$ is the well-defined, torsion-free, and metric-compatible covariant derivative operator on $(\M, g_{ab})$. In other words, if we explicitly know $\hat g_{ab}$ on the physical spacetime, we explicitly know $\nabla$ on the conformally completed spacetime. We will keep this fact in the back of our minds.\bigskip

\textbf{\underline{Introducing covariant derivatives on $\scrip$:}}\\
The boundary of the conformally completed spacetime, i.e., $\scrip$, is a three-dimensional hypersurface defined by $\Omega = 0$. It has a null normal $n^{a} := g^{ab}\nabla_b \Omega$ and an intrinsic metric $q_{ab} := \pb{g}_{ab}$. As usual, we choose a divergence-free conformal frame, which is characterized by the equation $\nabla_a n_b \equalhat 0$. Because~$q_{ab}$ is degenerate, we can \textit{not} use the Levi-Civita connection of $q_{ab}$ to introduce a covariant derivative on~$\scrip$. In fact, the Levi-Civita connection of this metric is not even defined because $q_{ab}$ possesses no unique inverse!

It seems, therefore, that we either have to work with a more general metric-affine geometry, or we find a way to circumvent this obstacle. We can actually do the latter, as the following observations suggest:
\begin{itemize}
    \item[1)] We have argued that the derivative operator $\hat\nabla$ defined on the physical spacetime induces a well-defined derivative operator $\nabla$ on the conformally completed spacetime. One possibility to define a covariant derivative on $\scrip$ would therefore be to pull back $\nabla$ to $\scrip$ and define $\D:= \pb{\nabla}$.
    \item[2)] Let $\D$ be a candidate covariant derivative operator for $\scrip$. There is no problem in imposing metric-compatibility and vanishing torsion. It is possible to find operators which do satisfy these conditions. The ``problem'' is only that there is more than one such operator, precisely because the metric $q_{ab}$ is not invertible. But is this really an obstacle?
\end{itemize}
We will see in this and in the next chapter that these two observations are related to each other and that having more than one derivative operator is not an obstacle. Rather, having more than one derivative operator is what allows us to distinguish between different asymptotically Minkowski spacetimes.

First, however, we have to address the question whether $\D:=\pb{\nabla}$ is actually a well-defined derivative operator on $\scrip$. It is clear that $\D$ acts like a directional derivative, i.e., that it satisfies~\eqref{eq:DirectDeriv} for scalars defined on $\scrip$. It inherits this property from $\nabla$, just as it inherits its linearity and the Leibniz property. What is less obvious, is whether it maps $(p,q)$ tensor fields \textit{on} $\mathbf{\scrip}$ to $(p+1,q)$ tensor fields.

Let $\omega_a$ be a $1$-form defined on the co-tangent space of $\scrip$. Such a $1$-form is tangent to $\scrip$ in the sense that $\omega_a n^{a} = 0$. Is $\D_a \omega_b$ a $(2,0)$ tensor field which is intrinsically defined on $\scrip$? To answer this question, we need to show that $T_{ab}:=\D_a \omega_b$ has no components perpendicular to $\scrip$. That is, we need to show that $n^{a} T_{ab}$ and $n^{b} T_{ab}$ vanish. The first case is clear, because we pull back $\nabla_a$ to $\scrip$ and the first index is thus intrinsically defined on $\scrip$. In particular, this means we can write
\begin{equation}\label{eq:Trade}
    t^{a}\pb{\nabla}_a \omega_b = t^{a} \nabla_a \omega_b,
\end{equation}
for any vector tangent to $\scrip$. This is particularly important when we consider the second case. Using Leibniz's rule, we find
\begin{align}
    t^{a} n^{b}\D_a \omega_b &= t^{a} n^{b} \nabla_a \omega_b\notag\\
    &= t^{a} \nabla_a\left(\omega_b n^{b}\right) - \omega_b t^{a}\nabla_a n^{b}\notag\\
    &= 0.
\end{align}
In the first step we used~\eqref{eq:Trade} in order to trade $\D$ for $\nabla$, while in the third step we used $\omega_b n^{b} = 0$ and $\nabla_a n^{b} = 0$. The latter equation follows from the equation of the divergence-free conformal frame, $\nabla_a n_b = 0$. 

We have thus succeeded in showing that $\D_a\omega_b$ is a tensor field which is intrinsically defined on $\scrip$. What remains to be done, is to show that it is sufficient to know how $\D$ acts on $1$-forms, even in the absence of a well-defined inverse metric. This is done in Exercise~\ref{ex:WellDefinedActionOnVectors}. 

Our idea of defining $\D$ as the pullback of $\nabla$ to $\scrip$ has panned out, in the sense that it has given us a well-defined derivative operator on $\scrip$. What properties does it have? Two immediate properties we can establish are metric-compatibility,
\begin{align}
    0 = \underleftarrow{\nabla_a g_{bc}} &= \pb{\nabla_a}q_{bc} = \D_a q_{bc},
\end{align}
and torsion-freeness,
\begin{equation}
    0 = \underleftarrow{\nabla_{[a}\nabla_{b]}f} = \D_{[a}\D_{b]} f,
\end{equation}
where $f$ is any scalar on $\M$. However, from these two properties we can \textit{not} conclude that $\D$ is the covariant derivative with respect to the Levi-Civita connection! The reason for this is, as we have alluded to in the introduction to this subsection, that $q_{ab}$ is degenerate. It thus lacks a unique inverse $q^{ab}$. However, is it still possible to define the inverse of $q_{ab}$ in some weaker sense? The answer is in the affirmative and we call \textit{any} tensor $q^{ab}$ which satisfies
\begin{align}\label{eq:PseudoInv}
    q^{ab} q_{ac} q_{bd} = q_{cd}
\end{align}
a \textbf{pseudo-inverse} of $q_{ab}$. As our choice of words suggest, there is more than one $q^{ab}$ which satisfies this equation. In fact, suppose we have found one tensor $q^{ab}$ which satisfies condition~\eqref{eq:PseudoInv}. Then it follows immediately that $q'^{ab} = q^{ab} + t^{(a} n^{b)}$ is \textit{also} a solution to~\eqref{eq:PseudoInv}, where $t^{a}$ is \textit{any} vector tangential to $\scrip$. That is, $t^{a}$ is any vector which satisfies $t^{a}n_a \equalhat 0$. The reason for this ambiguity is, of course, the degeneracy of $q_{ab}$ and the fact that $n^{a}$ is the null direction of $q_{ab}$ in the sense that $q_{ab}n^{b} = 0$.

Let us summarize the situation thus far. We have seen that $\D:=\pb{\nabla}$ is a well-defined covariant derivative operator on $\scrip$ and that it has the following properties
\begin{equation}\label{eq:D_Properties}
    \boxed{
    \begin{aligned}
        \D_{[a}\D_{b]}f &= 0, & && \D_a q_{ab} &= 0, & && \D_a n^{b} &= 0
    \end{aligned}
    }
\end{equation}
These properties are satisfied by \textit{any} covariant derivative operator on $\scrip$. Also, these properties strongly restrict the form of $\D$, but they do not completely fix it. How many $\D$'s are there with the above properties? How many conditions do we need to impose on $\D$ in order to completely determine its action?

To answer these questions, we will first study the action of $\D$ on $1$-forms $\omega_a$ which are transverse to~$\scrip$ and Lie dragged by $n^{a}$. Mathematically, these conditions can be expressed as
\begin{align}\label{eq:OmegaProperties}
    \omega_a n^{a} &= 0 & \textsf{and} && \lie_n \omega_a &= 0.
\end{align}
Our motivation for doing so is as follows: Recall that in Chapter~\ref{Chap4} we visualized $\scrip$ as being a cylinder, ruled by the integral lines of $n^{a}$. In that chapter, we also introduced the space of generators~$\bbG$ and a projector $\pi:\scrip\to\bbG$, which maps each integral line to a single point. From this property of~$\pi$, it follows that there is a one-to-one correspondence between $1$-forms $\underline{\omega}_a$ on $\bbG$ and $1$-forms $\omega_a$ on $\scrip$ which satisfy~\eqref{eq:OmegaProperties}. Furthermore, we know that $\bbG$ is equipped with a well-defined, non-degenerate metric $\underline{s}_{ab}$. Thus, we can introduce a unique torsion-free and metric-compatible covariant derivative operator $\underline{\D}$ on $\bbG$. Using $\pi$, we can pull back the tensor field $\underline{\D}_a \underline{\omega}_b$ to $\scrip$. Because of the one-to-one correspondence between $\underline{\omega}_a$ and $\omega_a$, we find that the pullback $\pi^*\left(\underline{\D}_a \underline{\omega}_b\right)$ is the same from section to section and equal to $\D_a \omega_b$. Because the derivative operator on $\bbG$ is unique, this means that $\D_a \omega_b = \D'_a \omega_b$. That is, the action of $\D$ on $1$-forms which are transverse to cross-sections of $\scrip$ and Lie dragged by $n^{a}$, is \textit{independent} of the choice of derivative operator on $\scrip$!

Let us explicitly show this, i.e., let us show that $\left(\D_a-\D'_a\right)\omega_b = 0$. To that end, let $\D$ be a derivative operator which satisfies~\eqref{eq:D_Properties} and let $\omega_a$ be a $1$-form which satisfies~\eqref{eq:OmegaProperties}. We can decompose $\D_a \omega_b$ into symmetric and anti-symmetric parts
\begin{align}\label{eq:SymAntiSymDecomp}
    \D_a \omega_b &= \D_{(a}\omega_{b)} + \D_{[a}\omega_{b]}\notag\\
    &=  \lie_{\tilde{\omega}}q_{ab} + \D_{[a}\omega_{b]},
\end{align}
where we have used that for a metric-compatible connection we can write $\lie_v q_{ab} = \D_{(a}v_{b)}$ for any vector field $v^{a}$. In the particular case at hand, we have defined $\tilde{\omega}^{a} := q^{ab}\omega_{b}$, where $q^{ab}$ is the pseudo-inverse introduced in~\eqref{eq:PseudoInv}. It seems that the action of $\D_a$ on $\omega_b$ depends on the particular choice of pseudo-inverse. However, we will now show that this is not the case. 

Recall that if $q^{ab}$ satisfies~\eqref{eq:PseudoInv}, then $q'^{ab} = q^{ab} + t^{(a}n^{b)}$ is also a solution. This leads to an ambiguity in the definition of the vector field $\tilde\omega^{a}$, which can be expressed as
\begin{align}
    \tilde\omega'^{ab} &:= q'^{ab}\omega_b = \tilde\omega^{a} + t^{(a}n^{b)}\omega_b\notag\\
    &\phantom{:}= \tilde\omega^{a} + n^{a} t^{b}\omega_b,
\end{align}
where we have used $\omega_a n^{a} = 0$. The contraction $t^{b}\omega_b$ is of course just a function, which we shall call $f$. This observation allows us to use 
\begin{align}
    \lie_{f n}q_{ab} = f\underbrace{\lie_n q_{ab}}_{=0} + \underbrace{q_{ac}n^{c}}_{=0}\left(\D_b f\right) + \underbrace{q_{cb}n^{c}}_{=0}\left(\D_a f\right) = 0.
\end{align}
Thus, we conclude that there is an ambiguity in defining the vector $\tilde{\omega}^{a}$, but this ambiguity ``washes out'' in $\lie_{\tilde{\omega}}q_{ab}$. No matter which pseudo-inverse we choose to define $\tilde{\omega}^{a}$, we always find the same $\lie_{\tilde{\omega}}q_{ab}$.

Next, let us have a closer look at the anti-symmetric part of~\eqref{eq:SymAntiSymDecomp}. We can either check by direct computation that $\D_{[a}\omega_{b]}$ does not depend on the connection or we can realize that $\D_{[a}\omega_{b]}$ is the coordinate expression of the exterior derivative of the $1$-form $\omega_a$. In either case, the anti-symmetric part does not depend on the connection used to define $\D$. Furthermore, we also know that the Lie derivative does not depend on the connection. Hence, the symmetric as well as the anti-symmetric part of~\eqref{eq:SymAntiSymDecomp} are \textit{independent of the connection}. Put differently, the tensor $\D_a \omega_b$ is independent of the choice of covariant derivative operator. 

This implies that if we have two covariant derivative operators, say $\D_a$ and $\D'_a$, which satisfy~\eqref{eq:D_Properties}, we must have
\begin{equation}
    \left(\D_a-\D'_a\right)\omega_b = 0
\end{equation}
for all $1$-forms which satisfy~\eqref{eq:OmegaProperties}. This is precisely what we wanted to prove.

\begin{mysidenote}{A qualitative picture}{}
    We have already understood geometrically, why $\left(\D_a-\D'_a\right)\omega_b = 0$ is true. Forms with the properties~\eqref{eq:OmegaProperties} really live in the space of generators $\bbG$ and that space is equipped with a \textit{unique} covariant derivative. Because this derivative operator is given by the Levi-Civita connection of $\underline{s}_{ab}$, we can qualitatively say that the action of $\D$ on such $1$-forms is determined by the information contained in $q_{ab}$. However, so far we have not said anything about forms perpendicular to cross-sections of $\scrip$. That is where the derivative operator $\D$ will reveal that it contains additional information compared to $q_{ab}$.
\end{mysidenote}
In Exercise~\ref{ex:D_on_1_Form}, it is shown that choosing $\{\ell_a, m_a, \bar{m}_a\}$  as basis of the co-tangent space to $\scrip$ allows us to determine the covariant derivative of any $1$-form $\alpha_a := f\, m_a + \bar{f}\,\bar{m}_a$, even when $\lie_n \alpha_a \neq 0$. Thus, the action of $\D$ is the same for all co-vectors which are transverse to $n^{a}$. In particular, this means
\begin{equation}
    \left(\D_a - \D'_a\right)\alpha_b = 0    
\end{equation}
for any choice of $\D$ and $\D'$ which satisfies~\eqref{eq:D_Properties} and any $1$-form  satisfying $\alpha_a n^{a} = 0$. However, we cannot say anything about $\D_a \ell_b$. Derivatives of $1$-forms in the direction perpendicular to cross-sections (recall that $\ell^{a} m_a = 0$) are \textit{not} determined by the pullback of $\underline{\D}$ from $\bbG$ to $\scrip$. We have thus arrived at an answer to our second questions below~\eqref{eq:D_Properties}. The question was, how many additional conditions we need in order to fix $\D$. The answer is one, because we need to prescribe what $\D_a\ell_b$ is. This completely fixes the operator $\D$.

We end this chapter with another qualitative comment. The $1$-form $\ell_a$ is not part of the universal structure. Thus, this element ``breaks'' the universality of the derivative operator $\D$. Of course, it never was universal. Only its action on $1$-forms with the properties~\eqref{eq:OmegaProperties} is universal. Not in the sense that it is the same for all spacetimes, but in the sense that it is determined by $q_{ab}$ (or, maybe more appropriately, by $\underline{s}_{ab}$). However, the fact that $\D_a\ell_b$ is \textit{not} determined by universal properties of $\scrip$ means that it is the derivative operator which allows us to distinguish between different asymptotically Minkowski spacetimes. Just as claimed in the introduction to this subsection. Furthermore, we see that also our idea of ``information transfer'' from $g_{ab}$ into $\D$ has panned out. Because $\ell_a$ does of course carry information about $g_{ab}$. 

Finally, we remark on a possible point of confusion. If we are given a concrete metric $g_{ab}$ of the conformally completed spacetime (think for instance of the completed Schwarzschild metric), we explicitly know $\nabla$. Thus, pulling back this derivative operator to $\scrip$ gives rise to a well-defined a fixed derivative operator~$\D$. There is no need to impose any conditions. In particular, we do not need to say something about $\D_a \ell_b$. Rather, the $\D$ we obtain by the pullback operation and the $\ell_a$ we obtain from $g_{ab}$ by performing the construction of the Newman-Penrose null tetrad, precisely tell us what $\D_a\ell_b$ is! That is, we obtain an explicit expression $\D_a\ell_b = ***$. 

In this subsection, however, we took a different perspective. Rather than starting from the bulk of spacetime and going to $\scrip$, we worked intrinsically, i.e., from within $\scrip$. We asked how many $\D$'s there are which satisfy the properties~\eqref{eq:D_Properties} and this revealed something about the structure of the $\D$'s and further properties they have, when we explicitly compute them from a given metric. Moreover, and this is a crucial point, even when we work with a specific metric $g_{ab}$ of the conformally completed spacetime, there is nothing special about the $\D$ we obtain. This is because we can always choose a \textit{different} conformal completion. Thus we will generally obtain \textit{different} $\D$'s \textit{for the same spacetime}, simply because we used different conformal completions. Therefore, in a sense, there is ``gauge redundancy'' in our description. This is a point which we will further explore in the next chapter and what we found here in this subsection will come in very handy.

Before doing so, however, we elaborate on an interesting connection between $\scrip$ and so-called non-expanding horizons.\bigskip

\subsubsection{Interlude: Non-Expanding Horizons}\label{subsubsec:NEH}
Capturing the essence of horizons is a delicate task and deserves more attention than we can dedicate to it in this brief interlude. What is important to us, is that there exists a notion of horizons which makes use of \textit{local} properties of null surfaces, as opposed to teleological notions such as the one encountered when defining \textbf{event horizons}\footnote{For readers not familiar with different notions of horizons and their subtleties: Event horizons require us to know \textit{the whole history of the spacetime under consideration} for their definition. We need to know what happened in the spacetime and what will happen in the far future. This is why we refer to it as a \textit{teleological} notion. The local notion we consider in this interlude is free of this disturbing and limiting property of having to know the whole history.}. The notion we are interested in is the one of a \textbf{non-expanding horizon}, or \textbf{NEH} for short: Let $(\M, g_{ab})$ be a spacetime endowed with the usual Levi-Civita connection $\nabla$. Furthermore, let $\N\subset\M$ be a co-dimension $1$ hypersurface defined by the equation
\begin{align}
    \Phi(x^a) = 0.
\end{align}
Define the normal $1$-form to this hypersurface as $\rho_a := \nabla_a\Phi$ and assume that it is null. That is, assume it satisfies $g^{ab}\rho_a \rho_b = 0$. This makes $\N$ a null hypersurface. 

Equally, we can define other normals by rescaling $\rho_a\mapsto\rho'_a=f\rho_a$ where $f$ is a non-vanishing function. Now let $\N$ be null. In this case, we denote the null normal $\rho_a \equiv n_a$. Covariant indices can be pulled back to $\N$ by restricting their action to vectors tangent to $\N$, e.g., a one form $\omega_a$ on $\M$ defines a one form $\pb{\omega}_a\in \N$ so that 
\begin{align}
    \pb{\omega}_a v^a = \omega_a v^a
\end{align}
for all $v^a\in \M$ tangent to $\N$. Note that then $\pb{n}_a = 0$ since $\pb{n}_a v^a = n_a v^a = 0$ for all $v^a$ tangent to $\N$. Note also that this definition implies that the pullback contains only a part of the original information. \\
We denote equality restricted to the submanifold $\N$, as we do throughout the whole script, using the symbol ``$\equalhat$''. In the particular case of NEH this can be understood as follows: $\pb{\omega}_a\equalhat\pb{v}_a$ implies equality only upon contraction with a vector tangent to $\N$ and $\omega_a\equalhat v_a$ upon contraction with an arbitrary vector. Given the above definitions it holds that
\begin{itemize}
    \item[1)] By definition, $n_a$ is hypersurface orthogonal, i.e., for tangent vectors $v^a,w^a$, it holds that
    \begin{align}
        v^a \omega^a \nabla_{[a}n_{b]} &\equalhat0  &\Longleftrightarrow &&   \underleftarrow{\nabla_{[a} n_{b]}} &\equalhat0.
    \end{align}
    \item[2)] Since $n_a n^a=0$, $n_a$ is tangent to $\N$. As such, $n^{a}$ is geodesic and satisfies the geodesic equation
    \begin{align}
        n^a \nabla_a n^b = \kappa_{(n)} n^b,
    \end{align}
    where $\kappa_{(n)}$ is the \textbf{surface gravity} of $\N$ with respect to the specific choice of $n^a$.\footnote{It is important to remember that when translated into a thermodynamical point of view, the surface gravity can be understood as an analogue of temperature. Thus, when entering the realm of thermodynamics, the freedom in the choice of $n^a$ has to be removed somehow as otherwise it would result in an ambiguous definition of surface gravity and hence temperature.}
\end{itemize}
We can further introduce a basis on each tangent space $T_p \N$ with $p\in \N$. For a given null normal $n^a$ we can always choose two tangent vectors $v,w$ such that $\{n,v,w\}$ form a basis of $T_p \N$ for all $ p$. Since $\N$ is null, every tangent vector to $\N$ must either be spacelike or null itself (we gave an intuition for this subtlety already in a side note in section~\ref{sn:TimetranslationBecomingNull}). Let us first chose both tangent vectors to be spacelike and normalized in a ``standard'' sense, i.e., $v^a w_a = 0$ and $v^a v_a = w^a w_a = 1$. Then, the spacelike subspace spanned by $\{v,w\}$ has two null normals, one of which is chosen to be $n^a$. We denote the other null normal by $\ell^a$ and normalize such that $n^a\ell_a=-1$. The term ``null normal'' has to be used with caution here. That is, we have to differentiate between null normals to $\N$ and null normals to the $2$-dimensional subspace spanned by $v,w$: The null surface $\N$ has only one null normal which we denoted as $n^a$ here, the vector $\ell^a$ is a null normal only with respect to the $\{v,w\}$-subspace. Note at this point that while $n^a$ is inward-pointing, $\ell^a$ is an outward-pointing null normal. Altogether, the vectors $\{n,\ell,v,w\}$ define a natural basis of $T_p \M$. For our purposes, it is convenient to choose a more ``familiar'' basis for $T_p \M$ namely a basis in which the tangent vectors to $\N$ are also null. This way we obtain the known null tetrad $\{n,\ell,m,\bar{m}\}$, in literature often named \textbf{Newman-Penrose basis}. All we have to do is define 
\begin{align}
    &m\equalhat\frac{1}{\sqrt{2}}(v+iw)&\text{and}&&\bar{m}\equalhat\frac{1}{\sqrt{2}}(v-iw).
\end{align}
It follows that $m$ and $\bar{m}$ automatically satisfy $m^am_a=\bar{m}^a\bar{m}_a=0$ and $m^a\bar{m}_a=\bar{m}^am_a=1$. It also holds that $\spn\{v,w\} = \spn\{m,\bar{m}\} $ and consequently $\{n,\ell,m,\bar{m}\}$ establishes an alternative basis for $T_p \M$.

Carrying on, we can define an intrinsic metric $q_{ab}$ of $\N$ as the pullback of the spacetime metric $g_{ab}$
\begin{align}
    q_{ab}\equalhat \pb{g}_{ab},
\end{align}
which is degenerate as $q_{ab} n^a \equalhat \pb{g}_{ab} n^a \equalhat \pb{n}_a = 0$, i.e., $n^a$ is the degenerate direction of $q_{ab}$. The signature of $q_{ab}$ is $(0,+,+)$ and its determinant vanishes. As discussed already in the context of $\scrip$, an inverse can be imposed here anyways by its defining equation 
\begin{align}\label{eq:InverseEquation}
    q^{ab} q_{ac} q_{bd} = q_{cd}.
\end{align}
For $\N$, we find a freedom of choice in the inverse given by $\tilde{q}^{ab} = q^{ab} + n^{(a} X^{b)}$ where $X^a$ is tangent to $\N$. Both $\tilde{q}^{ab}$ and $q^{ab}$ satisfy the inverse equation~\eqref{eq:InverseEquation}. \\
It is well known that the full spacetime metric of this interlude, $g_{ab}$, can be constructed solely from the tetrad we derived here. In fact, in  Exercises~\ref{ex:MetricInNPForm} it was shown that $g_{ab} = -2 \ell_{(a}n_{b)} - 2 m_{(a}m_{b)} = -2 \ell_{(a}n_{b)} + q_{ab}$, where $q_{ab}$ is the metric on our hypersurface $\N$. As $g_{ab}$ is non-degenerate, we are guaranteed to find a unique covariant derivative $\nabla$ on full spacetime such that $\nabla_a g_{bc}=0$ and $\nabla$ is torsion free. \\
This, unfortunately, is not the case for the null surface $\N$. Its intrinsic metric $q_{ab}$ is degenerate, prohibiting the uniqueness of a covariant derivative operator $\D$ on $\N$. It is impossible to pick out a preferred and well-defined derivative operator on the null surface if we do not impose additional conditions on $\N$.\footnote{``Well-defined'' here, as above, has to be equated with metric-compatible and torsion-free.}  \\
The difficulties of defining a derivative operator on $\N$ are exactly equivalent to the ones we are facing when encountering derivative operators defined at $\scri$. Furthermore, many definitions (such as the tetrads and the inverse metric) made here admit direct analogues in the context of $\scri$. For the null surface $\N$, we will solve issues arising with the definition of the derivative operator partially by using NEHs, and, as suggested above, we can do so for $\scrip$ as well. \\
Any null hypersurface defined in similar fashion as $\N$ can be cast into a NEH by imposing additional properties. For that purpose, we will keep the notation of the spacetime metric $g_{ab}$ and of the metric on the null hypersurface $q_{ab}$, as well as of the derivative on general spacetime $\nabla$. As the name suggests, we define NEH to be horizons in an \textbf{equilibrium state}. In the weakest form, this translates to 
\begin{align}\label{eq:ExpansionAndShear}
    \lie_n q _{ab} &\equalhat 0 &\Leftrightarrow && \underleftarrow{\nabla_{(a}  n_{b)}} &\equalhat 0
\end{align}
since $\lie_n q_{ab} \equalhat \lie_n \pb{g}_{ab}\equalhat 2\underleftarrow{\nabla_{(a}  n_{b)}}$. We can interpret the former statement as $\N$'s intrinsic geometry to be invariant under time translations which is the very definition of being ``static'' or in ``equilibrium''. This becomes more apparent when we rewrite the null normal of the hypersurface, $n^a$, and the Lie derivative in its direction, $\lie_n$, in terms of an affine parameter $u$ which, based on the coordinate chart $\{t,r,\theta,\phi\}$, can be identified with the retarded time coordinate. Note that we will \textit{not impose} equation \eqref{eq:ExpansionAndShear} for NEHs but we will show that it naturally follows from their definition. For completeness, note also that $\pb{\nabla_{a} \pb{n}_{b}}$ can be decomposed into a \textit{trace} which is commonly called \textbf{expansion} of $n^a$, here denoted as $\theta_{(n)}$, and a \textit{symmetric trace-free part} called \textbf{shear} of $n^a$, here $\sigma_{ab}$. They read
\begin{align}
    \theta_{(n)} &= q^{ab} \nabla_a n_b, & \sigma_{ab} &= \pb{\nabla_{(a} n_{b)}} -q_{ab} a^{cd} \nabla_c n_d.
\end{align}
Both vanish if $\pb{\nabla_{a}\pb n_{b}}$ is zero and are independent of the choice of the inverse metric, as will be shown in Exercise~\ref{ex:NEH}. As the trace of $\pb{\nabla_{a} \pb{n}_{b}}$ is part of the definition of NEHs we now have everything at hand to give a precise description.
\\

\textbf{Definition 6.1: Non-expanding horizon}\\
A three-dimensional hypersurface $\Delta\subset \M$ of a space time $(\M, g_{ab})$ is said to be a \textbf{non-expanding horizon} if it satisfies the following conditions:
\begin{itemize}
    \item[1)] $\Delta$ is topologically $\bbS^2\times \bbR$.
    \item[2)] The expansion $\theta_{(n)}$ vanishes for any null normal $n'^a=fn^a$.
    \item[3)] Einstein's field equations hold on $\Delta$.
    \item[4)] The energy-momentum  tensor $T_{ab}$ of external matter fields is such that at $\Delta$, the contraction $-T^a_b n^b$ is future directed for any null normal.
\end{itemize}
Let us take a moment to digest this definition and remark on a few points. First, from now on we focus on NEHs instead of ``normal'' null hypersurfaces $\N$, hence, we switch the notation of our hypersurface from $\N$ to $\Delta$. Secondly, note that in this framework with the above equilibrium condition, $n^a$ is hypersurface orthogonal. In Exercise~\ref{ex:NEHIII} it is shown that this has far reaching implications and one can conclude that every null normal of $\Delta$ Lie drags the metric $q_{ab}$ intrinsic to $\Delta$.\bigskip

Let us assume now the hypersurface we are considering in this interlude is indeed a NEH and all the properties we derived so far hold. We denote our null surface as $\Delta$ and construct two vectors $X^a, Y^a$ tangent to $\Delta$. For null surfaces in general, we do not have a natural way to pull back the full spacetime derivative operator $\nabla$ to obtain an derivative on the hypersurface. For NEHs, however, there is a well-defined way. What we mean by ``well-defined'' boils down to the core principles of the derivative operator. That is, we want to find an operator $\D=\pb \nabla$ satisfying the three axioms of a derivative operator
\begin{align}
    \D_a(X^b+Y^b)&=\D_a X^b+\D_a Y^b\notag,\\
    \D_a(f X^b)&=f\D_a X^b+X^b\D_a f\notag,\\
    X^a\D_a f &=\lie_X f.
\end{align}
Note that in the axioms we consider vectors $X,Y$ tangent to $\Delta$ as the derivative operator we aim for is intrinsic to $\Delta$. Hence, the tangent spaces of $\Delta$ are the vector spaces it is applied to. If we work through the axioms equipped with the machinery of the NEH-toolbox we find that the well-definiteness of the derivative operator boils down to the following statement: If $Y^a\nabla_aX^b$ is tangent to $\Delta$ then $\D_a=\pb{\nabla_a}$ is well-defined. Using the result of Exercise~\ref{ex:NEHIII}, namely that $n$ is twist-free, and $\pb{\nabla_{(a}}\pb n_{b)}\equalhat0$, we find $\pb{\nabla_{a}}\pb n_{b}\equalhat0$. Then, for $Y_an^a=X_an^a=0$, i.e., two vectors $X^a,Y^a$ tangent to $\Delta$ we find
\begin{align}
    Y^b\nabla_b(X^a n_a)&\equalhat Y^b X^a\nabla_b n_a + Y^b n_a \nabla_b X^a\notag\\
    &\equalhat Y^b X^a\pb{\nabla_b n_a} + n_a Y^b \nabla_b\notag\\
    &\equalhat l_a Y^b\nabla_b X^a\notag\\
    &\equalhat 0,
\end{align}
which means that $Y^a \nabla_a X^b$ is truly tangent to the hypersurface $\Delta$.
Consequently, $\D$ is defined by $\D_a X^b = \pb{\nabla_a}\tilde{X}^b$, where $\tilde{X}^b$ is an arbitrary expansion of $X^b$ to full spacetime. It is left as an exercise to show that the definition is independent of the extension. The above definition of the derivative operator intrinsic to $\Delta$ can be applied to co-vectors and, subsequently, tensors of arbitrary type on $\Delta$:
\begin{align}
    \D_e T\ud{a...b}{c...d}=\pb{\nabla_e \pb T\ud{a...b}{c...d}}.
\end{align}
Most importantly, the derivative operator $\D$ is metric-compatible with respect to the intrinsic metric of the NEH $\Delta$, i.e., $\D_aq_{bc}\equalhat \nabla_ag_{bc}$. \\

It seems like we reached the goal of this interlude. Before we turn back to our null infinity hypersurface $\scrip$, though, another remark is in order. The intrinsic derivative of our null normal $n^a$ can be shown to have a ``special'' form. To do so, we first notice that since $X^aY^b\nabla_an_b\equalhat0$ it follows that $X^a\nabla_an_b\sim n_b$. Capturing the proportionality to $X^a$ of the latter expression, we conclude that $X^a\nabla_an_b\equalhat X^a\omega_an_b$ where we define $\omega_a$ as the \textbf{induced normal connection} determined by $\D$ and $n^a$. Then, it holds that 
\begin{align}\label{ex6.2}
    \pb{\nabla_a}n_b\equalhat \D_an_b\equalhat \omega_an_b
\end{align}
and one can show that based on this definition, $\lie_n \omega_a = 0 = \lie_\ell \omega_a$. \\

Let us summarize the results so far. We found that a null surface has no unique derivative operator due to the degeneracy of the intrinsic metric and the pullback of the derivative operator from general spacetime to the hypersurface is only well-defined if we impose additional constraints on the null surface. That is, if we make it a NEH. As mentioned earlier, for us this is a very convenient observation as null infinity in a conformally completed spacetime is mathematically a NEH. Hence, we can use what we derived here regarding the operator $\D$ in order to obtain a well-defined derivative operator on $\scri$.\bigskip

\newpage
\subsection{Exercises}
\renewcommand{\thesection}{\arabic{section}}

\begin{Exercise}[]\label{ex:NEH}
    Let $(\N, q_{ab})$ be a co-dimension one null hypersurface embedded in $(\M, g_{ab})$ and let $n^{a}$ be the null normal to $\N$. Denote the torsion-free and metric-compatible covariant derivative on $(\M, g_{ab})$ by $\nabla$. The pullback $\pb{\nabla_{a}n_{b}}$ (from $\M$ to $\N$) can be decomposed into a trace and a symmetric trace-free part as
\begin{align*}
    \theta_{(n)} &:= q^{ab}\nabla_an_b, & \sigma_{ab}&:=\pb{\nabla_{(a}n_{b)}} -q_{ab}a^{cd}\nabla_cn_d,
\end{align*}
where $q^{ab}$ is a pseudo-inverse. Prove that $\theta_{(n)}$ and $\sigma^{(n)}_{ab}$ are independent of the choice of the pseudo-inverse, i.e., they are invariant under $q^{ab}\mapsto\tilde{q}^{ab}=q^{ab}+n^{(a}X^{b)}$, where $X^{b}$ is any vector tangent to~$\N$.
\end{Exercise}

\begin{Exercise}[]\label{ex:NEHII}
Let $\omega_a$ be the induced normal connection of a non-expanding horizon (NEH). Compute its behavior under transformations $n^a\mapsto n'^a = f n^a$ and find an explicit expression for the surface gravity $\kappa_{(n)}$ of a NEH in terms of $n^a$ and $\omega_a$.\bigskip
    
\textit{Hint: Use equation \eqref{ex6.2}.}
\end{Exercise}

\begin{Exercise}[]\label{ex:ConservedChargeI}
    Prove the transformation law \eqref{transprop} in Minkowski space for a displacement vector $d^a$ leading from an origin $O$ to $O'$.
\end{Exercise}

\begin{Exercise}[]\label{ex:WellDefinedActionOnVectors}
    Define $\D :=\pb{\nabla}$. In this chapter it was shown that the action of this operator on scalars $f$ and $1$-forms $\omega_a$ is well-defined. Show that if the action on $1$-forms is known, we also know how $\D$ acts on vector fields $v^{a}$ tangent to $\scrip$, even when there is no well-defined inverse metric to raise indices.\bigskip
    
    \textit{Hint: Consider $q_{bc}n^{c} \D_a v^{b}$ and $\omega_b \D_a v^{b}$.}
\end{Exercise}

\begin{Exercise}[]\label{ex:D_on_1_Form}
    Let $\D$ and $\D'$ be two differential operators which satisfy the properties~\eqref{eq:D_Properties}. Furthermore, let $m^{a}$ be part of a Newman-Penrose null tetrad. Show that 
    \begin{equation}
        \left(\D - \D'\right)\alpha_b = 0,
    \end{equation}
    where $\alpha_b = f\, m_a + \bar{f}\, \bar{m}_a$.
\end{Exercise}

\begin{Exercise}[]\label{ex:NEHIII}
Let $n^a$ be the null normal of a NEH as defined in the interlude on NEHs~\ref{subsubsec:NEH}. Prove that 
    \begin{align*}
       &\pb \nabla_{(a}\pb n_{b)}\equalhat 0 &&\text{and hence}&\lie_nq_{ab}=0.
    \end{align*}
    
    \textit{Hint: Prove first that $n^a$ is shear free using that it is hypersurface orthogonal and thus twist-free. Also, use the Raychaudhuri equation.}
\end{Exercise}

\begin{Exercise}[]\label{ex:LieDragNullNormal}
    Show that if $n^a$ is a null normal such that $q_{ab} n^{b}=0$ and $\lie_nq_{ab}=0$, then both equations hold for all null normals $n'^{a} = f n^{a}$, where $f$ is any smooth function.\bigskip
    
    \textit{Hint: Use the Cartan identity.}
\end{Exercise}

\newpage
\asection{7}{Radiative Modes in Full, Non-Linear General Relativity}\label{Chap7}
Let us briefly recapitulate the story so far. We have argued that the derivative operator on null infinity incorporates information about the radiative modes of the gravitational field in the bulk. We engaged on a search for a well-defined derivative operator on $\scrip$ and found that there is not one, but many such operators. In fact, any torsion-free and metric-compatible operator which satisfies $\D_a n^{b} = 0$ is admissible. 
In an attempt to quantify the ambiguity, we analyzed the operator's action on $1$-forms $\alpha_a$ transverse to $n^{a}$. In doing so, we found that all derivative operators on $\scrip$ have the same action on such $1$-forms. That is, any two derivative operators $D$, $D'$ which are torsion-free, metric compatible, and which satisfy $\D_a n^{b} = 0 = \D'_a n^{b}$ satisfy
\begin{equation}
    \left(\D_a-\D'_a\right)\alpha_b = 0.
\end{equation}
However, the actions of $\D$ and $\D'$ on $\ell_a$, which is a $1$-form which is \textit{not} transverse to $n^{a}$, are generally different. We have argued that this is where radiative modes will be revealed. It is this ``non-universality'' of the action of $\D$ on $\ell_a$ which allows us to distinguish between spacetimes with and without radiation. 

In this chapter, we will make this notion more precise by the introduction of equivalence classes of derivative operators. We will also ``count'' how many operators there are. We will find that each derivative operator possesses three degrees of freedom, in a sense which will become clear in the next subsection. One degree of freedom is pure ``gauge'' and corresponds to a choice of conformal completion. The remaining two degrees of freedom encode the radiative modes of the gravitational field. 
\bigskip

\subsection{Equivalence Classes of Derivative Operators on Null Infinity}\label{susec:DefiningClassesOfDerivatives}
Let $\D$ and $\D'$ be two distinct, torsion-free, and metric-compatible covariant derivative operators on $\scrip$ which satisfy $\D_a n^{b} = 0 = \D'_a n^{b}$. Furthermore, let $\alpha_b$ be \textit{any} $1$-form on the co-tangent space of $\scrip$. As is well-known from differential geometry, the difference between any two covariant derivative operators is a tensor. Thus, we can write
\begin{align}\label{eq:DifferenceInDerivativesI}
    (\D'_a - \D_a)\alpha_b = C\du{ab}{c} \alpha_c,
\end{align}
where $C\du{ab}{c}=C\du{(ab)}{c}$ is the tensor we alluded to above. Notice that it has to be symmetric because the connections used to construct $\D$ and $\D'$ are torsion-free. We know from the last chapter that the action of $\D$ and $\D'$ on $1$-forms $\omega_a$ which are transverse to $\scrip$, i.e., which satisfy $\omega_a n^{a}$, is universal in the sense that $\left(\D'_a-\D_a\right)\omega_b = 0$. From this we conclude
\begin{align}\label{7.3}
    \underbrace{(\D'_a - \D_a)\omega_b}_{=0} &= C\du{ab}{c}\omega_c &\Longleftrightarrow&&  C\du{ab}{c} = \Sigma_{ab} n^c.
\end{align}
In words: we can conclude that the tensor $C\du{ab}{c}$ is the product of a symmetric tensor $\Sigma_{ab}$ and the null normal $n^{c}$. By its very construction, this ensures that $C\du{ab}{c}\omega_c = 0$ for all transversal $1$-forms. In addition to that, we also know that $\D_a n^{b} = 0 = \D'_a n^{b}$, from which we conclude
\begin{align}
    \underbrace{(\D'_a - \D_a) n^b}_{=0} = C\du{ac}{b} n^c &= \left(\Sigma_{ac} n^c\right)n^b & \Longleftrightarrow&& \Sigma_{ac} n^c = 0.
\end{align}
It follows that $\Sigma_{ab}$ is transverse to $n^{a}$. From these findings we deduce that our freedom in choosing a derivative operator on $\scrip$ is reduced to choosing a symmetric tensor which is transverse to $\scrip$. Put differently, there are as many distinct covariant derivative operators on $\scrip$ as there are tensors of this type. How many such tensors are there?

Given that $\Sigma_{ab}$ is a rank-$2$ tensor which is defined on a three-dimensional manifold, it has $3\times 3$ components. However, it is symmetric and this reduces the number of independent components to $\frac{3(3+1)}{2} = 6$. Taking into account the transversality, which imposes the three constraint equations $\Sigma_{ab}n^{b} = 0$, finally leaves us with \textit{three} independent components which completely specify $\Sigma_{ab}$.

This is not the end of the story, though. We have to keep in mind that we work in a fixed divergence-free conformal frame $(q_{ab}, n^{a})$. However, there is nothing special about the choice we made and someone else might have chosen a \textit{different} divergence-free conformal frame $(q'^{ab}, n'^{a})$ for the same spacetime. We recall that any two such frames are related by a conformal transformation of the form $(q_{ab}, n^{a})\mapsto (q'^{ab}, n'^{a}) = (\omega^2 q^{ab}, \omega^{-1}\,n^{a})$, where $\omega$ is a smooth, nowhere vanishing function which is Lie dragged by $n^{a}$. How does this rescaling freedom affect our covariant derivative operator?
Notice that fixing a divergence-free conformal frame is akin of a partial ``gauge fixing'' in the following sense: We are free in choosing a conformal completion of the physical spacetime, $(\hatM, \hat g_{ab})\mapsto (\M, g_{ab}, \Omega)$. However, as we have spelled out in Chapter~\ref{Chap3}, there is a canonical choice for a conformal completion and we can always make that choice. Of course, we are talking about a divergence-free conformal frame. Hence, among all the possible ``gauge'' choices $\Omega$ we have selected one for which $\nabla_a n^{a}$ holds. However, this choice does not completely fix the ``gauge'' because there are infinitely many $\Omega$'s which lead to divergence-freeness and these $\Omega$'s are related by the residual rescaling freedom described above. Thus, we have not fixed a ``gauge'', but rather a ``gauge class''. This means that we actually have an equivalence class of derivative operators $\D$. Namely, given a $\D$ which satisfies all the conditions spelled out above, we can generate a new operator simply by a conformal rescaling, thus obtaining an equivalence class $[\D]$ in this sense that $\D \sim \D'$ if and only if they are related by a conformal rescaling. Observe that this is consistent with what we said in~\ref{Chap6D} about $\nabla$ being determined by $\hat g_{ab}$ (the \textit{physical} metric) and $\Omega$. Thus, in a sense, when we pull back $\nabla$ to $\scrip$ in order to define $\D$, this intrinsic derivative operator ``knows'' about $\Omega$. But this $\Omega$ is a ``gauge artifact''. Thus, we anticipate that one of the three degrees of freedom in $\Sigma_{ab}$ actually just represent the freedom to choose a conformal frame, leaving us with only two degrees of freedom which can be traced back to $\hat g_{ab}$. These are the two true degrees of freedom of the gravitational field and they represent the radiative modes!

This is the qualitative picture and it is somewhat hand-waving in certain places. So let us make everything precise and demonstrate that this picture is actually correct. To do so, we perform a conformal rescaling under the assumption that $\omega \equalhat 1$, but $\omega\neq 1$ off $\scrip$. It obviously follows that the universal structure of $\scrip$ is invariant under this particular transformation:
\begin{align}
    q'_{ab} &\equalhat q_{ab}, &\text{and} && n'^a &\equalhat n^a.\notag
\end{align}
However, as we have hopefully made clear in the last and in this chapter, the covariant derivative $\D$ is \textit{not} fully determined by the universal structure. In a sense, it probes the structure of $\scrip$ ``one level deeper''. More precisely, $\D$ is not oblivious to what happens in a neighborhood of $\scrip$ and therefore is affected by this particular transformation. Put in yet a different way, even when $\omega\equalhat 1$ this does \textit{not} imply that the derivative of $\omega$ off $\scrip$ is trivial (i.e., it does not vanish) and hence it \textit{does} affect the derivative operator $\D$.

To show that the action of $\D$ is altered by the above rescaling transformation, we step away from $\scrip$ into the conformally completed spacetime. There, we consider the derivative operator $\nabla$. As we know very well be now, this operator changes under a conformal rescaling to $\nabla'$, and the two operators are related by (see Exercise~\ref{ex:conformaltrans}):
\begin{align}
    (\nabla'_a - \nabla_a)\alpha_b &= C\du{ab}{c} \alpha_c &\text{with}&& C\du{ab}{c} &= -\omega^{-1} \left(2\delta\ud{c}{(a} \nabla_{b)}\omega-(\nabla^c\omega)g_{ab}\right).
\end{align}
We can use this result to relate the derivative operators $\D := \pb{\nabla}$ and $\D' := \pb{\nabla}'$ on $\scrip$. At this point, a word of caution is in order when performing this pullback operation.
\begin{mysidenote}{The meaning of ``pulling back free indices''}{PullBackIndices}
    When considering composite expressions with many indices, where some of them are fully contracted, one has to be careful with the pullback operation. In such a case we say, maybe somewhat unorthodoxically, that we ``pull back the free indices''.  This terminology is probably best explained with an example: Assume we are given a tensor $Q_{ab}$ which can be written as $(\nabla_c v^{c}) h_{ab}$. The pullback of this tensor is then given by $(\nabla_c v^{c}) \pb{h}_{ab}$, where it is understood that the scalar $\nabla_c v^{c}$ is restricted to the manifold onto which we are pulling back. But no additional pullback operation on individual tensors, i.e., something like $\pb{\nabla_c v^{c}}$ or $\nabla_c \pb{v^{c}}$, is necessary. In this sense, the dummy indices remain untouched and only the free indices $a$ and $b$ are pulled back.
\end{mysidenote}
With this clarification out of the way, we finally compute the pullback of $(\nabla'_a - \nabla_a)\alpha_b$ to $\scrip$. Taking care to only pull back the free indices $a$ and $b$, this leads to
\begin{align}
    \underleftarrow{(\nabla'_a-\nabla_a)\alpha_b}&\equalhat(\D'_a-\D_a)\pb{\alpha_b}\notag \\
    &\equalhat-\omega^{-1}\left(2 [\delta^{c}{}\underleftarrow{\prescript{}{(a}{\nabla}_{b)}\omega]}\alpha_c - (\nabla^c\omega)\alpha_c \,\pb{g_{ab}}\right)\notag \\
    &\equalhat-1\left(0 - (\nabla^c\omega)\alpha_c\, q_{ab}\right)\notag\\
    &=\left(\nabla^c\omega\right)\alpha_c\, q_{ab}.
\end{align}
In the second-to-last row we used that $\omega\equalhat1$ and we made use of the fact that
\begin{equation}
    \delta^{c}{}\underleftarrow{\prescript{}{(a}{\nabla}_{b)}\omega} \equalhat 0,
\end{equation} 
because we take the derivative of $\omega$ on $\scrip$, where $\omega$ is constant. Next, we use the fact that a scalar which is constant on a submanifold, has a gradient which is normal to that submanifold, i.e., $\left.\nabla^{a} \omega\right|_{\scrip} = f\, n^{a}$, where $f$ is some smooth function. This delivers the following result.
\begin{align}\label{eq:PullbackDerivativeDifference}
    (\D'_a - \D_a)\underleftarrow{\alpha_b} = f\,n^{c}\alpha_c\, q_{ab}.
\end{align}
Comparing this equations to equation~\eqref{eq:DifferenceInDerivativesI} finally implies that
\begin{equation}
    \Sigma_{ab} = f\, q_{ab}.
\end{equation}
This is a key result and we pause to put it into context: Performing a conformal rescaling with $\omega\equalhat 1$ leaves the universal structure invariant but changes the covariant derivative operator. The change in this operator is proportional to the intrinsic metric of $\scrip$. This proportionality function vanishes if and only if $\omega = 1$ everywhere, which means it is equivalent to applying the identity transformation to the universal structure. Thus, $f$ captures information about the conformal rescaling and represents the ``gauge artifact'' we described earlier. Furthermore, when the action of two derivative operators differs by $f\, q_{ab}$, we can conclude that the difference is due to the use of two different divergence-free conformal frames. Thus, the difference has nothing to do with physics, but with the arbitrary choice of our conformal frame. This observation leads us to introduce equivalence classes of covariant derivatives. We denote these equivalence classes by $[\D]$ and define equivalence as
\begin{align}
    &\D \sim \D' &\Longleftrightarrow && \left(\D'_a-\D_b\right)\alpha_b = f\, q_{ab}.
\end{align}
Ultimately, when computing physical quantities using $\D$, they have to be gauge-independent, meaning that they cannot depend on $f$. In yet other words, this means that physical quantities do not depend on our arbitrary choice of a conformal frame.

Our identification of $f$ with the rescaling freedom suggests that  $\Sigma_{ab}$ carries two physical degrees of freedom, by which we mean that $\Sigma_{ab}$  carries two degrees of freedom which are independent of the conformal completion we chose and encode information contained in the physical metric $\hat g_{ab}$. To see this mathematically, we first isolate the ``gauge artifact'' $f$ by computing the trace of $\Sigma_{ab}$,
\begin{equation}
    q^{ab}\Sigma_{ab} = \frac12 f.
\end{equation}
Observe that the trace is independent of the choice of pseudo-inverse because $\Sigma_{ab}$ is transverse to $n^{a}$. The above equation also implies that the trace of $\Sigma_{ab}$ is pure gauge and suggests that its trace-free part is the carrier of the physical degrees of freedom. The trace-free part of $\Sigma_{ab}$ shall be denoted by $\sigma_{ab}$ and it is explicitly given by
\begin{equation}
    \sigma_{ab} := \Sigma_{ab} - \frac12 q_{ab} q^{cd}\Sigma_{cd}.
\end{equation}
This tensor allows us to distinguish equivalence classes from each other and thus it allows us to distinguish physically distinct spacetimes from each other!

To clarify that, observe that if $\D$ and $\D'$ do \textit{not} differ by $f\, \Sigma_{ab}$ they belong to two different equivalence classes and necessarily differ by $\sigma_{ab}$. Thus, it follows that two equivalence classes are disjoint if and only if $\sigma_{ab}\neq 0$,
\begin{align}
    [\D]\cap [\D]' &= \emptyset &\Longleftrightarrow && \sigma_{ab} \neq 0.
\end{align}
where $\sigma_{ab}$ can be computed with \textit{any} representative of $[\D]$ and $[\D]'$. Observe that $\sigma_{ab}$ is gauge-invariant by construction and that it inherits the following properties from $\Sigma_{ab}$ and $q_{ab}$:
\begin{equation}
    \boxed{
    \begin{aligned}\label{eq:propertiesshear}
       \sigma_{ab} &= \sigma_{(ab)}, & && &&  \sigma_{ab} n^b &=0, & && && q^{ab} \sigma_{ab} &=0
\end{aligned}
    }
\end{equation}
In other words, $\sigma_{ab}$ is gauge-independent, symmetric, transverse, and trace-free. The last three properties imply that $\sigma_{ab}$ carries two independent degrees of freedom. In conjunction with its gauge-independence, this suggests that it carries two degrees of freedom of the physical metric $\hat g_{ab}$!

However, below we will give an intuition of how this connection arises before more rigorously deriving the connection in Chapter~\ref{Chap8}.\bigskip

\subsection{Radiative Modes and Geometry}\label{susec:RadiativeModesNGeometry}
In subchapter~\ref{Chap6D}, we have demonstrated that the covariant derivative operator~$\D$ of $\scrip$ is completely fixed if we impose one additional condition. Namely, if we know what $\D_a \ell_b$ is, then we completely know $\D$. We have also argued that this equation is what ``transfers information'' from the physical $\hat g_{ab}$ to the operator $\D$. In other words, it is this equation which leaves an imprint of the radiative modes in $\D$ and, as we have seen in the previous subsection, these radiative modes are potentially encoded in the two degrees of freedom of $\sigma_{ab}$. Hence, there should be a relation between $\D_a \ell_b$, the tensor $\sigma_{ab}$, and the radiative modes of the gravitational field, which we denote by $h_+$ and $h_\times$. We will show that this is indeed the case, although we will only sketch how $h_+$ and $h_\times$ appear in the formalism. A satisfactory derivation will be postponed to Chapter~\ref{Chap8}.

Let us begin by expressing the action of $\D$ on $\ell_a$ in terms of $\sigma_{ab}$. To that end, fix a divergence-free conformal frame $(q_{ab}, n^{a})$ and carry out the construction of a Newman-Penrose null tetrad $(\ell^{a}, n^{a}, m^{a}, \hat{m}^{a})$ as described in subsection~\ref{Chap3_NullTetrad} of Chapter~\ref{Chap3}. By defining $\D:=\pb{\nabla}$, we obtain an equivalence class of derivative operators $[\D]$ which all satisfy the properties~\eqref{eq:D_Properties}. Then, introduce a \textbf{fiducial derivative operator} $\Dcirc\in [\D]$, defined by the requirement
\begin{align}
    \Dcirc_a \ell_b \overset{!}{=} 0.
\end{align}
The relation between the fiducial derivative and any other derivative $\D\in [\D]$ is given by
\begin{align}\label{eq:ShearAndDerivativeEll}
    (\Dcirc_a-\D_a) \ell_b = \Sigma_{ab} n^c \ell_c = -\Sigma_{ab} = \D_a \ell_b,
\end{align}
where we used $n^{c} \ell_c = -1$. Thus, we have found
\begin{equation}
    \D_a \ell_b = -\Sigma_{ab}.
\end{equation}
The trace-free part of $\Sigma_{ab}$ can now be expressed in terms of $\D_a \ell_b$ and we introduce a new tensor, known as the \textbf{asymptotic shear} $\sigma^\circ_{ab}$:
\begin{align}\label{eq:RunningOutOfNames}
    \sigma_{ab}^\circ := -\D_a \ell_b + \frac12 q_{ab} q^{cd} (\D_c \ell_d).
\end{align}
This is an important result because it tells us that the gauge-invariant information captured by $\sigma_{ab}$ can be computed from $\D_a\ell_b$. This is precisely what our qualitative picture suggested! Namely that the radiative modes are captured by $\D_a \ell_b$, because this tensor represents information which is present in~$\D$ but absent in the universal structure (see discussion in~\ref{Chap6D}). Let us also recall that $\ell_a$ is \textit{not} part of the universal structure.

To continue our investigation of $\sigma^\circ_{ab}$, we choose a $u=u_0$ cross-section and expand it in a basis of that cross-section. Notice that the properties of the asymptotic shear tell us that on any $u=u_0$ cross-section it has the form
\begin{align}\label{eq:AsymptoticShear}
    \sigma_{ab}^\circ = -(\bar{\sigma}^\circ m_a m_b + \sigma^\circ \bar{m}_a \bar{m}_b).
\end{align}
The minus sign is convention and the complex function $\sigma^\circ$ is called the \textbf{shear}. Notice that it has spin weight $2$, while its complex conjugate $\bar{\sigma}^\circ$ has spin weight $-2$. 

We now jump ahead and show the relation between the shear and the radiative modes $h_+,h_\times$, but we postpone a proper derivation to Chapter~\ref{Chap8}. To that end, notice that it follows from~\eqref{eq:AsymptoticShear} that the shear is defined as
\begin{equation}
    \sigma^\circ(u,\theta,\phi) = - \lim_{r\to\infty} \left(m^{a}m^{b}\nabla_a \ell_b\right),
\end{equation}
where we have extended all fields into the bulk of spacetime. This limit can be explicitly computed and compared with the linearized theory, where $h_+$ and $h_\times$ encode the radiative modes. One then finds
\begin{equation}
    \sigma^\circ(u,\theta,\phi) = \frac12\left(h^\circ_+ + i\, h^\circ_\times\right)(u, \theta, \phi).
\end{equation}
We refer to $h^\circ_+$ and $h^\circ_\times$ as the \textbf{strains} of the gravitational wave and define them as
\begin{align}\label{eq:shearatscri}
    h^\circ_+ &:= \lim_{r\to\infty}r h_+ \notag\\ 
    h^\circ_\times &:= \lim_{r\to\infty}r h_\times.
\end{align}
Those unfamiliar with the strain of a gravitational wave, we refer to section~\ref{subsec:strain} for a brief introduction. The latter equations explicitly show that there is a relation between $\D_a\ell_b$ and $\sigma_{ab}$, and that $\sigma_{ab}$ does indeed carry information about the two radiative degrees of freedom, as we have sketched here. 

With this, we conclude our brief outlook on the connection between the strains of the linearized theory and the shear obtained in the non-linear theory. At this point, we wish to develop a deeper understanding of the interplay between geometry and physics. Given a torsion-free and metric-compatible derivative operator, the only other interesting tensor to consider is the curvature tensor $\R\du{abc}{d}$. Since we are on the three-dimensional hypersurface $\scrip$, we use the letter $\R$ for the curvature tensor of $\D$ in order to distinguish it from the Riemann tensor $R$ defined on $\M$.

With this comment out of the way, we can determine the curvature tensor $\R\du{abc}{d}$ from
\begin{align}
    2\D_{[a}\D_{b]}\alpha_c = \R\du{abc}{d} \alpha_d.
\end{align}
Keeping in mind that in three or less dimensions the Weyl tensor vanishes due to its symmetries, we can decompose $\R\du{abc}{d}$ solely in terms of the Shouten tensor, which leads to
\begin{align}
    \R\du{abc}{d}\alpha_d = \left(q_{c[a}S\du{b]}{d}+S_{c[a}\delta\du{b]}{d}\right)\alpha_d.
\end{align}
Notice that because $q_{ab}$ is degenerate, $S_{ab} = S\du{a}{d}q_{bd}$ loses some information with respect to $S\du{a}{d}$. Also, just as before, we are interested in the ``gauge-invariant'' part of the curvature, which means we have to study the conformally invariant part of it. To that end, notice that $S_{ab}$ as well as $S\du{a}{b}$ transform under the conformal rescaling $(q_{ab},n^a)\mapsto(\omega^2 q_{ab},\omega^{-1}n^a)$. In fact, one finds
\begin{align}
    S'\du{a}{b}&=\omega^{-1}S\du{a}{b}-\omega^{-3}\D_a(q^{bc}\D_c\omega)+4\omega^{-4}(\D_a\omega)q^{bc}\D_c\omega-\omega^{-4}\delta\du{a}{b}(q^{cd}\D_c\omega\D_d\omega) \notag\\
    S'_{ab}&=S_{ab}-2\omega^{-1}\D_a\D_b\omega+4\omega^2(\D_a\omega)(\D_b\omega)-\omega^2q_{ab}(q^{cd}\D_c\omega\D_d\omega).
\end{align}
In order to obtain the conformally invariant part of $S_{ab}$, it is convenient to work in a Bondi frame. Recall from subsection~\ref{susec:FluxandSupermomentum} that a Bondi frame is defined by having $q_{ab}$ as the metric of the unit $2$-sphere. In this particular frame, one finds that the trace-free part of $S_{ab}$ is also its conformally invariant part:
\begin{align}\label{eq:DefinitionBondi}
    N_{ab} = \underbrace{S_{ab}-\frac12 q^{cd} S_{cd} q_{ab}}_{\text{conformally invariant}}.
\end{align}
This particular combination of tensors is known in the literature as the \textbf{Bondi news tensor}.\footnote{Interesting historical side remark: The Bondi news $N$, to be introduced later, describes gravitational radiation in asymptotically flat spacetimes. Since gravitational waves contain information about the processes that created them, one can say that $N$ heralds the news that something interesting happened a long time ago and very far away. Back in the 1960's, a lot of research was geared toward finding this news brought to us by distant phenomena. For more mathematical details, we refer the reader to~\cite{Bondi:1962}} One can easily see that the Bondi news inherits the properties $N_{ab}=N_{(ab)}$, $N_{ab}n^b=0$ and $N_{ab}q^{ab}=0$ from the intrinsic metric and the Shouten tensor. 

Observe that the curvature tensor is constructed from the derivative operator and that we decomposed it in terms of the Shouten tensor. Thus, the Shouten tensor also depends on the derivative operator. With the Bondi news tensor, we have introduced a new tensor which depends on $\D$ in a ``gauge-invariant'' way. It is therefore natural to ask, what the relation between $N_{ab}$ and $\sigma_{ab}$ might be. In Exercise~\ref{ex:RiemannShear} it is shown that the two tensors are related to each other via
\begin{align}\label{eq:bondiI}
    N_{ab} = 2\lie_n \sigma_{ab}^\circ =: \dot{\sigma}_{ab}^\circ.
\end{align}
This is one of the most important results in these notes! What we derived here is a relationship between a tensor $N_{ab}$, which has been constructed using solely the geometry of the boundary $\scrip$, with the shear $\sigma_{ab}$, which encapsulates the radiative degrees of freedom of the bulk. This is a remarkable interplay between physics and geometry. 

Because the Bondi news tensor has the properties $N_{ab}=N_{(ab)}$, $N_{ab}n^b=0$ and $N_{ab}q^{ab}=0$, it is easy to see that it can be written as
\begin{align}
    N_{ab} = \bar{N}^\circ m_a m_b + N^\circ \bar{m}_a \bar{m}_b,
\end{align}
where $N^\circ$ is a complex function of spin weight $2$ which is simply called the \textbf{Bondi news}. This expansion closely resembles the one for the asymptotic shear given in~\eqref{eq:AsymptoticShear}. Qualitatively, the Bondi news tensor can also be regarded as the analogue of the Maxwell field strength tensor. Both tensors are conformally invariant and trace-free and, more importantly, the square of $N_{ab}$ is proportional to the energy-momentum flux across $\scrip$. We will see this in Chapter~\ref{Chap8}.

In summary, we found a useful and remarkable interplay between the geometry of $\scrip$ and the physics of gravitational waves. The conformally invariant part of the curvature of $\scrip$ is directly related to the strains $h_+$, $h_\times$, which contain information about gravitational waves produced by distant sources and violet physical phenomena such as the merger of two black holes. Of course, here we have just seen a glimpse of this latter relation between the shear and the strains. We will deepen this connection in Chapter~\ref{Chap8}. Moreover, we will revisit physical observables such as the energy carried by gravitational waves and see that the flux through $\scrip$ can be expressed in terms the Bondi news, or, equivalently, in terms of the strains.

Finally, we will also see how the Newman-Penrose scalars $\Psi^\circ_i$ can be expressed in terms of shear and strains. This is an important result which opens the door to comparing numerical relativity computations with observational data.

Before doing so, however, it is helpful to recall where the two strains $h_+, h_\times$ originate from. To that end, we briefly review linearized GR in an interlude.\bigskip

\subsection{Interlude: The Gravitational Wave's Strain}\label{subsec:strain}
Let us close this chapter with a brief excursion into the theory of gravitational waves in linearized GR. Here, we want to derive the strains $h_+,h_\times$ of a gravitational wave and demonstrate why they play an important role for the detection of such waves. All considerations in this subsection concern the \textit{physical} spacetime only and no conformal transformation is ever needed. We will therefore refrain from using hats on physical quantities. 

Let $(\M,g_{ab})$ be the physical spacetime and let $x^{a}$ be coordinates on the manifold. Assume there is a matter distribution described by an energy-momentum tensor $T^{ab}$ and assume Einstein's field equations,
\begin{align}
    R_{ab}-\frac12 g_{ab}R = 8\pi T_{ab},
\end{align}
hold. Now let us assume that an observer is located far away from a static matter distribution, whose contribution is contained in $T_{ab}$. Hence, it has an effect on spacetime and will be manifested in $g_{ab}$ through the field equations. Any change of $T_{ab}$, such as for instance rapid changes in the mass distribution, will induce a change in the gravitational field and thus lead to a change in the metric. For small changes $h_{ab}$, one can write (approximately)
\begin{align}
    \tilde{g}_{ab} = g_{ab}+h_{ab}.
\end{align}
As we will see later, the perturbations $h_{ab}$ can be interpreted as ripples on spacetime, more prominently known as \textit{gravitational waves}.

Assuming that the perturbations are much smaller than unity, $\left|h_{ab}\right|\ll1$, allows us to linearize the otherwise highly non-linear Einstein field equations. We shall further simplify our computations by assuming that the background metric $g_{ab}$ is flat, i.e., $\tilde{g}_{ab}=\eta_{ab}+h_{ab}$. Note that this technically corresponds to a vacuum background and $T_{ab}=0$. For this scenario, Einstein himself proved that the trace-free part of $h_{ab}$, that is
\begin{align}
    \tilde{h}_{ab} := h_{ab}-\frac12\eta_{ab}h\ud{a}{a},
\end{align}
indeed solves a wave equation which admits plane wave solutions similar to the ones of Maxwell's theory.

At this point, a word about gauge freedom: There are multiple procedures in order to partially or completely fix the gauge freedom that is inherent to GR. Here, we use a very particular gauge choice for the sake of being able to nicely display the form of the gravitational wave solution. In Chapter~\ref{Chap8}, on the other hand, we present two alternative gauge fixings, ultimately leading to the same result but taking different routes. The distinction lays in the detail and is of minor relevance for our objective. That said, without loss of generality, we can use the Hilbert gauge $\partial_a\tilde{h}^{ab}=0$ to show that Einstein's linearized field equations can be simplified to
\begin{align}
    \left(-\frac{\partial^2}{\partial t^2}+\nabla^2\right)\tilde{h}^{ab}\equiv \partial_a\partial^a\tilde{h}_{ab}=0.
\end{align}
Evidently, this is a wave equation and it does suggest that the perturbations around the background metric are indeed wavelike ripples in spacetime. We remind the reader at this point that under the Hilbert gauge, a transformation $x'^a=x^a+\xi^a$ induces a change $\tilde{h}'_{ab} =  \tilde{h}_{ab}-\partial_a\xi_b-\partial_b\xi_a+\eta_{ab}\partial_c\xi^c$ and it must hold that $\partial_a\partial^a\xi^b=0$ so that $\tilde{h}'_{ab}$ is again a solution of the above wave equation. The simplest solution to the above wave equation is of the form 
\begin{align}\label{eq:SolWaveEq}
    \tilde{h}^{ab}=\Re{A^{ab}e^{ik_ax^a}},
\end{align}
where $A^{ab}$ is a constant, symmetric tensor in which information about amplitude and polarization of the wave is encoded. The wave vector $k_a$  determines the direction of propagation and the wave's frequency. Moreover, it is a null vector. Note that the Hilbert gauge implies the constraint $A^{ab}k_b=0$, which reduces the number of independent components of $A$ to six. Also, note that in physical applications we are solely interested in the real part of $\tilde{h}_{ab}$, hence, we explicitly only considered the real part of \eqref{eq:SolWaveEq}.

Using the gauge freedom $\xi^a$, we can further reduce the independent components of $A^{ab}$ to two by means of a suitable gauge choice, such as the \textbf{transverse-traceless gauge}, or $\mathbf{TT}$ \textbf{gauge} for short.\footnote{This gauge choice will reappear in Chapter~\ref{Chap8} and is subtler than it appears at first sight.} In this gauge, only the spatial components of $\tilde{h}_{ab}$ are non-trivial. Note also that transversality here refers to transversality with respect to the direction of propagation, i.e., $\tilde{h}_{a0}=0$. Moreover, in the $TT$ gauge it holds that $\tilde{h}_{ab}=h_{ab}$. Therefore, to avoid confusion, we write $h^{TT}_{ab}$ whenever we use the transverse-traceless gauge.

Because $A^{ab}$ has precisely two independent components, we can rewrite it in terms of two dimensionless amplitudes $h_\times,h_+$, the \textbf{strain}, and real unit polarization tensors $\epsilon^{ab}_\times,\epsilon^{ab}_+$. Explicitly, we can write
\begin{align}
    A^{ab}=h_\times\epsilon^{ab}_\times+h_+\epsilon^{ab}_+.
\end{align}
The exact form of $\epsilon^{ab}_\times,\epsilon^{ab}_+$ depends on the direction of propagation and one often sees 
\begin{align}
    h_{ab}^{TT}=\begin{pmatrix}
        0&0&0&0\\
        0&h_+&h_\times&0\\
        0&h_\times&-h_+&0\\
        0&0&0&0
    \end{pmatrix}e^{ik_ax^a}.
\end{align}
Note that $(+)$ and $(\times)$ are two orthogonal polarization states. That is, it is impossible to construct $(+)$ from $(\times)$ and vice versa. 

Let us try to physically understand what the strains $h_\times,h_+$ effectively tell us. In the following, we sketch how the strains influence the motion of test masses. To that end, we consider the linearized Riemann tensor, which reads
\begin{align}
    R_{abcd}=\frac12 \left(\partial_a\partial_bh_{cb}+\partial_c\partial_bh_{ad}-\partial_a\partial_ch_{bd}-\partial_b\partial_dh_{ac}\right).
\end{align}
Some components drastically simplify in the $TT$ gauge and we find
\begin{align}
    R^{TT}_{i0j0} &=-\frac12 \PD{^2}{t^2}h^{TT}_{ij} & \text{for} i,j\in\{1,2,3\},
\end{align}
which looks similar to Newton's second law. This gives us a first hint at the role which these particular components of the Riemann tensor play in the $TT$ gauge. Indeed, in the Newtonian limit, one can show that
\begin{align}
     R^{TT}_{i0j0}\approx\frac{\partial^2\Phi}{\partial x^i\partial x^j},
\end{align}
where $\Phi$ is Newton's potential. In this context, the Riemann tensor has an interesting physical interpretation. Namely, it is similar to the tidal force field describing the relative acceleration between two test particles in free fall. Let us clarify this statement: Assuming we have two freely moving test masses in a detector moving on geodesics $x^a(\tau)$ and $x^a(\tau)+\zeta^a(\tau)$, respectively, we find a simplified form of the geodesic deviation equation,
\begin{align}\label{eq:GeodesicsDeviation}
    \frac{\dd^2\zeta^k}{\dd t^2}\approx-{R\ud{k}{0j0}}^{TT}\zeta^j.
\end{align}
Hence, we can express the tidal force $f^k$ as
\begin{align}
    f^k\approx-m\,{R\ud{k}{0j0}}^{TT}\zeta^j,
\end{align}
where $m$ is the mass of the particle. With this result at hand, let us look at a more interesting example. Consider the case of a detector with two test particles being ``hit'' by a gravitational wave propagating along the $z$-direction (of our coordinate system). For simplicity, we only consider one polarization and can thus writ
\begin{align}
    h_{ab}^{TT} = h_+ \epsilon_{ab}^{TT}\cos[\omega(t-z)],
\end{align}
where $\epsilon^{TT}_{ab}$ is an unspecified polarization tensor in the $TT$ gauge. Then, as the gravitational wave passes through the detector, the two test masses move relative to each other. This movement can be quantified by the variation $\delta\zeta^x$ with respect to an initial position $\zeta_0^x$ and one finds
\begin{align}
    \delta \zeta^x=-\frac12h_+\omega^2\cos[\omega(t-z)]\zeta^x_0
\end{align}
for the variation in $x$-direction and the same expression with a flipped sign for the $y$-direction. This follows from the geodesic deviation equation~\eqref{eq:GeodesicsDeviation}. The oscillation of the gravitational wave translates directly into a relative motion of the test masses. The force exerted onto the test masses by the passing wave is given by
\begin{align}
    &f^x\approx -\frac{m}{2}h_+\omega^2\cos[\omega(t-z)]\zeta^x_0 &\text{and}&& &f^y\approx \frac{m}{2}h_+\omega^2\cos[\omega(t-z)]\zeta^y_0.
\end{align}
The latter equations show that the strains $h_+,h_\times$ are directly linked to what is measured in gravitational wave detectors, even though we are considering a massively simplified example. They determine the amplitude of the displacement of the test masses. Therefore, linking the strain to the asymptotic shear and subsequently to the Bondi news tensor, with which we can calculate fluxes, resembles a powerful connection between observation and theory. This connection enables us to determine certain quantities based on measurements and do consistency checks with respect to numerical simulations of, for instance, wave form models. More of that will be revealed in the last chapter.

\newpage
\subsection{Exercises}
\renewcommand{\thesection}{\arabic{section}}

\begin{Exercise}[]\label{ex:conformaltrans}
    Show that in a conformally completed spacetime $(\M,g_{ab})$, the transformation $g_{ab}\mapsto g'_{ab} = \omega^2 g_{ab}$ implies that the metric-compatible, torsion-free derivative operator $\nabla$ is mapped to $\nabla'$ with 
    \begin{align*}
    (\nabla'_a-\nabla_a)\alpha_b &= C\du{ab}{c} \alpha_c &\text{with}&& C\du{ab}{c} &= -\omega^{-1} \left(2\delta\ud{c}{(a} \nabla_{b)}\omega-(\nabla^c\omega)g_{ab}\right).
    \end{align*}
\end{Exercise}

\begin{Exercise}[]\label{ex:shear}
The shear tensor $\sigma_{ab}$ is transverse, trace-less, and symmetric. That is, it satisfies $\sigma_{ab} n^{b} = 0$, $q^{ab}\sigma_{ab}$, and $\sigma_{[ab]} = 0$. Show that this implies that the shear is of the form
\begin{align*}
    \sigma_{ab} = -(\bar{\sigma}^\circ m_a m_b + \sigma^\circ \bar{m}_a \bar{m}_b).
\end{align*}
\end{Exercise}

\begin{Exercise}[]\label{ex:Riemanndecomp}
    Show that on a 3-dimensional manifold $(\scri,q_{ab})$ with vanishing Weyl contribution, the 3-dimensional Riemann tensor can be decomposed into another rank-2 tensor such that it holds that
    \begin{align*}
    R\du{abc}{d}\alpha_d = \left(q_{c[a}S\du{b]}{d}+S_{c[a}\delta\du{b]}{d}\right)\alpha_d
    \end{align*}
    on this manifold. Further, analyze the properties of $S\ud{a}{b}$ and show that under $(q_{ab},n^a)\mapsto(\omega^2 q_{ab},\omega^{-1}n^a)$ we find
    \begin{align*}
    S'\du{a}{b}&=\omega^{-1}S\du{a}{b}-\omega^{-3}\D_a(q^{bc}\D_c\omega)+4\omega^{-4}(\D_a\omega)q^{bc}\D_c\omega-\omega^{-4}\delta\du{a}{b}(q^{cd}\D_c\omega\D_d\omega)\\
    S'_{ab}&=S_{ab}-2\omega^{-1}\D_a\D_b\omega+4\omega^2(\D_a\omega)(\D_b\omega)-\omega^2q_{ab}(q^{cd}\D_c\omega\D_d\omega)
    \end{align*}
    Using these equations, show that $S_{ca}-\frac12 q^{mn} S_{mn} q_{ca}$ is invariant under the above transformation.\\
    
    \textit{Hint: Recapitulating the statements covered in section~\ref{subsec:PhysicalPropofasympflat} might be helpful.}
\end{Exercise}

\begin{Exercise}[]\label{ex:RiemannShear}
    Using what we know about the decomposition of the Riemann tensor $R\du{abc}{d} = \left(q_{c[a}S\du{b]}{d}+S_{c[a}\delta\du{b]}{d}\right)$ and the asymptotic shear $\sigma_{ab} = -\D_a \ell_b + \frac12 q_{ab} q^{cd} (\D_c \ell_d)$, show that 
    \begin{align*}
    N_{ab} = 2\lie_n \sigma_{ab}^\circ,
    \end{align*}
    where we defined $N_{ab}$ as the Bondi news tensor, which in a Bondi-conformal frame reads $N_{ab} = S_{ab} - \frac12 q^{cd} S_{cd} q_{ab}$.\\
    
    \textit{Hint: There are multiple ways of approaching this problem. One would be brute force calculation. However, it is useful to work in the Bondi gauge (see \cite{CompereBook} for details and support).}
\end{Exercise}

\newpage
\asection{8}{The Connection between Full and Linearized General Relativity}\label{Chap8}
The aim of this chapter, at least partially, is to draw a connection to what we discussed in the first few chapters. Specifically, our objective is to relate asymptotic shear and the Bondi news to the Newman-Penrose scalars of Chapter~\ref{Chap3}. Furthermore, we will also relate all these quantities with the strains $h_+, h_\times$ known from the theory of gravitational waves in linearized GR. As we have hinted at in the interlude~\ref{subsec:strain}, connecting shear, news, and strains is what opens the door to compare theory with observations. 
\bigskip

\subsection{Connecting the Newman-Penrose Scalars to Shear and Strain}
In order to draw a connection between the Newman-Penrose scalars and the Bondi news tensor, we need to look for an equation which relates the Shouten and the Weyl tensors. Such an equation is given by the Bianchi identities of the Riemann tensor on the conformally completed spacetime. One can show that these equations, in the limit $\Omega\to 0$, reduce to
\begin{align}\label{eq:BianchiforShoutenTensor}
    2\epsilon^{apq} \D_{[p}S_{q]}^b = K^{ab} \equiv \lim_{\Omega\to 0}\left( \Omega^{-1} C^{apbq} n_p n_q\right).
\end{align}
Recall from Chapter~\ref{Chap3} that the Weyl tensor vanishes on $\scrip$, but that $\Omega^{-1} C^{abcd}$ is \textit{not} zero. In fact, $\Omega^{-1} C^{abcd}$ is what we called the \textbf{asymptotic Weyl tensor} and we denoted it by $K^{abcd}$. Furthermore, we defined the Newman-Penrose scalars of the conformally completed spacetime with respect to $K^{abcd}$. We repeat these definition here for the convenience of the reader
\begin{align}\label{eq:DefNPGR}
	{\Psi}_4 &:= K_{abcd}\,n^{a} \bar{m}^{b} n^{c} m^{d}\notag\\
	{\Psi}_3 &:= K_{abcd}\,\ell^{a} n^{b} \bar{m}^c n^d\notag\\
	{\Psi}_2 &:= K_{abcd}\,\ell^{a} m^{b} \bar{m}^c n^d\notag\\
	{\Psi}_1 &:= K_{abcd}\,\ell^{a} n^{b} \ell^{c} m^{d}\notag\\
	{\Psi}_0 &:= K_{abcd}\, \ell^{a} m^{b}\ell^{c} m^{d}.
\end{align}
It follows from these definition, that the tensor $K^{ab}:= \Omega^{-1}C^{apbq}n_p n_q$ can be expressed in terms of $\Psi^\circ_4$, $\Psi^\circ_3$, and $\Im{\Psi^\circ_2}$. This are five out of ten components of the Weyl tensor. Moreover, because $S_{ab}$ is determined by $\D$, which in turn encodes the radiative modes, we can think of $\Psi^\circ_4$, $\Psi^\circ_3$, and $\Im{\Psi^\circ_2}$ as carrying information about gravitational radiation to $\scrip$. We also recall the Peeling Theorem from Chapter~\ref{Chap4}, which tells us that the \textit{physical} $\hat\Psi_4$ decays like $\hat\Psi_4 = \frac{\Psi^\circ_4}{r}$. This is the expected behavior for radiative modes and, in an analogy with electromagnetism, we call $\Psi^\circ_4$ the \textbf{radiation field}.

\begin{figure}[htb!]
	\centering
	\includegraphics[width=0.75\linewidth]{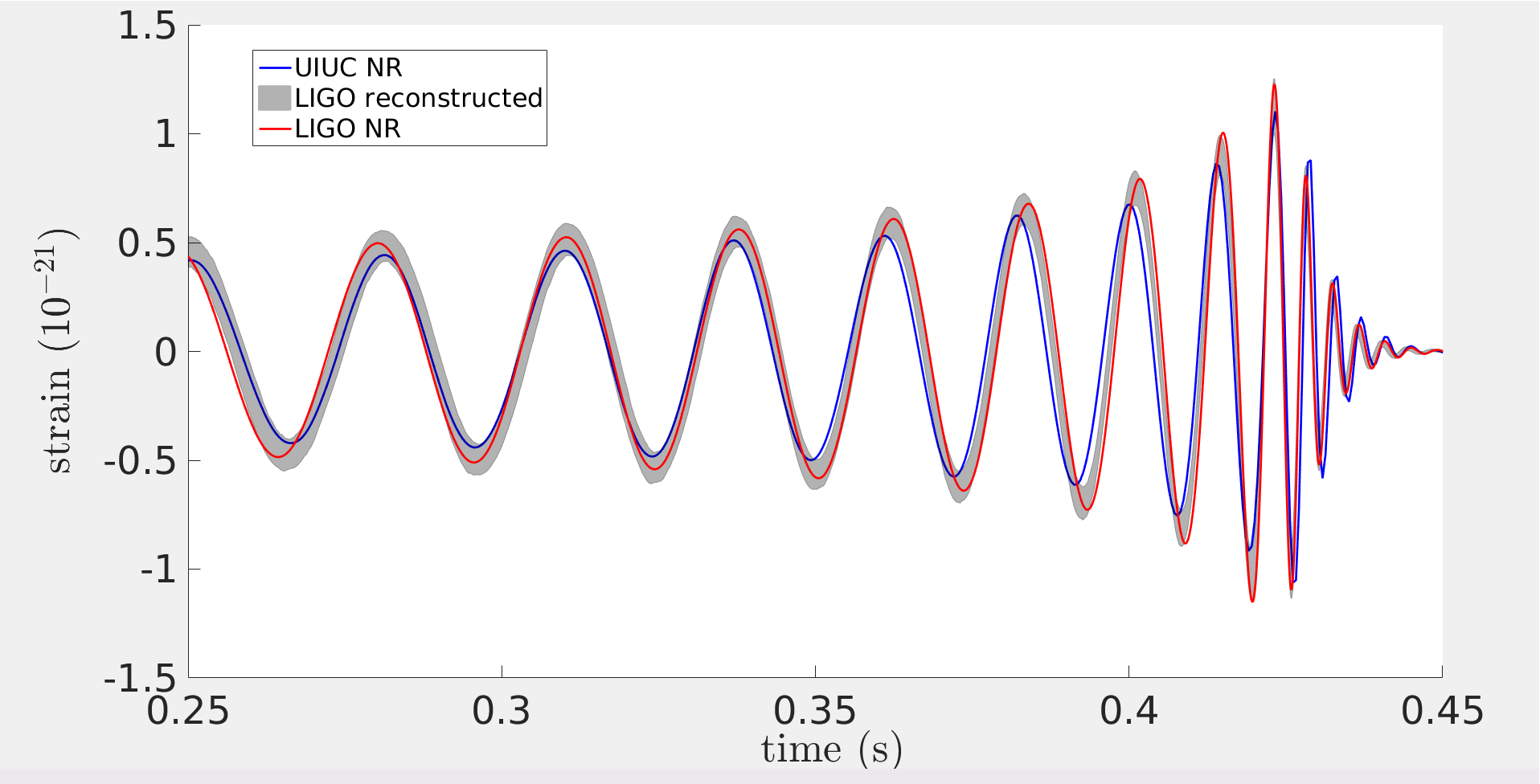}
	\caption{\textit{A waveform plot of the first official gravitational wave detection in 2015 by the LIGO and VIRGO collaborations~\cite{Ligo:2016}.}}
	\label{fig:waveform}
\end{figure}

The relation between the Newman-Penrose scalars and the shear emerges when we contract equation~\eqref{eq:BianchiforShoutenTensor} with the null tetrad components which are needed to construct $\Psi^\circ_4$, $\Psi^\circ_3$, and $\Psi^\circ_2$. For instance, contracting~\eqref{eq:BianchiforShoutenTensor} with $\bar{m}_a m_b$, and using the symmetries of the Weyl tensor, will make $\Psi_4$ appear. Contractions with $\ell_a \bar{m}_b$ and $\ell_a \bar{m}_b$ instead, will make $\Psi_3$ and $\Psi_2$ appear, respectively. We then use the Leibniz rule in order to make the derivative $\D$ act on the null tetrad (sine we know how it has to act on tetrads) and this finally makes the shear appear in our considerations. Explicitly carrying out these steps (see Exercise~\ref{ex:NPscalars}) leads to the three equations 
\begin{align}\label{eq:NPscalarsAsShear}
    \Psi_4^\circ &= -\ddot{\bar{\sigma}}^\circ \notag\\
    \Psi_3^\circ &= \eth\dot{\bar{\sigma}}^\circ \notag\\
    -2i\Im{\Psi_2^\circ} &= \sigma^\circ \dot{\bar{\sigma}}^\circ - \bar{\sigma}^\circ \dot{\sigma}^\circ + \eth^2\bar{\sigma}^\circ - \bar{\eth}^2 \sigma^\circ.
\end{align}
We recall that a dot on the shear symbolizes a derivative with respect to the retarded time coordinate $u$ and that $\eth$ is the angular derivative operator for spin-weighted function. This operator was introduced in Chapter~\ref{Chap2} (cf. equation~\eqref{eq:AlternativeAngularDeriv}). By making use of the relation between the shear and the strain, which we sketched in~\ref{subsec:strain} and which we will derive in the next subsection, one can write
\begin{align}\label{eq:Psi4inStrain}
    \Psi^\circ_4=-\frac12 \left(\Ddot{h}_+^\circ -i\Ddot{h}_\times^\circ\right),
\end{align}
where we defined
\begin{align}
    h^\circ_+(u,\theta,\phi) &:= \lim_{r\to\infty}r h_+ (u,r,\theta,\phi),\notag\\ 
    h^\circ_\times(u,\theta,\phi) &:= \lim_{r\to\infty}r h_\times(u,r,\theta,\phi).
\end{align}
Thus, we have established a connection between the Newman-Penrose scalar $\Psi^\circ_4$ and the strains of the gravitational wave we use in the linearized theory. The label ``radiation field'' is thus well-justified for $\Psi^\circ_4$. It is worth remarking that this links theory to observations and data analysis. In fact, $\Psi^\circ_4$ is a key quantity which is computed in Numerical Relativity and integrating it twice over $\dd u$, isolates the strains. This is what is ultimately used in waveform models and plotted in the famous waveform plots, such as the one shown in Figure~\ref{fig:waveform}. Further details can be found in the literature on waveform models such as \texttt{PHENOMD} or \texttt{SEOBNR}.

Let us now return to equation~\eqref{eq:NPscalarsAsShear} and observe that the scalars $\Psi^\circ_0$, $\Psi^\circ_1$, and $\Re{\Psi^\circ_2}$ are missing. In fact, it is easy to see that they cannot be isolated from equation~\eqref{eq:BianchiforShoutenTensor}. It is possible though to derive expressions for these remaining scalars, which show that they carry coulombic information, as we argued back in Chapter~\ref{Chap4}, as well as a description of longitudinal modes~\cite{Newman:1968}. We will, however, not go into details here and refer the reader to the literature instead.

Before concluding this subsection and presenting the derivation of~\eqref{eq:Psi4inStrain}, we briefly return to the flux of gravitational $4$-momentum. In Chapter~\ref{Chap6} we encountered mathematical expressions for the flux of energy and momentum through portions of $\scrip$ and mentioned that their derivation relies on an identification of the radiative modes in full, non-linear GR. Now that we have identified these modes and we know that they are encoded in the Bondi news tensor $N_{ab}$, we can show that these fluxes can also be expressed as
\begin{align}\label{eq:GWFluxes}
    \F_E(\Delta\scri) &= \int_{\Delta\scri} N_{ab} N_{cd} q^{ca} q^{bd}\, \dd u\,\dd^2\omega \notag\\
    \F_{P_i}(\Delta\scri) &= \int_{\Delta\scri} \alpha_i\, N_{ab} N_{cd} q^{ca} q^{bd}\,\dd u\,\dd^2\omega,
\end{align}
where $\alpha_i\in\{\sin\theta\cos\phi, \sin\theta\sin\phi, \cos\theta\}$. Here, $\F_E(\Delta\scri)$ denotes the flux of energy through a finite volume element of $\scrip$, while $\F_{P_i}(\Delta\scri)$ stands for the momentum flux. The right hand side of equation~\eqref{eq:GWFluxes} makes it manifest that these fluxes pertain solely to gravitational waves, as the integrand only depends on the news tensor. In the absence of gravitational waves, $N_{ab} = 0$ and consequently there is no flux at $\scrip$.

The fluxes represent a landmark in the discussion on the existence of gravitational waves, which culminated in the nineteen-sixties. Since the inception of gravitational waves in 1916 by Einstein, there has been much debate about whether they are a real physical phenomenon, or whether they are a mere coordinate artifact. Eventually, this dispute was settled by the mathematical rigorous framework presented here, as it provides a \textit{gauge-invariant} description of gravitational waves. In particular, it provides a gauge-invariant description of the flux of energy and momentum carried by gravitational waves.\bigskip

\subsection{Bridging between Linearized and Full General Relativity}
In this subsection, all computations are carried out in a physical spacetime and no conformal completion is ever introduced. Thus, in order to simplify the notation and have more readable equations, we refrain from putting hats on physical object. 

With this comment out of the way, let $(\M, g_{ab})$ be a physical spacetime. The sources of the gravitational field shall be assumed to have compact spatial support and to drop sufficiently fast for $r\to \infty$, such that we can assume  that $g_{ab} = \eta_{ab} + \O(1/r)$ is a good approximation for $r\to \infty$. We refer to $\eta_{ab}$ as the \textbf{background} spacetime and we expect that the vacuum field equations $G_{ab} = 0$ are satisfied for $r\to\infty$. Furthermore, we assume that the metric $g_{ab}$, which is the exact solution to the Einstein field equations, can be written as background metric plus perturbations. Concretely, we assume that
\begin{align}
   g_{ab} = \eta_{ab} + \lambda\, h_{ab},
\end{align}
where $\lambda$ is assumed to be a small parameter. Effectively, this ansatz turns the Einstein field equations into a $1$-parameter family of equations, $G_{ab}(\lambda)$. We can expand these equations in $\lambda$ around $\lambda = 0$. The zeroth order equation is rather trivial. However, at first order we find from
\begin{align}
    \left.\frac{\dd}{\dd\lambda}G_{ab}(\lambda)\right|_{\lambda = 0} = 0
\end{align}
a set of field equations for the perturbation $h_{ab}$,
\begin{align}\label{eq:LinearizedEinstein}
    0 = \left.\frac{\dd}{\dd\lambda}G_{ab}(\lambda)\right|_{\lambda = 0} = -\frac12 \Box\, \bar{h}_{ab} + \nabla^c \nabla_{(b}\bar{h}_{a)c} - \frac12\eta_{ab} \nabla^c \nabla^d \bar{h}_{cd}.
\end{align}
For notational compactness, we have introduced the \textbf{trace-free metric} $\bar{h}_{ab} := h_{ab} - \frac12 (\eta^{cd} h_{cd}) \eta_{ab}$ and the box symbol $\Box$ denotes the d'Alembert operator with respect to the background metric. More precisely, we have $\Box = \eta^{ab}\partial_a \partial_b$. Furthermore, $\nabla$ denotes the covariant derivative with respect to the Levi-Civita connection of $\eta_{ab}$.

We now wish to study~\eqref{eq:LinearizedEinstein}, which are known as the \textbf{linearized Einstein equations}. However, we should point out that these equation are \textbf{not} covariant with respect to arbitrary diffeomorphisms. Rather, these equations retain their form only with respect to so called \textbf{linearized diffeomorphisms}, which affect the perturbations as follows:
\begin{align}
    h_{ab}\,\mapsto\, h'_{ab} = h_{ab} + 2\nabla_{(a}\zeta_{b)} \equiv h_{ab} + \lie_\zeta\eta_{ab},
\end{align}
where $\zeta_a$ are four arbitrary functions. Effectively, this introduces a gauge freedom into the linearized theory.\footnote{A more accurate statement would be that the linearized theory inherits a particular gauge freedom from the full theory, whose gauge group is the group of spacetime diffeomorphisms.} As usual in gauge theories, we first need to fix a gauge in order to obtain deterministic results. There are several ways to achieve this. One option is the so-called \textbf{radiation gauge}, which imposes the conditions
\begin{align}
    h_{ab} t^b &= 0 &\text{and}&& h_{ab}\eta^{ab} &= 0
\end{align}
on the perturbations. Here, $t^a$ is a unit timelike vector on a constant-time-hypersurface in a coordinate chart  $(t,r,\theta,\phi)$. In this particular gauge, only the spatial components of $h_{ab}$ are non-zero and the perturbations are automatically  traceless. That is, $h_{a0}=0$ and $g_{ab}h^{ab}=0$, respectively. Another popular gauge choice is the \textbf{Lorenz gauge}, which is defined by the condition
\begin{align}
    \nabla^d \bar{h}_{cd} &=0 &\Longrightarrow && \Box\,\bar{h}_{ab} &= 0.
\end{align}
This is akin of the Lorenz gauge in electromagnetism\footnote{The Lorenz gauge in electromagnetism is named after Ludvig Lorentz, \textit{not} Hendrik Lorentz, who first discovered the Lorentz transformations.} and just like its electromagnetic counterpart, it only partially fixes the gauge. The residual gauge freedom can be used to make $h_{ab}$ trace-less and eliminate its time-time and space-time components, leaving only space-space components. Ultimately, the results we derive do not depend on the gauge choice. Moreover, in both gauges we have introduced, one is led to consider a perturbation which is trace-less and possesses only spatial components. Thus, it is convenient to introduce the concept of a trace-less, transverse tensor, which we denote by $h^{TT}_{ab}$. In both gauges we have $h_{ab} \equiv h^{TT}_{ab}$.

How many algebraically independent components does $h^{TT}_{ab}$ have? A symmetric rank-$2$ tensor in four dimensions has $\frac{4\times(4+1)}{2} = 10$ components. Transversality removes all components of the form $h_{0a}$, which is a total of seven components. Finally, the condition of a vanishing imposes an additional constraint on $h^{TT}_{ab}$ and thus removes one more component. This leaves us with $10-7-1 = 2$ algebraically independent components. This is in line with the fact that the full gravitational field has two physical degrees of freedom, while the remaining eight are pure gauge. Because we have fixed a gauge, we should expect to have at most two degrees of freedom left in $g_{ab}$ and consequently in $h_{ab}$.

Let us now introduce the shear. To be more precise, we consider the shear of a congruence of null geodesics on the spacetime $(\M, g_{ab})$. To that end, it is convenient to introduce a chart $(u, r, \theta, \phi)$, where $u$ is a null coordinate. Null geodesics are thus simply parametrized by $u=$ const. Furthermore, let us fix the null vector $\ell^{a}$ tangential to the congruence of null geodesics (i.e., tangential to every $u=$ const. surface). Simultaneously, this $\ell^{a}$ is an element of a Newman-Penrose null tetrad for $(\M, g_{ab})$. The shear is then defined as
\begin{align}
    \sigma_{ab} = -\nabla_a \ell_b + \frac12 s_{ab} s^{cd} (\nabla_c \ell_d),
\end{align}
where $\nabla$ is the covariant derivative with respect to $g_{ab}$ and where it is understood that this expression is evaluated on a $u=$ const. and $r=$ const. cross-section. By pulling back $g_{ab}$ to that cross-section, we obtain a metric $s_{ab}$ of signature $(++)$ and the tangent space to this cross-section is, as usual, spanned by $\{m^{a}, \bar{m}^{a}\}$.

Since the metric is essentially just the Minkowski metric plus a perturbation, we can expect that the shear can also be expanded into a background contribution and a perturbation (because it inherits a $\lambda$-dependence from $\nabla$). This allows us to define the linearized shear as
\begin{align}
   \delta\sigma &= \delta\sigma_{ab}m^am^b & \textsf{with}&& \delta\sigma_{ab} &= \frac{\dd}{\dd\lambda}\left.\left(-\nabla_a \ell_b + \frac12 s_{ab} s^{cd} (\nabla_c \ell_d)\right)\right|_{\lambda=0}.
\end{align}
Using the identity $2\nabla_{(a}\ell_{b)}=\lie_\ell g_{ab}$, which can be proven by expressing the metric in terms of the Newman-Penrose null tetrad, and using $s_{ab}m^{a} m^{b} = 0$ as well as  $s^{ab}\nabla_a\ell_b=\sqrt{2}/r$, it can be shown~\cite{Ashtekar:2017} that
\begin{align}\label{eq:LinearizedShear}
    \delta\sigma=\frac12m^am^b\lie_\ell h_{ab}-\frac{1}{2}m^am^bh_{ab}.
\end{align}
Next, we expand $h_{ab}^{TT}$ in powers of $\frac{1}{r}$, i.e.,
\begin{align}
   h_{ab}^{TT} = \frac{h^{\circ\, TT}_{ab}}{r} + \frac{h^{(1)\, TT}_{ab}}{r^2} + ...,
\end{align}
This is justified because we assumed that $\lim_{r\to\infty}g_{ab} = \eta_{ab}$, independently of $\lambda$. Plugging this expansion into~\eqref{eq:LinearizedShear} yields
\begin{align}\label{eq:FinalResult}
    \delta\sigma &=\frac{1}{2}m^am^b\frac{h_{ab}^{\circ\,TT}}{r^2}+\O\left(r^{-3}\right) &\text{with}&& \delta\sigma^\circ &= \frac12 \bar{m}^a\bar{m}^bh^{\circ\, TT}_{ab}.
\end{align}
For more details on this calculation, we refer the reader to~\cite{Ashtekar:2017}. Finally, this is the relation between shear and strains we sought. Using 
\begin{equation}
    h^{TT}_{ab} \propto \begin{pmatrix}
        0 & 0 & 0 & 0\\
        0 & h_+ & h_\times & 0 \\
        0 & h_\times & -h_+ & 0\\
        0 & 0 & 0 & 0 
    \end{pmatrix},
\end{equation}
which is the form of $h^{TT}_{ab}$ we encountered in~\ref{subsec:strain}, we obtain
\begin{align}\label{eq:linearizedshearandmetricTT}
    2\delta \sigma^\circ \equalhat h^{\circ\, TT}_{ab} \bar{m}^a \bar{m}^b\sim(h_+^\circ+ih_\times^\circ),
\end{align}
In this seemingly straight forward calculation we covered up one very important subtlety that requires some extra attention. Namely, the $h_{ab}^{TT}$ from equation~\eqref{eq:FinalResult} is not the same as in~\eqref{eq:TTmetric}. Why is that? In order to obtain a \textit{transverse traceless} version of our metric perturbation, one has to apply non-local operations such as, for instance, the inverse of the Laplacian operator. This in turn makes $h_{ab}^{TT}$ as in equation~\eqref{eq:TTmetric} \textit{non-local} itself. On the other hand, the shear is a very local quantity in physical spacetime. This becomes a bit more intuitive when we think of $h_{ab}^{TT}$ in terms of a metric and of the shear being related to the gravitational strain, or equally, the displacement of test masses in a detector. Hence, we would compare a local with a non-local quantity which is a troublesome venture. However, as mentioned, what we used in~\eqref{eq:FinalResult} is actually another form of transverse traceless that one commonly defines. In this equation, we refer to the local notion of transverse-traceless as in~\cite{Ashtekar:2017}. As there one can find a somewhat detailed explanation of the ambiguity in the notion of ``transverse-traceless'' in literature, we will not go into detail on this topic here. Note however that the use their \textit{local notion} in equation~\eqref{eq:FinalResult} leads to a consistent result regarding locality. This brings us to our conclusion on that matter: We were able to rigorously show how the shear is related to the strain fields of the gravitational wave. This simultaneously validates many assumptions form previous chapters regarding $\sigma_{ab}$. With this final remark, the circle between linearized theory and full GR is closed and we achieved our final goal. We close with a brief summary and outlook.\bigskip

\subsection{Summary}
In Chapter~\ref{Chap1} we started out with a discussion on the notion of radiation in Maxwell's theory and why isolating radiative modes from a generic source of electromagnetic fields is, in general, a non-trivial task. We were led to conclude that in order to extract information about radiative modes, it is helpful to move ``infinitely'' far away, as this naturally leads to non-radiative modes ``peeling off'' from the electromagnetic field. Qualitative arguments thus foreshadowed the Peeling Theorem. 

In order to make this idea of going ``infinitely far away'' mathematically precise, we introduced the notion of a \textit{conformally completed spacetime} $\M=\hatM\cup\scri$. The key feature of such a spacetime is its three-dimensional boundary $\scri$, which brings ``infinity'' to a finite distance and allows us to study the asymptotic region using methods from differential geometry, topology, and even group theory. We also introduced the Newman-Penrose tetrad in Minkowski space. The underlying idea was to introduce a tetrad basis which allows us to ``follow radiation all the way to infinity''. In particular, we showed that the tetrad is well defined on the boundary $\scrip$.

The fact that Maxwell's field strength tensor is conformally invariant facilitated, together with the well-defined tetrad, the construction of scalar quantities that exhibit useful properties in the asymptotic region. These quantities are the Newman-Penrose scalars and they enabled us to isolate the radiative and coulombic modes from the field strength tensor. What allowed to do so, is a key result known as the \textit{Peeling Theorem}. Furthermore, we defined fluxes of energy-momentum carried by electromagnetic waves in terms of the Newman-Penrose scalars. The machinery applied to electromagnetism, thus, enabled us to take the position of an observer located infinitely far away from sources and to calculate several physical observables based on this particular point of view.

For the transition to GR, we first had to restrict ourselves to a special class of spacetimes --- the class of \textit{asymptotically Minkowski spacetimes} introduced in Chapter~\ref{Chap3}. Once the mathematical framework was set, we described a general procedure to construct a Newman-Penrose null tetrad for any asymptotically Minkowski spacetime.

Once this was achieved, we analyzed the Riemann and Weyl tensors, confirming that these tensors vanish asymptotically for the class of spacetimes under consideration. In particular, the asymptotic properties of the Weyl tensor led to the introduction of Newman-Penrose scalars for GR and to a proof of the Peeling Theorem for gravity in Chapter~\ref{Chap4}. Most importantly, because of its asymptotic behavior, we speculated that $\Psi_4$ encodes gravitational radiation.

These results raised further questions. For instance, taking Maxwell's theory as a guideline, we asked how to define \textit{fluxes of $4$-momentum of the gravitational field} or we wondered what information about gravitational waves is actually encapsulated in the Newman-Penrose scalars. This drew our attention towards the \textit{universal structure} of asymptotically Minkowski spacetimes as, naturally, before being able to define fluxes of any sort, we had to properly identify a \textit{symmetry group}. A careful analysis of the universal structure and the subgroup of diffeomorphisms which leaves this structure invariant led us to the discussion of the Bondi-Metzner-Sachs group in Chapter~\ref{Chap5}.

We pointed out several similarities as well as crucial differences between the BMS and the Poincar\'e groups. In particular, we drew on some analogies to motivate how to define fluxes of $4$-momentum for the gravitational field in Chapter~\ref{Chap6}. We also discussed the limitation of this analogy and discussed in some detail, why the notion of angular momentum is highly ambiguous in GR. Towards the end of Chapter~\ref{Chap6} we caught a first glimpse of radiative modes in full, non-linear GR, when we presented an argument which motivated us to look for those modes in the covariant derivative operator on null infinity.

Our qualitative justification for this endeavor paid out and in Chapter~\ref{Chap7} we connected the well-defined derivative operator on $\scrip$ to the \textit{asymptotic shear} tensor, which in turn is related to the strain of gravitational waves. Thus, the covariant derivative does indeed encode the radiative modes of the gravitational field.

Furthermore, we used the derivative operator to compute the curvature of null infinity. We found that its trace-less transverse part (the Bondi news tensor) can also be related to the asymptotic shear, thus establishing a link between the geometry of null infinity and the physics of gravitational waves. 

Finally, in Chapter~\ref{Chap8}, using the Bondi news tensor, we reconsidered the fluxes related to the BMS group and expressed them in terms of the asymptotic shear and the Newman-Penrose scalars. This provided further evidence that the radiative modes are encoded in the Newman-Penrose scalar $\Psi^\circ_4$. Moreover, we investigated the relation between full, non-linear GR and the linearized theory. This finally led to the insight that the shear is indeed given by the strains $h_+,h_\times$. We came to the conclusion that it is fully consistent with the results of General Relativity at $\scri$ and that our identifications of previous chapters regarding shear and strain were well-justified.

\newpage
\subsection{Exercises}
\renewcommand{\thesection}{\arabic{section}}

\begin{Exercise}[]\label{ex:Bianchi}
Let us assume we are on $\scri$ and let $\D$ be the derivative operators on $\scri$ and $R\du{abc}{d}$ the corresponding Riemann tensor that can be decomposed into a sum of contractions of $S\ud{a}{b}$. Use the (symmetry) properties of these tensors together with the Bianchi identity of the full 4-dimensional conformally completed spacetime $(\M,g_{ab})$ to show that
\begin{align*}
    2\epsilon^{amn} \D_m S_n^b = 2\epsilon^{amn} \D_{[m}S_{n]}^b = \star K^{ab}
\end{align*}

    \textit{Hint: It can be useful to write the metric in terms of the Newman-Penrose tetrads and to split up the metric on $\scri$ explicitly such that $g_{ab}=\tilde g_{ab}+ q_{ab}$.}
\end{Exercise}

\begin{Exercise}[]\label{ex:NPscalars}
Derive equation~\eqref{eq:NPscalarsAsShear} using the strategy described in the main text. The definitions of asymptotic shear and Bondi news tensor are also relevant:
\begin{align*}
    \sigma_{ab}^\circ &= -(\bar{\sigma}^\circ m_a m_b + \sigma^\circ \bar{m}_a \bar{m}_b) \\
    N_{ab} &= 2\lie_n \sigma_{ab}^\circ = \dot{\sigma}_{ab}^\circ.
\end{align*}
Furthermore, the calculation simplifies if $N_{ab} = S_{ab}+\rho_{ab}$ is used. \bigskip

\textit{Hint: Use the properties of $\scri$ and the Bondi-frame regarding the metric and the contractions of the tetrads. Also, use the definition of the angular derivative operator $\eth$ in terms of derivatives along $\{m,\bar{m}\}$ (see Chapter~\ref{Chap2}).}
\end{Exercise}

\begin{Exercise}[]\label{ex:SpinWeights}
    Determine the spin weight of the Newman-Penrose scalars defined in~\eqref{eq:DefNPGR}. Check that in each equation in~\eqref{eq:NPscalarsAsShear}, the spin weight of the left hand side equals the spin weight of the right hand side.\bigskip
    
    \textit{Hint: Recall that $\eth$ increases the spin weight by one.}
\end{Exercise}

\begin{Exercise}[]\label{ex:linearizedMetric}
Let $(\M,\eta_{ab})$ be the spacetime manifold of flat physical spacetime. Assume now, there are small perturbations around the flat metric $\eta_{ab}$ such that we can expand in a perturbative parameter and the new full metric of physical spacetime becomes
\begin{align*}
    g_{ab}=\lambda^0\eta_{ab}+\lambda^1\gamma_{ab}+\O(\lambda^2).
\end{align*}
Here, $\gamma_{ab}$ is the linearized part of the metric and yields a non-trivial Einstein equation.\\
Focusing on its contribution, determine the linearized Einstein field equation 
\begin{align*}
    \left.\frac{\dd}{\dd\lambda}{G}_{ab}(\lambda)\right|_{\lambda = 0} = 0
\end{align*}
and show that this gives
\begin{align*}
    0  = -\frac12 \Box {\bar{\gamma}}_{ab} + {\nabla}^c {\nabla}_{(b}{\bar{\gamma}}_{a)c} - \frac12{\eta}_{ab} {\nabla}^c {\nabla}^d {\bar{\gamma}}_{cd}
\end{align*}
with ${\bar{\gamma}}_{ab} = {\gamma}_{ab} - \frac12 ({\eta}^{cd} {\gamma}_{cd}) {\eta}_{ab}$.
\end{Exercise}

\newpage
\appendix
\titleformat{\section}{\normalfont \Large \bfseries}
{Appendix\ \thesection:}{2.3ex plus .2ex}{}

\renewcommand{\thesubsection}{A.\arabic{subsection}}

\section{On Maxwell's Equations and the Theory of Partial Differential Equations}\label{App:A}
\bigskip

\subsection{Rewriting Maxwell's Equations in Terms of Newman-Penrose Scalars}\label{App:A1}
In this subsection we derive Maxwell's vacuum field equations in the Newman-Penrose formalism. All considerations concern quantities defined on the physical spacetime, but for notational simplicity we will abstain from using hats. In terms of Maxwell's $2$-form $F_{ab}$, the vacuum field equations are given by
\begin{align}\label{eq:MaxwellTraditional}
    \nabla_{[a} F_{bc]} &= 0 &\text{and} && \nabla_{[a}\prescript{\star}{}{ F}_{bc]} &= 0,
\end{align}
where $\prescript{\star}{}{F}_{ab} = \frac12 \epsilon_{abcd} {F}^{cd}$ is the Hodge dual of $F_{ab}$. In what follows, we will restrict ourselves to Minkowski space, rather than a generic background. Moreover, we work in outgoing Eddington-Finkelstein coordinates $(u, r, \theta, \phi)$. In this particular chart, the Newman-Penrose null tetrad $\{{\ell}^{a}, n^{a}, m^{a}, \bar{m}^{a}\}$ is explicitly given by
\begin{align}\label{eq:DefTetradMink}
    \ell^{a} &= \frac{1}{\sqrt{2}}\delta\ud{a}{r} \notag\\
    n^{a} &= \sqrt{2}\,\delta\ud{a}{u} - \frac{1}{\sqrt{2}}\delta\ud{a}{r} \notag\\
    m^{a} &= \frac{1}{\sqrt{2}\, r}\delta\ud{a}{\theta} + \frac{i}{\sqrt{2}\,r\,\sin\theta }\delta\ud{a}{\phi}.
\end{align}
The strategy for rewriting Maxwell's equations~\eqref{eq:MaxwellTraditional} in terms of Newman-Penrose scalars  consists of two steps:
\begin{itemize}
    \item[1)] First, construct all possible contractions of the field equations~\eqref{eq:MaxwellTraditional} with three elements of the null tetrad. Since the field equations are totally anti-symmetric, there are only four possible contractions for each equation.
    \item[2)] Secondly, express $F_{ab}$ in terms of the null tetrad and the Newman-Penrose scalars. After that, it is just a matter of simplifying the resulting equations.
\end{itemize}
As expressed in point 1, there are only four possible contractions. Explicitly, these are
\begin{align}\label{eq:MaxwellEqContractions}
    n^{a} m^{b} \bar{m}^{c} \nabla_{[a} F_{bc]} &= 0 & &&  n^{a} m^{b} \bar{m}^{c} \nabla_{[a}\prescript{\star}{}{F}_{bc]} &= 0 \notag\\
    \ell^{a} n^{b} m^{c} \nabla_{[a} F_{bc]} &= 0 & && \ell^{a} n^{b} m^{c} \nabla_{[a}\prescript{\star}{}{F}_{bc]} &= 0 \notag\\
    \ell^{a} n^{b} \bar{m}^{c} \nabla_{[a} F_{bc]} &= 0 & && \ell^{a} n^{b} \bar{m}^{c} \nabla_{[a}\prescript{\star}{}{F}_{bc]} &= 0 \notag\\
    \ell^{a} m^{b} \bar{m}^{c} \nabla_{[a} F_{bc]} &= 0  & && \ell^{a} m^{b} \bar{m}^{c} \nabla_{[a}\prescript{\star}{}{F}_{bc]} &= 0.
\end{align}
At this point, we recall the definitions of the electromagnetic Newman-Penrose scalars:
\begin{align}
    \Phi_0 &:= F_{ab} m^{a} \ell^{b}, & && \Phi_1 &:= \frac12 F_{ab} \left(m^{a} \bar{m}^{b} - \ell^{a} n^{b}\right), & && \Phi_2 &:= F_{ab} n^{a} \bar{m}^{b} .
\end{align}
Using these definitions, it can be shown that the Maxwell $2$-form and its dual can be expressed as (see Exercises~\ref{ex:Maxwell2FormInNPForm} and~\ref{ex:FIdentity} for a derivation)
\begin{align}
    \frac12 F_{ab} &= \Phi_0 n_{[a}\bar{m}_{b]} - \Phi_1 \left( n_{[a}\ell_{b]} - \bar{m}_{[a}m_{b]}\right) -\Phi_2\ell_{[a}m_{b]} + \text{c.c.} \notag\\
    \frac12 \prescript{\star}{}{F}_{ab} &= i\, \Phi_0 n_{[a}\bar{m}_{b]} - i\, \Phi_1 \left( n_{[a}\ell_{b]} - \bar{m}_{[a}m_{b]}\right) - i\, \Phi_2 \ell_{[a}m_{b]} + \text{c.c.},
\end{align}
where ``$\text{c.c.}$'' stands for ``complex conjugate''. Observe that $F_{ab}$ and $\prescript{\star}{}{F}_{ab}$ have a very similar structure, which suggests that the eight equations~\eqref{eq:MaxwellEqContractions} are \textit{not} linearly independent. In fact, one can easily verify that
\begin{align}\label{eq:LinearCombinationFs}
    \frac12\left(F_{ab} - i\, \prescript{\star}{}{F}_{ab}\right) &= \Phi_0 n_{[a}\bar{m}_{b]} - \Phi_1 \left( n_{[a}\ell_{b]} - \bar{m}_{[a}m_{b]}\right) -\Phi_2\ell_{[a}m_{b]}\notag\\
    \frac12\left(F_{ab} + i\, \prescript{\star}{}{F}_{ab}\right) &= \bar{\Phi}_0 n_{[a}m_{b]} - \bar{\Phi}_1 \left(n_{[a}\ell_{b]} - {m}_{[a}\bar{m}_{b]}\right) -\bar{\Phi}_2\ell_{[a}\bar{m}_{b]}.
\end{align}
Notice that the second line is just the complex conjugate of the first one. Thus, we only have a total of four complex equations, rather than eight. This is reassuring, since in the traditional formalism given by~\eqref{eq:MaxwellTraditional}, there are eight \textit{real} equations. Our four \textit{complex} equations can be rewritten as a set of eight real equations and thus the number of algebraically independent equations match. However, we will continue with the ``complexified'' version of the equations given by
\begin{align}\label{eq:MaxwellIndependent}
    n^{a} m^{b} \bar{m}^{c} \nabla_{[a}(F-i\, \prescript{\star}{}{F})_{bc]} &= 0 \notag\\
    \ell^{a} n^{b} m^{c} \nabla_{[a}(F-i\, \prescript{\star}{}{F})_{bc]} &= 0 \notag\\
    \ell^{a} n^{b} \bar{m}^{c} \nabla_{[a}( F-i\, \prescript{\star}{}{F})_{bc]} &= 0 \notag\\
    \ell^{a} m^{b} \bar{m}^{c} \nabla_{[a}(F-i\, \prescript{\star}{}{F})_{bc]} &= 0.
\end{align}
At this point we observe that once we substitute $(F-i\, \prescript{\star}{}{F})_{bc}$ by the right hand side expression of~\eqref{eq:LinearCombinationFs}, the derivative operator will hit not only the scalars, but also elements of the null co-tetrad. Thus, we need to work out explicitly the action of $\nabla$ on co-tetrad elements. To do so, we can use that the tetrad is normalized as $\ell^{a} n_a = -1$ and $m^{a} \bar{m}_a = 1$, while all other contractions vanish. From this we infer that
\begin{align}
    \ell^{a} n_a &= -1 & \Longrightarrow && \ell^{a}\nabla_b n_a &= -\left(\nabla_b \ell^{a}\right)n_a\notag\\
  \ell^{a} \ell_a &= 0 & \Longrightarrow && \ell^{a} \nabla_b \ell_a &= - \left(\nabla_b \ell^{a}\right)\ell_a
\end{align}
and so on for the remaining eight contractions. These relations allow us to rewrite the action of $\nabla$ on a co-tetrad element as the action of $\nabla$ on an element of the tetrad. Furthermore, we can use that the Newman-Penrose null tetrad is constructed such that (see construction procedure in Chapter~\ref{Chap3})
\begin{align}
    \ell^{a} \nabla_a \ell^{b} &= 0, & \ell^{a}\nabla_a n^{a} &= 0, & \ell^{a} \nabla_a m^{b} &= 0, & \ell^{a}\nabla_a \bar{m}^{b} &= 0\notag\\
    n^{a}\nabla_a \ell^{b} &= 0, & n^{a}\nabla_a n^{b} &= 0, & n^{a}\nabla_a m^{b} &= 0, & n^{a}\nabla_a \bar{m}^{b} &= 0.
\end{align}
These relations can also be checked by a direct computation using the Levi-Civita connection of the Minkowski metric in $(u, r, \theta, \phi)$ coordinates and the tetrad defined in~\eqref{eq:DefTetradMink}. What remains to be determined is the action of the operator $m^{a}\nabla_a$. The action of $\bar{m}^{a}\nabla_a$ can then be determined by complex conjugation. A direct computation yields the results
\begin{align}
    m^{a}\nabla_a \ell^{b} &= \frac{1}{\sqrt{2}\, r} m^{b} \notag\\
    m^{a}\nabla_a n^{b} &= -\frac{1}{\sqrt{2}\, r} m^{b} \notag\\
    m^{a}\nabla_a m^{b} &= \frac{\cot\theta}{\sqrt{2}\, r} m^{b} \notag\\
    m^{a}\nabla_a \bar{m}^{b} &= \frac{1}{\sqrt{2}\, r}\left(n^{b} - \ell^{b} - \cot\theta\, m^{b}\right).
\end{align}
This is all we need to simplify the equations~\eqref{eq:MaxwellIndependent} such that only derivatives of the Newman-Penrose scalars remain. We obtain the following equations.
\begin{align}\label{eq:MaxwellTetrad}
    n^{a}\nabla_a \Phi_1 - m^{a}\nabla_a\Phi_2 -\frac{\sqrt{2}}{r} - \frac{\cot\theta}{\sqrt{2}\, r}\Phi_2 &= 0 \notag\\
    n^{a}\nabla_a\Phi_0 - m^{a}\nabla_a \Phi_1 - \frac{1}{\sqrt{2}\, r}\Phi_0 &= 0 \notag\\
    \ell^{a}\nabla_a \Phi_2 - m^{a}\nabla_a\Phi_1 + \frac{1}{\sqrt{2}\, r}\Phi_2 &= 0 \notag\\
    \ell^{a}\nabla_a\Phi_1 - m^{a}\nabla_a\Phi_0 + \frac{\sqrt{2}}{r}\Phi_1 - \frac{\cot\theta}{\sqrt{2}\, r}\Phi_0 &= 0.
\end{align}
These equations can be further simplified using the angular derivative operator $\eth$ and its conjugate, which we introduced in Chapter~\ref{Chap1}. We recall that their action on a spin weight $s$ function $f_s$ is defined as
\begin{align}
    \eth f_s &:= \frac12 \left(\sin\theta\right)^s \left(\PD{}{\theta} + \frac{i}{\sin\theta}\PD{}{\phi}\right) \left(\sin\theta\right)^{-s}f_s = \frac12\left(\partial_\theta f_s + \frac{i}{\sin\theta}\partial_\phi f_s - s\, \cot\theta\, f_s\right)\notag\\
    \bar{\eth} f_s &:= \frac12 \left(\sin\theta\right)^{-s} \left(\PD{}{\theta} - \frac{i}{\sin\theta}\PD{}{\phi}\right)\left(\sin\theta\right)^s f_s = \frac12\left(\partial_\theta f_s - \frac{i}{\sin\theta}\partial_\phi f_s + s\, \cot\theta\, f_s\right).
\end{align}
Moreover, it is important to remember that $\eth$ increases the spin weight by one, while $\bar{\eth}$ decreases it by one. After plugging the explicit expressions~\eqref{eq:DefTetradMink} for the tetrad into~\eqref{eq:MaxwellTetrad}, we can use the angular derivative operators. This leads us to the result that Maxwell's equations can be written in the Newman-Penrose formalism as
\begin{equation}
\boxed{
\begin{aligned}\label{eq:MaxwellFinal}
    \left(2\PD{}{u} - \PD{}{r} - \frac{2}{r}\right)\Phi_1 - \frac{2}{r}\eth\Phi_2 &= 0 \\
    \left(2\PD{}{u} - \PD{}{r} - \frac{1}{r}\right)\Phi_0 -\frac{2}{r}\eth\Phi_1 &= 0 \\
    \left(\PD{}{r} + \frac{1}{r}\right)\Phi_2 - \frac{2}{r}\bar{\eth}\Phi_1 &= 0 \\
    \left(\PD{}{r} + \frac{2}{r}\right)\Phi_1 + \frac{1}{r}\bar{\eth}\Phi_0 &= 0
\end{aligned}
}
\end{equation}
We recall that $\Phi_0$, $\Phi_1$, and $\Phi_2$ have spin weight $1$, $0$, and $-1$, respectively. Using the properties of the angular derivative operators $\eth$ and $\bar{\eth}$, one can check that these equations have a consistent spin weight. From top to bottom, one finds spin weight $0$, $1$, $-1$, and $0$.

What we have derived are Maxwell's equations in terms of Newman-Penrose scalars in the \textit{bulk} of spacetime. However, the equations which were discussed in Chapter~\ref{Chap2} represented the limit of these equations to $\scrip$. To compute the limit, we use the Peeling Theorem, which states that 
\begin{align}
    \Phi_0(u, r, \theta, \phi) &= \frac{\Phi^\circ_0(u, \theta, \phi)}{r^3} + \O\left(r^{-4}\right) \notag\\
    \Phi_1(u, r, \theta, \phi) &= \frac{\Phi^\circ_1(u, \theta, \phi)}{r^2} + \O\left(r^{-3}\right) \notag\\
    \Phi_2(u, r, \theta, \phi) &= \frac{\Phi^\circ_2(u, \theta, \phi)}{r} + \O\left(r^{-2}\right).
\end{align}
For large $r$, we can use these relations to simplify~\eqref{eq:MaxwellFinal}, which will lead to a sum of terms which scale like $\frac{1}{r^n}$. To find the leading order behavior of the equations, we multiply each one with $r^{n}$, where $n$ is the smallest integer which occurs in the $\frac{1}{r^n}$ terms of that specific equation. This enables us to take the limit $r\to\infty$. Finally, we obtain the leading order Maxwell equations on $\scrip$ in the Newman-Penrose formalism.
\begin{equation}
\boxed{
\begin{aligned}
   & & \partial_u \Phi^\circ_1 &= \eth\Phi^\circ_2 & & \notag\\
   & & \partial_u \Phi^\circ_0 &= \eth\Phi^\circ_1 & &\notag\\
   & & \bar{\eth}\Phi^\circ_1 &= 0 & &\notag\\
   & & \bar{\eth}\Phi^\circ_0 &= 0 & &
\end{aligned}
}
\end{equation}
These are precisely the equations we encountered in Chapter~\ref{Chap2}. We conclude by remarking that it is evident from~\eqref{eq:MaxwellFinal} that there is no equation which determines $\Phi_2$. That is, no equation contains a term $\partial_u\Phi_2$, which means that Maxwell's equations fail to determine the dynamical behavior of $\Phi_2$. To solve the field equations~\eqref{eq:MaxwellFinal}, one has to specify initial data on a $u=$ const. surface \textit{and} one has to specify $\Phi_2$ \textit{everywhere} on spacetime. It thus follows that Maxwell's equations have lost their predictive power, since there are two field components which are not determined by the theory. 

As alluded to in Chapter~\ref{Chap2}, this loss of predictive power is a generic feature of partial differential equations when they are decomposed with respect to so-called \textit{characteristic surfaces}. In the next subsection, we give a self-contained introduction to the theory of partial differential equations, which will make these statements more precise and more comprehensible.\bigskip

\subsection{Basics of the Theory of Partial Differential Equations}\label{App:A2}
The purpose of this subsection is \textit{not} to give an in-depth treatment of the theory of partial differential equations. Rather, we pursue a much humbler goal: We want to find under which conditions first  order\footnote{The generalization to higher order equations follows the same reasoning and is straightforward. However, we limit ourselves to first order systems in order to keep the presentation simple and because this best serves our purposes.} systems of partial differential equations possess a unique solution. All we are interested in, is an existence criterion. We are \textit{not} looking for the general solution. This can be compared with systems of linear equations, where simple criteria tell us whether the system possesses a unique solution and if not, to what degree it is ambiguous (i.e., how many variables remain undetermined by the system of equations).  We will see that the situation for partial differential equations is, under certain weak assumptions, remarkably similar to systems of linear equations. Indeed, the existence criterion boils down to computing the determinant of certain matrices, just as in linear algebra!

Let us properly define the problem we wish to study: Let $\M$ be a differentiable manifold of dimension~$d$ and let $x = (x^1, \dots, x^d)$ be coordinates on the manifold. Let $y$ be a sufficiently regular (at least $C^{1}$) $m$-component field defined on a spacetime region $\Omega\subseteq\M$. Qualitatively,\footnote{Technically, one would have to talk about sections of fiber bundles and so on. But all these technicalities are irrelevant for our considerations. It is sufficient to view $y$ as a $m$-component vector made up of differentiable functions.} we can think of $y$ as being a ``vector-valued'' function $y:\Omega\to\bbR^m$, where each component is a differentiable function. Then we consider the following \textbf{first order initial value problem for $y$}
\begin{align}\label{eq:QLinPDE}
	\begin{cases}
		\displaystyle\sum_{i = 1}^{d} M^{(i)}(x)\partial_i y + N(x)\,y + h(x) = 0\\[15pt]
		\left.y\right|_\S = f
	\end{cases},
\end{align}
where $y:=(y^1, \dots, y^m)^\transpose$ are the $m$ fields to be ``solved for'', $M^{(i)}(x)$ and $N(x)$ are $m\times m$ matrices, and $h(x)$ is a $m$-dimensional vector. The index $i$ in $M^{(i)}$ ranges from $1$ to $d$ (the dimension of spacetime) and it \textit{labels} the matrices. Hence, there is a total of $d$ matrices $M^{(i)}$. Furthermore, $\S$ is a co-dimension one hypersurface embedded in the spacetime region $\Omega$. The notation $\left.y\right|_{\S} = f$ means that on $\S$ the field~$y$ is given by the \textbf{initial data} $f$. In particular this means that $f:\S\to\bbR^m$ is a \textit{freely specifiable} function. We can think of it as representing our knowledge, gathered through observations and measurements, of the field $y$ at \textit{a given instant of time}. The surface $\S$ represents that instant of time.

To make things more concrete, in electromagnetism we would set $d=4$, choose coordinates $x=(t, x^{1}, x^2, x^3)$, and the field $y$ is given by the six-component vector $y=(E_1, E_2, E_3, B_1, B_2, B_3)^\transpose$. The surface $\S$ would typically be a $t=$ const. surface and $f$ represents a snapshot of the electric and magnetic fields on that $t=$ const. surface.

Qualitatively, we can imagine that we know the field $y$ on a hypersurface which has one dimension less than spacetime, because $\left.y\right|_{\S} = f$. The differential equations then ``evolve'' the data $f$ off the surface~$\S$ and allow us to determine $y$ in the rest of spacetime. However, this procedure could also fail and the question we wish to answer is ``under which conditions can we determine~$y$ from the initial data and the differential equations?'' 

Looking at the initial value problem~\eqref{eq:QLinPDE}, we realize that the only things we can influence or choose are the coordinates $x$, the surface $\S$, and the initial data $f$. Our suspicion is that not every initial data surface $\S$ is a ``good'' surface, in the sense that not every surface allows us to evolve the data $f$ such that we gain knowledge about the field $y$ away from $\S$. 

To make this more precise, we fix a chart $x$ of $\Omega\subseteq\M$ and we represent $\S$ using the constraint equation~$\rchi(x) = 0$ with $\rchi:\Omega\to\bbR$. Furthermore, we assume that $\nabla\rchi(x)\neq 0$ for all $x\in\Omega$, where $\nabla$ represents the gradient of scalar functions. This is a reasonable requirement since it means that $\S$ has a well-defined normal vector and this vector can be used to describe the evolution of $y$ in the direction which points away from $\S$.

The next step is to perform a coordinate transformation to coordinates which are adapted to the surface~$\S$. Adapted means that in the new coordinates, we can represent the surface $\S$ as (the analogue of a) $t=$ const. surface. We also anticipate that these adapted coordinates bring about a greater computational simplicity because they reduce the evolution problem described in~\eqref{eq:QLinPDE} to its essential in the following sense: If we work in $d$ dimensions and consider $m$ fields, then~$\partial_i y$ represents $d\times m$ derivatives. Namely ``spatial'' and ``temporal'' derivatives. However, we ultimately only care about the ``time'' derivatives in order to determine the evolution of the fields. In other words, we are interested in the $m$ derivatives $\partial_t y$. Coordinates adapted to $\S$ help us in isolating precisely these derivatives, as we will see.

In order to perform the change of coordinates, we introduce the map $\phi:\Omega\to\Omega$ such that the new coordinates are $\phi(x) = (\phi^1(x), \dots, \phi^d(x))$. Furthermore, we demand that the coordinate $\phi^1$ is given by
\begin{align}
	\phi^1(x) &:= \rchi(x^1,\dots,x^d),	
\end{align}
where $\rchi$ is the scalar constraint which defines $\S$. Since coordinate transformations have to be invertible, we know that the Jacobian matrix $J$ for the change of coordinates (with components $\frac{\partial \phi^k}{\partial x^l}$ for $k,l\in\{1,\dots, d\}$) has a non-zero determinant throughout the whole domain of definition, $\det J\neq 0$ for all $x\in\Omega$.

Let us briefly pause to put the situation thus far into simple words: We can represent the initial value surface $\S$ through the constraint equation $\rchi(x) = 0$. The condition $\nabla\rchi(x)\neq 0$ ensures that $\S$ has a non-vanishing normal vector everywhere. In the new coordinates $(\phi^1,\dots,\phi^d)$, the initial value surface is simply located at 
\begin{equation}
    \S:\qquad \phi^1 = 0.
\end{equation}
This is the generalization of a $t=$ const. surface and we can think of the coordinates $(\phi^2,\dots, \phi^{d})$ to lie \textit{within} or to be \textit{parallel} to the surface $\S$. If we consider some function $F(\phi^1, \dots, \phi^d)$ and we vary the coordinate $\phi^1$, while keeping all the other coordinates fixed, we can visualize this as a ``movement'' of $F$ in the direction orthogonal to $\S$, i.e., a movement parallel to the normal vector  $\vec{n}:=\nabla\rchi(x)$. We refer to $\phi^1$ as the \textbf{evolution parameter} or as \textbf{time coordinate}, even though it has a priori nothing to do with a physical notion of time. This is just convenient language and derives from an analogy. Similarly, we refer to the other coordinates, $\phi^2, \dots, \phi^{d}$, as \textbf{spatial coordinates}. We reiterate that this is mere terminology and has, a priori, no physical significance.

Having established this simple geometric picture and introduced some useful terminology, we now proceed in rewriting the system~\eqref{eq:QLinPDE} in the new coordinate system. To that end, we need
\begin{equation}
	\PD{y}{x^{i}} = \PD{z}{\phi^j}\PD{\phi^j}{x^{i}} \equiv \left(J\cdot\nabla_\phi z\right)_i,
\end{equation}
where we introduced $z(\phi):= y(x(\phi))$. From this simple relation we immediately learn that our knowledge of the initial data $f$ on $\S$ allows us to determine the $m\times(d-1)$ (spatial) derivatives $\PD{z}{\phi^{j}}$ with $j\in\{2,\dots, d\}$ when evaluated on $\S$. In other words, given the initial data $f$ on the surface $\S$, i.e., given that we know $y$ on $\S$ as a function of $(\phi^2,\dots, \phi^{d})$, we can determine \textit{all spatial} derivatives within that surface. In electromagnetism this would mean that knowing $\vec{E}$ and $\vec{B}$ at an instant of time $t$ allows us to determine $\nabla\times\vec{E}$, $\nabla\cdot\vec{B}$ and other spatial derivatives (also higher order derivatives, if the fields are sufficiently regular) at that same instant of time. What we can \textit{not} do using only the data $f$ is determine how these fields change in time. That is, we do not know how to compute $\partial_t \vec{E}$, for instance, because on $\S$ we only know $\vec{E}$ as a function of $x^1, x^2, x^3$, not of $t$.

This is true in full generality: The only derivatives of $y$ we cannot determine from the data $f$ are $\PD{z}{\phi^1}$. That is, the derivatives of $z$ in the direction perpendicular to the surface $\S$. To determine those derivatives, we need to use the field equations. To prove this, we resort to our simple geometric picture and we recall that taking partial derivatives (on $\S$) simply means determining the limit
\begin{equation}\label{eq:limitingprocedure}
	\left.\PD{z}{\phi^{j}}\right|_{\S} := \lim_{\epsilon\to 0}\frac{z(0,\phi^2,\dots, \phi^j + \epsilon, \dots, \phi^d) - z(0,\phi^2,\dots, \phi^j, \dots, \phi^{d})}{\epsilon}\quad\text{for }j\in\{2,\dots, d\}.
\end{equation}
Since $z(0,\phi^2,\dots, \phi^j, \dots, \phi^d)$ as well as $z(0,\phi^1,\dots, \phi^j + \epsilon, \dots, \phi^d)$ are known function values on $\S$, and because $z$ is sufficiently regular, the limit itself is well-defined. Hence, we can determine all partial derivatives which are \textit{tangential} to $\S$. A different situation presents itself when we try to determine the partial derivatives in the direction $\phi^1$. To compute these $m$ derivatives, we would have to move in the direction orthogonal to $\S$. That is, we would have to consider terms like $z(\epsilon,\phi^2,\dots,\phi^{d})$ in the difference quotient. However, such terms are \textit{not} determined by the data on $\S$ because the point $(\epsilon, \phi^2,\dots,\phi^d)$ does not lie on the surface $\S$, which is described by $\phi^1=0$.

Using this little insight, we can rewrite the original system~\eqref{eq:QLinPDE} in the new coordinate system schematically as
\begin{equation}\label{eq:linearsystem}
	\sum_{i=1}^{d} \tilde{M}^{(i)} \PD{\rchi}{x^{i}}\PD{z}{\rchi}  = -\tilde{K} z - \tilde{h} + \text{terms known on }\S,
\end{equation}
where the quantities with tilde are simply the quantities without tilde expressed in the new coordinate system. Notice that we have achieved to rewrite the original system into a \textit{linear equation} for the vector~$\PD{z}{\rchi}$. That is, the vector $\PD{z}{\rchi}$ is multiplied from the left by the matrix $\sum_{i=1}^{d}\tilde{M}^{(i)}\PD{\rchi}{x^{i}}$ (recall that $i$ labels matrices and does not symbolize components of a vector or anything of-the-like). Thus, if we are able to invert this matrix, then we can express the vector $\PD{z}{\rchi}$ in terms of \textit{known functions}. This allows us, in principle, to integrate the equation in order to obtain the value of $z$ off the surface $\S$.

This last equation can also be rewritten in a more geometric fashion by recalling that $\vec{n} := \nabla\rchi \equiv \PD{\rchi}{x^{i}}\neq 0$ defines the normal vector to $\S$. Also, because we regard $\phi^1 = \rchi$ as a time coordinate, we can introduce the ``velocities'' $\dot z:= \PD{z}{\rchi}$. Then equation~\eqref{eq:linearsystem} takes the suggestive form
\begin{equation}\label{eq:LinearEquation}
	\sum_{i=1}^d \tilde{M}^{(i)} n_i \dot z = F,
\end{equation}
where $F$ stands as placeholder for known functions on $\S$. Hence, the right hand side is effectively under control because we control the initial data. Furthermore, we have managed to ``project'' our system of differential equations along the normal vector and rewrite them as evolution equations along that axis. The problem has been reduced to finding an expression for the $m$ velocities $\dot z$. If we manage to obtain an equation of the form $\dot z = G$, where $G$ is a function constructed from $F$ and the matrix $\sum_{i=1}^d \tilde{M}^{(i)} n_i$, then we can in principle integrate the equation and determine $z$ also off $\S$.

Determining whether we can write $\dot z = G$ is straightforward: As we have pointed out above, $\sum_{i=1}^{d}\tilde{M}^{(i)}n_i$ is simply a matrix. Thus, we can solve the equation for $\dot z$, provided $\sum_{i=1}^{d}\tilde{M}^{(i)}n_i$ is invertible! 
We have thus found a sufficient condition for the existence of a unique solution to the initial value problem~\eqref{eq:QLinPDE}. Namely,~\eqref{eq:QLinPDE} possesses a unique solution provided
\begin{equation}\label{eq:Condition}
	\det\left(\sum_{i=1}^{d} \tilde M^{(i)}(x)\,n_i(x)\right) \neq 0\quad\forall x\in\S.
\end{equation}
Conversely, if the determinant of this matrix is zero, we cannot solve for all velocities $\dot z$ and the initial data $f$ is \textit{not} sufficient to integrate the equations. More precisely, if the rank of the matrix $\sum_{i=1}^{d}\tilde{M}^{(i)}n_i$ is $r<m$, then we can solve for $r$ of the time derivatives $\PD{z}{\rchi}$. Hence, we only obtain a solution after \textit{specifying} the $m-r$ undetermined fields \textit{everywhere} on spacetime.\footnote{Since the matrix under consideration is a function of the coordinates $x$, it could happen that its rank is not constant all over spacetime. We are not aware of any physical system where this phenomenon occurs. Therefore, we will disregard this possibility in what follows.}

Observe that the existence criterion~\eqref{eq:Condition} effectively allows us to distinguish between ``good'' and ``bad'' initial value surfaces. To see this more explicitly, we introduce the so-called \textbf{characteristic equation}
\begin{equation}\label{eq:CharacteristicEquation}
	\det\left(\sum_{i=1}^{d} \tilde M^{(i)}(x)\,n_i(x)\right) = 0.
\end{equation}
This equation can be used to determine the \textbf{characteristic surface} $\S$, which is a ``bad'' surface in the sense that specifying initial data on $\S$ is not sufficient to determine $z$ everywhere in spacetime. Indeed, we can view~\eqref{eq:CharacteristicEquation} as a differential equation for $n = (\partial_1\rchi, \dots, \partial_d \rchi)^\transpose$. Thus, solutions $\rchi$ to this differential equation tell us which surfaces $\S$ are characteristic, as claimed.

Before applying the tools of this subsection to electrodynamics, let us illustrate these ideas and concepts with a simpler example.\bigskip

\subsubsection{Illustration of the Existence Criterion using the Beltrami Equation}
The Beltrami equation is a system of first order differential equations and it is given by
\begin{align}\label{eq:Beltrami}
	W \frac{\partial u}{\partial x} - b \frac{\partial v}{\partial x} - c \frac{\partial v}{\partial y} &= 0\notag\\
	W\frac{\partial u}{\partial y} + a \frac{\partial v}{\partial x} + b \frac{\partial v}{\partial y} &= 0,
\end{align}
where $W, a, b,$ and $c$ are known functions of $x$ and $y$ with $W\neq 0$. Moreover, the matrix
\begin{equation}\label{eq:PositiveMatrix}
	D:=\begin{pmatrix}
		a & b\\
		b & c
	\end{pmatrix}
\end{equation} 
is assumed to be \textit{positive definite}. We can bring this system of equations into the form~\eqref{eq:QLinPDE} by defining
\begin{equation}
	M^{(1)} := \begin{pmatrix}
		W & -b\\
		0 & a
	\end{pmatrix} \quad\text{and}\quad
	M^{(2)} := \begin{pmatrix}
		0 & -c\\
		W & b
	\end{pmatrix}.
\end{equation}
It then follows that the Beltrami equation can be written as
\begin{equation}
	M^{(1)}\begin{pmatrix}
		\frac{\partial u}{\partial x}\\ \frac{\partial v}{\partial x}
	\end{pmatrix}
	+ M^{(2)}
	\begin{pmatrix}
		\frac{\partial u}{\partial y} \\ \frac{\partial v}{\partial y}
	\end{pmatrix}
	=
	\begin{pmatrix}
		0 \\ 0
	\end{pmatrix}.
\end{equation}
According to the existence criterion~\eqref{eq:Condition}, this system is only solvable if we can find functions $n_1$ and $n_2$ such that
\begin{equation}
    \det\left(M^{(1)}n_1 + M^{(2)}n_2\right) \neq 0.
\end{equation}
Concretely, we find that
\begin{equation}
	\det\begin{pmatrix}
		W n_1 & -b n_1 - c n_2\\
		W n_2 & a n_1 + b n_2
	\end{pmatrix}
	= \underset{\neq 0}{\underbrace{W}}\underset{>0\text{ for }(n_1,n_2)^\transpose\neq(0,0)^\transpose}{\underbrace{\left(a n^2_1 + 2 b n_1 n_2 + c n^2_2\right)}} = W\, \vec{n}^\transpose D\,\vec{n} \neq 0.
\end{equation}
This is trivially true for any $\vec{n}:=(n_1,n_2)^\transpose\neq(0,0)^\transpose$ because $W\neq 0$ by assumption and because the matrix~\eqref{eq:PositiveMatrix} is positive definite by assumption. It thus follows that the Beltrami equation \textit{always} admits a unique solution. Or, in other words, the Beltrami equation possesses no characteristic surfaces.\bigskip

\subsubsection{The Existence Criterion for Maxwell's Equations}
We now turn to Maxwell's field equations in vacuum, which are given by
\begin{align}\label{eq:MaxwellEandB}
	\nabla\cdot\vec{E} &= 0 & \nabla\cdot\vec{B} &=0\notag\\
	\partial_t\vec{B}+ \nabla\times\vec{E} &= 0  & \partial_t\vec{E} - \nabla\times\vec{B} &= 0,
\end{align}
and we apply the existence criterion to this system. However, we first need to sort out a subtlety which we did not consider in the general treatment of first order partial differential equations. Namely, the two equations on the first line of~\eqref{eq:MaxwellEandB} are \textit{not} dynamical equations, they are \textit{constraints}. One can prove the following: If the constraints are satisfied for some initial data $f$ on $\S$, and if the other two equations are satisfied, then the constraints are satisfied \textit{everywhere and for all times}. Hence, we shall assume that we have chosen initial data which satisfies the constraints and we will focus on the two dynamical equations on the second line of~\eqref{eq:MaxwellEandB}. 

Our goal is to study the characteristic equation~\eqref{eq:CharacteristicEquation} and to determine which surfaces $\S$ are characteristic. Recall that choosing characteristic surfaces as initial value surfaces \textit{prevents} us from solving all dynamical field equations and therefore there will be \textit{undetermined} field components.

Let us begin by introducing coordinates $x=(t, x^1, x^2, x^3)$ and assuming that the initial value surface $\S$ is given by 
$\rchi(x) = 0$ with $n:=\nabla\rchi \neq 0$. Furthermore, the initial data $f=(\vec{E}_0, \vec{B}_0)$ satisfies the constraint equations. The two dynamical field equations can be written as
\begin{equation}
	M^{(0)}\begin{pmatrix}
		\partial_t \vec{E}\\ 
		\partial_t\vec{B}
	\end{pmatrix}
    +
	M^{(1)}\begin{pmatrix}
		\partial_1\vec{E} \\
		\partial_1 \vec{B}
	\end{pmatrix}
	+
	M^{(2)}\begin{pmatrix}
		\partial_2\vec{E} \\
		\partial_2 \vec{B}
	\end{pmatrix}
	+
	M^{(3)}\begin{pmatrix}
		\partial_3\vec{E} \\
		\partial_3 \vec{B}
	\end{pmatrix}
	=0,
\end{equation}
where the $6\times 6$ matrices $M^{(0)}$, $M^{(1)}$, $M^{(2)}$, and $M^{(3)}$ are explicitly given by
\begin{align}
	M^{(0)} &= \begin{pmatrix}
		1 & 0 & 0 & 0 & 0 & 0\\
		0 & 1 & 0 & 0 & 0 & 0\\
		0 & 0 & 1 & 0 & 0 & 0\\
		0 & 0 & 0 & 1 & 0 & 0\\
		0 & 0 & 0 & 0 & 1 & 0\\
		0 & 0 & 0 & 0 & 0 & 1
	\end{pmatrix}
	&
	M^{(1)} &= \begin{pmatrix}
		0 & 0 & 0 & 0 & 0 & 0\\
		0 & 0 & 0 & 0 & 0 & 1\\
		0 & 0 & 0 & 0 & -1 & 0\\
		0 & 0 & 0 & 0 & 0 & 0\\
		0 & 0 & -1 & 0 & 0 & 0\\
		0 & 1 & 0 & 0 & 0 & 0
	\end{pmatrix}\notag\\
	M^{(2)} &= \begin{pmatrix}
		0 & 0 & 0 & 0 & 0 & -1\\
		0 & 0 & 0 & 0 & 0 & 0\\
		0 & 0 & 0 & 1 & 0 & 0\\
		0 & 0 & 1 & 0 & 0 & 0\\
		0 & 0 & 0 & 0 & 0 & 0\\
		-1 & 0 & 0 & 0 & 0 & 0
	\end{pmatrix}
	&
	M^{(3)} &= \begin{pmatrix}
		0 & 0 & 0 & 0 & 1 & 0\\
		0 & 0 & 0 & -1 & 0 & 0\\
		0 & 0 & 0 & 0 & 0 & 0\\
		0 & -1 & 0 & 0 & 0 & 0\\
		1 & 0 & 0 & 0 & 0 & 0\\
		0 & 0 & 0 & 0 & 0 & 0
	\end{pmatrix}.
\end{align}
The characteristic equation can then be written as
\begin{equation}
	\det\left(M^{(0)} n_0 + M^{(1)}n_1 + M^{(2)} n_2 + M^{(3)} n_3\right) = \det\begin{pmatrix}
		n_0 & 0 & 0 & 0 & n_3 & -n_2\\
		0 & n_0 & 0 & -n_3 & 0 & n_1\\
		0 & 0 & n_0 & n_2 & -n_1 & 0\\
		0 & -n_3 & n_2 & n_0 & 0 & 0\\
		n_3 & 0 & -n_1 & 0 & n_0 & 0\\
		-n_2 & n_1 & 0 & 0 & 0 & n_0
	\end{pmatrix} \overset{!}{=} 0,
\end{equation}
where $n = (n_0, n_1, n_2, n_3)^\transpose = (\partial_t \rchi, \partial_1\rchi, \partial_2\rchi, \partial_3\rchi)^\transpose$. In other words, computing the determinant of this matrix leads to a differential equation for $\rchi$. If we solve this differential equation, we know which surfaces $\S$, described by $\rchi(x) = 0$, do \textit{not} allow us to solve for all time derivatives of our fields. In order to compute the determinant, observe that it can be written as
\begin{align}
    \det\begin{pmatrix}
        A & B\\
        -B & A
    \end{pmatrix} &= \det(A) \det(A + B A^{-1} B),
\end{align}
where we used the formula for the determinant of block matrices and where we introduced
\begin{align}\label{eq:AandBMatrics}
    A &:= n_0 \id_{3\times 3} &\text{and} && B &:= \begin{pmatrix}
        0 & n_3 & -n_2\\
        -n_3 & 0 & n_1\\
        n_2 & -n_1 & 0
    \end{pmatrix}.
\end{align}
This drastically simplifies the computation and one finds 
\begin{equation}\label{eq:CharacteristicED}
	\det\left(M^{(0)} n_0 + M^{(1)}n_1 + M^{(2)} n_2 + M^{(3)} n_3\right) = n_0^2\left(n_0^2 - n_1^2 - n_2^2 - n_3^2\right)^2 \overset{!}{=} 0.
\end{equation}
One obvious solution is $n_0 = 0$, which translates to $\partial_t\rchi = 0$. This is the less interesting solution and therefore we move to the second option, which is $n_0^2 - n_1^2 - n_2^2 - n_3^2 = 0$. In terms of $\rchi$, this can be written as
\begin{equation}
	(\partial_t\rchi)^2 - (\nabla_{\vec{x}}\rchi)^2 = 0.
\end{equation}
This is a wave equation for $\rchi$! Hence, we immediately know that this equation is solved by $\rchi = g(\vec{r}\cdot\vec{x}-t)$, where $g$ is an arbitrary function and  $\vec{r}$ is a unit vector in the direction of $\vec{x} = (x_1, x_3, x_3)^\transpose$. To better understand what this solution is telling us, i.e., what kind of surfaces it describes, let us first change to spherical coordinates, where $\vec{r}\cdot\vec{x}-t = r-t$ ($r$ being the radial coordinate, not the rank of a matrix). Now recall that the retarded time coordinate $u$ is defined as $u:= t-r$. Hence, in coordinates $(u, r, \theta, \phi)$, we can write the solution to the characteristic equation as $\rchi(u, r, \theta, \phi) = g(u)$. An interesting choice is $g(u) = u-u_0$, where $u_0$ is a constant. But this simply describes a light ray in Minkowski space! Hence, we find that Maxwell's equations do \textit{not} lead to a unique solution if we place ourselves on top of an outgoing light ray!

However, other choices than $g(u) = u-u_0$ are possible and all of them describe null surfaces. Let us prove that \textit{any} null surface leads  to \textit{under-determined} Maxwell equations. To that end, we just need to compute the normal vector $n:= \nabla \rchi(u) = \nabla g(u) = \left(\frac{\dd g(u)}{\dd u},0,0,0\right)^\transpose$, where, by assumption, $n\neq 0$. Since we are working in outgoing Eddington-Finkelstein coordinates, the Minkowski metric and its inverse are given by
\begin{align}
	\eta_{ab} &= \begin{pmatrix}
		-1 & -1 & 0 & 0\\
		-1 & 0 & 0 & 0\\
		0 & 0 & r^2 & 0\\
		0 & 0  & 0 & r^2 \sin^2\theta
	\end{pmatrix}
	& \Longleftrightarrow & &
	\eta^{ab} = \begin{pmatrix}
		0 & -1 & 0 & 0\\
		-1 & 1 & 0 & 0\\
		0 & 0 & \frac{1}{r^2} & 0\\
		0 & 0 & 0 & \frac{1}{r^2 \sin^2\theta}
	\end{pmatrix}
\end{align}
Using the inverse metric and $n=\left(\frac{\dd g(u)}{\dd u},0,0,0\right)^\transpose$, one sees that the norm of $n$ vanishes,
\begin{equation}
	\eta^{ab}n_a n_b = 0.
\end{equation}
In other words, $n$ is a null vector. Furthermore, we know that $n$ is also normal to $\S$, which is described by $\rchi(u) = g(u) = 0$. Hence, $\S$ is a null surface, and this proves our claim.\footnote{For completeness, we remark that there is a second solution to the wave equation, given by $\rchi = \tilde{g}(\vec{r}\cdot \vec{x} + t)$. This solution can be shown to be equal to $\tilde{g}(v)$, where $v:=t+r$ is the advanced time in ingoing Eddington-Finkelstein coordinates. Surfaces described by $\tilde{g}(v) = 0$ are also null surfaces.} 

Let us quickly return to the first solution to the characteristic equation~\eqref{eq:CharacteristicED}, which was $\partial_t \rchi = 0$. In Cartesian coordinates $(t, x^1, x^2 ,x^3)$, this translates into a normal vector $n:=\nabla\rchi(x) = (0, \nabla_{\vec{x}}\rchi)^\transpose$, where $\nabla_{\vec{x}}$ is the gradient with respect to the spatial coordinates. It follows that $n$ is a spacelike vector, making $\S$ consequently a timelike surface. 

In summary, we find that Maxwell's equations lead to a unique solution if the initial data $f$ is specified on a \textit{spacelike} surface $\S$. If we were to specify the initial data on a timelike or on a null surface $\S$, we would find that the data is \textit{not} sufficient to provide a unique solution to the field equations. The reason is that some components of the electric and magnetic field remain \textit{undetermined} by the equations, thus forcing us to provide them by hand. 

How many components remain undetermined? This question can readily be answered if we remember that in adapted coordinates, the first order system is reduced to $\sum_{i=1}^{d} \tilde{M}^{(i)}n_i \dot z = ...$, where the right hand side consists of known quantities on $\S$. Thus, the rank of $\sum_{i=1}^{d} \tilde{M}^{(i)}n_i$ provides us with the information how many velocities we can solve for and how many velocities are not determined by this linear equation. In order to determine the rank of the matrix in the case of electrodynamics, we use again the $A$ and $B$ matrices we encountered when computing the determinant. From definition~\eqref{eq:AandBMatrics}, we gather that
\begin{equation}
    B^\transpose = -B.
\end{equation}
Furthermore, let us define $\bbM := \sum_{i=1}^4 M^{(i)}n_i$ and show that this matrix is symmetric:
\begin{equation}
    \bbM = \begin{pmatrix}
      A & B\\
      -B & A
    \end{pmatrix}
    \qquad
    \Longrightarrow
    \qquad
    \bbM^\transpose = \begin{pmatrix}
        A^\transpose & -B^\transpose\\
        B^\transpose & A^\transpose
    \end{pmatrix}
    =
    \begin{pmatrix}
        A & B\\
        -B & A
    \end{pmatrix}
    = \bbM.
\end{equation}
From this we can conclude that the matrix $\bbM$ is diagonalizable. To compute its eigenvalues, we would have to solve the equation 
\begin{equation}
    \det\left(\bbM-\lambda\,\id_6\right) = 0.
\end{equation}
Observe that the diagonal of $\bbM$ is given by $\text{diag}(n_0, \dots, n_0)$ and that $n_0$ does not appear anywhere else. Thus, the eigenvalue equation can be obtained from the determinant~\eqref{eq:CharacteristicED} by the replacement $n_0 \mapsto n_0-\lambda$, resulting in
\begin{equation}
    \left(n_0-\lambda\right)^2\left(\left(n_0-\lambda\right)^2-\|\vec{n}\|^2\right) = 0,
\end{equation}
where we introduced the shorthand notation $\|\vec{n}\| := \sqrt{n_1^2+n_2^2+n_3^2}$. It follows that $\bbM$ has three eigenvalues, each with algebraic multiplicity two. Concretely, the eigenvalues are
\begin{align}
    \lambda_{1,2} &= n_0 + \|\vec{n}\| \notag\\
    \lambda_{3,4} &= n_0 - \|\vec{n}\| \notag\\
    \lambda_{5,6} &= n_0.
\end{align}
In its diagonalized form, the matrix $\bbM$ is thus given by
\begin{equation}
    \bbM = \textsf{diag}\left(n_0 + \|\vec{n}\|, n_0 + \|\vec{n}\|, n_0 - \|\vec{n}\|, n_0 - \|\vec{n}\|, n_0, n_0\right).
\end{equation}
In this form, we can simply read off the rank of $\bbM$. When $n_0 = 0$, the last two entries vanish and the remaining four are of course linearly independent. Thus,
\begin{equation}
    \textsf{rank}\left.\bbM\right|_{n_0 = 0} = 4.
\end{equation}
If we use the other solution to the characteristic equation, we need to choose a sign. The choice will not affect the end result and we arbitrarily choose $n_0 = +\|\vec{n}\|$ to be the second solution. This yield
\begin{equation}
    \textsf{rank}\left.\bbM\right|_{n_0 = \|\vec{n}\|} = 4.
\end{equation}
Hence, we conclude that both solutions to the characteristic equation yield a matrix $\bbM$ of rank $4$. Consequently, this means that precisely \textit{two} components of the electromagnetic field remain undetermined!

We repeat the meaning of this result: If we prescribe initial data on  a timelike or a null surface, not all field components will be determined by Maxwell's equations. Precisely two components will remain undetermined and we have to prescribe them by hand, if we wish to solve the field equations. There is no problem for spacelike surfaces. Any spacelike surface leads to unique solution.

In the last section we will prove that what remains undetermined in the case of null surfaces are precisely the radiative modes of the electromagnetic field.

Before doing so, let us conclude with a remark on the characteristic equation. It does not only tell us which surfaces are ``bad'', it also tells us which coordinate systems are ``bad''. In fact, the key step in deriving the characteristic equation was a coordinate transformation. Any coordinate transformation which fails the criterion~\eqref{eq:Condition} can be regarded as a ``bad'' because it prevents us from solving for what we regard as the ``velocities'' in the new system. In particular, as we have seen here, this is the case for coordinate systems based on the retarded time $u$. This explains why the Maxwell equations in the Newman-Penrose formalism lose their deterministic character: It is a ``bad'' choice of coordinates. At least, it is bad if we wish to actually solve the equations. However, this is not what try to do. Rather, we want to identify the radiative degrees of freedom. This leads us to the final section of this appendix.\bigskip

\subsection{A Proof that decomposing Maxwell's Equations with respect to a Null Surface leaves the Radiative Modes undetermined}\label{App:A3}
Given Maxwell's field equations, we can derive the uncoupled second order system of equations
\begin{align}
	\partial_t^2\vec{E} - \nabla^2\vec{E} &= 0 &  \partial_t^2\vec{B} - \nabla^2\vec{B} &= 0.
\end{align} 
These are the wave equations which describe electromagnetic radiation. Of course, we know that the electric and magnetic fields are not independent. Once one of these fields has been determined, we can compute the other one from that solution. Hence, we only need to study one of the wave equations. We choose to scrutinize the one for the electric field. Also, we know that this equation only propagates two degrees of freedom, not three. So let us now ask under which conditions the wave equation determines a unique solution. This time, we are facing a system of second order partial differential equations. The initial value problem of the above system has the form
\begin{align}
    \begin{cases}
        \sum_{i,j=1}^{d} M^{(ij)}(x) \partial_i\partial_j y = 0 \\
        \left.y\right|_{\S} = f \\
        \left.\dot y\right|_{\S} = g
    \end{cases}
\end{align}
where $M^{(ij)}$ are again matrices, but this time labeled by two indices (these indices to \textit{not} denote components of the matrices). Following the same logic as in~\ref{App:A2}, we can derive a characteristic equation of the form
\begin{equation}
    \det\left(\sum_{i,j}^{d}M^{(ij)}n_i n_j\right) \overset{!}{=} 0.
\end{equation}
In the case of the wave equation for $\vec{E}$, there are only four non-zero matrices and they have a very simple form:
\begin{align}
    M^{(00)} &= \id_3 &\text{and}&& M^{(11)} = M^{(22)} = M^{(33)} = -\id_3.
\end{align}
The characteristic equation thus becomes
\begin{align}
	\det&\left(M^{(00)}n_0^2 + M^{(11)}n_1^2 + M^{(22)}n_2^2 + M^{(33)}n_3^2 \right)\notag\\
	&= \det\begin{pmatrix}
		n_0^2 - n_1^2-n_2^2-n_3^2 & 0 & 0\\
		0 & n_0^2 - n_1^2-n_2^2-n_3^2 & 0\\
		0 & 0 & n_0^2 - n_1^2-n_2^2-n_3^2
	\end{pmatrix}\notag\\
	&= (n_0^2 - n_1^2-n_2^2-n_3^2)^3 \overset{!}{=} 0.
\end{align}
It follows that the equation describing electromagnetic waves does \textit{not} possess a unique solution when
\begin{equation}
	n_0^2 - n_1^2-n_2^2-n_3^2 \equiv (\partial_t\rchi)^2 - (\nabla_{\vec{x}}\rchi)^2 =0.
\end{equation}
That is, if we describe electromagnetic radiation from the point of view of a null surface, such as for instance from a light ray, we do not get a deterministic equation! This result had of course to be anticipated, since we derived the wave equation from the full set of Maxwell's equations. Notice, however, that the characteristic equation has only one solution, namely null surfaces. Timelike surfaces are no problem for electromagnetic waves. 

How many components remain undetermined if we work in null coordinates? To answer the question, we need to determine the rank of the above sum of matrices when evaluated on a characteristic surface. This is rather trivial, since in that case we obtain a matrix full of zeros and thus the rank is zero. This means that the wave equation does \textit{not determine any degrees of freedom} when we work in null coordinates. Thus, we find that the two radiative degrees of freedom of the electromagnetic field remain undetermined!

Recall that in the case of the full set of Maxwell's equations, we found that two field components remain undetermined. These are precisely the radiative modes. In the Newman-Penrose formalism, this fact becomes very transparent because the real field components $\vec{E} = (E_1, E_2, E_3)^\transpose$ and $\vec{B} = (B_1, B_2, B_3)^\transpose$ are mapped to the three complex fields $\Phi_0$, $\Phi_1$, and $\Phi_2$. However,  Maxwell's equations expressed in the chart $(u, r, \theta, \phi)$, as we have seen in~\ref{App:A1}, only provide evolution equations for $\Phi_0$ and $\Phi_1$. The two remaining components encoded in $\Phi_2$ are not determined by the equations and these components represent precisely the radiative degrees of freedom.

\newpage
\bibliographystyle{utcaps}
\bibliography{Bibliography}\addcontentsline{toc}{section}{Bibliography}

\providecommand{\href}[2]{#2}\begingroup\raggedright\begin{thebibliography}{10}

\bibitem{Ashtekar:2019YT}
A.~Ashtekar, ``Lectures on Gravitational Waves,'' 2019.
\newblock
  \url{https://www.youtube.com/playlist?list=PL9goACN_fIU_DoZnDA7OGgY-ZdFl-wPF7}.
  \\Date of last access: 2022-01-24.

\bibitem{AshtekarII:2020}
A.~Ashtekar, T.~De~Lorenzo, and N.~Khera, ``{Compact binary coalescences:
  Constraints on waveforms},''
  \href{http://dx.doi.org/10.1007/s10714-020-02764-1}{{\em Gen. Rel. Grav.}
  {\bfseries 52} no.~11, (2020) 107},
  \href{http://arxiv.org/abs/1906.00913}{{\ttfamily arXiv:1906.00913 [gr-qc]}}.

\bibitem{Ashtekar:2020}
A.~Ashtekar, T.~De~Lorenzo, and N.~Khera, ``{Compact binary coalescences: The
  subtle issue of angular momentum},''
  \href{http://dx.doi.org/10.1103/PhysRevD.101.044005}{{\em Phys. Rev. D}
  {\bfseries 101} no.~4, (2020) 044005},
  \href{http://arxiv.org/abs/1910.02907}{{\ttfamily arXiv:1910.02907 [gr-qc]}}.

\bibitem{JacksonBook}
J.~D. Jackson, {\em Classical Electrodynamics}.
\newblock Wiley, 1998.

\bibitem{WaldBook}
R.~M. Wald, {\em General Relativity}.
\newblock University Of Chicago Press, 1984.

\bibitem{Adamo:2016}
T.~Adamo and E.~T. Newman, ``{The Kerr-Newman metric: A Review},''
  \href{http://dx.doi.org/10.4249/scholarpedia.31791}{{\em Scholarpedia}
  {\bfseries 9} (2014) 31791}, \href{http://arxiv.org/abs/1410.6626}{{\ttfamily
  arXiv:1410.6626 [gr-qc]}}.

\bibitem{Alessio:2018}
F.~Alessio and G.~Esposito, ``{On the structure and applications of the
  Bondi\textendash{}Metzner\textendash{}Sachs group},''
  \href{http://dx.doi.org/10.1142/S0219887818300027}{{\em Int. J. Geom. Meth.
  Mod. Phys.} {\bfseries 15} no.~02, (2018) 1830002},
  \href{http://arxiv.org/abs/1709.05134}{{\ttfamily arXiv:1709.05134 [gr-qc]}}.

\bibitem{Ashtekar:1982}
A.~Ashtekar and A.~Magnon-Ashtekar, ``On the symplectic structure of general
  relativity,''
  \href{http://dx.doi.org/https://doi.org/10.1007/BF01205661}{{\em Commun.
  Math. Phys.} {\bfseries 86} (1982) 55--68}.

\bibitem{Ashtekar:1981}
A.~Ashtekar and M.~Streubel, ``Symplectic Geometry of Radiative Modes and
  Conserved Quantities at Null Infinity,''
  \href{http://dx.doi.org/http://www.jstor.org/stable/2397216}{{\em Proc. R.
  Soc. Lond. A} {\bfseries 376} no.~1767, (1981) 585--607}.

\bibitem{Sachs:1962}
R.~K. Sachs, ``Gravitational waves in general relativity VIII. Waves in
  asymptotically flat space-time,''
  \href{http://dx.doi.org/https://doi.org/10.1098/rspa.1962.0206}{{\em Proc. R.
  Soc. Lond. A} {\bfseries 270} (1962) 103 -- 126}.

\bibitem{Newman:1968}
E.~T. Newman, R.~Penrose, and H.~Bondi, ``New conservation laws for zero
  rest-mass fields in asymptotically flat space-time,''
  \href{http://dx.doi.org/https://royalsocietypublishing.org/doi/abs/10.1098/rspa.1968.0112}{{\em
  Proc. R. Soc. Lond. A} {\bfseries 305} no.~1481, (1968) 175--204}.

\bibitem{Mitman:2020}
K.~Mitman {\em et~al.}, ``{Adding gravitational memory to waveform catalogs
  using BMS balance laws},''
  \href{http://dx.doi.org/10.1103/PhysRevD.103.024031}{{\em Phys. Rev. D}
  {\bfseries 103} no.~2, (2021) 024031},
  \href{http://arxiv.org/abs/2011.01309}{{\ttfamily arXiv:2011.01309 [gr-qc]}}.

\bibitem{Bondi:1962}
H.~Bondi, M.~G.~J. Van~der Burg, and A.~W.~K. Metzner, ``Gravitational waves in
  general relativity, VII. Waves from axi-symmetric isolated system,''
  \href{http://dx.doi.org/http://dx.doi.org/10.1098/rspa.1962.0161}{{\em Proc.
  R. Soc. Lond. A} {\bfseries 269} no.~1336, (1962) 21--52}.

\bibitem{CompereBook}
G.~Comp{\`e}re,
  \href{http://dx.doi.org/10.1007/978-3-030-04260-8_3}{``Asymptotically Flat
  Spacetimes,''} in {\em Advanced Lectures on General Relativity}, pp.~81--102.
\newblock Springer International Publishing, Cham, 2019.

\bibitem{Ligo:2016}
{\bfseries LIGO Scientific, Virgo} Collaboration, B.~P. Abbott {\em et~al.},
  ``{Observation of Gravitational Waves from a Binary Black Hole Merger},''
  \href{http://dx.doi.org/10.1103/PhysRevLett.116.061102}{{\em Phys. Rev.
  Lett.} {\bfseries 116} no.~6, (2016) 061102},
  \href{http://arxiv.org/abs/1602.03837}{{\ttfamily arXiv:1602.03837 [gr-qc]}}.

\bibitem{Ashtekar:2017}
A.~Ashtekar and B.~Bonga, ``{On the ambiguity in the notion of transverse
  traceless modes of gravitational waves},''
  \href{http://dx.doi.org/10.1007/s10714-017-2290-z}{{\em Gen. Rel. Grav.}
  {\bfseries 49} no.~9, (2017) 122},
  \href{http://arxiv.org/abs/1707.09914}{{\ttfamily arXiv:1707.09914 [gr-qc]}}.

\end{thebibliography}\endgroup
\end{document}